\documentclass[12pt]{book}
\usepackage{amsmath,feynmf,bbm,fancyhdr,a4,graphicx,psfrag,umlaut,amssymb}
\usepackage[bf]{caption2}

\allowdisplaybreaks[1]

\begin{document}

\thispagestyle{empty}
\vspace*{3cm}
\begin{center}\LARGE Symmetry breaking in the Hubbard model\\\LARGE A bosonization approach\end{center}
\vspace{2cm}
\begin{center}\large Eike Bick\end{center}
\vspace{4cm}
\begin{align}
\text{Referees:}&\qquad\text{Prof. Dr. Christof Wetterich}\nonumber\\
&\qquad\text{Prof. Dr. Otto Nachtmann}\nonumber
\end{align}

\newpage
\thispagestyle{empty}
\hphantom{X}
\newpage

\thispagestyle{empty}
\begin{center}\Large
Dissertation\\
submitted to the\\
Combined Faculties for the Natural Sciences and Mathematics\\
of the Ruperto--Carola University of Heidelberg, Germany\\
for the degree of\\
Doctor of Natural Sciences
\end{center}
\vspace{6cm}
\begin{center}
\Large presented by
\end{center}
\begin{align}
\text{\Large Diplom--Physiker}&\qquad\text{\Large Eike Bick}\nonumber\\
\text{\Large born in}&\qquad\text{\Large Aachen}\nonumber
\end{align}
\vspace{0.5cm}
\begin{center}
\Large Oral examination: 3rd July, 2002
\end{center}

\newpage
\thispagestyle{empty}
\hphantom{X}
\newpage

\thispagestyle{empty}
\begin{center}
\sc Symmetriebrechung im Hubbardmodell\\
Ein Bosonisierungsansatz
\end{center}
\begin{center}
\sc Zusammenfassung
\end{center}
Die meisten bekannten Hochtemperatursupraleiter gehören zur Stoffklasse der Kuprate, die sich gut durch das zweidimensionale Hubbardmodell beschreiben lassen. Um das Zusammenspiel verschiedener Eigenschaften wie Antiferromagnetismus und Supraleitung zu verstehen, berechnet man das Phasendiagramm des Hubbardmodells als Funktion der Ladungsdichte und Temperatur. Für diese Rechnung eignen sich insbesondere exakte Renormierungsgruppengleichungen, die wir im Formalismus der mittleren effektiven Wirkung verwenden. Zu diesem Zweck leiten wir eine äquivalente Formulierung des Hubbardmodells her, die die Form einer Yukawatheorie besitzt und aus der Informationen über lang\-reich\-wei\-ti\-ge Ordnung in verschiedenen Kanälen durch die Berechnung bo\-so\-ni\-scher Erwartungswerte gewonnen werden können. Es gelingt uns, die wesentlichen Eigenschaften des Phasendiagramms von Hochtemperatursupraleitern zu reproduzieren. Außerdem zeigt unsere Untersuchung, wie das Mermin-Wagner Theo\-rem mit der Existenz antiferromagnetischer Ordnung bei nichtverschwindender Temperatur zu vereinbaren ist und wie sich die Berücksichtigung verschiedener bo\-so\-ni\-scher Fluktuationen auf das Phasendiagramm auswirkt. 

\vspace{1.5cm}

\begin{center}
\sc Symmetry breaking in the Hubbard model\\
A bosonization approach
\end{center}
\begin{center}
\sc Abstract
\end{center}
Almost all known high temperature superconductors are cuprates, which can be suitably modelled by the two dimensional Hubbard model. To understand the interplay of various long range properties as antiferromagnetism and superconductivity, one can calculate the phase diagram of the Hubbard model in the charge density-temperature plane. This analysis is conveniently carried out by means of exact renormalization group equations that we apply in the formalism of the effective average action. For this purpose, we derive an equivalent version of the Hubbard model that takes the form of a Yukawa theory. From this modified model long range order in various channels can be extracted by simple calculation of bosonic expectation values. We are able to reproduce the main features of the phase diagram of high temperature superconductors. Furthermore, our analysis shows how the Mermin-Wagner theorem can be reconciled with the existence of antiferromagnetic long range order at non vanishing temperature and how the inclusion of different kinds of bosonic fluctuations affect the phase diagram.  

\newpage

\tableofcontents
\listoffigures

\newpage
\pagenumbering{arabic}
\setcounter{page}{1}

\begin{fmffile}{fmfsave}

\chapter{Introduction}

\begin{quotation}\sl
 Tomorrow by the end of the day we shall come to a mountain of black stone hight the Magnet Mountain, for thither the currents carry us willy-nilly. As soon as we are under its lea, the ship's sides will open and every nail in plank will fly out and cleave fast to the mountain, for that Almighty Allah hath gifted the loadstone with a mysterious virtue and a love for iron, by reason whereof all which is iron traveleth toward it. And on this mountain is much iron, how much none knoweth save the Most High, from the many vessels which have been lost there since the days of yore.
\newline\newline
From: The Arabian Nights, The Third Kalandar's Tale, translated by Sir Richard Burton (1850) 
\newline\newline
\end{quotation}

The investigation of electromagnetic properties of condensed matter systems is one of the oldest branches of physics. Even the ancients knew about the mysterious magnetic force exerted by certain materials, and during the middle ages, magnetism was one of the most popular subjects of alchemical speculation. Over the last two centuries, the fast progress in physics greatly enriched our knowledge of possible electromagnetic properties of different materials. The long known magnetism is now interpreted as only one possible type of long range order, called ferromagnetism. Other long range structures, like antiferromagnetic or ferrimagnetic ordering were discovered. On the other hand, materials can be classified with respect to their conductivity: Conductors, semiconductors and insulators are known. More recently, the discovery of superconductors led to completely new developments --- both theoretically and experimentally with interesting applications.

\begin{figure}
\centering
\psfrag{AF}{AF}
\psfrag{SC}{SC}
\psfrag{Dotierungskonzentration}{Doping}
\psfrag{Temperatur}{Temperature}
\includegraphics{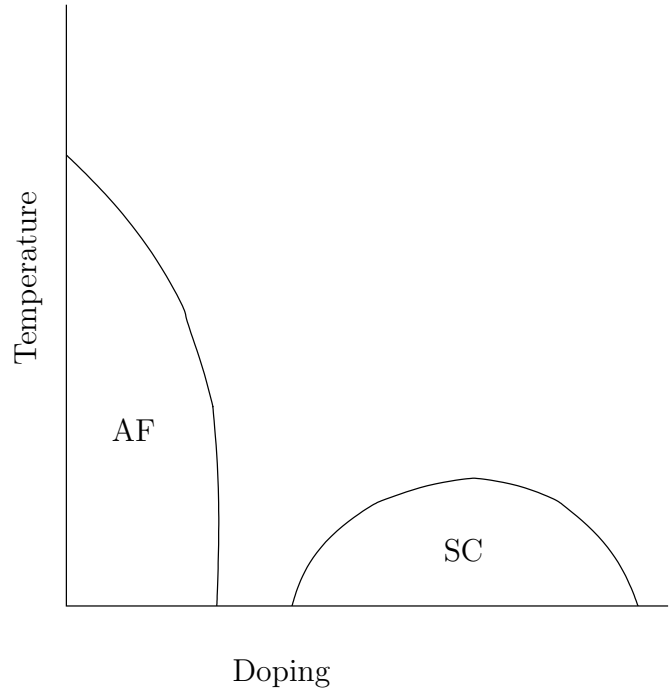}
\caption[The generic phase diagram for a p-doped cuprate superconductor]{The generic phase diagram for a p-doped cuprate superconductor. By {\rm AF} we denote regions of antiferromagnetic, by {\rm SC} regions of $d$-wave-superconducting behavior. The origin corresponds to zero temperature and no doping.}
\label{fig:cuprat}
\end{figure}

Although many of these properties are understood in principle, many unsolved problems remain. One of these problems is even the qualitative understanding of the phase diagram of high temperature superconductors (fig. (\ref{fig:cuprat})). This work is devoted to develop a formalism to deal with such systems exhibiting many different ways of symmetry breaking and --- in the framework of this formalism --- to shed some light on the origin of the different phases. 

As for all physical problems, we face two kinds of problems:

\begin{enumerate}
\item{\bf The question of modeling:} The typical high temperature superconductor possesses a rather complicated chemical structure. The question is how much information about this structure we are allowed to neglect without loosing anything significant giving rise to the phase diagram we want to explain. By reducing the amount of complexity, we achieve two goals: We hopefully end up with a model which can be treated by standard calculation methods and that furthermore contains the essential information about the actual system in a very condensed form. This should free us from complications obscuring our view on the true nature of the physical properties under investigation. Fortunately, such a model exists, the {\em Hubbard model}.   
\item{\bf The question of calculation technique:} Suppose the Hubbard model actually contains enough information to describe high temperature superconductors. However, it is far from clear how to extract this information, given the fact that for nearly forty years the Hubbard model has successfully resisted any attempt to being solved. The main obstacles in doing so are
\begin{enumerate}
\item the different nature of antiferromagnetic and superconducting behavior, both of which nevertheless should be treated on equal footing in our formalism. Besides, we see that for example ferromagnetic or $s$-wave-superconducting behavior is not present in the phase diagram of cuprates. Our formalism should not only provide an answer to the question why the antiferromagnetic and $d$-wave-superconducting phases are where they are, but also why other phases are not present. We attack this problem by a {\em bosonization procedure}. The idea is to artificially introduce additional ``particles'' into the description of the model corresponding to physical degrees of freedom like antiferromagnetic ordering etc. This allows to discuss antiferromagnetic, superconducting and other properties in terms of expectation values of bosonic fields --- a very convenient method to discuss these phenomena in one common language.
\item the strong coupling between the electrons. This prevents us from using perturbation theory to derive our results. Non-perturbative methods are needed. Particularly suitable for this kind of problem are {\em renormalization group techniques}, which we will apply in the setting of the effective average action. 
\end{enumerate}
\end{enumerate}
The focus of this work will be on the bosonization procedure. A lot of recent work has been devoted to analyze the Hubbard model in the framework of purely fermionic models. We hope to convince the reader that partial bosonization of the Hubbard model --- giving interesting physical degrees of freedom a particle interpretation --- greatly improves our physical intuition concerning this complicated system, which simplifies the motivation of approximation schemes.     

One key ingredient in the formalism to describe antiferromagnetism and superconductivity in the same formal language will be a certain viewpoint we adopt: We interprete both phenomena as spontaneously broken symmetries of the underlying model. Whereas this perspective is quite natural for antiferromagnetism, it deserves some explanation in the superconducting case. The usual textbook approach to superconductivity is by starting with a microscopic model, motivating it by calculating properties of the model and comparing them to experiment. To emphasize our point, we will show how the usual properties of a superconductor follow merely from the breakdown of $U(1)$-symmetry of some underlying model that we will not further specify. This topic will be covered in the following section of this introduction. The last two sections will discuss high temperature superconductors in general and our model for them, the Hubbard model. After a rather formal chapter introducing our starting point, the partition function of the Hubbard model, we will describe our bosonization procedure. A mean field calculation already reveals the main features of the phase diagram. After that, we describe our renormalization group procedure and investigate properties of the system beyond mean field --- particularly implications of the Mermin-Wagner theorem and the influence of charge density and antiferromagnetic fluctuations on the superconducting behavior of the model.

\section[Spontaneously broken symmetries]{Antiferromagnetism and superconductivity as spontaneously broken symmetries}
Electrons are conveniently described by field operators $\psi(x)$, where $\psi(x)$ is a basis for irreducible linear representations of
\begin{enumerate}
\item $SU(2)$ in the sense that $\psi(x)=(\psi_\uparrow(x),\psi_\downarrow(x))^T$ and the elements of $SU(2)$ are represented by $U(\vec{\theta})=\exp(i\vec{\sigma}\vec{\theta})$ and of
\item $U(1)$ in the sense that $\psi(x)=\psi_1(x)+i\psi_2(x)$ and the elements of $U(1)$ are represented by $U(\theta(x))=\exp(i\theta(x))$.
\end{enumerate}
Here $\vec{\theta}$ and $\theta(x)$ are used to parameterize the elements of the Lie groups and $\vec{\sigma}$ is the usual set of Pauli matrices. The two dimensional space spanned by $\psi_\uparrow(x)$ and $\psi_\downarrow(x)$ will be called the spinor space. Note that we consider global $SU(2)$-, but local $U(1)$-transformations. The reason will become clear below.

\subsection{The $SU(2)$-symmetry}
From rotational invariance in spinor space, we expect the Lagrangian of our theory to be composed of scalars with respect to $SU(2)$-transformations. This means that the Lagrangian itself is invariant under $SU(2)$-transformations. The $SU(2)$-symmetry is broken, if for example the (space dependent) expectation value of the operator $f(x)\psi^\dagger(x)\vec{\sigma}\psi(x)$ does not vanish, where $f(x)$ is an arbitrary non vanishing scalar function. In this case, one speaks of the spontaneous symmetry breakdown from $SU(2)$ to $U(1)$.

Ferromagnetic and antiferromagnetic behavior can be inferred from the space dependence of $f(x)$. Assume the electrons are strongly located at the sites of a quadratic or cubic lattice, so that the above operator expectation value $\langle f_i\psi^\dagger_i\vec{\sigma}\psi_i\rangle$ is taken at discrete lattice sites $i$. In the case that $\langle f_i\psi^\dagger_i\vec{\sigma}\psi_i\rangle$ is independent of $i$, we have ferromagnetic behavior if $f_i=f_j$ $\forall i,j$, and antiferromagnetic behavior if $f_i=-f_j$, where $i$ and $j$ label nearest neighbor lattice sites.

Since ferromagnetic or antiferromagnetic properties already follow directly from the operator expectation value breaking the {\em global} $SU(2)$-symmetry, we do not have to bother dealing with the complications of local symmetries.     

\subsection{The $U(1)$-symmetry}
In contrast to the $SU(2)$-case we here consider local transformations. This is necessary since the defining properties of a superconductor do not follow directly from the form of the symmetry breaking operator expectation value. Instead, many properties (like the Meissner effect) are connected to the $U(1)$-gauge field $A_\mu(x)$. We therefore assume our Lagrangian to be invariant under the $U(1)$-gauge transformation
\begin{equation}
\begin{aligned}
A_\mu(x) &\rightarrow A_\mu(x)+\partial_\mu\Lambda(x),\\
\psi(x)  &\rightarrow \exp(-ie\Lambda(x))\psi(x),
\end{aligned}
\end{equation}
where we replaced $\theta(x)$ by the more common $-e\Lambda(x)$ with the electron charge $-e$. $\Lambda(x)$ is an arbitrary function, but with $\Lambda(x)$ and $\Lambda(x)-2\pi n/e$, $n\in \mathbbm{Z}$ regarded as identical. We will now show how the typical properties of a superconductor can be derived by assuming spontaneous symmetry breaking of this $U(1)$-gauge symmetry. This discussion follows \cite{weinberg}.

The symmetry is broken if operators like $f(x)\psi^T(x)i\sigma_2\psi(x)$ develop a non vanishing expectation value (the $i\sigma_2$ between the electron fields is needed since due to the anticommutation rules for fermionic fields an operator like $f(x)\psi^T(x)\psi(x)$ would be equal to zero identically). In this case, the symmetry $U(1)$ is broken down to $Z_2$. $Z_2$ corresponds to the transformations $\psi(x)\rightarrow\pm\psi(x)$ (with $\Lambda(x)=0$ or $\Lambda(x)=-\pi/e$) which leave $f(x)\psi^T(x)i\sigma_2\psi(x)$ invariant. According to the spatial symmetry of the function $f(x)$ we distinguish $s$-wave-, $p$-wave-, $d$-wave, etc. $U(1)$-symmetry breaking. Instead of considering the transformation properties of $\psi_1(x)$ and $\psi_2(x)$ in $\psi(x)=\psi_1(x)+i\psi_2(x)$ it is more convenient to write
\begin{equation}
\psi(x)=\exp(-ie\phi(x))\rho(x).
\end{equation}
We define $\phi(x)$ to be periodic in $\pi/e$ ({\em not} in $2\pi/e$ as one would expect from this notation that resembles the decomposition into an absolute value and a phase factor). Then $\phi(x)$ serves as a basis for $U(1)/Z_2$ and $\rho(x)$ for $Z_2$ (if we had taken $\phi(x)$ to be periodic in $2\pi/e$, $\phi(x)$ would have corresponded to $U(1)$ and $\rho(x)$ to the trivial group). By this definition, we obtain two fields, one of which belongs to the broken $U(1)/Z_2$- and one to the unbroken $Z_2$-symmetry. For $U(1)/Z_2$ we have the transformation properties
\begin{equation}
A_\mu(x)\rightarrow A_\mu(x)+\partial_\mu\Lambda(x),\quad\phi(x)\rightarrow\phi(x)+\Lambda(x),\quad\rho(x)\rightarrow\rho(x)
\end{equation}  
($\phi(x)+\Lambda(x)$ is understood to be taken modulo $\pi/e$) and for $Z_2$
\begin{equation}
A_\mu(x)\rightarrow A_\mu(x)+\partial_\mu\Lambda(x),\quad\phi(x)\rightarrow\phi(x),\quad\rho(x)\rightarrow-\rho(x).
\end{equation}
Since no mass term $\psi^\dagger(x)\psi(x)=\rho^\dagger(x)\rho(x)$ in the Lagrangian involves $\phi(x)$, $\phi(x)$ is a massless mode of our theory, the well known Goldstone boson. If we are well within the symmetry breaking regime, the dynamics of the system is dominated by this Goldstone boson. In a first approximation, we may neglect the quantum fluctuation contributions of the massive modes altogether. Then $\rho(x)$ plays the role of some fixed external field and enters the Lagrangian as a parameter. In this case, we end up with a Lagrangian for the electromagnetic and Goldstone boson content
\begin{equation}
L=-\frac{1}{4}\int\,d^3xF_{\mu\nu}F^{\mu\nu}+L_s[A_\mu-\partial_\mu\phi],
\end{equation} 
which is valid in the region of symmetry breaking and not too close to the point where the broken symmetry becomes unbroken. The exact form of the functional $L_s$ is not known to us; however, the dependence on $A_\mu-\partial_\mu\phi$ is dictated by gauge invariance. Classically, $-L_s$ can be interpreted as a potential for our theory. We will assume that this potential possesses a minimum for vanishing external fields $A_\mu$ (i.e. the system is stable if external electromagnetic fields are absent) and vanishing Goldstone fields, which means that the minimum occurs in $A_\mu-\partial_\mu\phi=0$. This is all we need to derive the main properties of superconductors.

We see immediately that if the potential possesses a minimum in $A_\mu-\partial_\mu\phi=0$, we have $A_\mu=\partial_\mu\phi$, so that the magnetic field vanishes: $\vec{B}=\text{rot}\,\vec{A}=0$. This is the famous {\em Meissner effect}: Deep within a superconductor we have no magnetic field. Closer to the point where the broken symmetry becomes unbroken, i.e. closer to the spatial border of the region of superconductivity, $A_\mu-\partial_\mu\phi$ no longer vanishes. To describe the behavior of the superconductor near the border of the superconducting region, we may expand the energy to second order in $|\vec A-\vec{\nabla}\phi|$ around $|\vec A-\vec{\nabla}\phi|=0$. The linear term vanishes since we assumed the energy to possess a minimum at this point. The quadratic term has the form
\begin{equation}
\begin{aligned}
\Delta E_{\text{pen}}&=-\frac{1}{2}
&\approx |\vec A-\vec{\nabla}\phi|^2L^3/\lambda^2,
\end{aligned}
\end{equation}
where $L^3$ is the volume of the superconductor, $\lambda$ is some length depending on the material and in the second line $|\vec A-\vec\nabla\phi|^2$ is some average value of $|\vec A(\vec x)-\vec\nabla\phi(\vec x)|^2$ over the region of integration. $\Delta E_{\text{pen}}$ describes the energy cost for allowing a magnetic field to penetrate the superconductor. Since $ |\vec A-\vec{\nabla}\phi|$ is of order $BL$, $B$ being the magnetic field, we have
\begin{equation}
\Delta E_{\text{pen}}\approx \frac{B^2L^5}{\lambda^2}.
\end{equation}
On the other hand, the magnetic field carries an energy density of order $B^2$, thus the energy cost to {\em expel} the magnetic field from the superconductor is
\begin{equation}
\Delta E_{\text{ex}}\approx B^2L^3
\end{equation}
The magnetic field will be expelled from the superconductor, if the energy cost to expel the weak magnetic field from the superconductor is much smaller\footnote{``Much'' to be on the safe side with all our approximations.} than the energy cost we have to pay if the magnetic field is to penetrate the superconductor: $\Delta E_{\text{ex}}\ll \Delta E_{\text{pen}}$ or in other words $\lambda\ll L$. This means that in materials with small $\lambda$ the superconducting region from which the magnetic field is expelled is large and vice versa. For this reason, $\lambda$ is called the {\em penetration depth} of the superconductor.

Similarly we can see that superconductivity is destroyed, if the magnetic field $B$ is larger than some critical magnetic field $B_c$. The fact that some material becomes a superconductor means that the superconducting state is energetically favored in comparison to the normal state, say by the energy $L^3\Delta$, where $\Delta$ is the energy density gap between the superconducting and normal state. As we have argued above, the energy cost to expel a magnetic field from the superconductor is of order $B^2L^3$. If the energy cost to expel the magnetic field is larger than what we energetically win by favoring the superconducting state, $B>\sqrt{\Delta}$, the material will no longer remain to be a superconductor. The critical magnetic field is then given by $B_c\approx\sqrt{\Delta}$. However, note that this is only true for uniform superconductors. Especially high temperature superconductors are able to tolerate much larger magnetic fields than one would expect from these simple considerations without losing their superconducting properties. This is due to the fact that these materials form magnetic flux vortices, tiny tubes of non vanishing magnetic fields that traverse the superconductor. By this mechanism the energy cost for expelling the magnetic field is reduced, allowing the material to remain superconducting for large magnetic fields (``type II superconductor'').

We will now come to the most significant property of a superconductor, the fact that the resistance equals zero. Imagine a wire made of superconducting material with $L\gg\lambda$, where $L$ is the radial dimension of the wire. Bend the wire into a closed ring. Then we know that well inside the wire $|\vec A-\vec{\nabla}\phi|$ vanishes. We therefore can find a closed curve $C$ (following the linear dimension of the wire) along which $|\vec A-\vec{\nabla}\phi|$ always vanishes. Now start at some point $P$ on $C$ with the fields given at this point by $\vec{A}_P$ and $\phi_P$. Going around the ring following $C$ until we reach our starting point $Q=P$ of the closed curve, the fields are $\vec{A}_Q$ and $\phi_Q$. Since $P$ and $Q$ are equal, we should have $\vec{A}_Q=\vec{A}_P$. However, since $\phi(x)$ is periodic in $\pi/e$, we may have $\phi(x)_Q=\phi(x)_P+n\pi/e$, $n\in\mathbbm{Z}$, all of which are equivalent. Therefore the magnetic flux surrounded by our wire is
\begin{equation}
\int_F\vec{B}\vec{\hat n}\,dF=\oint_C\vec{A}\,d\vec{x}=\oint_C\vec{\nabla}\phi \,d\vec{x}=n\pi/e,
\end{equation}
where $F$ is the area surrounded by the wire, $\vec{\hat n}$ is a unit vector perpendicular to this area and $C$ is the closed curve connecting $P$ and $Q=P$. This result tells us that the magnetic flux is quantized. A given magnetic flux with $n\neq0$ is maintained by currents flowing in the superconductor. Since there is no way to smoothly change the magnetic flux, these currents cannot smoothly decay, which means that the resistance of the superconductor is zero. 

The last effect we would like to discuss occurs if two pieces of superconducting material $1$ and $2$ are brought together. Let $F$ be the area of the junction. Then the Lagrangian describing the system near the junction is 
\begin{equation}
L_j=\int_F\int_1^2 dx\,\tilde G[\vec{A}(x),\phi_1(x),\phi_2(x)].
\end{equation}
The integral over $x$ goes over some short line perpendicular to the surface of the junction, connecting two points $1$ and $2$ situated inside the two different materials. $\phi_1$ and $\phi_2$ are the Goldstone modes in the two materials. If we assume that no gradients of Goldstone fields and no components of magnetic fields parallel to the surface of the junction are present, we may simply write
\begin{equation}
L_j=F G[\vec{A},\phi_1,\phi_2],
\end{equation}
where we have absorbed the integration over $x$ into $G$. Gauge invariance tells us that $G=G[\Delta_A\phi=\int_1^2 dx\,\vec{\hat n}(\vec{\nabla}\phi-\vec{A})]$, $\vec{\hat n}$ being a unit vector perpendicular to the surface of the junction. The integral is necessary to guarantee the correct behavior in the case of vanishing vector potential. In this case
\begin{equation}
\Delta_{A=0}\phi=\int_1^2 dx\,\vec{\hat n}\vec{\nabla}\phi=\phi_2-\phi_1\equiv\Delta\phi
\end{equation}
so that we end up with a gauge invariant expression as it should be. We want to calculate the current flowing through the junction. The current density is given by
\begin{equation}
\vec{J}=\frac{\delta L_j}{\delta\vec{A}}=G'(\Delta_A\phi) F\frac{\delta\Delta_A\phi}{\delta\vec A}=-G'(\Delta_A\phi)\vec{\hat n}
\end{equation}
and in the case of vanishing vector potential
\begin{equation}
\label{eq:u1:J}
\vec{J}=-G'(\Delta\phi)\vec{\hat n}.
\end{equation}
The next step is to express $\Delta\phi$ by the voltage between the two materials. For this purpose, note that the charge density is given by
\begin{equation}
J^0(x)=\frac{\delta L_j}{\delta A_0(x)}=-\frac{\delta L_j}{\delta\dot\phi(x)},
\end{equation}
so that $-J^0(x)$ is the canonical conjugate to $\dot\phi(x)$. In the Hamiltonian formulation, this yields
\begin{equation}
\dot\phi(x)=\frac{\delta H_j}{\delta(-J^0(x))}.
\end{equation} 
The voltage $V(x)$ is nothing else than the change of energy density per change of charge density, so that
\begin{equation}
\dot\phi(x)=-V(x).
\end{equation}
As a side remark, note that this shows that for some superconductor in a stationary state for which $\dot\phi(x)=0$ we have $V(x)=0$ which is again the zero resistance property of a superconductor. If we now assume that our two superconductors are kept at a constant voltage and the voltage difference is given by $\Delta V$, we get
\begin{equation}
\Delta\phi=-\Delta V t+\text{const.}
\end{equation} 
Using this result in \eqref{eq:u1:J}, we have
\begin{equation}
\vec{J}=-G'(-\Delta V t+\text{const})\vec{\hat n}.
\end{equation} 
Since $\Delta\phi$ is periodic in $\pi/e$, this shows that the current oscillates with frequency
\begin{equation}
\nu=e\left|\Delta V\right|/\pi.
\end{equation} 
This is the {\em Josephson effect}. It allows high precision measurements of $e/\hbar$ (if we had bothered not taking $\hbar$ to be unity), since frequencies and voltages can be measured very accurately.

We would like to recall that all the results we derived in this section were solely based on the assumption of a broken $U(1)$-symmetry. No explicit dynamical model (as the Ginzburg-Landau- or BCS-Lagrangian) was needed to find the main properties of a superconductor. This point of view allows us to directly identify regions of broken $U(1)$-symmetry in the Hubbard model (which is the dynamical model we will use) with regions of superconducting behavior in much the same way as we naturally identify regions of broken $SU(2)$-symmetry with regions of ferromagnetic or antiferromagnetic behavior.

\section{High temperature superconductors}
We showed in the last section that superconducting properties can be derived by assuming $U(1)$-symmetry breaking of a gauge theory. In this section we review the history of superconductivity and especially that of the discovery of high temperature superconductors. For a recent and more complete overview, see \cite{supra}.

 In 1911, Heike Kamerlingh Onnes found the first material exhibiting superconductivity by cooling down a mercury wire to 4K. Nowadays we know that superconductivity can be observed for many conductors and semiconductors at low temperature. However, until 1986 the wide range of materials found to exhibit superconducting properties had in common that their critical temperature $T_c$ (the maximum temperature for which superconductivity occurs) did not exceed 20K. In 1986, the first high temperature superconductor $\rm(LaBa)_2CuO_4$ with a critical temperature of 35K was found by Bednorz and Müller \cite{bednorz}. The following years witnessed a series of records of critical temperatures for high temperature superconductors. The material with the currently highest critical temperature known is $\rm HgBa_2Ca_2Cu_3O_8$ with $T_c=134K$ \cite{schilling}.

Many of the recently found high temperature superconductors are so called {\em cuprates}, materials with a very anisotropic structure. In contrast to the superconductors known before and to what is usually discussed in dynamical models like the BCS-theory, the superconductivity in cuprates has $d_{x^2-y^2}$-wave-symmetry \cite{scalapino}. The cuprates consist of two dimensional $\rm CuO_2$-layers and La-, Sr-, Ba-atoms between these layers. For La-interlayer atoms, one effectively finds one electron per lattice site of the $\rm CuO_2$-layers. By replacing La bei Sr or Ba, one removes electrons from the $\rm CuO_2$-layers, which is called p-doping. Most of the electronic dynamics is constrained to the layers. We will exploit this fact by modeling a high temperature superconductor by a two dimensional model, neglecting the weak coupling between different layers.

 Experimentally, the phase diagram of a cuprate is qualitatively shown in fig. (\ref{fig:cuprat}). From inelastic neutron scattering experiments it is known that although the antiferromagnetic long range order disappears for strong doping, antiferromagnetic fluctuations are present even in the superconducting domain. There are speculations that these fluctuation have an important impact on the superconducting order. Furthermore, a whole variety of quantum fluctuations in different channels that do not correspond to any long ranged order is under discussion to explain the phase diagram. This discussion is additionally fed by the discovery of a so called {\em pseudo energy gap} below some temperature $T^*$. It is not known if this pseudo energy gap is connected to any kind of long range order (as the energy gap $\Delta$ discussed in the last section is connected to superconducting long range order), but it is strongly suspected that the key to an explanation of high temperature superconductivity lies in the understanding of this pseudo energy gap. 

All in all, at the present our understanding of the phase diagram is very limited. We hope to convince the reader that the bewildering variety of degrees of freedom discussed for the cuprates to explain their properties calls for a formalism which is able to include all these degrees of freedom in a transparent and unified way and that with our bosonized version of the Hubbard model we are able to provide such a formalism.  

\section{The Hubbard model}
The Hubbard model has been proposed independently by Hubbard, Kanamori and Gutzwiller \cite{hubbard} in 1963 as a model for strongly interacting electrons on a lattice. There is a wide range of electromagnetic properties of condensed matter systems that were or are under investigation by modeling them by the Hubbard model: Ferromagnetism, antiferromagnetism, conductor-insulator transitions and --- more recently --- high temperature superconductivity. The large spectrum of physical properties that are tried to be understood by means of this model is accompanied by an equally large spectrum of different calculational techniques used to approximately solve it. An exact solution of the model is only known in one dimension \cite{liebwu}. For two or more dimensions, in general approximations or numerical methods have to be used. Unfortunately, the results are not stable against choice of the method: A lot of contradicting results have been published during the last decades. This is the reason why exact solutions for particular values of the parameters of the model play an important role as tests that any reliable approximation has to pass.

The defining features of the Hubbard model are:
\begin{itemize}
\item The electrons are strongly located at the atoms of the lattice. This means that the electron field operator is given by $\psi_i$, where $i$ label the lattice sites, instead of some continuous operator $\psi(x)$. 
\item The Coulomb interaction between electrons at {\em different} lattice sites is neglected. Any electron interacts only with a possible second electron at the same lattice site. Due to the Pauli principle, only two electrons with opposite spin at one lattice site are allowed.
\item The electrons have the ability to hop between lattice sites. 
\end{itemize} 
For our purposes, we will additionally make the following assumptions:
\begin{itemize}
\item The lattice is two dimensional and quadratic. This is motivated by the actual chemical structure of the cuprates that we want to provide a model for. We completely neglect the weak interlayer coupling and the slight distortion of the lattice structure away from the ideal quadratic structure.
\item Electron hopping occurs only between nearest neighbor lattice sites. This should be the dominating effect, since the electron hopping amplitude becomes smaller with the distance of the lattice sites between which hopping may occur. Furthermore, we assume that the hopping amplitude is the same for all nearest neighbor pairs. 
\end{itemize}

With these preliminaries in mind, we can write down the Hamiltonian for the Hubbard model
\begin{equation}
\label{eq:hubb:ham1}
\hat H=\sum_{ij\sigma}{\cal T}_{ij}a_{i\sigma}^+a_{j\sigma}+\frac{1}{2}U\sum_{i\sigma}a_{i\sigma}^+a_{i\sigma}a_{i(-\sigma)}^+a_{i(-\sigma)}.
\end{equation}
$a_{i\sigma}^+$ and $a_{i\sigma}$ are the creation- and annihilation operators for an electron at lattice site $i$ with spin $\sigma$. The first term describes the hopping between different lattice sites. With our assumptions
\begin{equation}
{\cal T}_{ij}=\left\{\begin{aligned}-t&\quad\text{, if $i$ and $j$ are nearest neighbors}\\
                            0 &\quad\text{, else.}\end{aligned}\right.
\end{equation}
The sign in front of $t$ is purely conventional. In particular, we will not assume that $t>0$ (and indeed there is no simple argument to decide which is the correct sign of $t$). As we will see, we do not have to bother with this question since all our results only depend on $t^2$. The second term of the Hamiltonian describes the local Coulomb interaction between electrons at the same lattice site. We take $U>0$ to have a repulsive interaction ($U>0$ raises the energy of placing two electrons on the same lattice site, which corresponds to a repulsive force). 

The parameters of the model are obviously $U$ and $t$. If we decide to measure all quantities with respect to $U$, the Hamiltonian may be written as
\begin{equation}
\hat H/U=\sum_{ij\sigma}({\cal T}_{ij}/U)a_{i\sigma}^+a_{j\sigma}+\frac{1}{2}\sum_{i\sigma}a_{i\sigma}^+a_{i\sigma}a_{i(-\sigma)}^+a_{i(-\sigma)}.
\end{equation}
Introducing new variables that are dimensionless and measured with respect to $U$, we finally have
\begin{equation}
\label{eq:hubb:ham}
\hat H=\sum_{ij\sigma}{\cal T}_{ij}a_{i\sigma}^+a_{j\sigma}+\frac{1}{2}\sum_{i\sigma}a_{i\sigma}^+a_{i\sigma}a_{i(-\sigma)}^+a_{i(-\sigma)}.
\end{equation}
This transcription is unusual in the context of analyzing the Hubbard model in this form by means of the renormalization group, as in this case one is mostly interested in investigating the flow of the four fermion coupling constants. However, in our new approach that we present in this work we will not consider the flow of the four fermion coupling, so that this transcription is convenient.

Another parameter that should be fixed for the model is the number of electrons on the lattice $N_e$. If the number of lattice sites is $N_s$, we have $0\leq N_e\leq 2N_s$. The case $N_e=N_s$ is called half filling and it is especially interesting, since exact results are available for it (at least in the limit of large $U$). Furthermore, half filling corresponds to the undoped cuprate (where because of the chemical structure each atom provides one free electron to the system) whereas doping changes the number of electrons away from half filling. In contrast to $t$, the fixed electron number does not enter directly into the model as a parameter. We will have to include it by specifying a source for the electron charge density\footnote{This source is nothing else than the chemical potential.}, dynamically varying this source to keep the expectation value of the charge density constant. 

The last parameter is temperature. It enters our description when writing down the partition function for the Hubbard model as a statistical quantity. We will come to the details. 

As we mentioned before, rigorous results are important to test approximations. For a review, see e.g. \cite{tasaki}. One of these exact results is the fact that the Hubbard model has an antiferromagnetic ground state at temperature $T=0$, sufficiently large $U$ and for half filling in agreement with what is experimentally found for cuprates (cf. fig. (\ref{fig:cuprat})). For $T>0$ another exact result (the Mermin-Wagner theorem) forbids the existence of an antiferromagnetic ground state in two dimensions. This is somewhat disturbing, because we would like to predict antiferromagnetic order exactly in the region of the phase diagram where it is forbidden by the theorem. A possible explanation would be to argue that although the coupling between the layers of a cuprate is weak, it cannot be neglected when applying the Mermin-Wagner theorem --- for three dimensions, antiferromagnetic order {\em is} allowed by the theorem. One of the subjects of the last chapter, where we analyze the properties of the model using our formalism, is to show that it is possible to reconcile the Mermin-Wagner theorem with the occurrence of antiferromagnetic long range order for $T>0$ even in two dimensions. 

Over the last years, efforts have been made to investigate the properties of the Hubbard model numerically by renormalization group techniques \cite{renhub}. In all of these approaches the flow of various four fermion interactions was calculated, confirming the main symmetry breaking instabilities of antiferromagnetism and $d$-wave superconductivity in the Hubbard model. These instabilities are inferred from the divergence of the four fermion couplings. The divergence of couplings at the onset of spontaneous symmetry breaking prevents these approaches from following the flow into the broken phase and is one reason for constructing the alternative formalism presented in this work, which is more suitable for this task.     

\chapter{The partition function of the Hubbard model}

The starting point of this work will be the partition function of the Hubbard model. In this chapter the general method to derive the partition function once the Hamiltonian is given in second quantized form is presented. Much of the material in this chapter can be found in textbooks covering statistical field theory (cf. e.g. \cite{negele}) and will be known to the experienced reader. However, the last topic of this chapter, the formulation of the partition function via coherent states, deserves some explanation. We restrict ourselves to fermionic systems.

\section{Many particle systems}
The two ingredients for a quantum theory are states and operators. We will generalize these concepts from the one particle system to the many particle system in this section. First we cover the generalization of state kets.

Consider a system with $N$ identical fermionic particles. The Hilbert space for one particle be $\cal{H}$. Then the Hilbert space for the $N$-particle system is given by
\begin{equation}
{\cal H}_N={\underbrace{\cal{H}\otimes\cal{H}\otimes\ldots\otimes\cal{H}}_{\text{$N$ times}}}.
\end{equation}
If $\{\left|\alpha_i\right>\}$ is an orthonormal basis for the one particle Hilbert space $\cal{H}$ of particle $i$, we can define a basis for ${\cal H}_N$ by
\begin{equation}
\left|\alpha_1\ldots\alpha_N\right)=\left|\alpha_1\right>\otimes\ldots\otimes\left|\alpha_N\right>.
\end{equation} 
Orthonormality and completeness directly carry over from the one particle basis of $\cal{H}$ to this basis of ${\cal H}_N$:
\begin{equation}
\begin{aligned}
\left(\alpha_1\ldots\alpha_N\right|\left.\alpha_1'\ldots\alpha_N'\right)&=\left<\alpha_1\right|\left.\alpha_1'\right>\ldots\left<\alpha_N\right|\left.\alpha_N'\right>=\delta_{\alpha_1\alpha_1'}\ldots\delta_{\alpha_N\alpha_N'}\\
\sum_{\alpha_1\ldots\alpha_N}\left|\alpha_1\ldots\alpha_N\right)\left(\alpha_1\ldots\alpha_N\right|&=\sum_{\alpha_1}\left|\alpha_1\right>\left<\alpha_1\right|\ldots\sum_{\alpha_N}\left|\alpha_N\right>\left<\alpha_N\right|=1.
\end{aligned}
\end{equation}
For systems with identical fermions, any physical state has to be antisymmetric under particle exchange. We therefore define the totally antisymmetric basis
\begin{equation}
\label{eq:multpar:defanti}
\left|\alpha_1\ldots\alpha_N\right>=\frac{1}{\sqrt{N!}}\sum_{\cal P}\text{sgn}({\cal P})\left|\alpha_{{\cal P}(1)}\ldots\alpha_{{\cal P}(N)}\right),
\end{equation} 
where the sum runs over all permutations of the particles. The scalar product now reads
\begin{multline}
\label{eq:multpar:scalar}
\left<\alpha_1\ldots\alpha_N\right|\left.\alpha_1'\ldots\alpha_N'\right>=\sum_{\cal P}\text{sgn}({\cal P})\left<\alpha_1\right|\left.\alpha_{{\cal P}(1)}'\right>\ldots\left<\alpha_N\right|\left.\alpha_{{\cal P}(N)}'\right>\\
=\left\{\begin{aligned}1&\quad\text{, if the permutation transferring $\alpha_1\ldots\alpha_N$ into $\alpha_1'\ldots\alpha_N'$ is even,}\\
-1&\quad\text{, if the permutation transferring $\alpha_1\ldots\alpha_N$ into $\alpha_1'\ldots\alpha_N'$ is odd,}\\
0&\quad\text{, else.}\end{aligned}\right.
\end{multline}
The completeness relation is
\begin{equation}
\sum_{\alpha_1\ldots\alpha_N}\left|\alpha_1\ldots\alpha_N\right>\left<\alpha_1\ldots\alpha_N\right|=N!
\end{equation}

We now come to the second ingredient of a quantum theory, the operators. Suppose we are given a basis $\{\left|U_i\right>\}$ of a one particle (labeled with $i$) Hilbert space  which consists of eigenstates to some operator $\hat U_i$
\begin{equation}
\hat U_i\left|U_i\right>=U_i\left|U_i\right>,
\end{equation}
where $U_i$ is the eigenvalue to $\hat U_i$. A {\em one particle operator} $\hat U$ in the many particle system with a general basis $\{\left|\alpha_i\right>\}$ for the $i$th particle is then defined to be 
\begin{equation}
\hat U\left|\alpha_1\ldots\alpha_N\right)=\sum_{i=1}^N\hat U_i\left|\alpha_1\ldots\alpha_N\right),
\end{equation} 
where $\hat U_i$ only acts on the $\left|\alpha_i\right>$-part of $\left|\alpha_1\ldots\alpha_N\right)$. For example for non interacting particles, if we take $\hat U$ to be the energy operator and $\{\left|\alpha_i\right>\}$ to be the energy eigenbasis for the $i$th particle, this means that the energy of the many particle system is the sum of the single particle energies. The matrix elements of a one particle operator are given by
\begin{equation}
\label{eq:multpar:onepar}
\left(\alpha_1\ldots\alpha_N\right|\hat U\left|\beta_1\ldots\beta_N\right)=\sum_{i=1}^N\prod_{k\neq i}\left<\alpha_k\right|\left.\beta_k\right>\left<\alpha_i\right|\hat U\left|\beta_i\right>.
\end{equation}
Similarly, we define the {\em two particle operator} $\hat V$ by
\begin{equation}
\hat V\left|\alpha_1\ldots\alpha_N\right)=\frac{1}{2}\sum_{1\leq i,j\leq N, i\neq j}\hat V_{ij}\left|\alpha_1\ldots\alpha_N\right)
\end{equation}
with the matrix elements
\begin{equation}
\left(\alpha_1\ldots\alpha_N\right|\hat V\left|\beta_1\ldots\beta_N\right)=\frac{1}{2}\sum_{i\neq j}\prod_{k\neq i,j}\left<\alpha_k\right|\left.\beta_k\right>\left(\alpha_i\alpha_j\right|\hat V\left|\beta_i\beta_j\right).
\end{equation}

\section{Creation and annihilation operators}
Up to now we considered $N$-particle systems, where $N$ was some fixed number. In quantum field theory however, the number of particles may change. Instead of an $N$-particle Hilbert space ${\cal H}_N$ the underlying space is the {\em Fock space} $\cal F$, which is the direct sum of all $N$-particle Hilbert spaces
\begin{equation}
{\cal F}=\bigoplus_{N=0}^\infty{\cal H}_N.
\end{equation}
It is very convenient to introduce {\em creation and annihilation operators} on this Fock space and to express states and operators by means of these. Since basis kets belonging to Hilbert spaces with different $N$ are orthogonal, the completeness relation simply reads
\begin{equation}
\label{eq:cran:complet}
\left|0\right>\left<0\right|+\sum_{N=1}^\infty\frac{1}{N!}\sum_{\alpha_1\ldots\alpha_N}\left|\alpha_1\ldots\alpha_N\right>\left<\alpha_1\ldots\alpha_N\right|=1.
\end{equation}
The creation operator $a_\lambda^+$ is defined by
\begin{equation}
\label{eq:cran:defcr}
a_\lambda^+\left|\lambda_1\ldots\lambda_N\right>=\left|\lambda\lambda_1\ldots\lambda_N\right>,
\end{equation}  
transforming a state in ${\cal H}_N$ to one in ${\cal H}_{N+1}$. Remember that we only treat the fermionic case. For bosonic systems, additional factors appear in this definition to guarantee normalization. As a consequence, any state may be written as
\begin{equation}
\left|\alpha_1\ldots\alpha_N\right>=a_{\alpha_1}^+\ldots a_{\alpha_N}^+\left|0\right>,
\end{equation}
where $\left|0\right>$ is the vacuum state. Using \eqref{eq:multpar:defanti} and \eqref{eq:cran:defcr}, we find the anticommutation relation
\begin{equation}
\label{eq:cran:anticomcr}
\left\{a_\alpha^+,a_\beta^+\right\}=0.
\end{equation}
The annihilation operator $a_\alpha$ is defined by
\begin{equation}
\label{eq:cran:defan}
a_\alpha=(a_\alpha^+)^\dagger.
\end{equation}
From \eqref{eq:cran:anticomcr} we immediately find
\begin{equation}
\left\{a_\alpha,a_\beta\right\}=0.
\end{equation}
By using \eqref{eq:cran:complet}, \eqref{eq:cran:defan}, \eqref{eq:cran:defcr} and \eqref{eq:multpar:scalar} we show that
\begin{equation}
\label{eq:cran:acan}
a_\alpha\left|\alpha_1\ldots\alpha_N\right>=\left\{\begin{aligned}(-1)^{(i-1)}\left|\alpha_1\ldots\alpha_{i-1}\alpha_{i+1}\ldots\alpha_N\right>&\quad\text{, if the $i$th particle is in state $\left|\alpha\right>$},\\0&\quad\text{, if no particle is in state $\left|\alpha\right>$.}\end{aligned}\right.
\end{equation}
\eqref{eq:cran:defcr} and \eqref{eq:cran:acan} yield the last anticommutation relation
\begin{equation}
\left\{a_\alpha,a_\beta^+\right\}=\delta_{\alpha\beta}.
\end{equation}
In the same way we can show that the operator $\hat n_\alpha=a_\alpha^+a_\alpha$ counts the number of particles in the state $\alpha$:
\begin{equation}
\hat n_\alpha\left|\alpha_1\ldots\alpha_N\right>=\sum_{i=1}^N\delta_{\alpha\alpha_i}\left|\alpha_1\ldots\alpha_N\right>.
\end{equation} 
Of course for fermionic systems, this number is either $1$ or $0$. The operator 
\begin{equation}
\label{eq:cran:totop}
\hat N=\sum_\alpha \hat n_\alpha
\end{equation}
counts the total number of particles  in the system. 

Note that basis changes from a one particle basis $\{\left|\alpha\right>\}$ to another one particle basis $\{\left|\tilde\alpha\right>\}$ are easily implemented on the creation and annihilation operators. From
\begin{equation}
\left|\tilde\alpha\right>=\sum_\alpha\left<\alpha\right|\left.\tilde\alpha\right>\left|\alpha\right>
\end{equation}
we find
\begin{equation}
\label{eq:cran:basch}
\begin{aligned}
a_{\tilde\alpha}^+&=\sum_\alpha\left<\alpha\right|\left.\tilde\alpha\right>a_\alpha^+,\\
a_{\tilde\alpha}&=\sum_\alpha\left<\tilde\alpha\right|\left.\alpha\right>a_\alpha.
\end{aligned}
\end{equation}

One and two particle operators are usually expressed by creation and annihilation operators. For a one particle operator $\hat U$, we have with \eqref{eq:multpar:defanti} and \eqref{eq:multpar:onepar}
\begin{equation}
\left<\alpha_1\ldots\alpha_N\right|\hat U\left|\beta_1\ldots\beta_N\right>=\sum_{i=1}^NU_i\left<\alpha_1\ldots\alpha_N\right|\left.\beta_1\ldots\beta_N\right>,
\end{equation}
where we have assumed that $\{\left|\alpha_i\right>\}$ is an eigenbasis of $\hat U_i$. By using the identity
\begin{equation}
\sum_{i=1}^NU_i=\sum_\alpha U_\alpha n_\alpha,
\end{equation} 
where the sum over $\alpha$ goes over all possible one particle states and $n_\alpha$ is the number of particles present in the state $\alpha$, we conclude that 
\begin{equation}
\hat U=\sum_\alpha U_\alpha a_\alpha^+a_\alpha.
\end{equation}
In a general basis (not necessarily an eigenbasis of $\hat U_i$) we find by using \eqref{eq:cran:basch}
\begin{equation}
\hat U=\sum_{\lambda\mu}U_{\lambda\mu}a_\lambda^+a_\mu
\end{equation}
with $\lambda$, $\mu$ labeling basis kets of the general basis and 
\begin{equation}
U_{\lambda\mu}=\sum_\alpha\left<\lambda\right|\left.\alpha\right>U_\alpha\left<\alpha\right|\left.\mu\right>.
\end{equation}
In much the same way, but somewhat more involved, we can repeat these steps to derive the desired form of the two particle operator
\begin{equation}
\hat V=\frac{1}{2}\sum_{\lambda\mu\nu\rho}V_{\lambda\mu,\nu\rho}a_\lambda^+a_\mu^+a_\rho a_\nu.
\end{equation}

Recalling the Hamiltonian of the Hubbard model \eqref{eq:hubb:ham1}, we do now understand the specific form of the terms. The hopping term describes the one particle hopping from one lattice site to another, and $t$ is a matrix element giving the transition amplitude. The second term is a two particle Coulomb interaction term (therefore two creation and annihilation operators) and $U$ describes the strength of this interaction.

\section{Coherent states}
Up to now, most of our results do not depend on the choice of the basis. We will exploit this fact by specifying a very special basis that is useful to derive the partition function in the next section. This basis is composed of {\em coherent state kets}. A coherent state $\left|\psi\right>$ is defined to be an eigenstate of the annihilation operator:
\begin{equation}
a_\alpha\left|\psi\right>=\psi_\alpha\left|\psi\right>.
\end{equation}
$\psi_\alpha$ is the eigenvalue to the annihilation operator $a_\alpha$. Note that since the annihilation operators for fermions anticommute, the same is true for the eigenvalues. This means that the $\psi_\alpha$ are {\em Grassmann numbers}. If $\left|\psi\right>$ is an eigenket to $a_\alpha$, then $\left<\psi\right|$ is an eigenbra to $a_\alpha^+$. We call the corresponding eigenvalue $\psi_\alpha^*$:
\begin{equation}
\left<\psi\right|a_\alpha^+=\left<\psi\right|\psi_\alpha^*.
\end{equation}
Of course, the $\psi_\alpha^*$ are also Grassmann numbers, since the creation operators for fermions anticommute. Additionally, we {\em demand} the properties
\begin{equation}
\{\psi_\alpha,\psi_\beta^*\}=0,\quad\{\psi_\alpha,a_\beta\}=0.
\end{equation} 
The full set $\{\psi_\alpha,\psi_\alpha^*\}$ contains the elements of the Grassmann algebra of all eigenvalues of coherent states. Note that not only the coherent states $\left|\psi\right>$ no longer correspond to states of some definite particle number, but also these states do not belong to the Fock space introduced in the last section. Instead, a coherent state is a superposition of different kets from this Fock space {\em with Grassmann valued coefficients}.

We can now proceed to construct coherent states from the vacuum state, calculating scalar products, completeness relations and operator expectation values using coherent states as we did in the last section for ordinary Fock space states.

Coherent state kets can be constructed from the vacuum ket by
\begin{equation}
\label{eq:coher:ket}
\left|\psi\right>=\prod_\alpha(1-\psi_\alpha a_\alpha^+)\left|0\right>.
\end{equation} 
To prove that this is consistent, apply an annihilation operator to both sides:
\begin{equation}
\begin{aligned}
a_\alpha\left|\psi\right>&=a_\alpha\prod_\beta(1-\psi_\beta a_\beta^+)\left|0\right>\\
&=\prod_{\beta\neq\alpha}(1-\psi_\beta a_\beta^+)a_\alpha(1-\psi_\alpha a_\alpha^+)\left|0\right>\\
&=\prod_{\beta\neq\alpha}(1-\psi_\beta a_\beta^+)\psi_\alpha(1-\psi_\alpha a_\alpha^+)\left|0\right>\\
&=\psi_\alpha\prod_\beta(1-\psi_\beta a_\beta^+)\left|0\right>\\
&=\psi_\alpha\left|\psi\right>.
\end{aligned}
\end{equation} 
In the same way, we can show that we can construct coherent state bras from the vacuum bra by
\begin{equation}
\label{eq:coher:bra}
\left<\psi\right|=\left<0\right|\prod_\alpha(1+\psi_\alpha^*a_\alpha).
\end{equation}
It is now straightforward to calculate the {\em scalar product} of two coherent states
\begin{equation}
\label{eq:coher:scal}
\left<\psi\right|\left.\psi'\right>=\prod_\alpha(1+\psi_\alpha^*\psi_\alpha').
\end{equation}
The proof of the {\em completeness relation} 
\begin{equation}
\label{eq:coher:compl}
\int\prod_\alpha d\psi_\alpha^*d\psi_\alpha\,\prod_\beta(1-\psi_\beta^*\psi_\beta)\left|\psi\right>\left<\psi\right|=1
\end{equation}
is equally simple, but more lengthy. One proceeds by taking the matrix elements of both sides with respect to two $N$-particle states in the ordinary Fock space, expressing all states by annihilation and creation operators applied to the vacuum state. The integral is the usual one for Grassmann numbers
\begin{equation}
\int d\psi\,1=\int d\psi^*\,1=0,\quad\int d\psi\,\psi=\int d\psi^*\,\psi^*=1,
\end{equation} 
and for multiple integrals the innermost integration is performed first. 

For the partition function we will need the trace of an operator in coherent state representation. This trace is given by 
\begin{equation}
\label{eq:coher:trace}
\text{Tr}\,A=\int\prod_\alpha d\psi_\alpha^*d\psi_\alpha\prod_\beta(1-\psi_\beta^*\psi_\beta)\left<-\psi\right|A\left|\psi\right>.
\end{equation}
To prove this, start with the trace in some arbitrary orthonormal basis, insert the completeness relation for coherent states and use the completeness of the original basis. One also needs that
\begin{equation}
\int d\psi^* d\psi\,\left<\alpha\right|\left.\psi\right>\left<\psi\right|\left.\beta\right>=\int d\psi^* d\psi\,\left<-\psi\right|\left.\beta\right>\left<\alpha\right|\left.\psi\right>.
\end{equation}
The minus sign comes from the exchange of the integration variables hidden in the coherent state kets and bras (cf. \eqref{eq:coher:ket} and \eqref{eq:coher:bra}). 

The last equation we will need in the next section is the expectation value of a normal ordered operator $A(a_\alpha^+,a_\alpha)$
\begin{equation}
\label{eq:coher:expect}
\left<\psi\right|A(a_\alpha^+,a_\alpha)\left|\psi'\right>=\prod_\beta(1+\psi_\beta^*\psi_\beta')A(\psi_\alpha^*,\psi_\alpha').
\end{equation}
It follows immediately from the normal ordered form of $A$ and \eqref{eq:coher:scal}.

\section{The partition function}
The grand canonical partition function is
\begin{equation}
Z=\text{Tr}\,e^{-\beta(\hat H-\mu\hat N)}.
\end{equation}
$\beta$ is the inverse temperature $\beta=1/T$, $\mu$ is the chemical potential, $\hat N$ is the total number operator introduced in \eqref{eq:cran:totop}
\begin{equation}
\hat N=\sum_\alpha a_\alpha^+a_\alpha,
\end{equation}
and $\hat H$ is some Hamiltonian expressed by creation and annihilation operators. Again, we work on a lattice with sites labeled by $i$. With $\sigma$ being the spin-3-component of the electron ($\sigma\in\{\uparrow,\downarrow\}$), we use the collective index $\alpha$ with $\alpha=i\sigma$. We assume $\hat H$ to be given in normal ordered form. In particular, for the Hubbard model we have
\begin{equation}
\label{eq:part:hamhubb}
\hat H=\sum_{ij\sigma}{\cal T}_{ij}a_{i\sigma}^+a_{j\sigma}-\frac{1}{2}\sum_{i\sigma}a_{i\sigma}^+a_{i(-\sigma)}^+a_{i\sigma}a_{i(-\sigma)}.
\end{equation}
From \eqref{eq:coher:trace} we have\footnote{Recall that for Grassmann numbers $1-\sum_\gamma\psi_\gamma^*\psi_\gamma=\exp(-\sum_\gamma\psi_\gamma^*\psi_\gamma)$. The exponential notation is more convenient in the derivation of the partition function.}
\begin{equation}
Z=\int\prod_\alpha d\psi_\alpha^*d\psi_\alpha\,e^{-\sum_\gamma\psi_\gamma^*\psi_\gamma}\left<-\psi\right|e^{-\beta(\hat H-\mu\hat N)}\left|\psi\right>.
\end{equation}
We cannot apply \eqref{eq:coher:expect} directly, since the exponential is not normal ordered. To cure this problem, we proceed as usual in the derivation of path integral expressions by dividing $\beta$ into $M$ small ``time slices'' $\epsilon$, so that $\beta=M\epsilon$ and write
\begin{equation}
e^{-\beta(\hat H-\mu\hat N)}=\underbrace{e^{-\epsilon(\hat H-\mu\hat N)}\cdots e^{-\epsilon(\hat H-\mu\hat N)}}_{\text{$M$ times}}.
\end{equation} 
Between all of these factor we insert the completeness relation \eqref{eq:coher:compl} in the form
\begin{equation}
\int\prod_{\alpha'} d\psi_{\alpha',k}^*d\psi_{\alpha',k}\,e^{-\sum_{\gamma'}\psi_{\gamma',k}^*\psi_{\gamma',k}}\left|\psi_k\right>\left<\psi_k\right|=1,
\end{equation}
where $k=1,\ldots,M-1$ labels the inserted states. By setting $\psi_M=-\psi$, $\psi_0=\psi$, $\psi_M^*=-\psi^*$ and $\psi_0^*=\psi^*$, we have
\begin{equation}
\label{eq:part:Z1}
\begin{split}
Z&=\int\prod_\alpha d\psi_\alpha^*d\psi_\alpha\,e^{-\sum_\gamma\psi_\gamma^*\psi_\gamma}\int\left(\prod_{k=1}^{M-1}\prod_{\alpha'}d\psi_{\alpha',k}^*d\psi_{\alpha',k}\right)\\
&\quad\left(\prod_{k=1}^{M-1}e^{-\sum_{\gamma'}\psi_{\gamma',k}^*\psi_{\gamma',k}}\right)\prod_{k=1}^M\left<\psi_k\right|e^{-\epsilon(\hat H-\mu\hat N)}\left|\psi_{k-1}\right>\\
&=\int\left(\prod_{k=1}^{M}\prod_{\alpha}d\psi_{\alpha,k}^*d\psi_{\alpha,k}\right)\,e^{-\sum_{\gamma}\sum_{k=1}^M\psi_{\gamma,k}^*\psi_{\gamma,k}}\left(\prod_{k=1}^M\left<\psi_k\right|e^{-\epsilon(\hat H-\mu\hat N)}\left|\psi_{k-1}\right>\right).
\end{split}
\end{equation}
Since $\hat H$ and $\hat N$ are normal ordered, the same is true for $e^{-\epsilon(\hat H-\mu\hat N)}$, if $\epsilon$ is small. Then we can use \eqref{eq:coher:expect} to calculate the expectation value
\begin{equation}
\left<\psi_k\right|e^{-\epsilon(\hat H-\mu\hat N)}\left|\psi_{k-1}\right>=e^{\sum_\alpha\psi_{\alpha,k}^*\psi_{\alpha,k-1}}e^{-\epsilon(H(\psi_{\alpha,k}^*,\psi_{\alpha,k-1})-\mu\sum_\alpha\psi_{\alpha,k}^*\psi_{\alpha,k-1})},
\end{equation}
where 
\begin{equation}
H(\psi_{\alpha,k}^*,\psi_{\alpha,k-1})=\hat H(a_\alpha^+\rightarrow\psi_{\alpha,k}^*,a_\alpha\rightarrow\psi_{\alpha,k-1}).
\end{equation}
Inserting this result in \eqref{eq:part:Z1} yields 
\begin{equation}
\begin{split}
Z&=\int\left(\prod_{k=1}^{M}\prod_{\alpha}d\psi_{\alpha,k}^*d\psi_{\alpha,k}\right)\\
&\quad\exp\left(-\epsilon\sum_{k=1}^M\left(\sum_\alpha\psi_{\alpha,k}^*\left(\frac{\psi_{\alpha,k}-\psi_{\alpha,k-1}}{\epsilon}-\mu\psi_{\alpha,k-1}\right)+H(\psi_{\alpha,k}^*,\psi_{\alpha,k-1})\right)\right).
\end{split}
\end{equation}
Taking the continuum limit, we get a functional integral expression for the partition function:
\begin{equation}
\begin{split}
Z&=\int_{\psi_\alpha(\beta)=-\psi_\alpha(0),\,\psi^*_\alpha(\beta)=-\psi^*_\alpha(0)}{\cal D}\psi_\alpha^*(\tau){\cal D}\psi_\alpha(\tau)\\
&\quad\exp\left(-\int_0^\beta d\tau\,\left(\sum_\alpha\psi_{\alpha}^*(\tau)\left(\frac{\partial}{\partial\tau}-\mu\right)\psi_{\alpha}(\tau)+H(\psi_{\alpha}^*(\tau),\psi_{\alpha}(\tau))\right)\right).
\end{split}
\end{equation}
Particularly for the Hubbard model, we have
\begin{equation}
\begin{split}
Z&=\int_{\psi_\alpha(\beta)=-\psi_\alpha(0),\,\psi^*_\alpha(\beta)=-\psi^*_\alpha(0)}{\cal D}\psi_\alpha^*(\tau){\cal D}\psi_\alpha(\tau)\\
&\quad\exp\biggl(-\int_0^\beta d\tau\,\biggl(\sum_{ij\sigma}\psi_{i\sigma}^*(\tau)\left(\frac{\partial}{\partial\tau}-\mu+{\cal T}\right)_{ij}\psi_{j\sigma}(\tau)\\
&\quad\qquad-\frac{1}{2}\sum_{i\sigma}\psi_{i\sigma}^*(\tau)\psi_{i(-\sigma)}^*(\tau)\psi_{i\sigma}(\tau)\psi_{i(-\sigma)}(\tau)\biggr)\biggr),
\end{split}
\end{equation}
where
\begin{equation}
\left(\frac{\partial}{\partial\tau}-\mu\right)_{ij}=\left(\frac{\partial}{\partial\tau}-\mu\right)\delta_{ij}.
\end{equation}
Note that the ``derivative'' $\frac{\partial}{\partial\tau}$ is a purely formal transcription of the discrete version 
\begin{equation}
\frac{\partial\psi_\alpha(\tau)}{\partial\tau}=\lim_{\epsilon\rightarrow0}\frac{\psi_{\alpha}(\tau)-\psi_{\alpha}(\tau-\epsilon)}{\epsilon}
\end{equation}
since the difference between $\psi_{\alpha,k}$ and $\psi_{\alpha,k-1}$ (which are Grassmann valued) is not ``small'' in any sense. When we have to actually calculate such a ``derivative'', we will return to the discrete version.

A remarkable feature of the path integral expression for the partition function is the antiperiodic boundary condition $\psi_\alpha(\beta)=-\psi_\alpha(0)$. If we trace back our steps, we see that the anti-periodicity is caused by the minus sign in $\langle-\psi|$ in eq. \eqref{eq:coher:trace} that followed from the fact that the $\psi_\alpha$ are Grassmann valued. Therefore the anti-periodicity is typical for fermionic systems. If we repeat all the steps in this chapter for a bosonic system, we would find periodic boundary conditions for the functional integral. The anti-periodicity has important implications for the properties of a fermionic system and we will discuss it in more detail in the last section of this chapter. 

Finally, we clean up our notation by defining spinors
\begin{equation}
\psi_{i}(\tau)=\begin{pmatrix}\psi_{i\uparrow}(\tau)\\\psi_{i\downarrow}(\tau)\end{pmatrix},
\quad\psi_{i}^\dagger(\tau)=\begin{pmatrix}\psi_{i\uparrow}^*(\tau),&\psi_{i\downarrow}^*(\tau)\end{pmatrix}.
\end{equation}
Using
\begin{equation} 
\sum_\sigma\psi_{i\sigma}^*(\tau)\psi_{i(-\sigma)}^*(\tau)\psi_{i\sigma}(\tau)\psi_{i(-\sigma)}(\tau)=-\psi_{i}^\dagger(\tau)\psi_{i}(\tau)\psi_{i}^\dagger(\tau)\psi_{i}(\tau),
\end{equation}
the partition function becomes
\begin{equation}
\begin{split}
Z&=\int_{\psi_\alpha(\beta)=-\psi_\alpha(0),\,\psi^*_\alpha(\beta)=-\psi^*_\alpha(0)}{\cal D}\psi_\alpha^*(\tau){\cal D}\psi_\alpha(\tau)\\
&\quad\exp\biggl(-\int_0^\beta d\tau\,\biggl(\sum_{ij}\psi_{i}^\dagger(\tau)\left(\frac{\partial}{\partial\tau}-\mu+{\cal T}\right)_{ij}\psi_{j}(\tau)\\
&\quad\qquad+\frac{1}{2}\sum_{i}\psi_{i}^\dagger(\tau)\psi_{i}(\tau)\psi_{i}^\dagger(\tau)\psi_{i}(\tau)\biggr)\biggr).
\end{split}
\end{equation}

\section{Matsubara sums}
The anti-periodicity conditions $\psi_\alpha(0)=-\psi_\alpha(\beta)$ and  $\psi_\alpha^*(0)=-\psi_\alpha^*(\beta)$ tell us that $\psi_\alpha(\tau)$ and $\psi_\alpha^*(\tau)$ may be expanded as a series 
\begin{equation}
\begin{aligned}
\psi_\alpha(\tau)&=\sum_{n=-\infty}^\infty T\psi_{\alpha,n}\exp(i(2n+1)\pi T\tau),\\
\psi_\alpha^*(\tau)&=\sum_{n=-\infty}^\infty T\psi_{\alpha,n}^*\exp(-i(2n+1)\pi T\tau)
\end{aligned}
\end{equation}
with $\tau$-independent coefficients, where $T=1/\beta$ is the temperature and the factor $T$ in front of the expansion coefficients is conventional. These sums are called {\em Matsubara sums} and one usually introduces the {\em Matsubara frequencies}
\begin{equation}
\omega_n^F=(2n+1)\pi T
\end{equation}
so that 
\begin{equation}
\begin{aligned}
\psi_\alpha(\tau)&=\sum_{n=-\infty}^\infty T\psi_{\alpha,n}\exp(i\omega_n^F\tau),\\
\psi_\alpha^*(\tau)&=\sum_{n=-\infty}^\infty T\psi_{\alpha,n}^*\exp(-i\omega_n^F\tau).
\end{aligned}
\end{equation}
The index $F$ for the Matsubara frequencies indicates that these are fermionic frequencies. In the bosonic case, we would have had {\em periodic} boundary conditions for the functional integral and a Matsubara sum of the same form as for fermions, but with $\omega_n^F$ replaced by 
\begin{equation}
\omega_n^B=2n\pi T.
\end{equation}
One remarkable difference between fermionic and bosonic system is that for $T>0$ we have $\omega_n^F>0\,\,\forall n$, which is not the case for $\omega_n^B$, since $\omega_n^B=0$ if $n=0$. We will use this positivity property when specifying a regularization scheme for fermionic propagators.

By using
\begin{equation}
\int_0^\beta d\tau\,\exp(-i(\omega_n^F-\omega_m^F)\tau)=\beta\delta_{nm}
\end{equation}
and
\begin{equation}
\begin{aligned}
\int_0^\beta\psi_\alpha^*(\tau)\frac{\partial}{\partial\tau}\psi_\alpha(\tau)
&=\lim_{\epsilon\rightarrow0}\int_0^\beta d\tau\,\psi_\alpha^*(\tau)\frac{\psi_\alpha(\tau)-\psi_\alpha(\tau-\epsilon)}{\epsilon}\\
&=\lim_{\epsilon\rightarrow0}\sum_{n,m=-\infty}^\infty T^2\psi_{\alpha,n}^*\psi_{\alpha,m}\int_0^\beta d\tau\,e^{-i(\omega_n^F-\omega_m^F)\tau}\,\frac{1-\exp(-i\omega_m^F\epsilon)}{\epsilon}\\
&=\sum_{n=-\infty}^\infty T\psi_{\alpha,n}^*i\omega_n^F\psi_{\alpha,n},
\end{aligned}
\end{equation}
we find for the partition function of the Hubbard model
\begin{equation}
\label{eq:mats:parhub}
\begin{split}
Z&=\int_{\psi_\alpha(\beta)=-\psi_\alpha(0),\,\psi^*_\alpha(\beta)=-\psi^*_\alpha(0)}{\cal D}\psi_\alpha^*(\tau){\cal D}\psi_\alpha(\tau)\\
&\quad\exp\biggl(-\biggl(\sum_{n=-\infty}^\infty T\sum_{ij}\psi_{in}^\dagger\left(i\omega_n^F-\mu+{\cal T}\right)_{ij}\psi_{jn}\\
&\quad\qquad+\frac{1}{2}\sum_{n_1\ldots n_4=-\infty}^\infty T^4\,(\beta\delta_{n_1+n_2,n_3+n_4})\,\sum_{i}\psi_{in_1}^\dagger\psi_{in_3}\psi_{in_2}^\dagger\psi_{in_4}\biggr)\biggr).
\end{split}
\end{equation}

One important feature of this expression is the behavior in the high and low temperature limit. First note that the Matsubara frequency acts as some kind of mass term in the fermionic propagator of this theory. The mass term is proportional to temperature. This means that in the high temperature limit modes with large $n$ are suppressed. The dynamics of the system is then dominated by the Matsubara modes with $n=0$ and $n=-1$, which yield $\omega_n^F=\pm\pi T$. For our model, it follows that the system is completely two dimensional. As opposed to this, in the low temperature limit the modes tend to form a continuum around the zero mode. In this case, the Matsubara sum can be approximately replaced by an integral. This renders the system effectively three dimensional. The dimensional decrease from $T=0$ to $T>0$ is called {\em dimensional reduction} and has an important impact on the qualitative properties of the system. For example, the Mermin-Wagner theorem shows that in two dimensions no antiferromagnetic long range order can exist. However, in three dimensions this is not true. Because of dimensional reduction, this means that for $T=0$ antiferromagnetic long range order is allowed in our two dimensional model, and indeed there are rigorous results proving the existence of an antiferromagnetic phase for zero temperature in the two dimensional Hubbard model \cite{tasaki}.   
\chapter{Partial bosonization}

The partition function of the Hubbard model describes a purely fermionic model. Information about spontaneously broken symmetries is encoded in the renormalization group flow of the quartic couplings. In principle it is possible to extract this information by analyzing the size and momentum structure of these quartic couplings. However, the momentum dependence not only reflects interesting degrees of freedom, but also arises from complicated short range fluctuations, which we would like to ignore in a simple truncation scheme for solving the renormalization group equations. The aim of this chapter is to explicitly extract interesting quartic terms and the relevant momentum structure of their couplings by artificially introducing bosonic ``particles'' corresponding to them. We call this new model the ``colored Hubbard model''. The momentum dependence of the couplings in this new theory no longer contains essential information about the long range behavior of the system, and we will neglect this momentum dependence in calculations. We will introduce these approximations later on; in this chapter everything is exact, and the ``colored Hubbard model'' as introduced here is an equivalent transcription of the original Hubbard model.

\section{The Fermi surface}
\label{sec:fermisur}
\begin{figure}
\centering
\includegraphics[height=9cm,width=9cm]{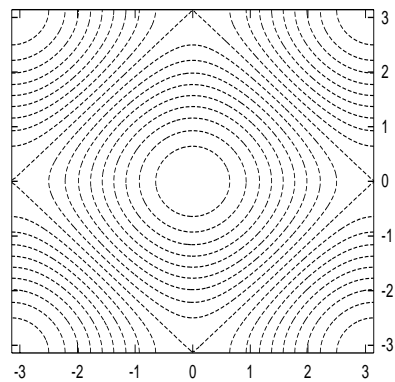}
\caption[The Fermi surface in the Hubbard model]{Equipotential lines of the function $-\cos q_1-\cos q_2$. The lines correspond to states of the same energy. This plot may be read as a plot of the Fermi energies for different values of the chemical potential.}
\label{fig:fermione}
\end{figure}

To gain a better understanding of the terms appearing in the partition function of the Hubbard model, consider the theory with no Coulomb interaction and at $T\ll1$\footnote{It is perfectly possible to actually set $T=0$. In this case the Matsubara sum in \eqref{eq:fermisurS} takes the form of an integral. However, we will never need this transcription explicitly, so that we do not bother to write it down here. For our purpose, it suffices to think of $T$ to be infinitesimally small.}. In this case, the action of the simplified model can be read off from \eqref{eq:mats:parhub}
\begin{equation}
\label{eq:fermisurS}
S\approx\sum_{n=-\infty}^\infty T\sum_{\boldsymbol{x}\boldsymbol{y}}\psi_{\boldsymbol{x}n}^\dagger\left(-\mu+{\cal T}\right)_{\boldsymbol{x}\boldsymbol{y}}\psi_{\boldsymbol{y}n},
\end{equation}
where we neglected $i\omega_n^F$ which is $\propto T$. We replaced the abstract index $i$ for the lattice sites by a two dimensional vector $\boldsymbol{x}$ labeling the lattice sites by two integer values for the two spatial directions. By Fourier transforming the fields 
\begin{equation}
\begin{aligned}
\psi_{\boldsymbol{x}n}&=\int_{-\pi}^\pi\frac{d^2 q}{(2\pi)^2}\psi_n(\boldsymbol{q})\exp(i\boldsymbol{x}\boldsymbol{q}),\\
\psi_{\boldsymbol{x}n}^\dagger&=\int_{-\pi}^\pi\frac{d^2 q}{(2\pi)^2}\psi_n^\dagger(\boldsymbol{q})\exp(-i\boldsymbol{x}\boldsymbol{q})
\end{aligned}
\end{equation}
we find
\begin{equation}
\label{eq:fermi1:S}
S=\sum_{n=-\infty}^\infty T\int_{-\pi}^\pi\frac{d^2 q}{(2\pi)^2}\psi_{n}^\dagger(\boldsymbol{q})\left(-\mu-2t(\cos q_1+\cos q_2)\right)\psi_n(\boldsymbol{q}).
\end{equation}
For $T\ll1$, we know that the chemical potential is equal to the Fermi energy $\mu=E_F$ and all energy states are filled up to this energy. The second term in \eqref{eq:fermi1:S} describes the one particle energies $E(\boldsymbol{q})$ (note that the model has a simple one band structure). We conclude that an electron state with energy $E(\boldsymbol{q})$ is occupied, if $-2t(\cos q_1+\cos q_2)<\mu$. A contour plot of $E(\boldsymbol{q})$ is given in fig. (\ref{fig:fermione}).
The equipotential line of quadratic shape corresponds to $-2t(\cos q_1+\cos q_2)=0$. As is evident from the plot, this is nothing else than the half filling case. We therefore identify $\mu=0$ with half filling. $\mu\neq0$ describes the doping of the system away from the undoped half filling state.

\section{The colored Hubbard model}
\label{sec:colhub}
The starting point of the transcription will be 
\begin{equation}
\begin{split}
Z&=\int_{\hat\psi_\alpha(\beta)=-\hat\psi_\alpha(0),\,\hat\psi^*_\alpha(\beta)=-\hat\psi^*_\alpha(0)}{\cal D}\hat\psi_\alpha^*(\tau){\cal D}\hat\psi_\alpha(\tau)\exp\left(-S_F-S_{coup}-S_j\right)
\end{split}
\end{equation}
with
\begin{equation}
\begin{aligned}
S_F&=\sum_{n=-\infty}^\infty T\sum_{\boldsymbol{x}\boldsymbol{y}}\hat\psi_{\boldsymbol{x}n}^\dagger\left(i\omega_n^F+{\cal T}\right)_{\boldsymbol{x}\boldsymbol{y}}\hat\psi_{\boldsymbol{y}n},\\
S_{coup}&=\frac{1}{2}\sum_{n_1\ldots n_4=-\infty}^\infty T^4\,(\beta\delta_{n_1+n_2,n_3+n_4})\,\sum_{\boldsymbol{x}}\hat\psi_{\boldsymbol{x}n_1}^\dagger\hat\psi_{\boldsymbol{x}n_3}\hat\psi_{\boldsymbol{x}n_2}^\dagger\hat\psi_{\boldsymbol{x}n_4}.
\end{aligned}
\end{equation}
We have introduced a ``hat'' $\,\hat{}\,$ to indicate fields. Symbols without $\,\hat{}\,$ will denote expectation values of these fields. Additionally, we have introduced a source term $S_j$ for fermion fields and fermion bilinears to be able to use $Z$ as a generating functional. We will specify $S_j$ later when we need it. Also note that the term involving the chemical potential (which has the form of a source term) is now included in $S_j$, serving as a source for the charge density. It is quite natural to do so, since we know that the charge density is essentially controlled by doping, rendering it a quantity controlled by external conditions on the system.
\begin{figure}
\centering
\psfrag{n1}{$n_1$}
\psfrag{n2}{$n_2$}
\includegraphics{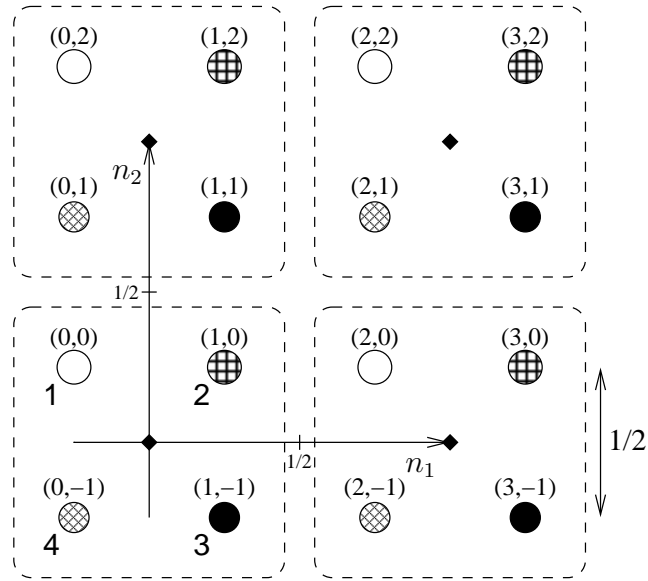}
\caption[Site labels in the colored Hubbard model]{Labeling of the sites in the colored Hubbard model. The labels of the original model are given in parentheses. The coarse lattice sites of the colored model are indicated by a $\blacklozenge$.}
\label{fig:colhub}
\end{figure}
The first step in the bosonization procedure is to realize that the most interesting degrees of freedom of the Hubbard model have to be implemented {\em non locally}. This means that if we want to decide whether a system exhibits e.g. antiferromagnetism, we have to compare electron spins at {\em different} lattice sites. In the same way, we are not able to decide whether a system exhibits $s$- or $d$-wave superconductivity, if we do not take into account the relative sign of electron pair expectation values at different lattice site pairs. The idea to deal with this complication is to introduce a coarse lattice which consists of plaquettes. Each plaquette contains four sites of the original lattice (cf. fig. \ref{fig:colhub}). The plaquettes --- or equivalently the lattice sites of the coarse lattice --- are labeled by a two dimensional vector $\boldsymbol n$, which takes integer values. The lattice sites belonging to a given plaquette are numbered clockwise. We will call these four labels {\em colors}. Instead of ${\hat\psi}_{{\boldsymbol x}n}$, where $ {\boldsymbol x}$ labels the sites of the original lattice, we have now ${\hat\psi}_{{\boldsymbol n}an}$, where ${\boldsymbol n}$ labels the sites of the coarse lattice, $a$ is the color label with $a\in\{1,2,3,4\}$ and in both cases $n$ is the Matsubara mode. Explicitly, we have
\begin{equation}
\begin{aligned}
{\hat\psi}_{{\boldsymbol n}1n}&={\hat\psi}_{(x_1,x_2)n}, & {\hat\psi}_{{\boldsymbol n}2n}&={\hat\psi}_{(x_1+1,x_2)n},\\
{\hat\psi}_{{\boldsymbol n}4n}&={\hat\psi}_{(x_1,x_2-1)n}, & {\hat\psi}_{{\boldsymbol n}3n}&={\hat\psi}_{(x_1+1,x_2-1)n}
\end{aligned}
\end{equation}
with $\boldsymbol{n}=\boldsymbol{x}/2$, $\boldsymbol{x}=(x_1,x_2)$.

The advantage of this transcription is that we can now write down fermion bilinears describing antiferromagnetic or superconducting behavior that are {\em local} on the coarse lattice. Before we do so, we repeat the discussion of sec. \ref{sec:fermisur} in this new language.

\subsection{Fourier transforms}
In a first step, we define Fourier transforms of the spinors as we did in sec. \ref{sec:fermisur}. The naive way is to simply define
\begin{equation}
\begin{aligned}
{\hat\psi}_{\boldsymbol{n}an}&=\int_{-\pi}^\pi\frac{d^2 q}{(2\pi)^2}{\hat\psi}_{an}(\boldsymbol{q})\exp(i\boldsymbol{n}\boldsymbol{q}),\\
{\hat\psi}_{\boldsymbol{n}an}^\dagger&=\int_{-\pi}^\pi\frac{d^2 q}{(2\pi)^2}{\hat\psi}_{an}^\dagger(\boldsymbol{q})\exp(-i\boldsymbol{n}\boldsymbol{q})
\end{aligned}
\end{equation}
in much the same way as before. However, in this case phase factors arise in the Fourier transformed expressions in the action, since we neglected the spatial dependence encoded in the color index in these transforms. The elegant way to perform the Fourier transforms is to define
\begin{equation}
\begin{aligned}
{\hat\psi}_{\boldsymbol{n}an}&=\int_{-\pi}^\pi\frac{d^2 q}{(2\pi)^2}{\hat\psi}_{an}(\boldsymbol{q})\exp(i(\boldsymbol{n}+\boldsymbol{z}_a)\boldsymbol{q}),\\
{\hat\psi}_{\boldsymbol{n}an}^\dagger&=\int_{-\pi}^\pi\frac{d^2 q}{(2\pi)^2}{\hat\psi}_{an}^\dagger(\boldsymbol{q})\exp(-i(\boldsymbol{n}+\boldsymbol{z}_a)\boldsymbol{q})
\end{aligned}
\end{equation}
with
\begin{equation}
\begin{aligned}
\boldsymbol{z}_1&=(-1/4,1/4) & \boldsymbol{z}_2&=(1/4,1/4)\\
\boldsymbol{z}_4&=(-1/4,-1/4) & \boldsymbol{z}_3&=(1/4,-1/4).
\end{aligned}
\end{equation}
Then no phase factors arise in the Fourier transformed expressions of the action. Note that the phase factors in this definition are no longer periodic in $2\pi$, which means that the same is true for $\hat\psi_{an}(\boldsymbol{q})$, since the integrand must be periodic in $2\pi$ as a whole.

This is a good place to clean up our notation with regard to Fourier transforms. We define
\begin{equation}
\begin{gathered}
Q=(\omega_n,\boldsymbol{q}),\quad X=(\tau,\boldsymbol{n}),\\
QX=\omega_n\tau+\boldsymbol{n}\boldsymbol{q},\\
\sum_X=\int_0^\beta d\tau\,\sum_{\boldsymbol{n}},\quad\sum_Q=T\sum_{n=-\infty}^\infty\int_{-\pi}^\pi\frac{d^2q}{(2\pi)^2},\\
\delta(Q-Q')=\beta\delta_{n,n'}(2\pi)^2\delta(\boldsymbol{q}-\boldsymbol{q'}),\\
\delta(X-X')=\delta(\tau-\tau')\delta_{\boldsymbol{n},\boldsymbol{n'}}.
\end{gathered}
\end{equation}
Note that $\delta(\boldsymbol{q}-\boldsymbol{q'})$ is periodic in $2\pi$ and that $\delta(\tau)$ is periodic in $\beta$ for bosons and antiperiodic for fermions. The definitions hold for $\omega_n$ in the bosonic ($\omega_n=\omega_n^B$) as well as in the fermionic ($\omega_n=\omega_n^F$) case. We can now write $\hat\psi_{\boldsymbol{n}an}=\hat\psi_a(X)$, $\hat\psi_{an}(\boldsymbol{q})=\hat\psi_a(Q)$ and similarly for $\hat\psi^*$. 

With these abbreviations, the complete Fourier transforms read
\begin{equation}
\label{eq:four:four}
\begin{aligned}
{\hat\psi}_a(X)&=\sum_Q{\hat\psi}_a(Q)\exp(i(QX+\boldsymbol{z}_a\boldsymbol{q})),\\
{\hat\psi}_a^\dagger(X)&=\sum_Q{\hat\psi}_a^\dagger(Q)\exp(-i(QX+\boldsymbol{z}_a\boldsymbol{q})).
\end{aligned}
\end{equation}

Furthermore, we will often use the notation $\hat\psi(Q)$ for the vector with components $\hat\psi_a(Q)$. Note that since the $\hat\psi_a(Q)$ themselves are two dimensional spinors, this means that the objects $\hat\psi(Q)$ live in the product space of spin and color and have $8$ components altogether.

Another notation we will use concerns infinite sums like $\sum_X1$. Note that (mathematicians hopefully forgive this)
\begin{equation}
\sum_X=\sum_Q\sum_X\delta(Q)\exp(iQX)=\sum_Q\delta(Q)\delta(Q)=\delta(Q=0).
\end{equation}
We define
\begin{equation}
{\cal V}=\sum_X=\delta(Q=0),
\end{equation}
which can be interpreted as the two dimensional volume of the system (that we assume to be large) divided by temperature.

\subsection{The Fermi surface}
We can now repeat the calculation of sec. \ref{sec:fermisur}, taking $U=0$, $T\ll1$ and all sources except $\mu$ equal to zero, so that we end up with
\begin{equation}
\begin{aligned}
\label{eq:fermi4:S}
S&=S_F+S_{coup}+S_j\\
&=\sum_Q\hat\psi^\dagger(Q)\left(-\mu-2t\left(\cos(q_1/2)A_1+\cos(q_2/2)B_1\right)\right)\hat\psi(Q).
\end{aligned}
\end{equation}
$A_1$ and $B_1$ are $4\times4$-matrices in color space and are defined in the appendix \ref{sec:matrices}. This result should be compared to \eqref{eq:fermi1:S}. The first difference to be observed is that the cosines are no longer periodic in $2\pi$, but in $4\pi$, which is a direct consequence of our Fourier transform. This means that all the cosines are positive in the interval of integration.

To see how the Fermi surface emerges in this picture, we temporarily switch to fields $\hat\Psi(Q)$, for which the fermionic propagator term in \eqref{eq:fermi4:S} becomes diagonal. For these fields, the action reads
\begin{equation}
S=\sum_Q\hat\Psi^\dagger(Q)D\hat\Psi(Q)
\end{equation}
with
\begin{equation}
D=\begin{pmatrix}-\mu+2t(c_1+c_2)&0&0&0\\0&-\mu+2t(c_1-c_2)&0&0\\0&0&-\mu-2t(c_1-c_2)&0\\0&0&0&-\mu-2t(c_1+c_2)\end{pmatrix},
\end{equation}
\begin{equation}
c_i=\cos(q_i/2).\nonumber
\end{equation}
\begin{figure}
\centering
\psfrag{pip}{$\pi$}
\psfrag{pim}{$-\pi$}
\psfrag{q1}{$q_1$}
\psfrag{q2}{$q_2$}
\includegraphics{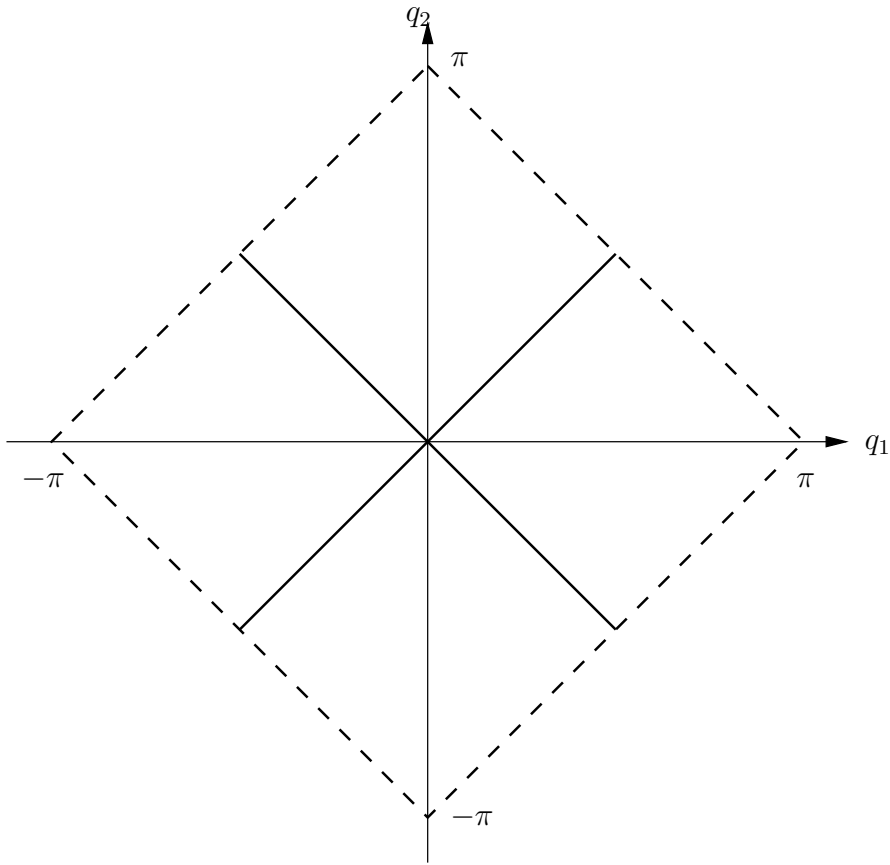}
\caption[The Fermi surface in the colored Hubbard model]{The Fermi surface for half filling in the Hubbard (dashed line) and the colored Hubbard model (full line).}
\label{fig:fermi}
\end{figure}
We can immediately read off the (cross shaped) Fermi surface for half filling from this result (fig. \ref{fig:fermi}).

The reason why we are so interested in the shape of the Fermi surface is the following. Note that the matrix $D$ is nothing else than the propagator matrix of four distinct fermion modes (keep in mind that the chemical potential will be absorbed in a source term and is not regarded as part of the propagator). Zeroes of the propagator are a well known problem in any quantum field theory calculation for massless particles, since they lead to divergencies. These zeroes appear in our case on the Fermi surface. The usual way to deal with the divergencies is to define a regularization scheme. This is not that difficult in simple theories where the propagator (in a Euclidean formulation) is proportional to the squared momentum --- any positive mass like term added to the propagator will cure the divergency problem. However, given the complicated (cross shaped) momentum structure of the Fermi surface we face in our formalism, momentum cutoffs become tedious to define and to work with. We will pursue a different way by noting that if we do not neglect $i\omega_n^F$ in the action above, we end up with a propagator of the form $i\omega_n^F+D$, which does not vanish for all $T>0$. We exploit this fact by using a regularization scheme that uses temperature as a flowing cutoff function. By defining some (unphysical) temperature $T_k$ as a function of a parameter $k$, $\lim_{k\rightarrow\infty}T_k=\infty$, $\lim_{k\rightarrow0}T_k=T$, we can lower $T_k$ starting with some large $k$ and letting $k\rightarrow0$ in a controlled way until we reach the physical temperature $T$ of the system. The flow of the system with $k$ will be described by exact renormalization group equations. We will come to this later.  

Another aspect concerns the motivation of approximations. For the bosonic propagators (introduced in the next chapter), we will be able to expand trigonometric functions to quadratic order. This is very convenient, since one achieves formal agreement with known theories (e.g. the propagator of a boson $2(2-\cos(q_1)-\cos(q_2))$ becomes $\approx\boldsymbol{q}^2$ in quadratic order). However, it is not possible, even in principle, to expand the trigonometric functions in the fermionic case for low temperature without loosing significant information, since the dynamics is dominated by modes with energy close to the Fermi surface, not just by modes with zero momentum. The main complications of calculations we are about to attack are that we are forced to keep these trigonometric functions in the fermionic sector.  

\subsection{Symmetries}
\label{sec:origsymm}
In this section we discuss the various symmetries of the colored Hubbard model. 

As already discussed in the introduction, we have the {\em $U(1)$-symmetry}
\begin{equation}
\label{eq:symm:U1}
\psi(X)\rightarrow\exp(i\theta)\psi(X),\quad\psi^\dagger(X)\rightarrow\psi^\dagger(X)\exp(-i\theta)
\end{equation}
and the {\em $SU(2)$-symmetry}
\begin{equation}
\psi(X)\rightarrow\exp(i\vec{\sigma}\vec{\theta})\psi(X),\quad\psi^\dagger(X)\rightarrow\psi^\dagger(X)\exp(-i\vec{\sigma}\vec{\theta}).
\end{equation}
Note that now we only consider global $U(1)$-transformations, since no gauge bosons are present in our theory (the reason being that these are ``integrated out'' under the assumption of negligible interaction of electrons at different lattice sites, giving rise to the Hubbard model as an effective purely fermionic model).

Furthermore, the model possesses the symmetries of the underlying lattice. These may be composed from translations, rotations and reflections. We restrict ourselves to the translation $T_x$ by one lattice site in the positive $1$-direction, the counterclockwise rotation $R$ by $90^\circ$ around the origin (at the center of a plaquette!) and the reflection $I$ at the the $2$-axis containing the origin. All other lattice symmetries can be built up by products of these three symmetries (for example, the translation $T_y$ along the $2$-direction can be composed by a translation $T_x$ and a rotation $R$).

In position space, these symmetries act as
\begin{equation}
\label{eq:symm:symm}
\begin{aligned}
T_x&:\left\{\begin{aligned}\hat\psi_{(n_1,n_2)1n}&\rightarrow\hat\psi_{(n_1,n_2)2n}\\
                        \hat\psi_{(n_1,n_2)2n}&\rightarrow\hat\psi_{(n_1+1,n_2)1n}\\
                        \hat\psi_{(n_1,n_2)3n}&\rightarrow\hat\psi_{(n_1+1,n_2)4n}\\
                        \hat\psi_{(n_1,n_2)4n}&\rightarrow\hat\psi_{(n_1,n_2)3n},
\end{aligned}\right.\\
R&:\left\{\begin{aligned}\hat\psi_{(n_1,n_2)1n}&\rightarrow\hat\psi_{(-n_2,n_1)4n}\\
                        \hat\psi_{(n_1,n_2)2n}&\rightarrow\hat\psi_{(-n_2,n_1)1n}\\
                        \hat\psi_{(n_1,n_2)3n}&\rightarrow\hat\psi_{(-n_2,n_1)2n}\\
                        \hat\psi_{(n_1,n_2)4n}&\rightarrow\hat\psi_{(-n_2,n_1)3n},
\end{aligned}\right.\\
I&:\left\{\begin{aligned}\hat\psi_{(n_1,n_2)1n}&\rightarrow\hat\psi_{(-n_1,n_2)2n}\\
                        \hat\psi_{(n_1,n_2)2n}&\rightarrow\hat\psi_{(-n_1,n_2)1n}\\
                        \hat\psi_{(n_1,n_2)3n}&\rightarrow\hat\psi_{(-n_1,n_2)4n}\\
                        \hat\psi_{(n_1,n_2)4n}&\rightarrow\hat\psi_{(-n_1,n_2)3n},
\end{aligned}\right.
\end{aligned}
\end{equation} 
The same applies for $\hat\psi^*$. Fourier transforming yields
\begin{equation}
\begin{aligned}
T_x&:\left\{\begin{aligned}\hat\psi(Q)&\rightarrow A_1\hat\psi(Q)\exp(iq_1/2)\\\hat\psi^\dagger(Q)&\rightarrow\hat\psi^\dagger(Q)A_1\exp(-iq_1/2)\end{aligned}\right.,\\
R&:\left\{\begin{aligned}\hat\psi_n(q_1,q_2)&\rightarrow\begin{pmatrix}0&0&0&1\\1&0&0&0\\0&1&0&0\\0&0&1&0\end{pmatrix}\hat\psi_n(-q_2,q_1)\\
\hat\psi^\dagger_n(q_1,q_2)&\rightarrow\hat\psi^\dagger_n(-q_2,q_1)\begin{pmatrix}0&1&0&0\\0&0&1&0\\0&0&0&1\\1&0&0&0\end{pmatrix}\end{aligned}\right.,\\
I&:\left\{\begin{aligned}\hat\psi_n(q_1,q_2)&\rightarrow A_1\hat\psi_n(-q_1,q_2)\\\hat\psi^\dagger_n(q_1,q_2)&\rightarrow\hat\psi^\dagger_n(-q_1,q_2)A_1\end{aligned}\right..\end{aligned}
\end{equation}
Again $A_1$ is one of the matrices defined in the appendix \ref{sec:matrices}.

It is interesting to consider a transformation like 
\begin{equation}
\label{eq:relabelingsym}
L_{A_1}:\left\{\begin{aligned}\hat\psi(Q)&\rightarrow A_1\hat\psi(Q)\\\hat\psi^\dagger(Q)&\rightarrow\hat\psi^\dagger(Q)A_1\end{aligned}\right.
\end{equation}
which is also a symmetry of the action. This transformation does not possess an interpretation in position space by means of lattice symmetries as one would expect. Instead, it reflects our freedom in choosing how to label the sites. For example, formally translating this transformation back to position space, we end up with e.g. $\psi_{(n_1,n_2)1n}\rightarrow\psi_{(n_1-1/2,n_2)2n}$. We may read this as the identity transformation, written by merely choosing another origin of the coarse lattice, shifted one lattice site of the original lattice to the left. This class of symmetries is therefore new in the colored formulation and corresponds to the unity transformation in the original theory. We call this kind of symmetry {\em relabeling symmetry}. In the same way we can define the relabeling symmetry transformations $L_{A_0}$, $L_{B_0}$ and $L_{B_1}$, which are derived from $L_{A_1}$ by replacing the matrix $A_1$ by $A_0$, $B_0$ and $B_1$ respectively. These four relabeling symmetries correspond to the four different ways to assign color labels by choosing a different origin of the coarse lattice. As in the case of $L_{A_1}$ we see that by multiplying with appropriate momentum phase factors these additional relabeling symmetries correspond to a translation in the $2$-direction for $L_{B_1}$, a translation along the diagonal connecting color sites $1$ and $3$ for $L_{B_0}$ and no translation at all for $L_{A_0}$. 

The last symmetries we mention are reminiscent of time reversal symmetries. They are realized by
\begin{equation}
\label{eq:timereversal}
\hat\psi_n(\boldsymbol{q})\to M_i\hat\psi_{-n}(\boldsymbol{q}),\quad\hat\psi_n^*(\boldsymbol{q})\to-M_i\hat\psi_{-n}^*(\boldsymbol{q}),\quad\mu\to-\mu,
\end{equation}
where $M_i\in\{A_2,B_2,B_3\}$ and are denoted by $T_{A_2}$, $T_{B_2}$ and $T_{B_3}$.

\section{Partial bosonization}

\subsection{Definitions of fermion bilinears }
Using the color notation, we are now able to define fermion bilinears corresponding to interesting degrees of freedom in a simple way. In particular, we want to include bilinears describing the charge density, antiferromagnetic order as well as $s$- and $d$-wave superconductivity. After defining these bilinears, we try to decompose the four fermion action of the colored Hubbard model with respect to these bilinears. We see that this decomposition is not possible until a whole set of additional bilinears is added. We define this set in this section and discuss the decomposition of the four fermion interaction of the Hubbard model with respect to this set in the next section.

We start by defining
\begin{equation}
\begin{aligned}
\tilde\sigma_{ab}(X)&=\hat\psi^\dagger_b(X)\hat\psi_a(X)\\
\vec{\tilde\varphi}_{ab}(X)&=\psi_b^\dagger(X)\vec\sigma\hat\psi_a(X)\\
\tilde\chi_{ab}(X)&=\hat\psi^T_b(X)i\sigma_2\hat\psi_a(X)\\
\tilde\chi_{ab}^*(X)&=-\hat\psi_b^\dagger(X)i\sigma_2\hat\psi_a^*(X).
\end{aligned}
\end{equation}
The operators $\tilde\sigma_{ab}(X)$ are uncharged spin singlet operators, $\vec{\tilde\varphi}_{ab}(X)$ uncharged spin triplet operators and $\tilde\chi_{ab}(X)$, $\tilde\chi_{ab}^*(X)$ charged spin singlet operators. Note that $\tilde\chi_{ab}(X)=\tilde\chi_{ba}(X)$ and $\tilde\chi_{ab}^*(X)=\tilde\chi_{ba}^*(X)$. For sake of simplicity, we do not take into account charged operators in the spin triplet. 

Suppressing the $X$-dependence, we now define the composite bilinears
\begin{equation}
\label{eq:defbil:defbil}
\begin{aligned}
\tilde\rho&=\tilde\sigma_{11}+\tilde\sigma_{22}+\tilde\sigma_{33}+\tilde\sigma_{44}
&\qquad\vec{\tilde m}&=\vec{\tilde\varphi}_{11}+\vec{\tilde\varphi}_{22}+\vec{\tilde\varphi}_{33}+\vec{\tilde\varphi}_{44}\\
\tilde p&=\tilde\sigma_{11}-\tilde\sigma_{22}+\tilde\sigma_{33}-\tilde\sigma_{44}
&\qquad\vec{\tilde a}&=\vec{\tilde\varphi}_{11}-\vec{\tilde\varphi}_{22}+\vec{\tilde\varphi}_{33}-\vec{\tilde\varphi}_{44}\\
\tilde{q}_y&=\tilde\sigma_{11}+\tilde\sigma_{22}-\tilde\sigma_{33}-\tilde\sigma_{44}
&\qquad\vec{\tilde g}_y&=\vec{\tilde\varphi}_{11}+\vec{\tilde\varphi}_{22}-\vec{\tilde\varphi}_{33}-\vec{\tilde\varphi}_{44}\\
\tilde{q}_x&=\tilde\sigma_{11}-\tilde\sigma_{22}-\tilde\sigma_{33}+\tilde\sigma_{44}
&\qquad\vec{\tilde g}_x&=\vec{\tilde\varphi}_{11}-\vec{\tilde\varphi}_{22}-\vec{\tilde\varphi}_{33}+\vec{\tilde\varphi}_{44}\\ \\
\tilde{s}&=\tilde\chi_{11}+\tilde\chi_{22}+\tilde\chi_{33}+\tilde\chi_{44}
&\qquad\tilde e&=\tilde\chi_{12}+\tilde\chi_{23}+\tilde\chi_{34}+\tilde\chi_{41}\\
\tilde{c}&=\tilde\chi_{11}-\tilde\chi_{22}+\tilde\chi_{33}-\tilde\chi_{44}
&\qquad\tilde d&=\tilde\chi_{12}-\tilde\chi_{23}+\tilde\chi_{34}-\tilde\chi_{41}\\
\tilde{t}_y&=\tilde\chi_{11}+\tilde\chi_{22}-\tilde\chi_{33}-\tilde\chi_{44}
&\qquad\tilde{v}_y&=\tilde\chi_{12}-\tilde\chi_{34}\\
\tilde{t}_x&=\tilde\chi_{11}-\tilde\chi_{22}-\tilde\chi_{33}+\tilde\chi_{44}
&\qquad\tilde{v}_x&=\tilde\chi_{23}-\tilde\chi_{41}\\ \\
\tilde{s}^*&=\tilde\chi_{11}^*+\tilde\chi_{22}^*+\tilde\chi_{33}^*+\tilde\chi_{44}^*
&\qquad\tilde e^*&=\tilde\chi_{12}^*+\tilde\chi_{23}^*+\tilde\chi_{34}^*+\tilde\chi_{41}^*\\
\tilde{c}^*&=\tilde\chi_{11}^*-\tilde\chi_{22}^*+\tilde\chi_{33}^*-\tilde\chi_{44}^*
&\qquad\tilde d^*&=\tilde\chi_{12}^*-\tilde\chi_{23}^*+\tilde\chi_{34}^*-\tilde\chi_{41}^*\\
\tilde{t}_y^*&=\tilde\chi_{11}^*+\tilde\chi_{22}^*-\tilde\chi_{33}^*-\tilde\chi_{44}^*
&\qquad\tilde{v}_y^*&=\tilde\chi_{12}^*-\tilde\chi_{34}^*\\
\tilde{t}_x^*&=\tilde\chi_{11}^*-\tilde\chi_{22}^*-\tilde\chi_{33}^*+\tilde\chi_{44}^*
&\qquad\tilde{v}_x^*&=\tilde\chi_{23}^*-\tilde\chi_{41}^*.
\end{aligned}
\end{equation}

The bilinears that we wanted to include in  our formalism are the charge density $\tilde\rho$, the antiferromagnetic spin density $\vec{\tilde a}$, the superconducting $s$-wave $\tilde s$ and $d_{x^2-y^2}$-wave $\tilde d$. In order to make the before mentioned decomposition of the four fermion interaction into these bilinears possible, the rest of the bilinears has to be additionally included. This set is minimal and cannot be further reduced. However, by concentrating on $\tilde\rho$, $\vec{\tilde a}$ and $\tilde d$ (that is, dropping $\tilde s$ from the list of bilinears we want to include), we may reduce this set by dropping $\tilde s$, $\tilde c$ and $\tilde{t}_{x/y}$. 

Many of the bilinears we considered only in order to be able to perform the decomposition in the next section have a simple physical interpretation. For example $\vec{\tilde m}$ describes the ferromagnetic spin density, and the various charged composite bilinears correspond to $d_{xy}$-waves ($\tilde c$), extended $s$-wave ($\tilde e$) or $p$-waves ($\tilde v_{x/y}$).

\subsection{Decomposition of the four fermion interaction}
The four fermion interaction in the colored Hubbard model reads
\begin{equation}
\begin{aligned}
S_{coup}&=\frac{1}{2}\sum_X\sum_a\hat\psi^\dagger_a(X)\hat\psi_a(X)\hat\psi^\dagger_a(X)\hat\psi_a(X)\\
&=\frac{1}{2}\sum_X\sum_a\tilde\sigma_{aa}(X)^2.
\end{aligned}
\end{equation}
$S_{coup}$ may be decomposed into our fermion bilinears by use of the identities
\begin{equation}
\label{eq:decom:decom}
\begin{aligned}
4\sum_a\tilde\sigma_{aa}^2&=\tilde\rho^2+\tilde p^2+\tilde q_x^2+\tilde q_y^2\\
-12\sum_a\tilde\sigma_{aa}^2&=\vec{\tilde a}^2+\vec{\tilde m}^2+\vec{\tilde g}_x^2+\vec{\tilde g}_y^2\\
8\sum_a\tilde\sigma_{aa}^2&=\tilde s^*\tilde s+\tilde c^*\tilde c+\tilde t_x^*\tilde t_x+\tilde t_y^*\tilde t_y\\
0&=\frac{1}{2}(-\tilde\rho^2+\tilde p^2+\vec{\tilde m}^2-\vec{\tilde a}^2)+\tilde d^*\tilde d+\tilde e^*\tilde e+2(\tilde v_x^*\tilde v_x+\tilde v_y^*\tilde v_y).
\end{aligned}
\end{equation}
To prove this, note that
\begin{equation}
\begin{aligned}
\tilde\chi^*_{ab}\tilde\chi_{cd}&=\tilde\sigma_{ca}\tilde\sigma_{db}+\tilde\sigma_{cb}\tilde\sigma_{da}\\
\vec{\tilde\varphi}_{ab}\vec{\tilde\varphi}_{cd}&=-\tilde\sigma_{ab}\tilde\sigma_{cd}-2\tilde\sigma_{ad}\tilde\sigma_{cb}.
\end{aligned}
\end{equation}
The clue of \eqref{eq:decom:decom} is that this list of possibilities to write down combinations of fermion bilinears to either give a multiple of $\sum_a\tilde\sigma_{aa}^2$ or $0$ is {\em exhaustive}. No other independent combinations of fermion bilinears can be found to give $\sum_a\tilde\sigma_{aa}^2$ or $0$. 

\subsection{The partial bosonization}
Consider the fermionic partition function of the colored Hubbard model
\begin{equation}
\label{eq:parbos:fermpar}
\begin{aligned}
Z&=\exp\left(-\frac{2\pi^2{\cal V}\mu^2}{h_\rho^2}\right)\int{\cal D}\hat\psi^*{\cal D}\hat\psi\,\exp(-S_F-S_{coup}-S_j)\\
S_F&=\sum_Q\hat\psi^\dagger(Q)\left(i\omega_n^F-2t\left(\cos(q_1/2)A_1+\cos(q_2/2)B_1\right)\right)\hat\psi(Q)\\
S_{coup}&=\frac{1}{2}\sum_X\sum_a\hat\psi^\dagger_a(X)\hat\psi_a(X)\hat\psi^\dagger_a(X)\hat\psi_a(X)=\frac{1}{2}\sum_X\sum_a\tilde\sigma_{aa}(X)^2\\
S_j&=-\sum_X\biggl(\sum_\beta\left(j_\beta^*(X)\tilde u_\beta(X)+j_\beta(X)\tilde u_\beta^*(X)+\frac{4\pi^2}{h_\beta^2}j_\beta^*(X)j_\beta(X)\right)\\
&\quad+\sum_\gamma\left(l_\gamma(X)\tilde w_\gamma(X)+\frac{2\pi^2}{h_\gamma^2}l_\gamma(X)^2\right)\biggr)+S_j^F.
\end{aligned}
\end{equation}
We already know the kinetic and the coupling term. Additionally, now we have specified an explicit form of the source term. In this term, we included sources 
\begin{equation}
\begin{aligned}
l_\gamma&\in\{l_\rho=l_\rho'+\mu,l_p,l_{q_{x/y}},l_{\vec m},l_{\vec a},l_{\vec{g}_{x/y}}\}\\
j_\beta&\in\{j_s,j_c,j_{t_{x/y}},j_e,j_d,j_{v_{x/y}}\}\\
j_\beta^*&\in\{j_s^*,j_c^*,j_{t_{x/y}}^*,j_e^*,j_d^*,j_{v_{x/y}}^*\}
\end{aligned}
\end{equation}
for all the bilinears introduced in \eqref{eq:defbil:defbil} with
\begin{equation}
\begin{aligned}
\tilde w_\gamma&\in\{\tilde\rho,\tilde p,\tilde q_{x/y},\vec{\tilde m},\vec{\tilde a},\vec{\tilde g}_{x/y}\}\\
\tilde u_\beta&\in\{\tilde s,\tilde c,\tilde t_{x/y},\tilde e,\tilde d,\tilde v_{x/y}\}\\
\end{aligned}
\end{equation}
so that $\tilde w_\gamma$ denote the uncharged and $\tilde u_\beta$ the charged bilinears. We also added terms quadratic in the sources. Since physical properties are not affected by these quadratic terms (which only give rise to a field independent factor to the partition function), we have the freedom to do so. Note that the chemical potential is now part of one of the sources, as we discussed at the beginning of section \ref{sec:colhub}. In the case of vanishing sources except for the chemical potential $\mu$, we demand the only term to survive to be the one linear in $\mu$. This demand gives rise to the quadratic term in $\mu$ we added (and wrote as an exponential factor in front of the partition function) to cancel the contribution quadratic in $\mu$ from the term $\frac{2\pi^2}{h_\gamma^2}l_\gamma(X)^2$. The quantities $h_\beta$, $h_\gamma$ are arbitrary at the moment. We will come to them soon. Additionally, $S_j$ contains a source term $S_j^F$ for the fermions.

We define the {\em partially bosonized partition function} by
\begin{equation}
\label{eq:parbos:bospar}
\begin{aligned}
Z&=\exp\left(-\frac{2\pi^2{\cal V}\mu^2}{h_\rho^2}\right)\int{\cal D}\hat\psi^*{\cal D}\hat\psi{\cal D}\hat u^*{\cal D}\hat u{\cal D}\hat w\,\exp(-S_F-S_{B}-S_Y-S_J)\\
S_F&=\sum_Q\hat\psi^\dagger(Q)\left(i\omega_n^F-2t\left(\cos(q_1/2)A_1+\cos(q_2/2)B_1\right)\right)\hat\psi(Q)\\
S_B&=\sum_X\left(4\pi^2\sum_\beta\hat u_\beta^*(X)\hat u_\beta(X)+2\pi^2\sum_\gamma\hat w_\gamma(X)^2\right)\\
S_Y&=-\sum_X\left(\sum_\beta h_\beta(\hat u^*_\beta(X)\tilde u_\beta(X)+\hat u_\beta(X)\tilde u_\beta^*(X))+\sum_\gamma h_\gamma\hat w_\gamma(X)\tilde w_\gamma(X)\right)\\
S_J&=-\sum_X\left(\sum_\beta\frac{4\pi^2}{h_\beta}(j_\beta^*(X)\hat u_\beta(X)+j_\beta(X)\hat u_\beta^*(X))+\sum_\gamma\frac{4\pi^2}{h_\gamma}l_\gamma(X)\hat w_\gamma(X)\right)+S_j^F.
\end{aligned}
\end{equation}
In this partition function, we have a fermionic kinetic term $S_F$ which coincides with the corresponding term in the Hubbard model. In the remaining terms we introduced {\em bosonic} fields $\hat u$, $\hat u^*$ and $\hat w_\gamma$, one for each fermionic bilinear $\tilde u$, $\tilde u^*$ and $\tilde w_\gamma$. $S_B$ is a mass term for these fields. $S_Y$ describes a Yukawa like coupling of the bosonic fields to the corresponding fermionic bilinears with Yukawa couplings $h_\beta$, $h_\gamma$. The source term now provides sources for the bosonic fields. 

The next step is to prove that this partially bosonized partition function is equivalent to \eqref{eq:parbos:fermpar} for appropriate values of the Yukawa couplings. We realize that since the action is quadratic in the bosons, the bosonic functional integral can be performed as a simple Gaussian integral. Recall that Gaussian integrals can be evaluated by evaluating the exponent at its stationary value. Since the bosonic propagators are mass like, we can neglect the constant pre-factor altogether and only have to insert the stationary values
\begin{equation}
\label{eq:parbos:backtrans}
\begin{aligned}
\hat u_\beta(X)&=\frac{h_\beta}{4\pi^2}\tilde u_\beta(X)+\frac{j_\beta(X)}{h_\beta}\\
\hat u^*_\beta(X)&=\frac{h_\beta}{4\pi^2}\tilde u^*_\beta(X)+\frac{j_\beta^*(X)}{h_\beta}\\
\hat w_\gamma(X)&=\frac{h_\gamma}{4\pi^2}\tilde w_\gamma(X)+\frac{l_\gamma(X)}{h_\gamma}
\end{aligned}
\end{equation}     
into the exponential of the partially bosonized partition function in order to perform the bosonic functional integrals. We see that the source term reduces to the source term of \eqref{eq:parbos:fermpar}. $S_B+S_Y$ gives a quartic fermionic coupling term of the form
\begin{equation}
S_{int}=-\sum_X\left(\sum_\beta\frac{h_\beta^2}{4\pi^2}\tilde u_\beta^*(X)\tilde u_\beta(X)+\sum_\gamma\frac{h_\gamma^2}{8\pi^2}\tilde w_\gamma(X)^2\right).
\end{equation} 
If \eqref{eq:parbos:fermpar} and \eqref{eq:parbos:bospar} are equivalent, we must have
\begin{equation}
\label{eq:parbos:condyuk}
-\sum_X\left(\sum_\beta\frac{h_\beta^2}{4\pi^2}\tilde u_\beta^*(X)\tilde u_\beta(X)+\sum_\gamma\frac{h_\gamma^2}{8\pi^2}\tilde w_\gamma(X)^2\right)=\frac{1}{2}\sum_X\sum_a\tilde\sigma_{aa}(X)^2.
\end{equation}
This equation should be read as a condition on the Yukawa couplings $h_\beta$, $h_\gamma$. At this point the identities \eqref{eq:decom:decom} come in handy. We can use them to parameterize the solutions of \eqref{eq:parbos:condyuk}. The general solution is further restricted by the fact that we demand the couplings to be real. Then the general solution is (with $h_\beta^2=\frac{\pi^2}{3}H_\beta$, $h_\gamma^2=\frac{\pi^2}{3}H_\gamma$)
\begin{equation}
\label{eq:parbos:hubyuk}
\begin{aligned}
H_\rho&=3(\lambda_2-\lambda_3) & H_{\vec m}&=2\lambda_1+\lambda_2+3\lambda_3+1\\
H_p&=3(\lambda_2+\lambda_3) & H_{\vec a}&=2\lambda_1+\lambda_2-3\lambda_3+1\\
H_{q_{x/y}}&=3\lambda_2 & H_{\vec g_{x/y}}&=2\lambda_1+\lambda_2+1,
\end{aligned}
\end{equation}
\begin{equation}
H_s=H_c=H_{t_{x/y}}=\frac{3}{2}\lambda_1,\quad 2H_e=2H_d=H_{v_{x/y}}=6\lambda_3.
\end{equation}
The parameters $\lambda_i$ obey
\begin{equation}
\begin{aligned}
\lambda_i&>0\quad\forall i\\
\lambda_2&>\lambda_3\\
2\lambda_1+\lambda_2+1&>3\lambda_3
\end{aligned}
\end{equation}
to guarantee that the condition $h_\beta,h_\gamma\in\mathbbm{R}$ is fulfilled.

For any choice of the parameters $\lambda_i$ meeting these conditions, the partially bosonized partition function is equivalent to the fermionic partition function of the Hubbard model we started from. This means that the choice of the Yukawa couplings contains a lot of arbitrariness --- which does not matter in the exact transcription we used here, since all choices are equivalent to the original Hubbard model. However, if we use approximations in calculations, the results can and will depend on the initial choices of the couplings. This can serve as a test for approximation schemes --- an approximation is regarded as well justified, if the results do not depend on the initial choice of the couplings. However, this problem will remain a disturbing one and is the weakness of our theory that we traded in for the possibility to investigate the properties of a system by direct calculation of expectation values of bosonic fields.

Note that is not possible to bosonize the theory without taking into account the spin triplet bilinears (due to the signs in \eqref{eq:parbos:condyuk} and \eqref{eq:decom:decom}). 

\subsection{Symmetries}
Most of the symmetries discussed in sec. \ref{sec:origsymm} can be easily implemented in the partially bosonized version of the colored Hubbard model. As we know how the fermionic fields transform under the symmetry transformations, we can derive the transformation behavior of the fermion bilinears involved in the Yukawa coupling terms in \eqref{eq:parbos:bospar}. Since we know that \eqref{eq:parbos:bospar} and \eqref{eq:parbos:fermpar} are equivalent, they should have the same symmetries and we define the behavior of the bosonic fields under symmetry transformations such that the partially bosonized action becomes invariant. 

In particular, we find for the $U(1)$-symmetry \eqref{eq:symm:U1}
\begin{equation}
\hat w_\gamma\rightarrow \hat w_\gamma,\quad\hat u_\beta\rightarrow\exp(2i\theta)\hat u_\beta,\quad\hat u_\beta^*\rightarrow\exp(-2i\theta)\hat u_\beta^*.
\end{equation}
Similarly, for the $SU(2)$-symmetry we obviously have invariance of all spin singlet bosons. The spin triplet bosons transform as three dimensional vectors under $SO(3)$-rotations around the $\vec\theta$-axis with rotation angle $2|\vec\theta|$. 

The rotation $R$, reflection $I$ and translation $T_x^2$ are also implemented in an obvious way. However, simple translations $T_x$ cannot be defined in our present formulation, since these translations correspond to shifting the field by {\em half} a lattice site of the {\em coarse} lattice and the bosons are defined on the coarse lattice only. To preserve invariance under translations $T_x$, we define a color label $a$ for the bosons and bilinears and set
\begin{equation}
\begin{aligned}
\hat w_{\gamma1}(X)&=T_yT_x^{-1}\hat w_\gamma(X), & \hat w_{\gamma2}(X)&=T_y\hat w_\gamma(X),\\
\hat w_{\gamma3}(X)&=\hat w_\gamma(X), & \hat w_{\gamma4}(X)&=T_x^{-1}\hat w_\gamma(X),
\end{aligned}
\end{equation} 
\begin{equation}
\begin{aligned}
\tilde w_{\gamma1}(X)&=T_yT_x^{-1}\tilde w_\gamma(X), & \tilde w_{\gamma2}(X)&=T_y\tilde w_\gamma(X),\\
\tilde w_{\gamma3}(X)&=\tilde w_\gamma(X), & \tilde w_{\gamma4}(X)&=T_x^{-1}\tilde w_\gamma(X),
\end{aligned}
\end{equation} 
and similarly for the bosons $\hat u_\beta$, $\hat u_\beta^*$ and the boson and bilinear sources. The translations $T_x$ are now simply implemented by e.g. $T_x\hat w_{\gamma1}(X)=\hat w_{\gamma2}(X)$, $T_x\hat w_{\gamma2}(X)=\hat w_{\gamma1}(X+\boldsymbol{e}_1)$ etc. Note that with these definitions, the lattice symmetry operations may be written in a way completely analogous to \eqref{eq:symm:symm}, since the bosonic fields also live on a coarse lattice with the same lattice spacing as the fermions, with four color labeled fields attached to each coarse lattice site. One consequence is that for the same reason as in the fermionic case we introduce an additional relabeling symmetry that acts in the same way on the bosonic color space as it did on the fermionic color space. Again, these relabeling transformations all correspond to the identity transformation in the uncolored formulation. 

To implement bosonic color into our partially bosonized partition function, we write  
\begin{align}
\label{eq:symm:bospar}
Z&=\exp\left(-\frac{2\pi^2{\cal V}\mu^2}{h_\rho^2}\right)\int{\cal D}\hat\psi^*{\cal D}\hat\psi{\cal D}\hat u^*{\cal D}\hat u{\cal D}\hat w\,\exp(-S_F-S_{B}-S_Y-S_J)\nonumber\\
S_F&=\sum_Q\hat\psi^\dagger(Q)\left(i\omega_n^F-2t\left(\cos(q_1/2)A_1+\cos(q_2/2)B_1\right)\right)\hat\psi(Q)\nonumber\\
S_B&=\frac{1}{4}\sum_a\sum_X\left(4\pi^2\sum_\beta\hat u_{\beta a}^*(X)\hat u_{\beta a}(X)+2\pi^2\sum_\gamma\hat w_{\gamma a}(X)^2\right)\nonumber\\
S_Y&=-\frac{1}{4}\sum_a\sum_X\left(\sum_\beta h_\beta(\hat u^*_{\beta a}(X)\tilde u_{\beta a}(X)+\hat u_{\beta a}(X)\tilde u_{\beta a}^*(X))+\sum_\gamma h_\gamma\hat w_{\gamma a}(X)\tilde w_{\gamma a}(X)\right)\nonumber\\
S_J&=-\frac{1}{4}\sum_a\sum_X\left(\sum_\beta\frac{4\pi^2}{h_\beta}(j_{\beta a}^*(X)\hat u_{\beta a}(X)+j_{\beta a}(X)\hat u_{\beta a}^*(X))+\sum_\gamma\frac{4\pi^2}{h_\gamma}l_{\gamma a}(X)\hat w_{\gamma a}(X)\right)+S_j^F.
\end{align}
instead of \eqref{eq:parbos:bospar} and realize that the same calculation that we performed to show that \eqref{eq:parbos:bospar} and \eqref{eq:parbos:fermpar} are equivalent also goes through here.  

\subsection{The colored partition function}
We finally summarize our results for the partition function. In position space, we have 
\begin{align}
\label{eq:colpar:parpos}
Z&=\exp\left(-\frac{2\pi^2{\cal V}\mu^2}{h_\rho^2}\right)\int{\cal D}\hat\psi^*{\cal D}\hat\psi{\cal D}\hat u^*{\cal D}\hat u{\cal D}\hat w\,\exp(-S_F-S_{B}-S_Y-S_J)\nonumber\\
S_F&=\sum_Q\hat\psi^\dagger(Q)\left(i\omega_n^F-2t\left(\cos(q_1/2)A_1+\cos(q_2/2)B_1\right)\right)\hat\psi(Q))\nonumber\\
S_B&=\sum_a\sum_X\left(\pi^2\sum_\beta\hat u_{\beta a}^*(X)\hat u_{\beta a}(X)+\frac{\pi^2}{2}\sum_\gamma\hat w_{\gamma a}(X)^2\right)\nonumber\\
S_Y&=-\frac{1}{4}\sum_a\sum_X\left(\sum_\beta h_\beta(\hat u^*_{\beta a}(X)\tilde u_{\beta a}(X)+\hat u_{\beta a}(X)\tilde u_{\beta a}^*(X))+\sum_\gamma h_\gamma\hat w_{\gamma a}(X)\tilde w_{\gamma a}(X)\right)\nonumber\\
S_J&=-\sum_a\sum_X\left(\sum_\beta(J_{\beta a}^*(X)\hat u_{\beta a}(X)+J_{\beta a}(X)\hat u_{\beta a}^*(X))+\sum_\gamma L_{\gamma a}(X)\hat w_{\gamma a}(X)\right)+S_j^F.
\end{align}
We have redefined the sources by setting
\begin{equation}
J_{\beta a}(X)=\frac{\pi^2}{h_\beta}j_{\beta a}(X),\quad J_{\beta a}^*(X)=\frac{\pi^2}{h_\beta}j_{\beta a}^*(X),\quad L_{\gamma a}(X)=\frac{\pi^2}{h_\gamma}l_{\gamma a}(X).
\end{equation}

The partition function can be used as a generating functional for the bosonic $n$-point functions. In particular, we have
\begin{equation}
\label{eq:colpar:expec}
\begin{aligned}
u_{\beta a}&\equiv\left<\hat u_{\beta a}\right>=\frac{\delta}{\delta J_{\beta a}^*}\ln Z\\
u_{\beta a}^*&\equiv\left<\hat u_{\beta a}^*\right>=\frac{\delta}{\delta J_{\beta a}}\ln Z\\
w_{\gamma a}&\equiv\left<\hat w_{\gamma a}\right>=\frac{\delta}{\delta L_{\gamma a}^*}\ln Z
\end{aligned}
\end{equation}
for the expectation values of the bosonic fields\footnote{Note that $L^*_{\gamma a}(X)=L_{\gamma a}(X)$, but $L^*_{\gamma a}(Q)=L_{\gamma a}(-Q)$ in momentum space. If we write \eqref{eq:colpar:expec} as we did here, the equations hold true both in position and momentum space.}. If we rewrite \eqref{eq:parbos:fermpar} using colored bilinears and insert the resulting partition function in the right hand side of \eqref{eq:colpar:expec}, we find the relation between expectation values of bilinears and bosons
\begin{equation}
\begin{aligned}
u_{\beta a}&=\frac{h_\beta^2}{4\pi^2}\left<\tilde u_{\beta a}\right>+\frac{1}{\pi^2}J_{\beta a}\\
u_{\beta a}^*&=\frac{h_\beta^2}{4\pi^2}\left<\tilde u_{\beta a}^*\right>+\frac{1}{\pi^2}J_{\beta a}^*\\
w_{\gamma a}&=\frac{h_\gamma^2}{4\pi^2}\left<\tilde w_{\gamma a}\right>+\frac{1}{\pi^2}L_{\gamma a}.
\end{aligned}
\end{equation}
These relations show that for vanishing sources the expectation values of the bilinears and the corresponding bosons are equal up to a factor. This is the reason why our formalism makes sense: We know (by construction) that the bilinears describe interesting properties of the fermionic system. The expectation value of the bilinears tell us whether the system exhibits e.g. antiferromagnetic behavior. This is the case if $\langle\vec{\tilde a}_a\rangle$ is non vanishing, which in turn means that also the bosonic expectation value $\vec a_a$ does not vanish. We therefore can analyze the properties of the system by merely calculating expectation values of bosonic fields in a Yukawa like theory. In principle, all calculations in this work are dedicated to do exactly this, and to interprete the results by means of the underlying fermionic theory. 

Again we stress the special role of the charge density $\tilde\rho_a$. The source of the charge density contains the chemical potential and will in general not vanish. The bosonic expectation value of $\hat\rho_a$ is
\begin{equation}
\rho_a=\frac{h_\rho}{4\pi^2}\langle\tilde\rho_a\rangle+\frac{\mu}{h_\rho}.
\end{equation}   

Using the Fourier transforms \eqref{eq:four:four} for the fermions and defining $\hat u_\beta$ and $\hat w_\gamma$ to have the same Fourier transform as $\hat\psi$, and $\hat u_\beta^*$ to have the same Fourier transform as $\hat\psi^*$, we can write down the partition function in momentum space
\begin{align}
\label{eq:colpar:parmom}
Z&=\exp\left(-\frac{2\pi^2{\cal V}\mu^2}{h_\rho^2}\right)\int{\cal D}\hat\psi^*{\cal D}\hat\psi{\cal D}\hat u^*{\cal D}\hat u{\cal D}\hat w\,\exp(-S_F-S_{B}-S_Y-S_J)\nonumber\\
S_F&=\sum_Q\hat\psi^\dagger(Q)\left(i\omega_n^F-2t\left(\cos(q_1/2)A_1+\cos(q_2/2)B_1\right)\right)\hat\psi(Q)\nonumber\\
S_B&=\sum_a\sum_Q\left(\pi^2\sum_\beta\hat u_{\beta a}^*(Q)\hat u_{\beta a}(Q)+\frac{\pi^2}{2}\sum_\gamma\hat w_{\gamma a}(-Q)\hat w_{\gamma a}(Q)\right)\nonumber\\
S_Y&=-\sum_{abc}\sum_{QQ'Q''}\delta(Q-Q'-Q'')\nonumber\\
&\qquad\biggl(\sum_\beta \left(\hat u^*_{\beta c}(Q)\hat\psi_a^T(Q')V_{ab,c}^{u_\beta^*}(Q',Q'')\hat\psi_b(Q'')+\hat u_{\beta c}(Q)\hat\psi_a^\dagger(Q')V_{ab,c}^{u_\beta}(Q',Q'')\hat\psi_b^*(Q'')\right)\nonumber\\
&\qquad+\sum_\gamma \hat w_{\gamma c}(Q)\hat\psi_a^\dagger(Q')V_{ab,c}^{w_\gamma}(Q',-Q'')\hat\psi_b(-Q'')\biggr)\nonumber\\
S_J&=-\sum_a\sum_Q\left(\sum_\beta(J_{\beta a}^*(Q)\hat u_{\beta a}(Q)+J_{\beta a}(Q)\hat u_{\beta a}^*(Q))+\sum_\gamma L_{\gamma a}(-Q)\hat w_{\gamma a}(Q)\right)+S_j^F.
\end{align}
The vertex factors $V_{ab,c}$ in the coupling term are given in appendix \ref{sec:vertexfac}.
\chapter{A mean field calculation}
\label{sec:meanf}

Before continuing to improve our formalism, we want to give a first impression of the power of our formalism even in a very simple mean field like approximation. The results of this chapter have been published in \cite{bick}. The main ingredients of this approximation are the following:
\begin{itemize}
\item The bosonic fields act as constant background fields, so that the bosonic functional integrals can be trivially performed by simply setting all bosonic fields to these constant background values.
\item The bosonic background fields are homogeneous in the sense that they do not possess any spatial dependence.
\item All background fields except the charge density $\hat\rho$, the antiferromagnetic spin density $\vec{\hat a}$ and the superconducting $d$-wave $\hat d$ vanish. 
\end{itemize} 
We proceed by deriving an expression for the effective action in this approximation, which depends on the values of the three non vanishing background fields. For given temperature and charge density, we then look for minima of the effective potential with respect to the expectation values of $\vec{\hat a}$ and $\hat d$. If the minimum occurs at non vanishing expectation values of the antiferromagnetic or superconducting fields, we conclude that the system exhibits antiferromagnetic or superconducting behavior at the given temperature and charge density.  

\section{The effective potential}
\label{sec:meanf:effpot}
Using \eqref{eq:colpar:expec}, we define the {\em effective action} as the Legendre transform of the log of the partition function
\begin{multline}
\label{eq:effpot:effact}
\Gamma[u_{\beta a},u^*_{\beta a},w_{\gamma a}]=-\ln Z[J_{\beta a},J^*_{\beta a},L_{\gamma a}]\\
+\sum_a\sum_Q\left(\sum_\beta(J_{\beta a}^*(Q) u_{\beta a}(Q)+J_{\beta a}(Q) u_{\beta a}^*(Q))+\sum_\gamma L_{\gamma a}(-Q) w_{\gamma a}(Q)\right)+S_j^F,
\end{multline}
where the sources on the right hand side are functionals of the expectation values $u_{\beta a}$, $u_{\beta a}^*$, $w_{\gamma a}$, $\psi$ and $\psi^*$. The usual properties of a Legendre transform tell us that
\begin{equation}
\label{eq:effpot:acprin}
\frac{\delta\Gamma}{\delta u_{\beta a}}=J_{\beta a}^*,\quad\frac{\delta\Gamma}{\delta u_{\beta a}^*}=J_{\beta a},\quad\frac{\delta\Gamma}{\delta w_{\gamma a}}=L_{\gamma a}^*.
\end{equation}
For vanishing sources these equations are formally nothing else than the classical action principle --- therefore the name effective action. They may be regarded as the equations of motion for the field expectation values, taking quantum corrections into account. Assuming that the fermions have been integrated out (we will do this explicitly below) and that the bosonic field expectation values do not have any spatial dependence, the effective action may be written as
\begin{equation}
\label{eq:effpot:Udef}
\Gamma[u_{\beta a},u^*_{\beta a},w_{\gamma a}]={\cal V}U[u_{\beta a},u^*_{\beta a},w_{\gamma a}],
\end{equation}
where $U$ serves as the {\em effective potential} for our theory. $U$ is finite (apart from a $T$-dependent additive constant) and position independent. \eqref{eq:effpot:acprin} tells us that --- for vanishing sources --- the system favors the state for which the effective potential as a function of the expectation values becomes stationary, and it can be shown that not only stationary, but even minimal \cite{weinberg1}.   

To realize our mean field conditions, we introduce a factor
\begin{equation}
\prod_{\beta\gamma a}\,\delta(\hat u_{\beta a}(Q)-u_{\beta a}\delta(Q))\,\delta(\hat u^*_{\beta a}(Q)-u_{\beta a}^*\delta(Q))\,\delta(\hat w_{\gamma a}(Q)-w_{\gamma a}\delta(Q))
\end{equation}
under the functional integral of \eqref{eq:colpar:parmom}. This sets the bosonic fields to constant background fields, neglecting all bosonic fluctuations. The momentum $\delta$-functions implement our condition that these constant background fields should be homogeneous. Our partition function then becomes
\begin{align}
\label{eq:effpot:Zmf}
Z_{mf}&=\exp\left(-\frac{2\pi^2{\cal V}\mu^2}{h_\rho^2}\right)\int{\cal D}\hat\psi^*{\cal D}\hat\psi\,\exp(-S_F-S_{B}-S_Y-S_J)\nonumber\\
S_F&=\sum_Q\hat\psi^\dagger(Q)\left(i\omega_n^F-2t\left(\cos(q_1/2)A_1+\cos(q_2/2)B_1\right)\right)\hat\psi(Q)\nonumber\\
S_B&={\cal V}\sum_a\left(\pi^2\sum_\beta u_{\beta a}^*u_{\beta a}+\frac{\pi^2}{2}\sum_\gamma w_{\gamma a}w_{\gamma a}\right)\nonumber\\
S_Y&=-\sum_{abc}\sum_{Q}\biggl(\sum_\gamma w_{\gamma c}\hat\psi_a^\dagger(Q)V_{ab,c}^{w_\gamma}(Q,Q)\hat\psi_b(Q)\nonumber\\
&\qquad+\sum_\beta \left(u^*_{\beta c}\hat\psi_a^T(Q)V_{ab,c}^{u_\beta^*}(Q,-Q)\hat\psi_b(-Q)+u_{\beta c}\hat\psi_a^\dagger(Q)V_{ab,c}^{u_\beta}(Q,-Q)\hat\psi_b^*(-Q)\right)\nonumber\biggr)\\
S_J&=-\sum_a\left(\sum_\beta(J_{\beta a}^*(0)u_{\beta a}+J_{\beta a}(0)u_{\beta a}^*)+\sum_\gamma L_{\gamma a}(0)w_{\gamma a}\right)+S_j^F.
\end{align}
Using the same approximations in \eqref{eq:effpot:effact} and inserting \eqref{eq:effpot:Zmf}, by using \eqref{eq:effpot:Udef} we arrive at
\begin{equation}
U=\sum_a\left(\pi^2\sum_\beta u_{\beta a}^*u_{\beta a}+\frac{\pi^2}{2}\sum_\gamma w_{\gamma a}w_{\gamma a}\right)+\frac{2\pi^2}{h_\rho^2}\mu^2+\Delta U
\end{equation}
with
\begin{align}
\label{eq:effpot:U1}
\Delta U&=-\frac{1}{\cal V}\ln\int{\cal D}\hat\psi^*{\cal D}\hat\psi \exp(-S_\Delta)\nonumber\\
S_\Delta&=\sum_Q\hat\psi^\dagger(Q)\left(i\omega_n^F-2t\left(\cos(q_1/2)A_1+\cos(q_2/2)B_1\right)\right)\hat\psi(Q)\nonumber\\
&\quad-\sum_{abc}\sum_{Q}\biggl(\sum_\gamma w_{\gamma c}\hat\psi_a^\dagger(Q)V_{ab,c}^{w_\gamma}(Q,Q)\hat\psi_b(Q)\\
&\quad+\sum_\beta \left(u^*_{\beta c}\hat\psi_a^T(Q)V_{ab,c}^{u_\beta^*}(Q,-Q)\hat\psi_b(-Q)+u_{\beta c}\hat\psi_a^\dagger(Q)V_{ab,c}^{u_\beta}(Q,-Q)\hat\psi_b^*(-Q)\right)\nonumber\biggr)\nonumber.
\end{align}
As it must be, $U$ can be written as the classical potential terms (which are pure mass terms in our theory) and a correction $\Delta U$ describing the influence of fermionic fluctuations. We must now calculate the functional integral in $\Delta U$. The easiest way to do so is to define the vector
\begin{equation}
\tilde\psi(Q)=\begin{pmatrix}\hat\psi(Q)\\\hat\psi^*(-Q)\end{pmatrix}.
\end{equation}
Note that this vector has $16$ components ($2$ explicitly, $4$ in color and $2$ in spinor space). By aid of this vector, we may rewrite $S_\Delta$ in the form
\begin{equation}
S_\Delta=\frac{1}{2}\sum_Q\tilde\psi^T(-Q)P(Q)\tilde\psi(Q)
\end{equation}
and perform the Gaussian integration to obtain
\begin{equation}
\label{eq:delUP}
\Delta U=-\frac{1}{2}\sum_Q\ln\det P(Q).
\end{equation}
In general, $P(Q)$ contains contributions from all bosonic fields. From now on, we will set all fields except $\rho$, $\vec a$ and $d$ equal to zero. For the remaining fields, we set
\begin{gather}
w_{\rho 1}=w_{\rho 2}=w_{\rho 3}=w_{\rho 4}=\rho\nonumber\\
w_{\vec a 1}=-w_{\vec a 2}=w_{\vec a 3}=-w_{\vec a 4}=\vec a\nonumber\\
u_{d 1}=u_{d 2}=u_{d 3}=u_{d 4}=d\nonumber\\
u_{d 1}^*=u_{d 2}^*=u_{d 3}^*=u_{d 4}^*=d^*.
\end{gather}
The explicit expression for $P(Q)$ follows by inserting the vertex factors from appendix \ref{sec:vertexfac} in \eqref{eq:effpot:U1} and collecting terms. One obtains
\begin{align}
P(Q)&=\begin{pmatrix}0&i\omega_n^F+2t\tilde T\\i\omega_n^F-2t\tilde T&0\end{pmatrix}-h_\rho\rho\begin{pmatrix}0&-A_0\\A_0&0\end{pmatrix}-h_a\vec a\begin{pmatrix}0&-A_3\otimes\vec\sigma^T\\A_3\otimes\vec\sigma&0\end{pmatrix}\nonumber\\
&-h_d(\cos(q_1/2)A_1-\cos(q_2/2)B_1)\otimes\begin{pmatrix}d^*&0\\0&-d\end{pmatrix}\otimes i\sigma_2
\end{align}
with $\tilde T=\cos(q_1/2)A_1+\cos(q_2/2)B_1$. By using the identity
\begin{equation}
\ln\det A=\frac{1}{2}\ln\det\left(A\begin{pmatrix}0&1\\1&0\end{pmatrix}A^T\begin{pmatrix}0&1\\1&0\end{pmatrix}\right)
\end{equation}
it is possible to diagonalize $P(Q)$ in the two dimensional space of the two components of $\tilde\psi$, so that the determinant reduces to one over a matrix of dimension $8$:
\begin{align}
\Delta U&=-\frac{1}{2}\sum_Q\ln{\det}_8\bigl((\omega_n^F)^2+(2t\tilde T+h_\rho\rho)^2+h_a^2\vec a^2\nonumber\\
&\qquad+2h_\rho h_a\rho\vec a A_3\otimes\vec\sigma+h_d^2d^*d(\cos(q_1/2)A_1-\cos(q_2/2)B_1)^2\bigr).
\end{align}
The determinant can be further diagonalized in spinor space, yielding
\begin{equation}
\vec a\vec\sigma\rightarrow \left|\vec a\right|\sigma_3.
\end{equation}
With $\{\tilde T,A_3\}=0$ and defining $\delta=d^*d$, $\alpha=\vec a^2$, we find
\begin{align}
\Delta U&=-\frac{1}{2}\sum_Q\ln\det\bigl((\omega_n^F)^2+(2t\tilde T+h_\rho\rho+h_a\sqrt{\alpha}A_3\otimes\sigma_3)^2\nonumber\\
&\qquad+h_d^2\delta\,(\cos(q_1/2)A_1-\cos(q_2/2)B_1)^2\bigr).
\end{align}
We proceed by evaluating the remaining determinant by brute force. In the resulting expression, the Matsubara sum can be performed using the product expansion \cite{fischer}
\begin{equation}
\cosh(\pi z)=\prod_{n\in\mathbbm N_0}\left(1+\frac{4z^2}{(2n+1)^2}\right),\quad z\in\mathbbm{C}.
\end{equation}
Up to a temperature dependent divergent constant, the final result for the effective potential is
\begin{align}
\label{eq:effpot:U}
U&=2\pi^2\rho^2+2\pi^2\alpha+4\pi^2\delta+\frac{2\pi^2\mu^2}{h_\rho^2}+\Delta U\\
\Delta U&=-2T\int_{-\pi}^\pi\frac{d^2q}{(2\pi)^2}\sum_{\epsilon_1,\epsilon_2\in\{-1,1\}}\nonumber\\
&\ln\cosh\left(\frac{1}{2T}\sqrt{\left( h_\rho\rho+\epsilon_2\sqrt{4t^2(c_1+\epsilon_1c_2)^2+h_a^2\alpha}\right)^2+h_d^2\delta(c_1-\epsilon_1c_2)^2}\right)\nonumber
\end{align}
with $c_i=\cos(q_i/2)$.

\section{Discussion of the effective potential}
We will now discuss our result for the effective potential \eqref{eq:effpot:U} and calculate the phase diagram. 

First note that for large temperature $\Delta U$ vanishes and $U$ is given by $U=2\pi^2\rho^2+2\pi^2\alpha+4\pi^2\delta+\frac{2\pi^2\mu^2}{h_\rho^2}$. In this case, the minimum of $U$ with respect to $\alpha$ and $\delta$ occurs at $\alpha=\delta=0$ for all $\rho$. Therefore for large temperature, no symmetry breaking, i.e. no antiferromagnetism or superconducting behavior is present. If $T$ is lowered, the minimum may be destabilized by the contribution of $\Delta U$. However, note that (as can be seen by expanding $\Delta U$ for $\alpha\gg1$) $U$ grows as $\alpha$ for large $\alpha$, which means that even for low temperature the minimum always occurs at {\em finite} $\alpha$. The same argument holds for $\delta$. 

A sufficient condition for the minimum to occur at non vanishing $\alpha$ or $\delta$ is that the masses 
\begin{equation}
M_a^2=\left.2\frac{\partial U}{\partial\alpha}\right|_{\alpha=\delta=0},\quad M_d^2=\left.\frac{\partial U}{\partial\delta}\right|_{\alpha=\delta=0}
\end{equation}
become negative. If e.g. the mass $M_a^2$ becomes negative, the effective potential possesses a local maximum in $\alpha=0$. Since for large $\alpha$ the potential $U$ increases $\propto\alpha$, the minimum must occur in $\alpha>0$ and the symmetry is spontaneously broken. Again, the same argument holds for $\delta$. The masses can be calculated from \eqref{eq:effpot:U} and we find
\begin{align}
M_a^2&=4\pi^2-h_a^2\int_{-\pi}^\pi\frac{d^2q}{(2\pi)^2}\sum_{\epsilon_1,\epsilon_2\in\{-1,1\}}\frac{\tanh\left(\frac{1}{2T}(h_\rho\rho+2t\epsilon_2(c_1+\epsilon_1c_2))\right)}{2t\epsilon_2(c_1+\epsilon_1c_2)}\nonumber\\
&=4\pi^2-h_a^2\int_{-\pi}^\pi\frac{d^2q}{(2\pi)^2}\sum_{\epsilon_1\in\{-1,1\}}\frac{\sinh\frac{2t(c_1+\epsilon_1c_2)}{T}}{t(c_1+\epsilon_1c_2)\left(\cosh\frac{h_\rho\rho}{T}+\cosh\frac{2t(c_1+\epsilon_1c_2)}{T}\right)}\nonumber\\
M_d^2&=4\pi^2-\frac{1}{2}h_d^2\int_{-\pi}^\pi\frac{d^2q}{(2\pi)^2}\sum_{\epsilon_1,\epsilon_2\in\{-1,1\}}\tanh\left(\frac{1}{2T}(h_\rho\rho+2t\epsilon_2(c_1+\epsilon_1c_2))\right)\nonumber\\
&\quad\qquad\qquad\cdot\frac{(c_1-\epsilon_1c_2)^2}{h_\rho\rho+2t\epsilon_2(c_1+\epsilon_1c_2)}.
\end{align}
In the expression for $M_a^2$ we performed the sum over $\epsilon_2$ to show that the right hand side is finite for $c_1+\epsilon_1c_2\to0$ and $T>0$. We see that for $\delta$, the mass correction arising from $\Delta U$ always tends to destabilize the symmetric minimum, since its contribution to the mass is always negative. The same holds true for $\alpha$. Whether the mass actually becomes negative or not depends on the choice of the Yukawa couplings. By increasing the strength of the couplings, we necessarily find negative masses and therefore spontaneous symmetry breaking. It is clear from this qualitative point of view that our numerical results for the phase diagrams in this approximation will crucially depend on the initial choice of the couplings. In the frame of the mean field approximation, there is no way to decide which choice is the correct one (remember that without any approximations, any choice of the couplings respecting \eqref{eq:parbos:hubyuk} will be viable and physically equivalent). 

We stress that the condition of vanishing masses is sufficient, but not necessary for a phase transition to take place.    
\begin{figure}
\centering
\psfrag{Tgg1}{$T\gg1$}
\psfrag{U}{$U$}
\psfrag{alpha}{$\alpha$}
\psfrag{PHI}{PH I}
\psfrag{PHII}{PH II}
\includegraphics[scale=0.5]{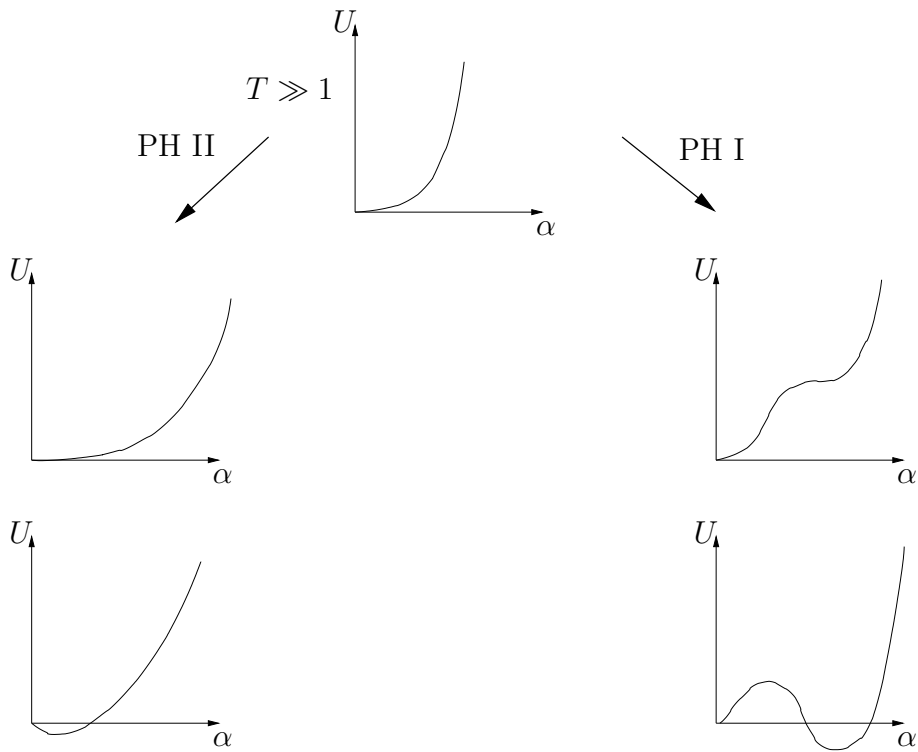}
\caption[Phase transitions of first and second order]{For high temperature, the potential minimum occurs in $\alpha=0$. By lowering the temperature, phase transitions of first or second order can appear. Only for phase transitions of second order the temperature where the mass changes sign coincides with the temperature of the phase transition. }
\label{fig:mass}
\end{figure}
To understand this, we consider the change of the effective potential with temperature as a function of $\alpha$ for fixed $\delta$ (fig. \ref{fig:mass}). For high temperature, we know that the minimum occurs in $\alpha=0$. Suppose for low temperature the symmetry is broken. The phase transition into the state of broken symmetry can take place in two different ways:
\begin{itemize}
\item The minimum of the potential remains at $\alpha=0$, until the mass $M_a^2$ changes its sign. Then, as a function of temperature, the minimum moves away from $\alpha=0$ to some finite $\alpha$. The order parameter $\alpha$ for this kind of phase transition starts at zero in the symmetric phase and continuously moves away from zero in the broken phase. The phase transition is therefore continuous or of second order.
\item At some temperature, the potential builds up a local minimum at some finite $\alpha$. The value of the effective potential at this minimum decreases, until the minimum becomes global. $M_a^2$ remains positive during this process. In this case the order parameter $\alpha$ changes discontinuously from zero to some finite value, so that we face a discontinuous phase transition (or equivalently a phase transition of first order).   
\end{itemize} 
This discussion shows that one has to be careful when analyzing the properties of the effective potential. For example, if we have found that say $M_\delta^2$ is negative and $M_\alpha^2$ is positive, we cannot immediately conclude that the system exhibits  superconducting behavior, since it is possible that the global minimum occurs in $\alpha>0$, $\delta=0$, if we happen to have the case of a first order phase transition in the antiferromagnetic channel. The main difficulties of numerically finding the minima of \eqref{eq:effpot:U} are rooted in these possibilities of first order phase transitions. However, even without explicitly calculating the minima of the effective potential, it is possible to gain further insight into the possibility of first order phase transitions. First consider the case that the minimum occurs in $\alpha=0$, $\delta>0$. Then this minimum must obey
\begin{equation}
\left.\frac{\partial U}{\partial\delta}\right|_{\alpha=0}=0.
\end{equation}
This equation may be rewritten in the form
\begin{align}
M_d^2&=\frac{1}{2}h_d^2\int_{-\pi}^\pi\frac{d^2q}{(2\pi)^2}\sum_{\epsilon_1,\epsilon_2}(c_1-\epsilon_1c_2)^2\\
&\quad\biggl(\frac{\tanh\left(\frac{1}{2T}\sqrt{(h_\rho\rho+2t\epsilon_2(c_1+\epsilon_1c_2))^2+h_d^2\delta(c_1-\epsilon_1c_2)^2}\right)}{\sqrt{(h_\rho\rho+2t\epsilon_2(c_1+\epsilon_1c_2))^2+h_d^2\delta(c_1-\epsilon_1c_2)^2}}\nonumber\\
&\qquad-\frac{\tanh\left(\frac{1}{2T}(h_\rho\rho+2t\epsilon_2(c_1+\epsilon_1c_2))\right)}{h_\rho\rho+2t\epsilon_2(c_1+\epsilon_1c_2)}\biggr)\nonumber
\end{align}
By using
\begin{equation}
\frac{\tanh a}{a}>\frac{\tanh\sqrt{a^2+x^2}}{\sqrt{a^2+x^2}}\quad\forall x\neq0,a,
\end{equation}
we see that the right hand side is strictly negative. Therefore solutions with $\delta>0$ are possible only for $M_d^2<0$, which in turn means that the phase transition is of second order. In other words: The phase transition between the symmetric phase and the superconducting phase is always of second order. The same argument holds for $\alpha$, if $\rho$ is not too large. For large $\rho$, we may (and will) encounter first order phase transitions into the antiferromagnetic phase.

The parameters of our theory are $T$, $\rho$, $t$ and the Yukawa couplings. The chemical potential has been removed from the list of free parameters by the Legendre transform \eqref{eq:effpot:effact} and can be inferred from \eqref{eq:effpot:acprin} to be
\begin{equation}
\label{eq:murho}
\mu=\frac{h_\rho}{4\pi^2}\frac{\partial U}{\partial\rho}.
\end{equation}
Particularly, we explicitly see by inserting \eqref{eq:effpot:U} that $\rho=0$ gives $\mu=0$, so that $\rho=0$ corresponds to the case of half filling as it must. 

\section{Numerical results}
In what follows, we set $t=1$, fix the value of the Yukawa couplings and minimize $U$ numerically with respect to $\alpha$ and $\delta$ for a large number of $(\rho,T)$-pairs, which yields a complete picture of the phase diagram in this approximation. We present results for four different choices of the Yukawa couplings.
\begin{figure}
\centering
\psfrag{AF}{AF}
\psfrag{SC}{SC}
\psfrag{SYM}{SYM}
\includegraphics[scale=0.4]{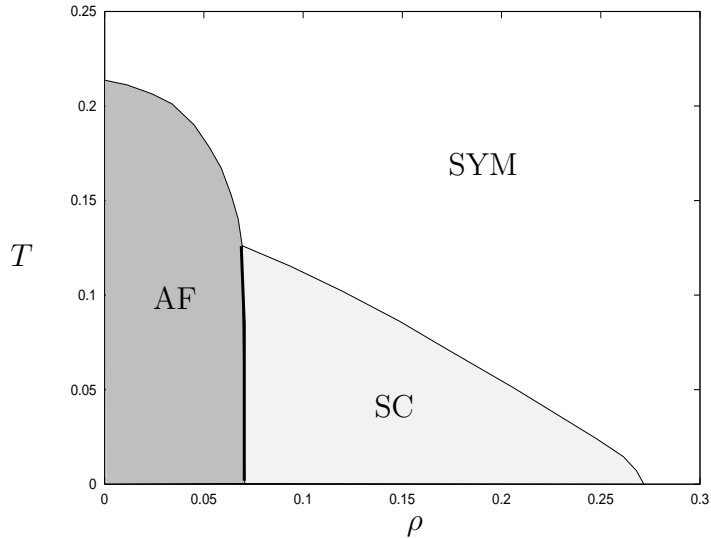}
\put(-120,0){$\rho$}
\put(-270,100){$T$}
\caption[The $\rho$-$T$ phase diagram for $h_\rho^2=h_a^2=h_d^2=10$]{The $\rho$-$T$ phase diagram for $h_\rho^2=h_a^2=h_d^2=10$ with symmetric (SYM), antiferromagnetic (AF) and superconducting (SC) phase. In the region marked by the bold line the phase transition into the antiferromagnetic phase is of first order; all other phase transitions are of second order.}
\label{fig:rho_T.1}
\end{figure}
\begin{figure}
\centering
\psfrag{AF}{AF}
\psfrag{SC}{SC}
\psfrag{SYM}{SYM}
\includegraphics[scale=0.4]{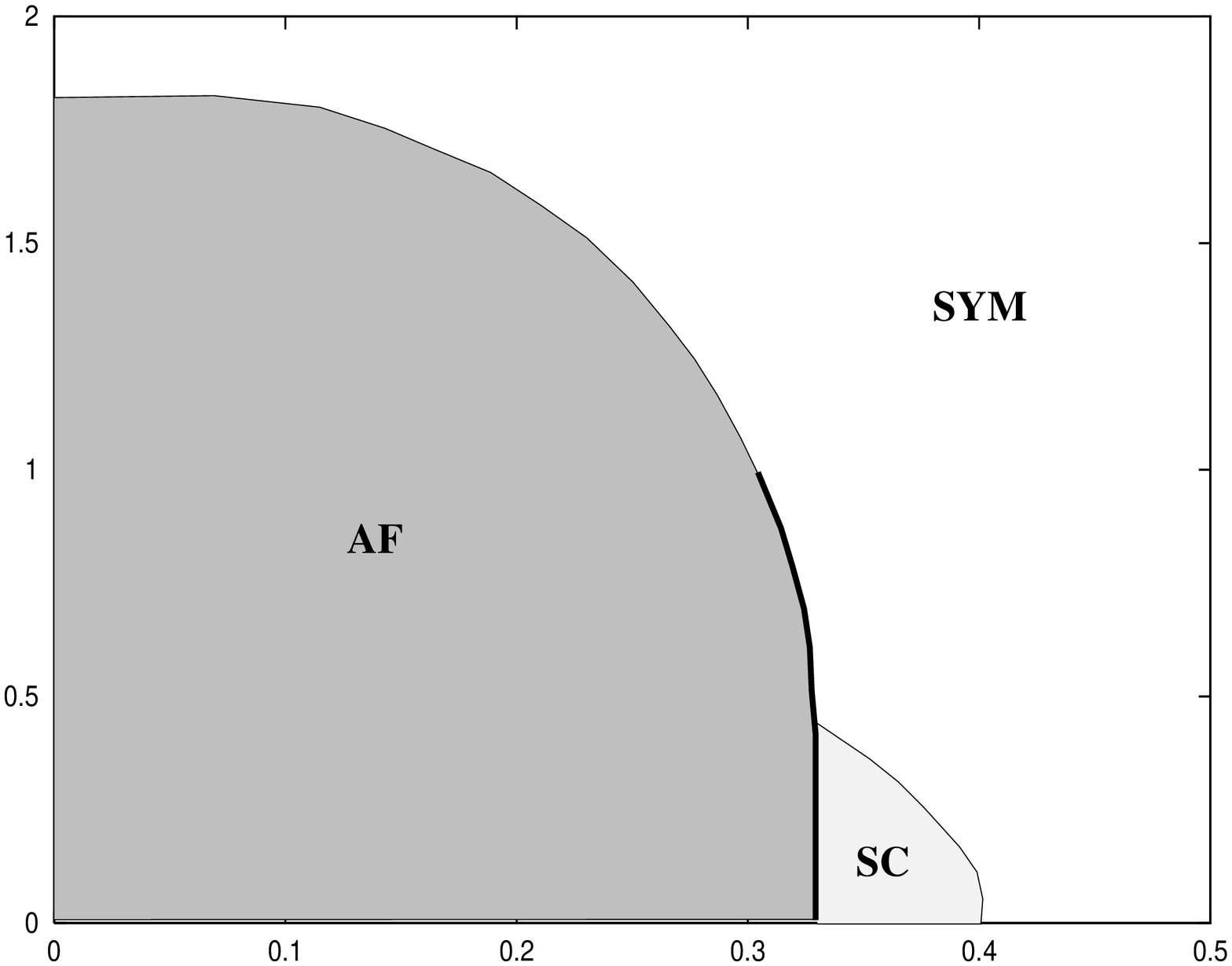}
\put(-120,0){$\rho$}
\put(-270,100){$T$}
\caption[The $\rho$-$T$ phase diagram for $h_\rho^2=h_a^2=h_d^2=40$]{The $\rho$-$T$ phase diagram for $h_\rho^2=h_a^2=h_d^2=40$ with symmetric (SYM), antiferromagnetic (AF) and superconducting (SC) phase. In the region marked by the bold line the phase transition into the antiferromagnetic phase is of first order; all other phase transitions are of second order.}
\label{fig:rho_T.2}
\end{figure}
\begin{figure}
\centering
\psfrag{AF}{AF}
\psfrag{SC}{SC}
\psfrag{SYM}{SYM}
\includegraphics[scale=0.4]{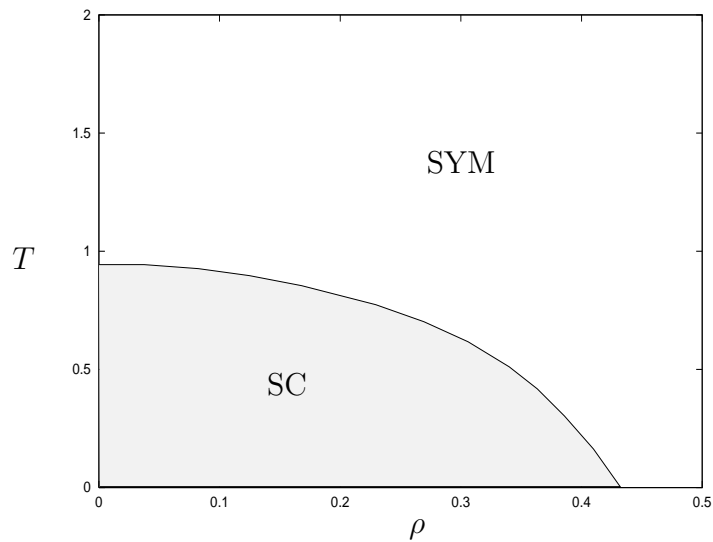}
\put(-120,0){$\rho$}
\put(-270,100){$T$}
\caption[The $\rho$-$T$ phase diagram for $h_\rho^2=h_d^2=40$, $h_a^2=10$]{The $\rho$-$T$ phase diagram for $h_\rho^2=h_d^2=40$, $h_a^2=10$ with symmetric (SYM) and superconducting (SC) phase. All phase transitions are of second order.}
\label{fig:rho_T.1.2}
\end{figure}
\begin{figure}
\centering
\psfrag{AF}{AF}
\psfrag{SC}{SC}
\psfrag{SYM}{SYM}
\includegraphics[scale=0.4]{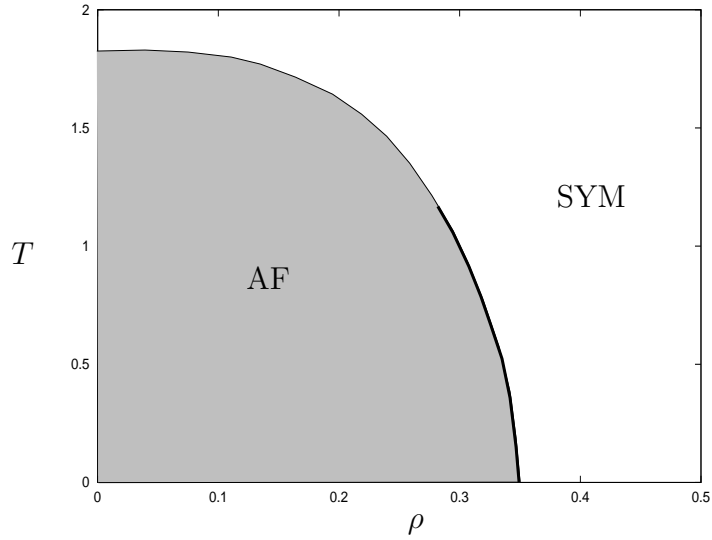}
\put(-120,0){$\rho$}
\put(-270,100){$T$}
\caption[The $\rho$-$T$ phase diagram for $h_\rho^2=h_a^2=40$, $h_d^2=10$]{The $\rho$-$T$ phase diagram for $h_\rho^2=h_a^2=40$, $h_d^2=10$ with symmetric (SYM) and antiferromagnetic (AF) phase. In the region marked by the bold line the phase transition into the antiferromagnetic phase is of first order; all other phase transitions are of second order.}
\label{fig:rho_T.2.1}
\end{figure}
\begin{figure}
\centering
\psfrag{rho}{$\rho$}
\psfrag{T}{$T$}
\psfrag{delta}{$\delta$}
\includegraphics{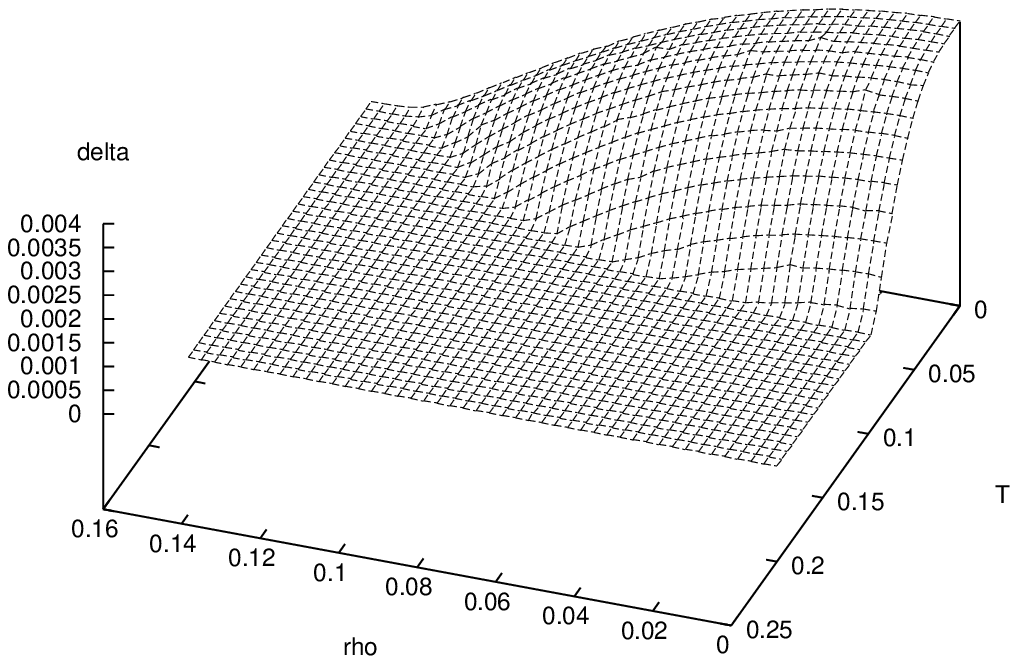}
\caption[$\alpha(\rho,T)$ where $U(\alpha,\delta)$ becomes minimal with $h_\rho^2=h_d^2=40$, $h_a^2=10$]{The $\rho$-$T$ phase diagram for $h_\rho^2=h_d^2=40$, $h_a^2=10$.}
\label{fig:rho_T_hill.1.2}
\end{figure}
\begin{figure}
\centering
\psfrag{rho}{$\rho$}
\psfrag{T}{$T$}
\psfrag{alpha}{$\alpha$}
\includegraphics{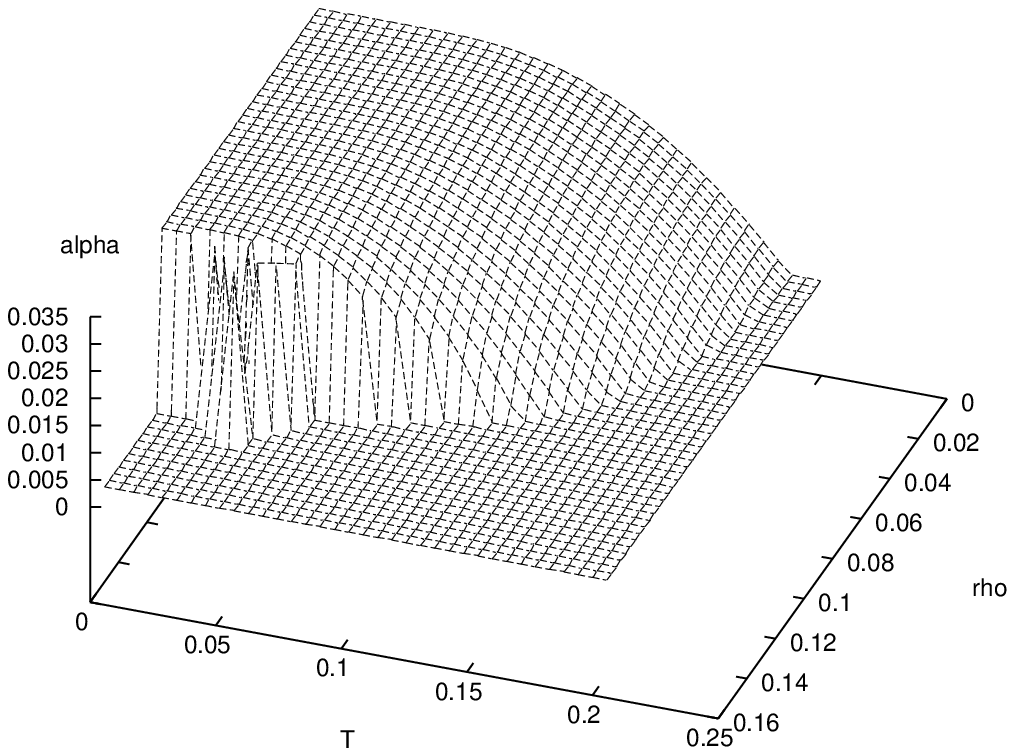}
\caption[$\delta(\rho,T)$ where $U(\alpha,\delta)$ becomes minimal with $h_\rho^2=h_a^2=40$, $h_d^2=10$]{The $\rho$-$T$ phase diagram for $h_\rho^2=h_a^2=40$, $h_d^2=10$.}
\label{fig:rho_T_hill.2.1}
\end{figure}

Some remarks concerning the numerics: To evaluate the momentum integrals in the effective potential numerically, we used a very simple self written Riemannian sum like method on a grid. This works well for the kind of functions encountered here and is faster than more sophisticated methods. The search for the minimum is performed by using the Broyden-Fletcher-Goldfarb-Shanno variant of the Davidon-Fletcher-Powell minimization. This method is implemented in the Numerical Recipes routine dfpmin (cf. \cite{numrec:mini}). To be sure that the method does not return a local minimum, we repeat the calculation several times for each point of the phase diagram with random initial values of $\alpha$ and $\delta$. To identify phase transitions of first order, we additionally calculate $M_a^2$ in every step and check whether for non vanishing minimum $M_a^2$ is positive or negative.

The results for different choices of the Yukawa couplings are presented in the figures. Note that the choice of $h_\rho$ does not change the qualitative shape of the phases, since $h_\rho$ only enters as a factor for $\rho$ and therefore does nothing else than to rescale the $\rho$-axis. For equal values of $h_d$ and $h_a$ (figs. \ref{fig:rho_T.1} and \ref{fig:rho_T.2}), we find phase diagrams that already resemble the phase diagram fig. \ref{fig:cuprat} of a high temperature superconductor. It is interesting that no region exists where the minimum of the effective actions occurs for both $\alpha>0$ and $\delta>0$. An expectation value of the antiferromagnetic field tends to suppress spontaneous symmetry breaking in the superconducting channel and vice versa. Additionally, we find a first order phase transition into the antiferromagnetic phase for large $\rho$, which is in agreement with our discussion on analytical grounds in the last section. 

If we increase either the antiferromagnetic coupling $h_a$ or the superconducting coupling $h_d$, the antiferromagnetic or superconducting phase dominates respectively. This feature is illustrated in figs. \ref{fig:rho_T.1.2} and  \ref{fig:rho_T.2.1}. In fig. \ref{fig:rho_T_hill.1.2} we plotted the value of $\delta$ where the effective potential becomes minimal as a function of $\rho$ and $T$. The choice for the Yukawa couplings is the same as in fig. \ref{fig:rho_T.1.2} ($h_d^2=40$, $h_a^2=10$), so that only the superconducting phase is present. We did the same in fig. \ref{fig:rho_T_hill.2.1} for $\alpha$ with Yukawa couplings as in fig. \ref{fig:rho_T.2.1}, in which case symmetry breaking always takes place in the antiferromagnetic channel. This plot clearly shows the occurrence of a first order phase transition into the antiferromagnetic phase for large $\rho$.

We conclude that even the mean field approximation presented here gives a rough picture of the phase diagram of high temperature superconductors. However, the drawback is also apparent: Different choices of the Yukawa couplings lead to different phase diagrams, although the original theory was invariant under this choice. The strong dependence on the couplings is unphysical, and we are not able to remove this dependence in the framework of the mean field approximation. To resolve this problem and to build up a more reliable picture of the phase diagram, we have to include bosonic fluctuations that we completely neglected in our mean field approach. However, the inclusion of bosonic fluctuations is highly non trivial and can no longer be treated by a simple calculation of an effective potential as we did here. The method of choice to deal with them is a renormalization group analysis of the effective potential and the couplings. In the next chapter we will present the renormalization group formalism that we will use in this work. After that, we further transform our partition function of the colored Hubbard model into a form more suitable for a renormalization group analysis.  

\chapter{Renormalization group equations and the effective average action}
\label{ch:renor}
One loop calculations in quantum field theory are usually plagued by divergencies, which appear as unbounded loop momentum integrals. Two sources of these divergencies are possible. For theories defined on a spatial continuum (as the usual theories defining the standard model), we face ultraviolet divergencies which occur because the integrals over the loop momenta extend to arbitrarily large momenta. This kind of divergencies is not present in our case, since the spacing of the underlying lattice provides a physical UV-cutoff for momentum integrals, constraining them to some finite interval (in our case to the interval $[-\pi,\pi]$). Another source of divergencies is the presence of massless modes. These infrared divergencies do emerge in our theory on the lattice and have to be regularized. Whatever regularization scheme we use, this regularized theory will depend on some unphysical regularization parameter --- an artificially introduced mass, a momentum cutoff, or whatever. The aim of any regularization procedure is to be able to calculate the loop integrals in the regularized theory and to remove the regulator afterwards to arrive at physical results. Renormalization group equations are a well established tool to describe the change of a given theory including quantum fluctuations with some flow parameter $k$, where $k$ parameterizes the regulator. We will assume that $k=0$ corresponds to a vanishing regulator. The ultimate goal then is to solve the flow equation, which gives the regularized theory as a function of $k$ and to go to the limit $k\rightarrow 0$ yielding the physical theory including quantum fluctuations.

In this chapter we describe the renormalization group formalism that we will use to further investigate the properties of the colored Hubbard model. However, the treatment of the formalism presented here is rather general and applies to any system with fermionic and bosonic degrees of freedom. The concept of investigating the flow of the effective average action has been introduced in \cite{renorm} and has been used to treat a wide range of problems in quantum field theory and statistical physics (for a review, see \cite{renormrev}). 

\section{Generalized fields and regularization}
We consider a theory with a fermionic field $\hat\psi^*$, $\hat\psi$, a real bosonic field $\hat w$ and a complex bosonic field $\hat u$, $\hat u^*$. The generalization to several fields is straightforward. We use generalized matrix notation by regarding the fields as vectors with a discrete label $\alpha$ (e.g. color, spin) and a position label $X$ or momentum label $Q$. In the same way, we introduce generalized matrices. We then use generalized matrix multiplication e.g. in the form 
\begin{align}
\hat\psi^\dagger A\hat\psi&=\sum_{Q,Q'}\sum_\alpha\hat\psi^*_\alpha(Q)A_{\alpha\beta}(Q,Q')\hat\psi_\beta(Q')\nonumber\\
(AB)_{\alpha\beta}(Q_1,Q_2)&=\sum_{Q}\sum_\gamma A_{\alpha\gamma}(Q_1,Q)B_{\gamma\beta}(Q,Q_2)
\end{align}
Similarly, Tr denotes a generalized trace.

Now define the generalized fields
\begin{equation}
\hat\chi=\begin{pmatrix}\hat u\\\hat u^*\\\hat w\\-\hat\psi\\\hat\psi^*\end{pmatrix},\quad\hat\chi^\dagger=\begin{pmatrix}\hat u^\dagger,&\hat u^T,&\hat w^\dagger,&\hat\psi^\dagger,&\hat\psi^T\end{pmatrix}
\end{equation} 
and the generalized sources
\begin{equation}
K=\begin{pmatrix}J\\J^*\\L\\\eta\\\eta^*\end{pmatrix},\quad K^\dagger=\begin{pmatrix}J^\dagger,&J^T,&L^\dagger,&-\eta^\dagger,&\eta^T\end{pmatrix}.
\end{equation}
$\eta$ and $\eta^*$ are fermionic sources (these sources enter $S_j^F$ in \eqref{eq:colpar:parmom}, if we had bothered to explicitly write it out). Note that in general $\hat w\neq\hat w^*$, although we introduced $\hat w$ as a real boson. But since our notation applies to both position and momentum space, we have to take care of the fact that $\hat w^*(X)=\hat w(X)$, but $\hat w^*(Q)=\hat w(-Q)$! The same applies to the real source $L$.  

The general partition function we want to consider is
\begin{equation}
Z={\cal N}(T)\int{\cal D}\hat\psi^*{\cal D}\hat\psi{\cal D}\hat u^*{\cal D}\hat u{\cal D}\hat w\,\exp(-S-S_j),
\end{equation}
where ${\cal N}(T)$ is some temperature dependent constant, $S$ is the action without sources and $S_j$ is a source term, which can be written in our matrix notation as  
\begin{align}
S_j&=-J^\dagger\hat u-J^T\hat u^*-L^\dagger\hat w-\eta^\dagger\hat\psi-\eta^T\hat\psi^*\nonumber\\
&=-K^\dagger\hat\chi=-K^T\hat\chi^*.
\end{align}

We regularize the theory by introducing a cutoff function $\Delta S_k[\hat\chi]$ so that
\begin{equation}
S[\hat\chi]\rightarrow S[\hat\chi]+\Delta S_k[\hat\chi].
\end{equation}
In the simple case of a theory with IR-divergencies only, the cutoff function could be a correction to the propagator terms in $S$, giving a $k$-dependent mass to the troublesome massless mode. For $k\rightarrow0$, we would then demand that this $k$-dependent mass vanishes, so that in this limit we recover the physical theory described by $S$ and a possible source term only. Specifically, we define the cutoff function by
\begin{equation}
\label{eq:genfi:delsk}
\Delta S_k[\hat\chi]=\frac{1}{2}\text{Tr}\,(R_k\hat\chi\hat\chi^\dagger).
\end{equation}
$R_k$ is a cutoff matrix, and again we stress that this equation has to be read in the sense of generalized matrix notation. Note that the right hand side is not equal to $\frac{1}{2}\text{Tr}\,(\hat\chi^\dagger R_k\hat\chi)$ due to the fermionic fields contained in $\hat\chi$. We assume that $R_k$ is diagonal in the space of fields, so that we can set
\begin{equation}
R_k=\text{diag}\,(R_k^C,(R_k^C)^T,R_k^R,R_k^F,(R_k^F)^T).
\end{equation} 
$R_k^C$ serves as a cutoff matrix for the complex bosonic field, $R_k^R$ for the real bosonic field and $R_k^F$ for the fermionic field. With this simplification, we can write \eqref{eq:genfi:delsk} as
\begin{equation}
\Delta S_k[\hat\chi]=\text{Tr}\,\left(\hat u^\dagger R_k^C\hat u+\frac{1}{2}\hat w^\dagger R_k^R\hat w+\hat\psi^\dagger R_k^F\hat\psi\right).
\end{equation}

\section{The effective average action}
The regularized partition function now becomes
\begin{equation}
\label{eq:effavac:Zk}
Z_k={\cal N}(T)\int{\cal D}\hat\psi^*{\cal D}\hat\psi{\cal D}\hat u^*{\cal D}\hat u{\cal D}\hat w\,\exp(-S[\hat\chi]-\Delta S_k[\hat\chi]+K^\dagger\hat\chi)
\end{equation}
and we define
\begin{equation}
W_k=\ln Z_k.
\end{equation}
The ($k$-dependent) expectation values of the fields are then given by
\begin{equation}
\chi=\langle\hat\chi\rangle=\frac{\delta W_k}{\delta K^\dagger},\quad\chi^\dagger=\langle\hat\chi^\dagger\rangle=\frac{\delta W_k}{\delta K}.
\end{equation}
We can now define 
\begin{align}
\tilde\Gamma_k[\chi]&=-W_k+K^\dagger\chi\nonumber\\
\Gamma_k[\chi]&=-W_k+K^\dagger\chi-\Delta S_k[\chi].
\end{align}
Note that these definitions are completely analogous to those we presented in \eqref{sec:meanf:effpot} except the last for $\Gamma_k[\chi]$. The reason why we subtracted $\Delta S_k[\chi]$ in the last of these definitions will become clear in a moment. For $\tilde\Gamma_k[\chi]$, which is the Legendre transform of $W_k$, we have
\begin{equation}
\label{eq:effavac:sourc}
\frac{\delta\tilde\Gamma_k[\chi]}{\delta\chi}=K^\dagger M,\quad\frac{\delta\tilde\Gamma_k[\chi]}{\delta\chi^\dagger}=MK
\end{equation}
with $M=\text{diag}\,(1,1,1,-1,-1)$. 

Some remarks are in order concerning the interpretation of $\Gamma_k[\chi]$. This quantity is called the {\em effective average action}. We will now assume that 
\begin{equation}
\label{eq:effavac:Rkconstr}
\lim_{k\rightarrow0}R_k=0,\quad\lim_{k\rightarrow\Lambda}R_k=\infty,
\end{equation}
where $\Lambda$ is either a natural UV-cutoff of the theory (in our case something proportional to the inverse lattice spacing) or $\infty$ for theories with UV-divergencies. It is clear that then
\begin{equation}
\lim_{k\rightarrow0}\Gamma_k[\chi]=\Gamma[\chi],
\end{equation}
where $\Gamma[\chi]$ is the full effective action, since by letting $R_k\rightarrow0$, we remove the cutoff function that causes $\Gamma_k$ to differ from $\Gamma$. In \eqref{sec:meanf:effpot} we already discussed the interpretation of the full effective action: The effective action yields --- by use of the classical action principle --- the equations of motion for the expectation values of the fields of the theory with all quantum fluctuations included. 

We now turn our attention to the limit $k\rightarrow\Lambda$. We start by rewriting \eqref{eq:effavac:Zk} in the form
\begin{align}
\label{eq:effavac:expgam}
\exp(-\Gamma_k[\chi])&={\cal N}(T)\int{\cal D}\hat\psi^*{\cal D}\hat\psi{\cal D}\hat u^*{\cal D}\hat u{\cal D}\hat w\nonumber\\
&\exp\left(-K^\dagger\chi+\Delta S_k[\chi]-S[\hat\chi]+K^\dagger\hat\chi-\frac{1}{2}\text{Tr}\,(R_k\hat\chi\hat\chi^\dagger)\right),
\end{align} 
where on the right hand side the sources are understood to be functions of $\chi$. Explicitly, we find by starting from \eqref{eq:effavac:sourc} that
\begin{equation}
K^\dagger=\frac{\delta\Gamma_k}{\delta\chi}M+\chi^\dagger R_kM.
\end{equation}
Inserting this in \eqref{eq:effavac:expgam} and rearranging terms, we find
\begin{align}
\exp(-\Gamma_k[\chi])&={\cal N}(T)\int{\cal D}\hat\psi^*{\cal D}\hat\psi{\cal D}\hat u^*{\cal D}\hat u{\cal D}\hat w\nonumber\\
&\exp\left(-S[\hat\chi]+\frac{\delta\Gamma_k}{\delta\chi}M(\hat\chi-\chi)\right)\exp\left(-\frac{1}{2}(\hat\chi^\dagger-\chi^\dagger)R_kM(\hat\chi-\chi)\right).
\end{align}
The second exponential vanishes for $R_k\rightarrow\infty$ unless $\hat\chi=\chi$, so that it effectively acts as a $\delta$-function and we obtain
\begin{equation}
\lim_{k\rightarrow\Lambda}\exp(-\Gamma_k[\chi])=\text{const.}\cdot{\cal N}(T)\exp(-S[\chi]).
\end{equation}
Up to some irrelevant constants, we find that $\Gamma_k[\chi]$ approaches the classical action $S[\chi]$ with the fields replaced by their expectation values as $k\to\Lambda$. Note that to achieve this nice property of $\Gamma_k[\chi]$ it was necessary to  subtract $\Delta S_k[\chi]$ in the definition of $\Gamma_k[\chi]$. 

In conclusion, the effective average action interpolates between the known classical action and the unknown effective action. Neither the initial value $\Gamma_{k\to\Lambda}$ nor the final value $\Gamma_{k\to0}$ depends on how we chose to define $R_k$ --- it suffices that $R_k$ meets the constraints \eqref{eq:effavac:Rkconstr}. However, it is clear that the interpretation of $\Gamma_k$ for some finite $k$ depends on the specific form of $R_k$. 

One very illuminating choice of the cutoff function is a simple sharp momentum cutoff $R_k\sim k^2\theta(k^2-q^2)$. This ansatz meets the conditions \eqref{eq:effavac:Rkconstr} for $\Lambda\to\infty$. This kind of cutoff does not influence the momentum modes with $q^2>k^2$, but gives a mass $k^2$ to the momentum modes with $q^2<k^2$. The propagation of these low momentum modes is suppressed by the mass, so that in $\Gamma_k$ quantum fluctuations of the high momentum modes only are integrated out, yielding an effective theory at scale $k$ for the propagation of the low momentum modes. In position space, this means that by lowering $k$ we average over larger and larger regions, integrating out the short range fluctuations and building up a an effective theory for long range fluctuations only. 

It is a well known feature of the effective action that it preserves the symmetries of the original action in the sense that the same symmetry operations acting on the fields in the original action applied to the field expectation values leave the effective action invariant, if these symmetry transformations are {\em linear}. This is the case for all the symmetries we consider here, so that we expect the effective action to respect the same symmetries we discussed before for the action of the Hubbard model. However, whether $\Gamma_k[\chi]$ respects these symmetries depends on the choice of $\Delta S_k[\chi]$. It is convenient to choose $\Delta S_k[\chi]$ such that it also respects the symmetries of the theory. Then $\Gamma_k[\chi]$ is invariant under the symmetry transformations for all $k$. The advantage of $\Gamma_k$ being invariant under the symmetry transformations is that we will have to write down an ansatz for $\Gamma_k$ to solve the flow equation calculated in the next section. During the flow an infinite number of terms contributing to $\Gamma_k$ will be generated, so that we need some guidance in selecting terms that we include in our truncation. The fact that only terms respecting the symmetries of the original theory are allowed in $\Gamma_k$ constrains the number of terms we may or may not include in our ansatz, which makes the motivation of truncation schemes much easier. Another reason for demanding invariance of $\Gamma_k$ is that we are often interested in symmetry breaking properties of the theory {\em at some scale $k$} and not only in the limit $k\to0$. Of course, if $\Gamma_k$ breaks the symmetries at finite $k$ explicitly, spontaneous symmetry breaking properties are completely obscured. 

\section{The flow equation for the effective average action}
In this section we derive an exact flow equation for $\Gamma_k[\chi]$ that can be used to calculate the effective average action.

We start by noting that
\begin{equation}
\frac{\delta^2W_k}{\delta K^\dagger\delta K}=\left<\hat\chi\hat\chi^\dagger\right>-\left<\hat\chi\right>\left<\hat\chi^\dagger\right>,
\end{equation}
which is the connected 2-point function, and
\begin{equation}
\label{eq:renor:invprop}
\frac{\delta^2W_k}{\delta K^\dagger\delta K}=M\left(\frac{\delta^2\tilde\Gamma_k}{\delta\chi^\dagger\delta\chi}\right)^{-1}.
\end{equation}
Recall that in our notation, expressions like $\frac{\delta^2W_k}{\delta K^\dagger\delta K}$ are matrices. The derivation of the flow equation is now straightforward:
\begin{align}
\frac{d}{dk}\tilde\Gamma_k&=\frac{d}{dk}(-W_k+K^\dagger\chi)=-\partial_kW_k-\frac{d K^\dagger}{dk}\frac{\delta W_k}{\delta K^\dagger}+\frac{d K^\dagger}{dk}\chi=-\partial_kW_k\nonumber\\
&=\left<\partial_k\Delta S_k[\chi]\right>=\frac{1}{2}\text{Tr}\,((\partial_kR_k)\left<\hat\chi\hat\chi^\dagger\right>)\nonumber\\
&=\frac{1}{2}\text{Tr}\,\left((\partial_kR_k)\left(\frac{\delta^2W_k}{\delta K^\dagger\delta K}+\chi\chi^\dagger\right)\right)
\end{align}
so that
\begin{equation}
\label{eq:masterflow}
\frac{d}{dk}\Gamma_k=\frac{1}{2}\text{Tr}\,\left((\partial_kR_k)M\left(\frac{\delta^2\Gamma_k}{\delta\chi^\dagger\delta\chi}+R_k\right)^{-1}\right).
\end{equation}
This is our master equation for deriving all the flow equations we need. The ultimate goal is to solve this equation with an initial condition given by the classical action. In general it is not possible to find an exact solution to the flow equation. However, the fact that we know that $\Gamma_k$ equals the classical action at the beginning of the flow $k\to\Lambda$ allows to motivate sensible truncation schemes for possible solutions. 

We have to stress that up to this point no approximations entered the calculation, so that the result is exact. This is particularly interesting if one compares it to the perturbative one loop calculation of the effective action, which yields
\begin{equation}
\frac{d}{dk}\Gamma_k=\frac{1}{2}\text{Tr}\,\left((\partial_kR_k)M\left(\frac{\delta^2S[\chi]}{\delta\chi^\dagger\delta\chi}+R_k\right)^{-1}\right).
\end{equation}
This is completely the same, except that the exact propagator is replaced by the classical propagator here. The simple replacement of the classical propagator by the exact propagator in the one loop correction to the effective average action renders this equation exact, including arbitrary high loop orders and genuinely non perturbative effects. The formal similarity to perturbative one loop expressions allows to use well known calculation techniques and leads to results that can be simply interpreted in a diagrammatic language. For further discussion, applications and references, see \cite{renormrev}.

\chapter{Diagonalization of the propagator matrix}

At this point, we are in principle done with the preparations to extract the properties of the Hubbard model. We have derived a partition function \eqref{eq:colpar:parmom}, which contains the interesting degrees of freedom in an explicit way, and we have presented the formalism of renormalization group equations for the effective average action, which we can use to derive the effective action. However, as we already stated, \eqref{eq:masterflow} cannot be solved exactly. Instead, we have to invent some truncation for the effective average action, which consists of terms respecting the symmetries of our theory containing $k$-dependent variables (masses, couplings, wave function renormalization constants etc.). Since our theory is non renormalizable, it is clear that one cannot avoid approximations at this point. The question is: Which terms should be included in our truncation without making the truncation error too large? As we mentioned in the course of discussing symmetries of $\Gamma_k$, it is clear that the smaller the number of terms allowed by exact symmetries, the smaller the truncation error that we make by discarding terms. 

Unfortunately, in the present formulation of our theory \eqref{eq:colpar:parmom}, a huge number of terms in the effective action are allowed, all of which are of the same order of magnitude. Particularly, consider the propagator terms of the form
\begin{align}
\sum_{QQ'}u_{\beta a}^*(Q)P_{\beta\beta',ab}(Q,Q')u_{\beta'b}(Q'),\\
\sum_{QQ'}w_{\gamma a}(Q)P_{\gamma\gamma',ab}(Q,Q')w_{\gamma'b}(Q').
\end{align}
We know that --- due to $U(1)$-invariance --- propagator terms mixing complex and real bosons will not occur in the effective average action. Similarly, the spin singlet bosons and spin triplet bosons will not mix due to $SU(2)$-invariance. To keep things simple, we will set $\lambda_1\to0$, which because of \eqref{eq:parbos:hubyuk} is equivalent to discarding the bosons $\hat s$, $\hat c$ and $\hat t_{x/y}$ from our theory. Then we have three sets of bosons (real bosons in the spin singlet, real bosons in the spin triplet and complex bosons) with four boson species each. Every boson species occurs in $4$ different colors. This means that we face $3(4\cdot4)^2=768$ boson propagator matrix entries, non of which vanish due to symmetries. Of course, it is completely hopeless to include so many terms in any useful truncation scheme, but on the other hand, by considering only some of the terms and neglecting all others, we introduce large truncation errors.

The problem that we face here is rooted in the fact that we are not able to make use of the additional lattice symmetries (translations, rotations and reflections) to narrow down the number of allowed terms, since the bosons mix under these transformations. If we were given a set of bosons belonging to different inequivalent representations of these lattice symmetries, it would be clear that mixing between these different bosons could not happen. In this case, most of the terms of the propagator matrix vanish due to exact symmetries. The idea of this chapter is to find linear combinations of the existing bosons to make the corresponding states eigenstates of the symmetry transformations. Truncation schemes are then proposed for a theory which is written in terms of these new bosons.

The simplest way we found to attack this problem is to explicitly calculate the oneloop corrections to the bosonic propagators perturbatively and to diagonalize the resulting one loop improved propagator matrix. As a byproduct, we gain some information about the propagation of bosonic modes that will be useful when we define truncation schemes. Although we diagonalized the perturbative expressions only, it will turn out that the propagator matrix remains diagonal to all orders in perturbation theory and that in fact this transcription leads to bosonic states which are eigenstates of symmetries of our theory. Finally, we will be able to write down a partition function of our model with a new set of bosons that all belong to different inequivalent representations of translational symmetries and therefore do not mix. Much of the material in this chapter has been published in \cite{bick1}. 

\section{The diagonalization procedure}

\subsection{The oneloop calculation}

Our starting point is \eqref{eq:colpar:parmom}. We want to calculate the one loop corrections to the bosonic propagators. It is clear that these one loop corrections correspond to the diagrams
\[
\setlength{\unitlength}{1mm}
\begin{fmfgraph*}(80,50)
\fmfleft{i}
\fmfright{o}
\fmf{scalar,tension=3.0,label=$K$}{v1,i}
\fmf{scalar,tension=3.0,label=$K$}{o,v2}
\fmf{fermion,right,label=$Q$}{v1,v2}
\fmf{fermion,right,label=$K+Q$}{v2,v1}
\fmfv{label=$V$,label.angle=0}{v1}
\fmfv{label=$V$,label.angle=180}{v2}
\fmfv{label=$w$, label.angle=180}{i}
\fmfv{label=$w$, label.angle=0}{o}
\fmfdot{v1,v2}
\end{fmfgraph*}
\]
and 
\[
\setlength{\unitlength}{1mm}
\begin{fmfgraph*}(80,50)
\fmfleft{i}
\fmfright{o}
\fmf{scalar,tension=3.0,label=$K$}{v1,i}
\fmf{scalar,tension=3.0,label=$K$}{o,v2}
\fmf{fermion,left,label=$Q$}{v2,v1}
\fmf{fermion,right,label=$K-Q$}{v2,v1}
\fmfv{label=$V$,label.angle=0}{v1}
\fmfv{label=$V$,label.angle=180}{v2}
\fmfv{label=$u$, label.angle=180}{i}
\fmfv{label=$u^*$, label.angle=0}{o}
\fmfdot{v1,v2}
\end{fmfgraph*}
\]
where the solid lines denote fermions and the dashed lines bosons. Thus only fermionic fluctuations enter these one loop corrections. To calculate these, we expand the fields around their expectation values
\begin{gather}
\hat u_{\beta a}(Q)\rightarrow u_{\beta a}(Q),\quad\hat u_{\beta a}^*(Q)\rightarrow u_{\beta a}^*(Q),\quad\hat w_{\gamma a}(Q)\rightarrow w_{\gamma a}(Q)\nonumber\\
\hat\psi(Q)\rightarrow\psi(Q)+\delta\psi(Q),\quad\hat\psi^*(Q)\rightarrow\psi^*(Q)+\delta\psi^*(Q)
\end{gather}
and find for the action $S=S_F+S_B+S_Y+S_J$
\begin{equation}
S[\hat u,\hat u^*,\hat w,\hat\psi,\hat\psi^*]=\underbrace{S[u,u^*,w,\psi,\psi^*]}_{=S_0}+\underbrace{S[u,u^*,w,\delta\psi,\delta\psi^*]}_{=S_2}.
\end{equation}
The effective action (with bosonic fluctuations neglected) is then given by
\begin{equation}
\label{eq:oneloop:gamma1}
\Gamma[u,u^*,w,\psi,\psi^*]=\frac{2\pi^2{\cal V}\mu^2}{h_\rho^2}+S_0-S_J-\ln\int{\cal D}\delta\psi^*{\cal D}\delta\psi\exp(-S_2).
\end{equation}
Note the similarity to the meanfield calculation of chapter \ref{sec:meanf}. The only difference is that our bosonic background fields are not assumed to be homogeneous, and we will only consider corrections to the propagator (which, in the mean field case, would correspond to expanding the effective potential up to quadratic order in the fields). A similar transcription as in chapter \ref{sec:meanf}
\begin{equation}
\tilde\psi(Q)=\begin{pmatrix}\delta\psi(Q)\\\delta\psi^*(Q)\end{pmatrix}
\end{equation}
allows to perform the functional integral in \eqref{eq:oneloop:gamma1} and we get
\begin{equation}
\label{eq:oneloop:gamma2}
\Gamma[u,u^*,w,\psi,\psi^*]=\frac{2\pi^2{\cal V}\mu^2}{h_\rho^2}+S_0-S_J\underbrace{-\frac{1}{2}\ln\det\tilde P[u,u^*,w]}_{=\Delta\Gamma}
\end{equation}
with
\begin{gather}
\tilde P[u,u^*,w](Q,Q')=\underbrace{\begin{pmatrix}0&-(P^\psi)^T\\P^\psi&0\end{pmatrix}}_{=\tilde P_0}-\underbrace{\begin{pmatrix}C&-A^T\\A&B\end{pmatrix}}_{=\Delta\tilde P}\\
P^\psi(Q,Q')=(i\omega_n-2t(\cos(q_1/2)A_1+\cos(q_2/2)B_1))\delta(Q-Q')\\
\begin{aligned}
A&=\sum_\gamma\sum_cV^{w_\gamma}_{,c}(Q,Q')w_{\gamma c}(Q-Q')\\
B&=2\sum_\beta\sum_cV^{u_\beta}_{,c}(Q,Q')u_{\beta c}(Q+Q')\\
C&=2\sum_\beta\sum_cV^{u_\beta^*}_{,c}(Q,Q')u_{\beta c}^*(Q+Q').
\end{aligned}
\end{gather}
Recall that the vertex factors $V_{,c}$ are matrices in color and spinor space with components $V_{ab,c}$ in color space.  

We can now expand $\Delta\Gamma$ in numbers of bosonic fields
\begin{align}
\Delta\Gamma&=-\frac{1}{2}\text{Tr}\,\ln\tilde P=-\frac{1}{2}\text{Tr}\,\ln(\tilde P_0(1-\tilde P_0^{-1}\Delta\tilde P))\nonumber\\
&=-\frac{1}{2}\left(\text{Tr}\,\ln\tilde P_0-\text{Tr}(\tilde P_0^{-1}\Delta\tilde P)-\frac{1}{2}\text{Tr}\,(\tilde P_0^{-1}\Delta\tilde P)^2+\cdots\right).
\end{align}
Only the third term of this expansion contributes to the propagator corrections we want to calculate. Note that the trace involves summation over color, spin and generalized momentum.

More explicitly, we have for $\Delta\Gamma_2=1/4\text{Tr}(\tilde P_0^{-1}\Delta\tilde P)^2$
\begin{eqnarray}
  \label{eq:Gam_korr}
  \Delta\Gamma_2&=&\sum_{KQ}\Big[\sum_{\gamma\gamma'}\sum_{cc'}\frac{1}{2}w_{\gamma c}(-K)
  \,\text{Tr}\Big\{(P^\psi)^{-1}(Q)V_{,c}^{w_\gamma}(Q,K+Q)(P^\psi)^{-1}(K+Q)\nonumber\\ 
    &&\qquad V_{,c'}^{w_{\gamma'}}(K+Q,Q)\Big\}
  w_{\gamma' c'}(K)\nonumber\\
  &&-2\sum_{\beta\beta'}\sum_{cc'}u^*_{\beta c}(K)
  \,\text{Tr}\Big\{(P^\psi)^{-1}(Q)V_{,c}^{u_\beta}(Q,K-Q)((P^\psi)^{-1})^T(K-Q)\nonumber\\ 
  &&\qquad V_{,c'}^{u^*_{\beta'}}(K-Q,Q)\Big\}u_{\beta' c'}(K)\Big].
\end{eqnarray}
Note that we have written out the momentum sums, so that the trace only sums over color and spin.

From \eqref{eq:renor:invprop}  we know that the inverse propagator is simply given by the second derivative of $\Gamma$ with respect to the field expectation values. The one loop corrections to the inverse propagators are therefore given by
\begin{eqnarray}
  \label{eq:delta_GammaB2}
    \lefteqn{\Delta\Gamma^{(2)}_{w_{\gamma c}w_{\gamma'c'}}(K)}\nonumber\\
&=&\sum_Q\mbox{Tr}\left\{(P^\psi)^{-1}(Q)V_{,c}^{w_\gamma}(Q,K+Q)(P^\psi)^{-1}(K+Q)V_{,c'}^{w_{\gamma'}}(K+Q,Q)\right\}\nonumber\\
    \lefteqn{\Delta\Gamma^{(2)}_{u_{\beta c}u_{\beta'c'}}(K)}\\
&=&-2\,\sum_Q\mbox{Tr}\left\{(P^\psi)^{-1}(Q)V_{,c}^{u_\beta}(Q,K-Q)((P^\psi)^{-1})^T(K-Q)V_{,c'}^{u^*_{\beta'}}(K-Q,Q)\right\}\nonumber
\end{eqnarray}
where $\Delta\Gamma^{(2)}_{w_{\gamma c}w_{\gamma'c'}}(K)$ denotes the second derivative of $\Delta\Gamma_2$ with respect to the field expectation values.

The one loop calculations will be simplified if we make a slight change to our bosonization prescription. We replace the fermion bilinears $\tilde e$ and $\tilde d$ by
\begin{equation}
\label{eq:exeydef}
\tilde e_y=\frac{1}{2}(\tilde e+\tilde d),\quad\tilde e_x=\frac{1}{2}(\tilde e-\tilde d)
\end{equation}
The bosonization procedure gives the same result as before with the following modifications: Replace $\hat e$, $h_e$ by $\hat e_y$, $h_{e_y}$ and $\hat d$, $h_d$ by $\hat e_x$, $h_{e_x}$ everywhere. In \eqref{eq:parbos:hubyuk} we have to replace $2H_e=2H_d=H_{v_{x/y}}=6\lambda_3$ by $H_{e_{x/y}}=H_{v_{x/y}}=6\lambda_3$. The vertex factors are given by $V^{e_y}=1/2(V^e+V^d)$ and $V^{e_x}=1/2(V^e-V^d)$.

In principle one could now insert the vertex factors and fermionic propagators and proceed by calculating the traces for all possible $768$ propagator matrix entries (we actually did that, but it is not very illuminating to present it here). However, a simple transformation renders the propagator matrix diagonal in color space. We will discuss this transformation in the following section and present the results for the one loop corrections afterwards.

\subsection{Diagonalization in color space}
The idea to diagonalize the propagator matrix in color space is to note that for homogeneous fields, translations can be easily applied by color transformations. Particularly, we have
\begin{equation}
T_x b_a=\sum_b(A_1)_{ab}b_b,\quad T_y b_a=\sum_b(B_1)_{ab}b_b,
\end{equation}
where $b_a$ is any of our bosons with color $a$. It is clear that $T_{x/y}^2=1$ in this case and $A_1$ and $B_1$ commute. The group of transformations consisting of  $1$, $T_x$, $T_y$ and $T_xT_y$ is therefore isomorphic to $G=Z_2\times Z_2$, where $Z_2$ is the cyclic group of order $2$. $G$ is Abelian and possesses therefore $4$ classes. The number of inequivalent irreducible representations of any group is equal to the number of classes of the group, so that we have four inequivalent irreducible representations, which are necessarily one dimensional. A basis for these irreducible representations is easy to define. Simply take
\begin{align}
\label{eq:diagcol:barbb}
\bar b_1&=\frac{1}{4}(b_1+b_2+b_3+b_4),&\quad\bar b_2&=\frac{1}{4}(b_1-b_2+b_3-b_4),\nonumber\\
\bar b_3&=\frac{1}{4}(b_1+b_2-b_3-b_4),&\quad\bar b_4&=\frac{1}{4}(b_1-b_2-b_3+b_4).
\end{align} 
The representations of the elements of $G\cong\{1,T_x,T_y,T_xT_y\}$ in this basis are $\{1,1,1,1\}$ for $\bar b_1$, $\{1,-1,-1,1\}$ for $\bar b_2$, $\{1,1,-1,-1\}$ for $\bar b_3$ and $\{1,-1,1,-1\}$ for $\bar b_4$. Since we demand all terms in the effective action to respect the translation symmetries, it follows that all propagator matrix entries mixing different colors have to vanish. For example, 
\begin{equation}
\bar b^*_1P^{\bar b_1^*\bar b_2}\bar b_2=(T_x\bar b^*_1)P^{\bar b_1^*\bar b_2}(T_x\bar b_2)=-\bar b^*_1P^{\bar b_1^*\bar b_2}\bar b_2=0,
\end{equation}  
if $P^{\bar b_1^*\bar b_2}$ is the propagator matrix element coupling the bosons $\bar b_1^*$ and $\bar b_2$.

If we write \eqref{eq:diagcol:barbb} in the form $\bar b_a=\frac{1}{4}\sum_b M_{ab}b_b$, we have the inversion $b_a=\sum_b M_{ab}\bar b_b$. Inserting this in \eqref{eq:colpar:parmom}, we find that as functionals of the new bosons $\bar b$, the bosonic terms become
\begin{align}
S_B&=\sum_a\sum_Q\left(4\pi^2\sum_\beta\hat{\bar u}_{\beta a}^*(Q)\hat{\bar u}_{\beta a}(Q)+2\pi^2\sum_\gamma\hat{\bar w}_{\gamma a}(-Q)\hat{\bar w}_{\gamma a}(Q)\right)\nonumber\\
S_Y&=-\sum_{abc}\sum_{QQ'Q''}\delta(Q-Q'-Q'')\nonumber\\
&\qquad\biggl(\sum_\beta \left(\hat{\bar u}^*_{\beta c}(Q)\hat\psi_a^T(Q')V_{ab,c}^{\bar u_\beta^*}(Q',Q'')\hat\psi_b(Q'')+\hat{\bar u}_{\beta c}(Q)\hat\psi_a^\dagger(Q')V_{ab,c}^{\bar u_\beta}(Q',Q'')\hat\psi_b^*(Q'')\right)\nonumber\\
&\qquad+\sum_\gamma \hat{\bar w}_{\gamma c}(Q)\hat\psi_a^\dagger(Q')V_{ab,c}^{\bar w_\gamma}(Q',-Q'')\hat\psi_b(-Q'')\biggr),
\end{align}
where the new vertex factors $V^{\bar b}_{,a}$ are given by $V^{\bar b}_{,a}=\sum_bM_{ab}V^{b}_{,b}$.

The above argument only applies to homogeneous fields. In general, the propagator matrix elements for different colors will not vanish for non vanishing momentum. However, by choosing the Fourier transforms for the original bosonic fields as in \eqref{eq:app:fourtrans}, we can show --- by explicitly calculating the propagator matrix entries --- that indeed the transition to the fields $\bar b$ renders the propagator matrix diagonal in color space for arbitrary bosonic momentum.

\subsection{Color diagonal one loop results}
We present our results for the one loop corrections to the bosonic propagators of the new bosons $\bar b$ which can be derived from \eqref{eq:delta_GammaB2}. 

Consider the $12$ sets $\{\bar\rho,\bar p,{\bar q}_y,{\bar q}_x\}_a$, $\{{\vec{\bar a}},{\vec{\bar m}},{\vec{\bar g}_y},{\vec{\bar g}_x}\}_a$ and $\{\bar e_y,\bar e_x,{\bar v_y},{\bar v_x}\}_a$, where $a=1,\ldots,4$ is the color index. All propagator matrix elements connecting bosons belonging to different sets vanish (due to $U(1)$- and $SU(2)$-invariance and diagonalization in color space). The full propagator matrix is then block diagonal with the blocks given by $12$ $4\times4$ matrices. We call the one loop corrections to the $4\times4$-propagator matrix blocks corresponding to the four colors of $\{\bar\rho,\bar p,{\bar q}_y,{\bar q}_x\}_a$ $(\Delta\Gamma^R)_a$, $(\Delta\Gamma^S)_a$ for $\{{\vec{\bar a}},{\vec{\bar m}},{\vec{\bar g}_y},{\vec{\bar g}_x}\}_a$ and $(\Delta\Gamma^\chi)_a$ for $\{\bar e_y,\bar e_x,{\bar v_y},{\bar v_x}\}_a$. 

For the sets of real bosons in the spin singlet, we find for $(\Delta\Gamma^R)^{(2)}_a(K)$, $K=(\omega_m^B,\boldsymbol{k}))$
\begin{align}
\label{eq:coldigr}
(\Delta\Gamma^R)^{(2)}_1(K)&=T\int_{-\pi}^\pi\frac{d^2q}{(2\pi)^2}S^r(k_1,k_2)(g_-^r(+,+)+g_-^r(-,-)),\nonumber\\
(\Delta\Gamma^R)^{(2)}_2(K)&=-T\int_{-\pi}^\pi\frac{d^2q}{(2\pi)^2}S^r(k_1+2\pi,k_2+2\pi)(g_+^r(+,+)+g_+^r(-,-)),\nonumber\\
(\Delta\Gamma^R)^{(2)}_3(K)&=T\int_{-\pi}^\pi\frac{d^2q}{(2\pi)^2}S^r(k_1,k_2+2\pi)(g_-^r(+,-)+g_-^r(-,+)),\nonumber\\
(\Delta\Gamma^R)^{(2)}_4(K)&=-T\int_{-\pi}^\pi\frac{d^2q}{(2\pi)^2}S^r(k_1+2\pi,k_2)(g_+^r(+,-)+g_+^r(-,+)).
\end{align}
$S^r(k_1,k_2)$ is a $4\times4$-matrix given by
\begin{equation}
\label{eq:coldig:Sr}
S^r(k_1,k_2)=\begin{pmatrix}4h_\rho^2\text{cs}_1^2\text{cs}_2^2&h_\rho h_p\bar s_1\bar s_2&-2ih_\rho h_{q_y}\text{cs}_1^2\bar s_2&2ih_\rho h_{q_x}\bar s_1\text{cs}_2^2\\
h_\rho h_p\bar s_1\bar s_2&4h_p^2\text{sn}_1^2\text{sn}_2^2&-2ih_ph_{q_y}\bar s_1\text{sn}_2^2&2ih_ph_{q_x}\text{sn}_1^2\bar s_2\\
2ih_\rho h_{q_y}\text{cs}_1^2\bar s_2&2ih_ph_{q_y}\bar s_1\text{sn}_2^2&4h_{q_y}^2\text{cs}_1^2\text{sn}_2^2&-h_{q_y}h_{q_x}\bar s_1\bar s_2\\
-2ih_\rho h_{q_x}\bar s_1\text{cs}_2^2&-2ih_ph_{q_x}\text{sn}_1^2\bar s_2&-h_{q_y}h_{q_x}\bar s_1\bar s_2&4h_{q_x}^2\text{sn}_1^2\text{cs}_2^2
\end{pmatrix}
\end{equation}
with $\text{cs}_i=\cos(k_i/4)$, $\text{sn}_i=\sin(k_i/4)$ and $\bar s_i=\sin(k_i/2)$. The functions $g$ are Matsubara sums ($m,n\in\mathbb{Z}$, $\epsilon_i\in\{1,-1\}$):
\begin{gather}
  \label{eq:oneloop:gfunc}
  \begin{split}
  g^{r,c}_{\epsilon_3}(\epsilon_1,\epsilon_2) &= \sum_{n}
  \frac{(c_1+\epsilon_1 c_2)(c_1'+\epsilon_2 c_2')+\epsilon_3\omega\omega'}
  {[(c_1+\epsilon_1 c_2)^2+\omega^2][(c_1'+\epsilon_2 c_2')^2+{\omega'}^2]} \\
  &= (c_1+\epsilon_1 c_2)(c_1'+\epsilon_2 c_2')\frac{S_1(m,a_{\epsilon_1}, b_{\epsilon_2})}{(\pi T)^4}
  \pm \epsilon_3 \frac{S_2(m,a_{\epsilon_1}, b_{\epsilon_2})}{(\pi T)^2},
  \end{split}
\end{gather}
which depend on $m$, $\boldsymbol{k}$ and $\boldsymbol{q}$. The upper sign applies to real bosons (r) and the lower sign for complex bosons (c) (we will use it later when we write down the one loop results for the complex bosons). The sums $S_1$ and $S_2$ are
\begin{eqnarray}
  \label{eq:oneloop:sums}
  S_1\left(m,a,b\right) 
  := \sum_{n\in\mathbbm{Z}}\frac{1}{[(2n+1)^2+a^2][(2(n+m)+1)^2+b^2]} \nonumber\\
  = \frac{\pi}{2}
  \frac{b(4m^2-a^2+b^2)\tanh(\frac{\pi a}{2})+
    a(4m^2+a^2-b^2)\tanh(\frac{\pi b}{2})}
  {ab[4m^2+(a+b)^2][4m^2+(a-b)^2]},\\[.5cm]
  S_2\left(m,a,b\right) 
  := \sum_{n\in\mathbbm{Z}}\frac{(2n+1)(2(n+m)+1)}{[(2n+1)^2+a^2][(2(n+m)+1)^2+b^2]} \nonumber\\
  = \frac{\pi}{2}
  \frac{a(4m^2+a^2-b^2)\tanh(\frac{\pi a}{2})+
    b(4m^2-a^2+b^2)\tanh(\frac{\pi b}{2})}
  {[4m^2+(a+b)^2][4m^2+(a-b)^2]}.
\end{eqnarray}
The frequencies $\omega$, $\omega'$ appearing in the definition of $g$ read
\begin{equation}
  \omega = (2n+1)\pi T, \quad 
  \omega' = \left\{\begin{array}{ll} (2(m+n)+1)\pi T  &\text{for real bosons}\\ 
      (2(m-n)-1)\pi T & \text{for complex bosons.} \end{array}\right.
\end{equation}
For the arguments of $S_{1,2}$ we use the abbreviations ($\epsilon_i\in\{1,-1\}$)
\begin{eqnarray}
  a_{\epsilon_i} &=& (c_1 +\epsilon_ic_2 )/(\pi T)\nonumber\\
  b_{\epsilon_i} &=& (c'_1+\epsilon_ic'_2)/(\pi T),
\end{eqnarray}
where $c_i$ and $c_i'$ are given by
\begin{eqnarray}
  \label{cicsi}
  c_i &=& 2t\cos(q_i/2), \quad 
  c'_i = \left\{\begin{array}{ll} 2t\cos((k_i+q_i)/2) &\text{for real bosons}\\ 
      2t\cos((k_i-q_i)/2) & \text{for complex bosons}.
    \end{array}\right.
\end{eqnarray}

For the real bosons in the spin triplet, we get the same results with $h_\rho\to h_m$, $h_p\to h_a$ and $h_{q_{x/y}}\to h_{g_{x/y}}$.

The result for the complex bosons is similar. Define
\begin{equation}
\label{eq:coldig:Sc}
S^c(\boldsymbol{q},\boldsymbol{q}')=\begin{pmatrix}
-2h_{e_y}^2c_{1-}^2c_{2+}^2&\frac{1}{2}h_{e_y}h_{e_x}\bar c_1\bar c_2&ih_{e_y}h_{v_y}c_{1-}^2s_{2+}&-\frac{1}{2}ih_{e_y}h_{v_x}\bar s_1\bar c_2\\
\frac{1}{2}h_{e_y}h_{e_x}\bar c_1\bar c_2&-2h_{e_x}^2c_{1+}^2c_{2-}^2&-\frac{1}{2}ih_{e_x}h_{v_y}\bar c_1\bar s_2&ih_{e_x}h_{v_x}s_{1+}^2c_{2-}^2\\
-ih_{e_y}h_{v_y}c_{1-}^2s_{2+}&\frac{1}{2}ih_{e_x}h_{v_y}\bar c_1\bar s_2&-2h_{v_y}^2c_{1-}^2s_{2+}^2&\frac{1}{2}h_{v_x}h_{v_y}\bar s_1\bar s_2\\
\frac{1}{2}ih_{e_y}h_{v_x}\bar s_1\bar c_2&-ih_{e_x}h_{v_x}s_{1+}^2c_{2-}^2&\frac{1}{2}h_{v_x}h_{v_y}\bar s_1\bar s_2&-2h_{v_x}^2s_{1+}^2c_{2-}^2
\end{pmatrix}
\end{equation}
with $c_{i\pm}=\cos((q_i\pm q_i')/4)$, $s_{i\pm}=\sin((q_i\pm q_i')/4)$, $\bar c_i=\cos q_i+\cos q_i'$ and $\bar s_i=\sin q_i+\sin q_i'$. Then we have for $(\Delta\Gamma^\chi)_a^{(2)}(K)$
\begin{align}
\label{eq:coldigc}
(\Delta\Gamma^\chi)_1^{(2)}(K)&=T\int_{-\pi}^\pi\frac{d^2q}{(2\pi)^2}\left(S^c(\boldsymbol{q},\boldsymbol{k}-\boldsymbol{q})g_-^c(-,-)+A_3S^c(\boldsymbol{q},\boldsymbol{k}-\boldsymbol{q})A_3g_-^c(+,+)\right)\nonumber\\
(\Delta\Gamma^\chi)_2^{(2)}(K)&=-T\int_{-\pi}^\pi\frac{d^2q}{(2\pi)^2}(S^c(\boldsymbol{q}+2\pi(\boldsymbol{e}_1+\boldsymbol{e}_2),\boldsymbol{k}-\boldsymbol{q}-2\pi(\boldsymbol{e}_1+\boldsymbol{e}_2))g_+^c(-,-)\nonumber\\
&\qquad+A_3S^c(\boldsymbol{q}+2\pi(\boldsymbol{e}_1+\boldsymbol{e}_2),\boldsymbol{k}-\boldsymbol{q}-2\pi(\boldsymbol{e}_1+\boldsymbol{e}_2))A_3g_+^c(+,+))\nonumber\\
(\Delta\Gamma^\chi)_3^{(2)}(K)&=T\int_{-\pi}^\pi\frac{d^2q}{(2\pi)^2}(S^c(\boldsymbol{q}+2\pi\boldsymbol{e}_2,\boldsymbol{k}-\boldsymbol{q}-2\pi\boldsymbol{e}_2)g_-^c(+,-)\nonumber\\
&\qquad+A_3S^c(\boldsymbol{q}+2\pi\boldsymbol{e}_2,\boldsymbol{k}-\boldsymbol{q}-2\pi\boldsymbol{e}_2)A_3g_-^c(-,+))\nonumber\\
(\Delta\Gamma^\chi)_4^{(2)}(K)&=-T\int_{-\pi}^\pi\frac{d^2q}{(2\pi)^2}(S^c(\boldsymbol{q}+2\pi\boldsymbol{e}_1,\boldsymbol{k}-\boldsymbol{q}-2\pi\boldsymbol{e}_1)g_+^c(+,-)\nonumber\\
&\qquad+A_3S^c(\boldsymbol{q}+2\pi\boldsymbol{e}_1,\boldsymbol{k}-\boldsymbol{q}-2\pi\boldsymbol{e}_1)A_3g_+^c(-,+)).
\end{align}

\subsection{Discussion of the one loop expressions}
\label{sec:disoneloop}
\begin{figure}
\centering
\psfrag{delgam2}{\hspace*{-1cm}$(\Delta\Gamma^{\bar{a}_2})^{(2)}_{m=0}(\boldsymbol{k})$}
\psfrag{0}{$0$}
\psfrag{pi}{$\pi$}
\psfrag{mpi}{$-\pi$}
\psfrag{2pi}{$2\pi$}
\psfrag{m2pi}{$-2\pi$}
\psfrag{k1}{$k_1$}
\psfrag{k2}{$k_2$}
\includegraphics[scale=0.9]{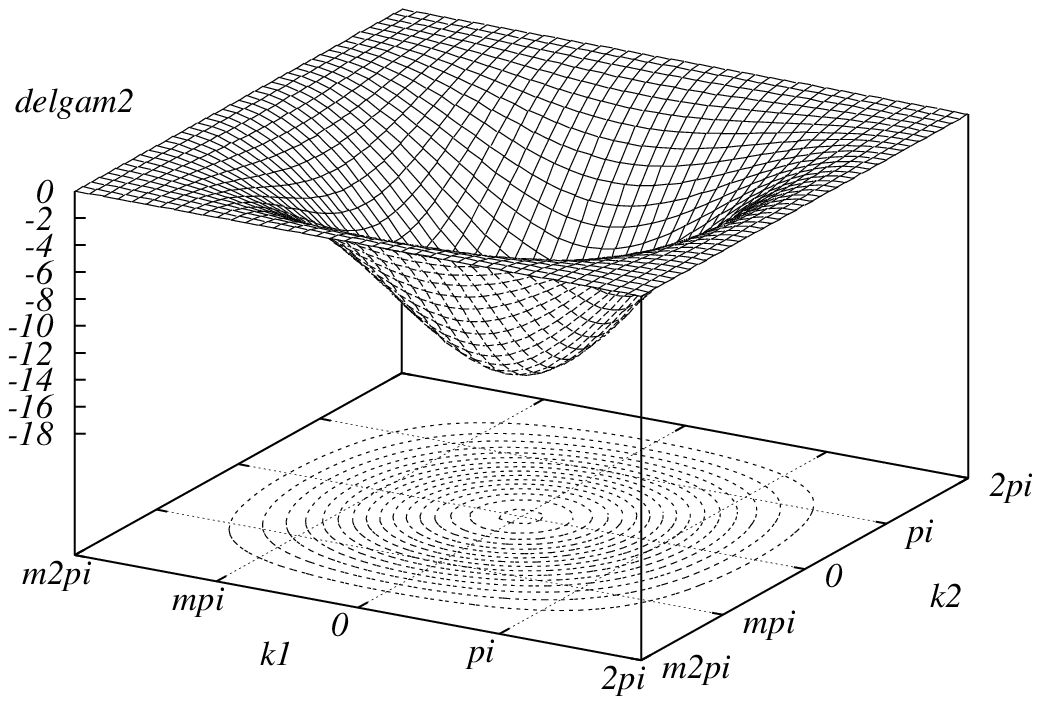}
\caption[The one loop correction to the bosonic propagator of $\overline{\vec a}_2$ for high temperature]{The second derivative of the one loop correction to the bosonic kinetic term in the effective action of $\overline{\vec a}_2$ for high temperature $T=1$. We have set $t=1$; the Yukawa couplings are at their Hubbard model values \eqref{eq:yukhubb} with $h_a^2=10$. We have plotted the Matsubara mode $m=0$.}
\label{fig:propR2hightemp}
\end{figure}
\begin{figure}
\centering
\psfrag{delgam2}{\hspace*{-1cm}$(\Delta\Gamma^{\bar{a}_2})^{(2)}_{m=0}(\boldsymbol{k})$}
\psfrag{0}{$0$}
\psfrag{pi}{$\pi$}
\psfrag{mpi}{$-\pi$}
\psfrag{2pi}{$2\pi$}
\psfrag{m2pi}{$-2\pi$}
\psfrag{k1}{$k_1$}
\psfrag{k2}{$k_2$}
\includegraphics[scale=0.9]{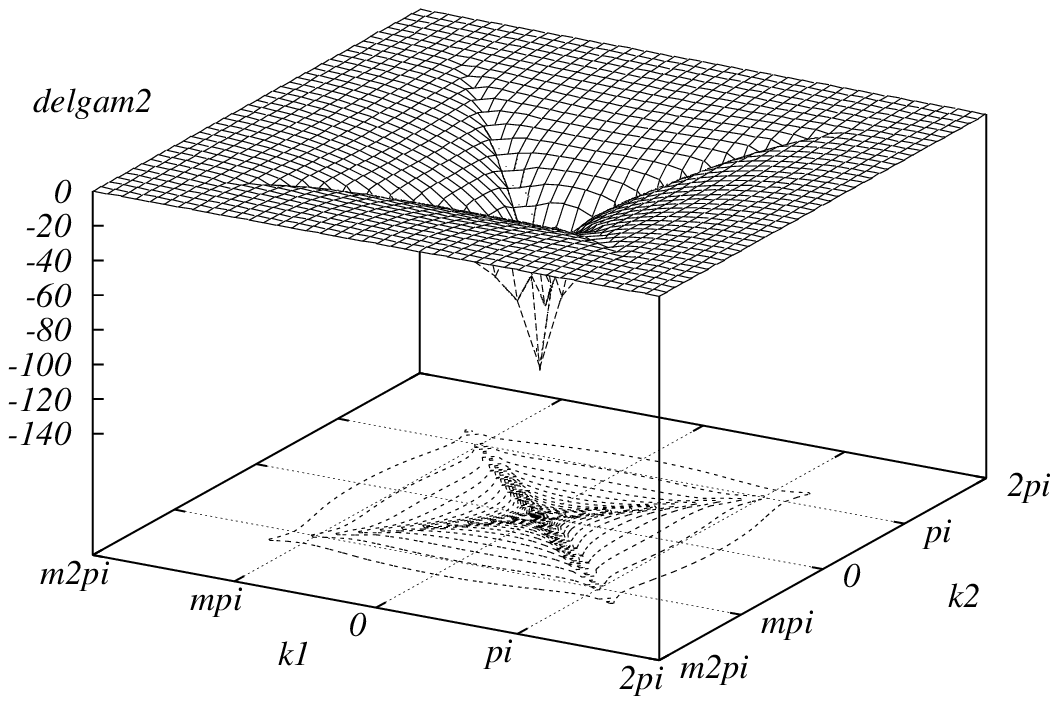}
\caption[The one loop correction to the bosonic propagator of $\overline{\vec a}_2$ for low temperature]{The second derivative of the one loop correction to the bosonic kinetic term in the effective action of $\overline{\vec a}_2$ for low temperature $T=0.01$. We have set $t=1$; the Yukawa couplings are at their Hubbard model values \eqref{eq:yukhubb} with $h_a^2=10$. We have plotted the Matsubara mode $m=0$.}
\label{fig:propR2lowtemp}
\end{figure}
As a byproduct of the diagonalization procedure of the propagator matrix we have at hand the explicit one loop corrections to the bosonic kinetic terms. We here discuss the generalized momentum dependence of the one loop expressions given by \eqref{eq:coldigr} and \eqref{eq:coldigc}. As the generic case, we show plots of these expressions as functions of the bosonic momentum for high and low temperature for the boson $\overline{\vec a}_2$. The most remarkable features of the one loop expressions are the following:
\begin{itemize}
\item The one loop expressions are periodic in $4\pi$, not in $2\pi$ as one would naively expect from the boundaries of the integrations. In fact, even if one takes into account the periodicity properties of the bosonic fields, it turns out that no single bosonic propagator term in the effective action is periodic in $2\pi$. This leads to interesting consequences discussed in the last section of this chapter.
\item The plots show that the one loop correction becomes minimal at $\boldsymbol{k}=0$, which holds true for any temperature. Since the classical propagators are momentum independent, the momentum dependence of the one loop corrected propagator is given by the one loop correction term and our result means that the propagation of momentum modes near vanishing momentum is facilitated in comparison to the higher momentum modes. The zero momentum mode corresponds to a spatially homogeneous field. We conclude that an approximation of the effective action that mainly keeps the dependence on spatially homogeneous fields will be justified, as the dynamics of the system is dominated by these homogeneous fields.
\item For high temperature the momentum dependence is quite simple. It is mainly given by $-(\cos(k_1/4)\cos(k_2/4))^2$ in the case of $\overline{\vec a}_2$. For low temperature, the momentum dependence becomes more complicated. Note the appearance of the cross shaped region in fig. \eqref{fig:propR2lowtemp}, where the momentum dependence becomes non analytical for low temperature. This cross exactly corresponds to the  Fermi surface in the colored Hubbard model, which makes perfect sense.
\end{itemize}

The high temperature limit of the one loop expression can be easily calculated analytically. To do this, we can expand the one loop expressions with respect to $1/T$. Using
\begin{equation}
\label{eq:hightemplim}
g_{\epsilon_3}^r(\epsilon_1,\epsilon_2)=\frac{\epsilon_3}{(2T)^2}\delta_{m0}+{\cal O}(1/T^4),\quad g_{\epsilon_3}^c(\epsilon_1,\epsilon_2)=-\frac{\epsilon_3}{(2T)^2}\delta_{m0}+{\cal O}(1/T^4)
\end{equation}
we see that in this limit the momentum dependence of the propagators is completely dominated by the $m=0$-Matsubara mode. For low temperature, where we expand the sums $S_1$ and $S_2$  with respect to $T$, we note that $m$ enters only via the product $(mT)^2$, so that the first $m$-dependent term in the expansion with respect to $T$ is $\propto m^2$.  

\subsection{The final diagonalization step}
\label{sec:finaldig}
Although we have greatly reduced the number of non vanishing propagator matrix entries, we can do even better by diagonalizing the remaining $4\times4$-matrices \eqref{eq:coldig:Sr} and \eqref{eq:coldig:Sc}. Define the transformation matrices
\begin{equation}
U^r(k_1,k_2)=\begin{pmatrix}h_\rho\text{cs}_1\text{cs}_2&-h_p\text{sn}_1\text{sn}_2&ih_{q_y}\text{sn}_2&-ih_{q_x}\text{sn}_1\\
h_p\text{sn}_1\text{sn}_2&h_\rho\text{cs}_1\text{cs}_2&0&0\\
ih_{q_y}\text{cs}_1\text{sn}_2&0&h_\rho\text{cs}_2&0\\
-ih_{q_x}\text{sn}_1\text{cn}_2&0&0&h_\rho\text{cs}_1\end{pmatrix}
\end{equation}
and
\begin{equation}
U^c(k_1,k_2)=\begin{pmatrix}h_{e_y}\text{cs}_2&0&ih_{v_y}\text{sn}_2&0\\
0&h_{e_x}\text{cs}_1&0&ih_{v_x}\text{sn}_1\\
ih_{v_y}\text{sn}_2&0&h_{e_y}\text{cs}_2&0\\
0&ih_{v_x}\text{sn}_1&0&h_{e_x}\text{cs}_1\end{pmatrix}.
\end{equation}
Then the transformations
\begin{align}
U_1^r(k_1,k_2)&=U^r(k_1,k_2)&U_1^c(k_1,k_2)&=U^c(k_1,k_2)\nonumber\\
U_2^r(k_1,k_2)&=\left.U^r(k_1,k_2)\right|_{\begin{subarray}{ll}\text{cs}_1\to-i\,\text{sn}_1\\\text{cs}_2\to i\,\text{sn}_2\\\text{sn}_1\to i\,\text{cs}_1\\\text{sn}_2\to-i\,\text{cs}_2\end{subarray}}&U_2^c(k_1,k_2)&=\left.U^c(k_1,k_2)\right|_{\begin{subarray}{ll}\text{cs}_j\to i\,\text{sn}_j\\\text{sn}_j\to-i\,\text{cs}_j\end{subarray}}\nonumber\\
U_3^r(k_1,k_2)&=\left.U^r(k_1,k_2)\right|_{\begin{subarray}{ll}\text{cs}_2\to i\,\text{sn}_2\\\text{sn}_2\to-i\,\text{cs}_2\end{subarray}}&U_3^c(k_1,k_2)&=\left.U^c(k_1,k_2)\right|_{\begin{subarray}{ll}\text{cs}_2\to i\,\text{sn}_2\\\text{sn}_2\to-i\,\text{cs}_2\end{subarray}}\nonumber\\
U_4^r(k_1,k_2)&=\left.U^r(k_1,k_2)\right|_{\begin{subarray}{ll}\text{cs}_1\to-i\,\text{sn}_1\\\text{sn}_1\to i\,\text{cs}_1\end{subarray}}&U_4^c(k_1,k_2)&=\left.U^c(k_1,k_2)\right|_{\begin{subarray}{ll}\text{cs}_1\to i\,\text{sn}_1\\\text{sn}_1\to-i\,\text{cs}_1\end{subarray}}
\end{align}
render the one loop propagator matrices diagonal for the real bosons and block diagonal for the complex bosons. More explicitly, we have 
\begin{multline}
\bar S^r(k_1,k_2)=U_1^r(-k_1,-k_2)^TS^r(k_1,k_2)U_1^r(k_1,k_2)=\\
\text{diag}\left(4\left(\text{cs}_1^2(h_\rho^2\text{cs}_2^2+h_{q_y}^2\text{sn}_2^2)+\text{sn}_1^2(h_p^2\text{sn}_2^2+h_{q_x}^2\text{cs}_2^2)\right)^2,0,0,0\right),\nonumber
\end{multline}
and
\begin{align}
\label{eq:oneloopSr}
U_2^r(-k_1,-k_2)^TS^r(k_1+2\pi,k_2+2\pi)U_2^r(k_1,k_2)&=\bar S^r(k_1+2\pi,k_2+2\pi)\nonumber\\
U_3^r(-k_1,-k_2)^TS^r(k_1,k_2+2\pi)U_3^r(k_1,k_2)&=\bar S^r(k_1,k_2+2\pi)\nonumber\\
U_4^r(-k_1,-k_2)^TS^r(k_1+2\pi,k_2)U_4^r(k_1,k_2)&=\bar S^r(k_1+2\pi,k_2)
\end{align}
for the real bosons and
\begin{multline}
\bar S^c(\boldsymbol{q},\boldsymbol{k}-\boldsymbol{q})=U_1^c(k_1,k_2)^\dagger S^c(\boldsymbol{q},\boldsymbol{k}-\boldsymbol{q})U_1^c(k_1,k_2)=\\
\begin{pmatrix}
-2c_{1-}^2\left(h_{e_y}^2\text{cs}_2^2+h_{v_y}^2\text{sn}_2^2\right)^2&\begin{aligned}2c_{1-}c_{2-}\left(h_{e_y}^2\text{cs}_2^2+h_{v_y}^2\text{sn}_2^2\right)\\\left(h_{e_x}^2\text{cs}_1^2+h_{v_x}^2\text{sn}_1^2\right)\end{aligned}&0&0\\
\begin{aligned}2c_{1-}c_{2-}\left(h_{e_y}^2\text{cs}_2^2+h_{v_y}^2\text{sn}_2^2\right)\\\left(h_{e_x}^2\text{cs}_1^2+h_{v_x}^2\text{sn}_1^2\right)\end{aligned}&-2c_{2-}^2\left(h_{e_x}^2\text{cs}_1^2+h_{v_x}^2\text{sn}_1^2\right)^2&0&0\\
0&0&0&0&\\0&0&0&0&
\end{pmatrix},\nonumber
\end{multline}
\begin{multline}
U_2^c(k_1,k_2)^\dagger S^c(\boldsymbol{q}+2\pi(\boldsymbol{e}_1+\boldsymbol{e}_2),\boldsymbol{k}-\boldsymbol{q}-2\pi(\boldsymbol{e}_1+\boldsymbol{e}_2))U_2^r(k_1,k_2)=\\
\bar S^c(\boldsymbol{q}+2\pi(\boldsymbol{e}_1+\boldsymbol{e}_2),\boldsymbol{k}-\boldsymbol{q}-2\pi(\boldsymbol{e}_1+\boldsymbol{e}_2))\nonumber
\end{multline}
\begin{multline}
U_3^c(k_1,k_2)^\dagger S^c(\boldsymbol{q}+2\pi\boldsymbol{e}_2,\boldsymbol{k}-\boldsymbol{q}-2\pi\boldsymbol{e}_2)U_3^r(k_1,k_2)=\\\bar S^c(\boldsymbol{q}+2\pi\boldsymbol{e}_2,\boldsymbol{k}-\boldsymbol{q}-2\pi\boldsymbol{e}_2)\nonumber
\end{multline}
\begin{multline}
\label{eq:oneloopSc}
U_4^c(k_1,k_2)^\dagger S^c(\boldsymbol{q}+2\pi\boldsymbol{e}_1,\boldsymbol{k}-\boldsymbol{q}-2\pi\boldsymbol{e}_1)U_4^r(k_1,k_2)=\\\bar S^c(\boldsymbol{q}+2\pi\boldsymbol{e}_1,\boldsymbol{k}-\boldsymbol{q}-2\pi\boldsymbol{e}_1)
\end{multline}
for the complex bosons. 

The next step is to define the set of bosons corresponding to the transformations we performed in order to diagonalize the propagator matrix and to express the partition function with respect to these new bosons. Define the new bosons $R_a$, $\vec s_a$ and $\chi_a$ by
\begin{align}
\begin{pmatrix}\bar\rho,&\bar p,&\bar q_y,&\bar q_x\end{pmatrix}_a^T(K)&=U_a^r(K)R_a(K)\nonumber\\
\begin{pmatrix}\vec{\bar m},&\vec{\bar a},&\vec{\bar g}_y,&\vec{\bar g}_x\end{pmatrix}_a^T(K)&=U_a^r(K)\vec s_a(K)\nonumber\\
\begin{pmatrix}\bar e,&\bar d,&\bar v_y,&\bar v_x\end{pmatrix}_a^T(K)&=U_a^c(K)\chi_a(K).
\end{align}
Here the new bosons $R_a$, $\vec s_a$ and $\chi_a$ are vectors with four components {\em for each color} $a$. We have written these transformation rules for the expectation values, but the same applies for the fields $\hat R_a$, $\hat{\vec s}_a$ and $\hat\chi_a$. Then the one loop correction to the propagator matrix for these new bosons has the form calculated above. In analogy to the color transformation in the last section, the vertex factors of the theory with the new bosons are given by linear combinations of the old vertex factors
\begin{equation}
\begin{pmatrix}V_1^{R_a}\\V_2^{R_a}\\V_3^{R_a}\\V_4^{R_a}\end{pmatrix}=(U_a^r)^T\begin{pmatrix}V_{,a}^{\bar\rho}\\V_{,a}^{\bar p}\\V_{,a}^{\bar q_y}\\V_{,a}^{\bar q_x}\end{pmatrix}
\end{equation}   
and similarly for $\hat{\vec s}_a$ and $\hat\chi_a$. Again, be aware of the notation: For any fixed color $a$, the old vertex factors $V_{,a}^{\bar\rho}$, etc. are $4\times4$-matrices in color space. The different vertex factors for $\bar\rho$, $\bar p$ and $\bar q_{y/x}$ are linearly combined by the matrix $(U_a^r)^T$ to give the new vertex factors $V_i^{R_a}$, which are again $4\times4$-matrices. The vertex factors can be calculated explicitly, and it turns out that
\begin{equation}
V_2^{R_a}=V_3^{R_a}=V_4^{R_a}=V_2^{\vec s_a}=V_3^{\vec s_a}=V_4^{\vec s_a}=V_3^{\chi_a}=V_4^{\chi_a}=0\quad\forall a.
\end{equation}
This leads to a great simplification of our formalism, since all the bosons with vanishing vertex factors decouple from the fermionic sector of the theory and can be integrated out! The theory depends only on the remaining set of bosons, i.e. on four real bosons in the spin singlet, four real bosons in the spin triplet and eight complex bosons. For each vector $R_a$ with fixed $a$, one boson remains that we again call $R_a$, but $R_a$ now understood to represent exactly one boson. In the same way, $\vec s_a$ will denote one boson for each $a$, and $\chi_a$ denotes a vector with two bosons for each $a$. We will present the complete expression for the partition function with the new bosons in the following section.

\section{The final form of the partition function}
In the last section we diagonalized the propagator matrix in two steps, first in color space, and then in the spaces of boson species blocks for each color. We now present our final form of the partition function that we will actually use for renormalization group calculations. Since the stepwise diagonalization procedure is very error prone, we show that by starting with our final form of the partition function, we can reproduce the Hubbard model by integrating out the bosons and inserting the Hubbard model values for the Yukawa couplings. The rest of this section will discuss the partition function and the one loop expressions.

\subsection{The partition function}
Our final version of the partition function is
\begin{equation}
\label{eq:partfunc}
Z = {\cal N}(T)\int D\hat\psi^* D\hat\psi D\hat\chi^* D\hat\chi D\hat R\, D\hat{\vec s}\exp\left\{-(S_{kin}^F+S_{kin}^B+S_Y+S_j)\right\}.
\end{equation}
The real fields $\hat R$, $\hat{\vec s}$ as well as the complex fields 
\begin{equation}
\hat\chi^*=\begin{pmatrix}\hat\chi^*_x\\ \hat\chi^*_y\end{pmatrix},\qquad
\hat\chi  =\begin{pmatrix}\hat\chi_x  \\ \hat\chi_y\end{pmatrix}
\end{equation}
are understood to carry a color-index $a=1\ldots4$. The terms of the action read
\begin{equation}
\label{eq:action}
\begin{split}
S_{kin}^F &= \sum_Q\sum_{ab}\hat\psi_a^\dagger(Q)P_{ab}^\psi(Q)\hat\psi_b(Q),\\
S_{kin}^B &= \sum_K\sum_{ab} \Big[\frac{1}{2}\hat R_a(-K)P_{ab}^R(K)\hat R_b(K)+\frac{1}{2}\hat{\vec s}_a(-K)P_{ab}^{\vec s}(K)\hat{\vec s}_b(K)\\
          &                       \hspace{2cm}+\hat\chi^\dagger_a(K)P_{ab}^\chi(K)\hat\chi_b(K)\Big],\\
S_Y       &= -\sum_{KQQ'}\sum_{abc}\Big[\delta(K-Q+Q')\Big(\hat R_c(K)\hat\psi_a^\dagger(Q)V^R_{ab,c}(K)\hat\psi_b(Q')\\
          &                                           \hspace{5.5cm}+\hat{\vec s}_c(K)\hat\psi_a^\dagger(Q)V^{\vec s}_{ab,c}(K)\hat\psi_b(Q')\Big)\\
          &                             \hspace{2cm}+\delta(K-Q-Q')\Big(\hat\chi^\dagger_c(K)\hat\psi_a^T(Q)V^{\chi^*}_{ab,c}(Q,Q')\hat\psi_b(Q')\\
          &                                           \hspace{5.5cm}+\hat\chi^T_c(K)\hat\psi_a^\dagger(Q)V^\chi_{ab,c}(Q,Q')\hat\psi^*_b(Q')\Big)\Big],\\
S_j       &= -\sum_Q\sum_c\Big[L^R_c(-Q)\hat R_c(Q)+L^{\vec s}_c(-Q)\hat{\vec s}_c(Q)+{L_c^{\chi^*}}^\dagger(Q)\chi_c(Q)+{L_c^{\chi}}^T(Q)\chi^*_c(Q)\\
          &               \hspace{2cm}+\eta_c^*(Q)\hat\psi_c(Q)+\eta_c(Q)\hat\psi^*_c(Q)\Big].
\end{split}
\end{equation}
The propagator matrices are ($B$ stands for $R$, $\vec s$ or $\chi$)
\begin{equation}
  \label{eq:propagators}
  \begin{split}
    P_{ab}^\psi(Q)                 &= \left[i\omega_n-2t\left(\cos(q_1/2)A_1
        +\cos(q_2/2)B_1\right)\right]_{ab},\\
    P_{ab}^B(K)                    &= (2\pi)^2P^B_a(K)\delta_{ab}
  \end{split}
\end{equation}
with
\begin{equation}
  \label{eq:massmatrixreal}
  \begin{split}
    P^R_1(K)        &= h_\rho^2\cos^2(k_1/4)\cos^2(k_2/4)+h_p^2\sin^2(k_1/4)\sin^2(k_2/4)\\
    &  \quad+h_{q_y}^2\cos^2(k_1/4)\sin^2(k_2/4)+h_{q_x}^2\sin^2(k_1/4)\cos^2(k_2/4)\\
    P^{\vec s}_1(K) &= h_{m}^2\cos^2(k_1/4)\cos^2(k_2/4)+h_{a}^2\sin^2(k_1/4)\sin^2(k_2/4)\\
    &  \quad+h_{g_y}^2\cos^2(k_1/4)\sin^2(k_2/4)+h_{g_x}^2\sin^2(k_1/4)\cos^2(k_2/4)\end{split}
\end{equation}
for the real bosons and
\begin{equation}
  \label{eq:massmatrixcomplex}
  \begin{aligned}
    P^\chi_a(K) &= \begin{pmatrix} P_a^y(K) & 0 \\ 0 & P_a^x(K)\end{pmatrix},\\
    P_1^x(K)    &= h_{e_x}^2\cos^2(k_1/4)+h_{v_x}^2\sin^2(k_1/4),\\
    P_1^y(K)    &= h_{e_y}^2\cos^2(k_2/4)+h_{v_y}^2\sin^2(k_2/4)
  \end{aligned}
\end{equation}
for the complex bosons. $P^R_a(K)$, $P_a^x(K)$ and $P_a^y(K)$ for $a=2\ldots 4$ are given by the expressions for $a=1$ with the replacements $\sin(k_i/4)\leftrightarrow\cos(k_i/4)$ for $a=2$, $\sin(k_2/4)\leftrightarrow\cos(k_2/4)$ for $a=3$ and $\sin(k_1/4)\leftrightarrow\cos(k_1/4)$ for $a=4$.

The vertices for the real bosons read
\begin{equation}
  \label{eq:vertexreal}
  \begin{aligned}
    V^R_{ab,1}(K) &= P^R_1(K)(A_0)_{ab}\otimes\sigma_0^{spin}, & V^R_{ab,2}(K) &= P^R_2(K)(A_3)_{ab}\otimes\sigma_0^{spin},\\
    V^R_{ab,3}(K) &= P^R_3(K)(D_0)_{ab}\otimes\sigma_0^{spin}, & V^R_{ab,4}(K) &= P^R_4(K)(D_3)_{ab}\otimes\sigma_0^{spin},\\
    V^{\vec s}_{ab,c}(K) &= \left.V^R_{ab,c}(K)\right|_{\sigma_0^{spin}\rightarrow\vec\sigma^{spin},\,P^R\to P^{\vec s}}
  \end{aligned}
\end{equation}
For the complex bosons, the vertices are
\begin{equation}
  \begin{aligned}
    \label{eq:vertexcomplex}
    V^{\chi^*}_{ab,1}(Q,Q') &= \frac{1}{2}\begin{pmatrix}\cos((q_1-q_1')/4)P_1^y(Q+Q')(A_1)_{ab}\\ \cos((q_2-q_2')/4)P_1^x(Q+Q')(B_1)_{ab}\end{pmatrix}\otimes (i\sigma_2),\\
    V^{\chi^*}_{ab,2}(Q,Q') &= \frac{1}{2}\begin{pmatrix}\sin((q_1-q_1')/4)P_2^y(Q+Q')(A_2)_{ab}\\ \sin((q_2-q_2')/4)P_2^x(Q+Q')(B_2)_{ab}\end{pmatrix}\otimes (i\sigma_2),\\
    V^{\chi^*}_{ab,3}(Q,Q') &= \frac{1}{2}\begin{pmatrix}-\cos((q_1-q_1')/4)P_3^y(Q+Q')(D_1)_{ab}\\ \sin((q_2-q_2')/4)P_3^x(Q+Q')(C_1)_{ab}\end{pmatrix}\otimes (i\sigma_2),\\
    V^{\chi^*}_{ab,4}(Q,Q') &= \frac{1}{2}\begin{pmatrix}-\sin((q_1-q_1')/4)P_4^y(Q+Q')(D_2)_{ab}\\ -\cos((q_2-q_2')/4)P_4^x(Q+Q')(C_2)_{ab}\end{pmatrix}\otimes (i\sigma_2),\\
    V^{\chi}_{ab,c}(Q,Q') &= -V^{\chi^*}_{ab,c}(-Q,-Q').
  \end{aligned}
\end{equation}
The matrices $A_\mu$, $B_\mu$, $C_\mu$ and $D_\mu$ are defined in the appendix \eqref{sec:matrices}.

\subsection{Equivalence to the Hubbard model}
In this section we will show that our new partition function \eqref{eq:partfunc} is equivalent to the one of the Hubbard model, if we set the Yukawa couplings to their Hubbard model values according to 
\begin{equation}
\label{eq:yukhubb}
  \begin{gathered}
    h_b^2=\frac{\pi^2}{3}H_b U\\
    \begin{aligned}
      H_\rho &= 3(\lambda_2-\lambda_3) & H_{\vec m} &= \lambda_2+3\lambda_3+1\\
      H_p    &= 3(\lambda_2+\lambda_3) & H_{\vec a} &= \lambda_2-3\lambda_3+1\\
      H_{q_x} = H_{q_y} &= 3\lambda_2  & H_{\vec g_x} = H_{\vec g_y} &= \lambda_2+1\\
    \end{aligned}\\
    H_{e_x}=H_{e_y}=H_{v_x}=H_{v_y}=6\lambda_3.
  \end{gathered}
\end{equation}
In this case, the propagators simplify to
\begin{equation}
\label{eq:prophubb}
  \begin{gathered}
    \begin{aligned}
      P_1^R(K) &= f^R(K,1,1) & P_1^{\vec s} &= f^{\vec s}(K,1,1)\\
      P_2^R(K) &= f^R(K,-1,1) & P_2^{\vec s} &= f^{\vec s}(K,-1,1)\\
      P_3^R(K) &= f^R(K,1,-1) & P_3^{\vec s} &= f^{\vec s}(K,1,-1)\\
      P_4^R(K) &= f^R(K,-1,-1) & P_4^{\vec s} &= f^{\vec s}(K,-1,-1)
    \end{aligned}\\
    P_a^\chi(K) = 2\pi^2\lambda_3\mathbbm{1}_2
  \end{gathered}
\end{equation}
with
\begin{equation}
  \begin{aligned}
    f^R(K,\epsilon_1,\epsilon_2)        &= \pi^2\left(\lambda_2-\epsilon_1\frac{\lambda_3}{2}\Big(\cos(k_1/2)+\epsilon_2\cos(k_2/2)\Big)\right)\\
    f^{\vec s}(K,\epsilon_1,\epsilon_2) &= \pi^2\left(\frac{1}{3}+\frac{\lambda_2}{3}+\epsilon_1\frac{\lambda_3}{2}\Big(\cos(k_1/2)+\epsilon_2\cos(k_2/2)\Big)\right).
  \end{aligned}
\end{equation}

The solution of the field equation for $\hat R_c(K)$ is
\begin{equation}
  \label{fieldeq_R}
  \hat R_c(K)=\frac{1}{4\pi^2P_c^R(K)}\sum_{QQ'}\sum_{ab}\delta(K+Q-Q')\hat\psi_a^\dagger(Q)V_{ab,c}^R(K)\hat\psi_b(Q')+\frac{L_c^R(K)}{4\pi^2P_c^R(K)}.
\end{equation}
Inserting this in the action, we see that only the source independent part of this solution contributes to the four fermion term in the purely fermionic theory. We will therefore set the sources equal to zero from now on (it is a simple task to check that the source dependent terms produce the correct source dependent terms of \eqref{eq:parbos:fermpar}). The contribution from the $R$-bosons to the four fermion term in the action is
\begin{equation}
  \label{eq:s4r}
  \begin{split}
    S_4^R &= \frac{1}{2}\sum_{QQ'\bar Q\bar Q'}\delta(Q-Q'+\bar Q-\bar Q')\Big(-\lambda_2([1111]+[2222]+[3333]+[4444])\\
    &  \quad+\lambda_3\cos((q_1-q_1')/2)([1122]+[3344])+\lambda_3\cos((q_2-q_2')/2)([1144]+[2233])\Big)
  \end{split}
\end{equation}
where we use the short hand notation
\begin{equation}
  \label{eq:abcd}
  [abcd]=\hat\psi_a^\dagger(Q)\hat\psi_b(Q')\hat\psi_c^\dagger(\bar Q)\hat\psi_d(\bar Q').
\end{equation}
We proceed similarly for the $\vec s$-bosons. To bring $S_4^{\vec s}$ into a form which can easily be compared with (\ref{eq:s4r}), we use the identity
\begin{equation}
  \begin{split}
    &\hat\psi_a^\dagger(Q)\vec\sigma\hat\psi_b(Q')\hat\psi_c^\dagger(\bar Q)\vec\sigma\hat\psi_d(\bar Q')\\
    &=-\hat\psi_a^\dagger(Q)\hat\psi_b(Q')\hat\psi_c^\dagger(\bar Q)\hat\psi_d(\bar Q')-2\hat\psi_a^\dagger(Q)\hat\psi_d(\bar Q')\hat\psi_c^\dagger(\bar Q)\hat\psi_b(Q')\\
    &=-[abcd]-2[adcb]_{24}.
  \end{split}
\end{equation} 
Here $[abcd]$ is defined as in (\ref{eq:abcd}) and $[abcd]_{ij}$ is equal to $[abcd]$ with the momenta of the $i$-th and $j$-th field exchanged. We find 
\begin{equation}
  \label{eq:s4s}
  \begin{split}
    S_4^{\vec s} &= \frac{1}{2}\sum_{QQ'\bar Q\bar Q'}\delta(Q-Q'+\bar Q-\bar Q')\Big((1+\lambda_2)([1111]+[2222]+[3333]+[4444])\\
    &  \quad+\lambda_3\cos((q_1-q_1')/2)([1122]+[3344]+2([1221]_{24}+[3443]_{24}))\\
    &  \quad+\lambda_3\cos((q_2-q_2')/2)([1144]+[2233]+2([1441]_{24}+[2332]_{24}))\Big)
  \end{split}
\end{equation}
For the $\chi$-bosons, we use the identity
\begin{equation}
  \begin{split}
    &\hat\psi_a(Q)i\sigma_2\hat\psi_b(Q')\hat\psi_c^\dagger(\bar Q)i\sigma_2\hat\psi_d^*(\bar Q')\\
    &=-\hat\psi_d^\dagger(\bar Q')\hat\psi_b(Q')\hat\psi_c^\dagger(\bar Q)\hat\psi_a(Q)-\hat\psi_c^\dagger(\bar Q)\hat\psi_b(Q')\hat\psi_d^\dagger(\bar Q')\hat\psi_a(Q)
  \end{split}
\end{equation} 
to calculate
\begin{equation}
  \label{eq:s4chi}
  \begin{split}
    S_4^{\chi} &= \frac{1}{2}\sum_{QQ'\bar Q\bar Q'}\delta(Q-Q'+\bar Q-\bar Q')\\
    &  \quad\Big(-2\lambda_3\cos((q_1-q_1')/2)([1122]+[3344]+[1221]_{24}+[3443]_{24})\\
    &  \qquad-2\lambda_3\cos((q_2-q_2')/2)([1144]+[2233]+[1441]_{24}+[2332]_{24})\Big)
  \end{split}
\end{equation}
From this we find
\begin{equation}
  S_4^R+S_4^{\vec s}+S_4^\chi = \frac{1}{2}\sum_{QQ'\bar Q\bar Q'}\delta(Q-Q'+\bar Q-\bar Q')([1111]+[2222]+[3333]+[4444])
\end{equation}
which is exactly the four fermion Coulomb term of the Hubbard model in the form of \eqref{eq:parbos:fermpar}.

\section{Discussion}
The last two sections were very formal and it is time to get some physical insight into the properties of our new formulation. In this section we will address the following issues:
\begin{itemize}
\item When first partially bosonizing the Hubbard model, we introduced the original bosons to be able to express physical degrees of freedom by expectation values of bosonic fields. Although we simplified our formalism analytically, we seem to have lost the intuitive grip on the physical significance of the bosons. However, we will show that in the case of homogeneous fields our new bosons really describe the physical degrees of freedom we want to investigate.
\item The new formalism is based on a perturbative one loop calculation. It is not yet clear that the full propagator matrix with all quantum fluctuations included remains diagonal beyond one loop. We will show that this is indeed the case and what we actually have done is to switch to a set of bosons that belong to different irreducible representations of translations, so that mixing of the bosons is prevented by translational invariance.
\item The one loop expressions offer a first insight into the way quantum fluctuations are included into bosonic propagators. They give some valuable information for the motivation of truncation schemes, e.g.: Which momentum modes are favored in propagation? How does the bosonic Matsubara frequency enter the propagators? How do the propagators behave as functions of temperature?
\item The last topic of this section will be the periodicity properties of the new fields, which will also turn out to play an important role in the definition of truncation schemes.    
\end{itemize}

\subsection{Homogeneous fields}
The renormalization group analysis we present in the following chapter will be focused on the investigation of the properties of the effective potential. The effective potential is a function of constant bosonic fields, $b(X)=b=\text{const.}$ in position space or respectively $b(Q)=\delta(Q)b$ in momentum space. The interpretation of our new bosonic fields in this limit is therefore of great interest to understand what the effective potential tells us about the physical degrees of freedom of the theory. 

For homogeneous fields, the bosonic kinetic term reads
\begin{align}
\frac{1}{\cal V}S_{kin}^B&=2\pi^2(h_\rho^2\hat R_1^2+h_p^2\hat R_2^2+h_{q_y}^2\hat R_3^2+h_{q_x}^2\hat R_4^2)\nonumber\\
&\quad+2\pi^2(h_m^2\hat{\vec s}_1^2+h_a^2\hat{\vec s}_2^2+h_{g_y}^2\hat{\vec s}_3^2+h_{g_x}^2\hat{\vec s}_4^2)\nonumber\\
&\quad+4\pi^2(\hat\chi_1^\dagger\text{diag}(h_{e_y}^2,h_{e_x}^2)\hat\chi_1+\hat\chi_2^\dagger\text{diag}(h_{v_y}^2,h_{v_x}^2)\hat\chi_2\nonumber\\
&\qquad\quad+\hat\chi_3^\dagger\text{diag}(h_{v_y}^2,h_{e_x}^2)\hat\chi_3+\hat\chi_4^\dagger\text{diag}(h_{e_y}^2,h_{v_x}^2)\hat\chi_4).
\end{align}
The Yukawa coupling term becomes
\begin{align}
\label{eq:couplhom}
S_Y&=-\sum_Q(\hat\psi^\dagger(Q)(\hat R_1(h_\rho^2A_0)+\hat R_2(h_p^2A_3)+\hat R_3(h_{q_y}^2D_0)+\hat R_4(h_{q_x}^2D_3)\hat\psi(Q)\nonumber\\
&\qquad\quad+\hat\psi^\dagger(Q)(\hat{\vec s}_1(h_m^2A_0\vec\sigma)+\hat{\vec s}_2(h_a^2A_3\vec\sigma)+\hat{\vec s}_3(h_{g_y}^2D_0\vec\sigma)+\hat{\vec s}_4(h_{g_x}^2D_3\vec\sigma))\hat\psi(Q)\nonumber\\
&\quad-\frac{1}{2}\sum_Q\biggl(\hat\psi^\dagger(-Q)\biggl(\hat\chi_1^\dagger\begin{pmatrix}h_{e_y}^2\cos(q_1/2)A_1\\h_{e_x}^2\cos(q_2/2)B_1\end{pmatrix}i\sigma_2
+\hat\chi_2^\dagger\begin{pmatrix}h_{v_y}^2\sin(q_1/2)A_2\\h_{v_x}^2\sin(q_2/2)B_2\end{pmatrix}i\sigma_2\nonumber\\
&\qquad\quad+\hat\chi_3^\dagger\begin{pmatrix}-h_{v_y}^2\cos(q_1/2)D_1\\h_{e_x}^2\sin(q_2/2)C_1\end{pmatrix}i\sigma_2
+\hat\chi_4^\dagger\begin{pmatrix}-h_{e_y}^2\sin(q_1/2)D_2\\-h_{v_x}^2\cos(q_2/2)C_2\end{pmatrix}i\sigma_2\biggr)\hat\psi(Q)\nonumber\\
&\qquad+(\hat\psi\to\hat\psi^*,\sin\to-\sin,i\sigma_2\to-i\sigma_2,\hat\chi^\dagger\to\hat\chi^T)\biggr).
\end{align}
We see that in this limit the bosons $\hat R$ and $\hat{\vec s}$ couple to the {\em same} fermionic bilinears as the original bosons $\hat\rho$, $\hat p$ etc. This means that for homogeneous fields, we can interprete $\hat R_1$ as the charge density, $\hat{\vec s}_1$ as the magnetic spin density, $\hat{\vec s}_2$ as the antiferromagnetic spin density etc. 

The interpretation of the complex bosons is not so simple. To understand the coupling terms of the complex bosons, we translate them to position space. For example, this yields for the first component of $\hat\chi_1$
\begin{multline}
-\frac{1}{2}h_{e_y}^2\hat\chi_{1_1}^*\sum_X(\hat\psi_1(X)i\sigma_2\hat\psi_2(X)+\hat\psi_1(X)i\sigma_2\hat\psi_2(X-\boldsymbol{e}_1)\nonumber\\
\qquad+\hat\psi_3(X)i\sigma_2\hat\psi_4(X)+\hat\psi_3(X-\boldsymbol{e}_1)i\sigma_2\hat\psi_4(X)).
\end{multline}  
\begin{figure}
\centering
\psfrag{1}{$1$}
\psfrag{2}{$2$}
\psfrag{3}{$4$}
\psfrag{4}{$3$}
\psfrag{+}{$+$}
\psfrag{-}{$-$}
\psfrag{c11}{$\hat\chi^*_{1_1}$}
\psfrag{c12}{$\hat\chi^*_{1_2}$}
\psfrag{c21}{$\hat\chi^*_{2_1}$}
\psfrag{c22}{$\hat\chi^*_{2_2}$}
\psfrag{c31}{$\hat\chi^*_{3_1}$}
\psfrag{c32}{$\hat\chi^*_{3_2}$}
\psfrag{c41}{$\hat\chi^*_{4_1}$}
\psfrag{c42}{$\hat\chi^*_{4_2}$}
\includegraphics[scale=0.7]{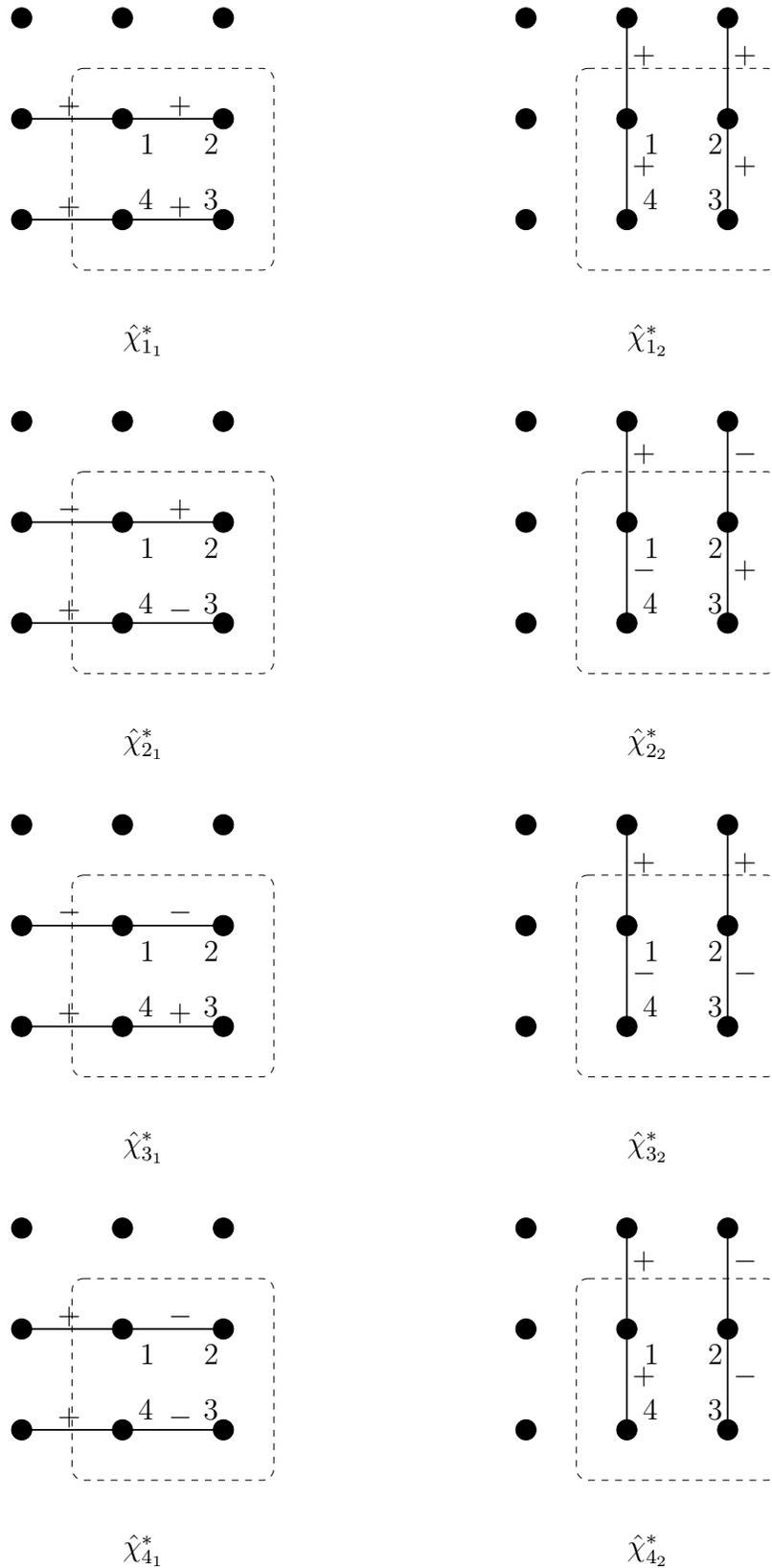}
\caption[Charged fermion bilinears in position space]{The complex bosons couple to fermion bilinears that have a position space structure as shown.}
\label{fig:linkpairs}
\end{figure}
We represent this result as a pictorial expression in fig. \ref{fig:linkpairs}. The plaquette with position label $X$ is indicated by a dashed line. The results for the other complex bosons are given in the same way by the remaining diagrams in fig. \ref{fig:linkpairs}. Particularly we see that $\hat\chi_{1_1}-\hat\chi_{1_2}$ describes a $d$-wave. Similarly, bosons with different spatial symmetry properties can be built up by simple inspection of these diagrams.

In conclusion, we have found that despite the transformations we performed to diagonalize the propagator matrix, the physical interpretation of the boson is as simple as for the original bosons in the most interesting case of homogeneous fields. 

\subsection{Symmetries}
Since our final form of the partition function is equivalent to the original Hubbard model for the Hubbard values of the Yukawa couplings, it is clear that it should respect the lattice symmetries as well as $U(1)$- and $SU(2)$-symmetry of the original model. In principle we could infer the symmetry transformation properties of the new bosons from the known transformation behavior of the bilinears (under the assumption that the action should be invariant). However, there is no need to do this with one important exception: The behavior with respect to translations. As we have mentioned in the discussion following \eqref{eq:relabelingsym}, the translations $T_x$, $T_y$ and $T_xT_y$ act as
\begin{equation}
\begin{pmatrix}1\\T_x\\T_y\\T_xT_y\end{pmatrix}\hat\psi(Q)=\begin{pmatrix}A_0\\A_1\exp(iq_1/2)\\B_1\exp(iq_2/2)\\B_0\exp(i(q_1+q_2)/2)\end{pmatrix}\hat\psi(Q).
\end{equation}
For completeness, we also wrote down the identity transformation $1$. Then, to guarantee translational invariance, we have
\begin{align}
\begin{pmatrix}1\\T_x\\T_y\\T_xT_y\end{pmatrix}\hat R_1(Q)&=\begin{pmatrix}1\\\exp(iq_1/2)\\\exp(iq_2/2)\\\exp(i(q_1+q_2)/2)\end{pmatrix}\hat R_1(Q)\nonumber\\
\begin{pmatrix}1\\T_x\\T_y\\T_xT_y\end{pmatrix}\hat R_2(Q)&=\begin{pmatrix}1\\-\exp(iq_1/2)\\-\exp(iq_2/2)\\\exp(i(q_1+q_2)/2)\end{pmatrix}\hat R_2(Q)\nonumber\\
\begin{pmatrix}1\\T_x\\T_y\\T_xT_y\end{pmatrix}\hat R_3(Q)&=\begin{pmatrix}1\\\exp(iq_1/2)\\-\exp(iq_2/2)\\-\exp(i(q_1+q_2)/2)\end{pmatrix}\hat R_3(Q)\nonumber\\
\begin{pmatrix}1\\T_x\\T_y\\T_xT_y\end{pmatrix}\hat R_4(Q)&=\begin{pmatrix}1\\-\exp(iq_1/2)\\\exp(iq_2/2)\\-\exp(i(q_1+q_2)/2)\end{pmatrix}\hat R_4(Q).
\end{align}
The same transformations apply to $\hat{\vec s}_a$ and $\hat\chi_a$. This tells us that {\em the bosons $\hat R_a$ belong to inequivalent irreducible one dimensional representations of the translation group}! The same reasoning holds for $\hat{\vec s}_a$ and $\hat\chi_a$. In other words, to preserve translational invariance the full propagator matrix (not only the one loop corrected one) has to be diagonal in the same sense in which the one loop corrected propagator matrix is diagonal (which means diagonal except for the $2\times2$-blocks for the complex bosons). In conclusion, $U(1)$-invariance tells us that the real and complex bosons cannot mix, $SU(2)$-invariance forbids mixing of bosons from the spin singlet and triplet, and finally translational invariance excludes mixing of bosons with different $a$. Note that these invariance arguments do not only hold for the propagator matrix, but can also be used to narrow down the possible form of arbitrary $n$-point functions. We will exploit this when writing down a truncation of the effective action in the following chapter. 

Translations leave each term of the action separately invariant. This is not the case for rotations. Although we know that the action as a whole is invariant under rotations, this is not the case for single terms in the action. For example, if we only consider the terms in the original action containing the boson $\hat q_y$ and translate these terms back into the purely fermionic theory (as we did in \eqref{eq:parbos:backtrans}), we find that these terms give rise to a four fermion interaction term $\sim(\hat\psi_1^\dagger\hat\psi_1+\hat\psi_2^\dagger\hat\psi_2-\hat\psi_3^\dagger\hat\psi_3-\hat\psi_4^\dagger\hat\psi_4)^2$. This term breaks rotational invariance. However, the conditions for the Yukawa couplings guarantee that these symmetry breaking terms cancel each other, but only, if all terms are taken into account. To preserve rotational invariance during the flow, we must keep in mind that during the renormalization group flow the individual terms we wrote down in our action will not develop wave function renormalization constants or couplings that are independent from each other. For example, the condition that $h_{q_x}=h_{q_y}$ at the beginning of the flow remains true during the whole flow due to rotational invariance. In the same way, the ratio of wave function renormalization constants for say $R_1$ and $R_i$, $i\in\{2,3,4\}$ will always be constant, if we truncate the wave function renormalizations to be momentum independent. However, symmetry considerations do not keep mass terms or couplings in the effective potential from flowing independently, since the effective potential is a function of homogeneous fields. For homogeneous fields, the new bosons coincide with the original ones, and symmetry transformations do no longer mix different terms.       

The last symmetry we want to discuss is the ``time reversal'' \eqref{eq:timereversal}. In the limit of spatially homogeneous bosonic fields $\hat R_{an}(\boldsymbol{q}=0)$ etc. the corresponding symmetry transformations for the bosons are simply
\begin{align}
T_{A_2}\hat R_{n}(0)&=\text{diag}(-1,1,-1,1)\hat R_{-n}(0)\nonumber\\
T_{B_2}\hat R_{n}(0)&=\text{diag}(-1,1,1,-1)\hat R_{-n}(0)\nonumber\\
T_{B_3}\hat R_{n}(0)&=\text{diag}(-1,-1,1,1)\hat R_{-n}(0).
\end{align}
Particularly,
\begin{equation}
\label{eq:Tsymm}
T_{A_2}T_{B_2}T_{B_3}\hat R_{an}(0)=-\hat R_{a(-n)}(0)\quad\forall a.
\end{equation}
The last equation also holds if $\hat R$ is replaced by $\hat{\vec s}$ or $\hat\chi$. We can use this symmetry to argue that in any term of the effective action that only depends on homogeneous fields $R_{a(n=0)}(\boldsymbol{q}=0)$, etc. the number of bosonic fields must be even.  

\subsection{Discussion of the momentum dependence}
\begin{figure}
\centering
\psfrag{delgam2}{\hspace*{-1cm}$(\Delta\Gamma^{\chi_{1_1}})^{(2)}_{m=0}(\boldsymbol{k})$}
\psfrag{0}{$0$}
\psfrag{pi}{$\pi$}
\psfrag{mpi}{$-\pi$}
\psfrag{2pi}{$2\pi$}
\psfrag{m2pi}{$-2\pi$}
\psfrag{k1}{$k_1$}
\psfrag{k2}{$k_2$}
\includegraphics[scale=0.9]{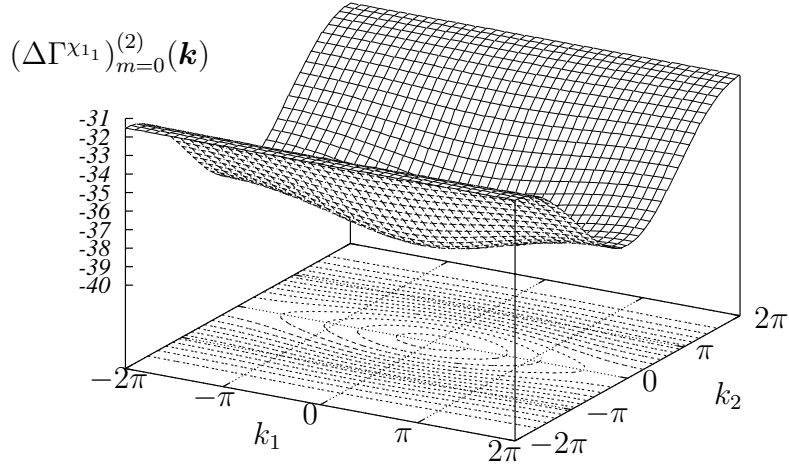}
\caption[The one loop correction to the bosonic propagator of $\chi_{1_1}$ for high temperature]{The second derivative of the one loop correction to the bosonic kinetic term in the effective action of $\chi_{1_1}$ for high temperature $T=1$. We have set $t=1$; the Yukawa couplings are at their Hubbard model values \eqref{eq:yukhubb} with $h_{e_y}^2=h_{e_x}^2=10$. We have plotted the Matsubara mode $m=0$.}
\label{fig:chi11hightemp}
\end{figure}
\begin{figure}
\centering
\psfrag{delgam2}{\hspace*{-1cm}$(\Delta\Gamma^{\chi_{2_2}})^{(2)}_{m=0}(\boldsymbol{k})$}
\psfrag{0}{$0$}
\psfrag{pi}{$\pi$}
\psfrag{mpi}{$-\pi$}
\psfrag{2pi}{$2\pi$}
\psfrag{m2pi}{$-2\pi$}
\psfrag{k1}{$k_1$}
\psfrag{k2}{$k_2$}
\includegraphics[scale=0.9]{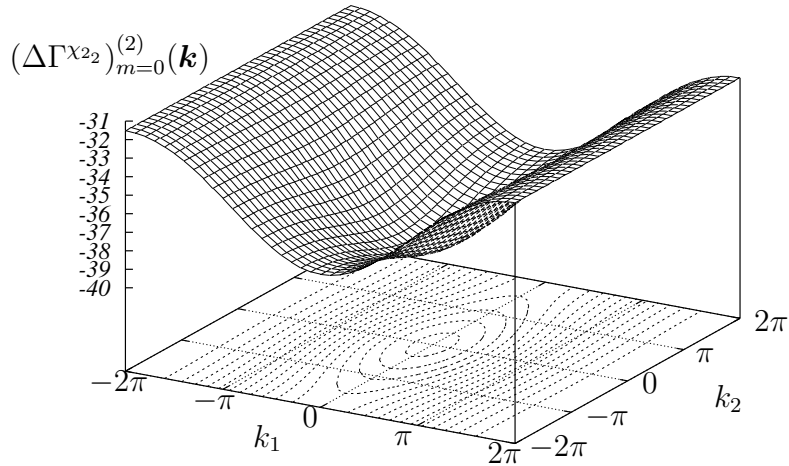}
\caption[The one loop correction to the bosonic propagator of $\chi_{2_2}$ for high temperature]{The second derivative of the one loop correction to the bosonic kinetic term in the effective action of $\chi_{2_2}$ for high temperature $T=1$. We have set $t=1$; the Yukawa couplings are at their Hubbard model values \eqref{eq:yukhubb} with $h_{e_y}^2=h_{e_x}^2=10$. We have plotted the Matsubara mode $m=0$.}
\label{fig:chi22hightemp}
\end{figure}
Before our final diagonalization step carried out in sec. \ref{sec:finaldig}, the bosonic propagator terms in the action (as functionals of the original bosonic fields or of the fields $\bar b_i$) were pure mass terms. The momentum dependence emerged at one loop level as discussed in sec. \ref{sec:disoneloop}. However, in our final form of the theory, already the classical propagators \eqref{eq:massmatrixreal}, \eqref{eq:massmatrixcomplex} are momentum dependent. For high temperature, this momentum dependence of the classical propagators dominates over the momentum dependence induced by quantum corrections.  

In the truncation for the effective action that we will use in this work, we will keep the Yukawa couplings constant at their initial values. If we insert the initial Hubbard values for the Yukawa couplings, the propagators reduce to \eqref{eq:prophubb}. Then the momentum dependence of the propagators for $R_1$ and $\vec s_2$ goes as $-\cos(k_1/2)-\cos(k_2/2)$, which takes its minimum at $\vec k=0$. The propagator matrix for $\chi_1$ becomes diagonal and momentum independent. To see how the momentum dependence emerges for $\chi_1$, we investigate the one loop correction to the diagonal elements of the propagator matrix of $\chi_1$ in the case that the Yukawa couplings are set to their Hubbard model values (we will neglect the off diagonal elements in our truncation). The results are shown in figs. \ref{fig:chi11hightemp}, \ref{fig:chi22hightemp}. In both cases, the minimum of the propagator correction again occurs in $\vec k=0$. Thus we find that the dominating momentum dependence of the bosonic propagators of our final theory exhibits minima at $\vec k=0$ for the bosons describing the charge density, antiferromagnetism and $d$-wave superconductivity in the limit of homogeneous fields. We will use this fact to argue that the propagation of momentum modes close to $\vec k=0$ is facilitated, so that we can approximate these fields homogeneously by their $\vec k=0$-mode.     

\subsection{Periodicities}
The bosonic integrals extend over the range $[-\pi,\pi]$. This range originated in the periodicity of the integrand in our definitions of the Fourier transforms. It is interesting to analyze the periodicity properties of our new bosons and to see how the periodicity of the integrand is maintained.

The periodicity behavior of the new bosons can be inferred from the known behavior of the original bosons, which follows directly from the form of the Fourier transforms. By applying all the transformations we performed to arrive at our new bosons, we find after a tedious but straightforward calculation
\begin{align}
\label{eq:period:neven}
\hat R(K+2\pi n\boldsymbol{e}_i)&=(-1)^{n/2}\hat R(K)\nonumber\\
\hat \chi(K+2\pi n\boldsymbol{e}_i)&=(-1)^{n/2}\hat\chi(K)
\end{align}
for $n$ even, $i\in\{1,2\}$ and
\begin{align}
\hat R(K+2\pi n\boldsymbol{e}_1)&=i(-1)^{(n+1)/2}B_1\hat R(K),\nonumber\\
\hat R(K+2\pi n\boldsymbol{e}_2)&=-i(-1)^{(n+1)/2}B_0\hat R(K),\nonumber\\
\hat \chi_{a_1}(K+2\pi n\boldsymbol{e}_1)&=\sum_b(B_1)_{ab}\hat \chi_{b_1}(K)\nonumber\\
\hat \chi_{a_1}(K+2\pi n\boldsymbol{e}_2)&=i(-1)^{(n+1)/2}\sum_b(B_0)_{ab}\hat \chi_{b_1}(K)\nonumber\\
\hat \chi_{a_2}(K+2\pi n\boldsymbol{e}_1)&=i(-1)^{(n+1)/2}\sum_b(B_1)_{ab}\hat \chi_{b_2}(K)\nonumber\\
\hat \chi_{a_2}(K+2\pi n\boldsymbol{e}_2)&=\sum_b(B_0)_{ab}\hat \chi_{b_2}(K)
\end{align}
for $n$ odd. The equations for $\hat R$ also hold respectively for $\hat{\vec s}$. Apart from a possible phase factor, the fields are periodic in $4\pi$. They are not periodic in uneven multiples of $2\pi$, but transform into each other. Thus it is clear that no single term in the bosonic part of the action is periodic in $2\pi$. However, periodicity of the integrand is restored when considering sums of terms. For example, whereas
\begin{equation}
\frac{1}{2}\hat R_1(-K)P_{11}^R\hat R_1(K)
\end{equation}
is not periodic in $2\pi$, the sum
\begin{equation}
\frac{1}{2}\sum_{ab}\hat R_a(-K)P_{ab}^R\hat R_b(K)
\end{equation}
is periodic in $2\pi$. 

The interesting thing about this behavior is that it is possible to write terms as higher momentum modes of other terms. For example, we can write
\begin{equation}
\frac{1}{2}T\sum_n\int_{-\pi}^{\pi}\frac{d^2k}{(2\pi)^2}\sum_{ab}\hat R_a(-K)P_{ab}^R\hat R_b(K)=\frac{1}{2}T\sum_n\int_{-2\pi}^{2\pi}\frac{d^2k}{(2\pi)^2}\hat R_c(-K)P_{cc}^R\hat R_c(K)
\end{equation}
for {\em any} $c\in\{1,2,3,4\}$. We have shown that the same property holds for the propagator terms of the complex bosons, as well as for the coupling terms. One possible equivalent transcription of the bosonic terms in \eqref{eq:action} that we will use is
\begin{equation}
\label{eq:action1}
\begin{split}
S_{kin}^B &= \sum_K\Big[\frac{1}{2}\hat R_1(-K)P_{11}^R(K)\hat R_1(K)+\frac{1}{2}\hat{\vec s}_2(-K)P_{22}^{\vec s}(K)\hat{\vec s}_2(K)\\
          &                       \hspace{2cm}+\hat\chi^\dagger_1(K)P_{11}^\chi(K)\hat\chi_1(K)\Big],\\
S_Y       &= -\sum_{KQQ'}\sum_{ab}\Big[\delta(K-Q+Q')\Big(\hat R_1(K)\hat\psi_a^\dagger(Q)V^R_{ab,1}(K)\hat\psi_b(Q')\\
          &                                           \hspace{5.5cm}+\hat{\vec s}_2(K)\hat\psi_a^\dagger(Q)V^{\vec s}_{ab,2}(K)\hat\psi_b(Q')\Big)\\
          &                             \hspace{2cm}+\delta(K-Q-Q')\Big(\hat\chi^\dagger_1(K)\hat\psi_a^T(Q)V^{\chi^*}_{ab,1}(Q,Q')\hat\psi_b(Q')\\
          &                                           \hspace{5.5cm}+\hat\chi^T_1(K)\hat\psi_a^\dagger(Q)V^\chi_{ab,1}(Q,Q')\hat\psi^*_b(Q')\Big)\Big],
\end{split}
\end{equation}
where it is understood that bosonic momentum integrals extend over the interval $[-2\pi,2\pi]$ (that is, $\sum_K=T\sum_n\int_{-2\pi}^{2\pi}\frac{d^2k}{(2\pi)^2}$, if $K$ denotes the momentum of a boson) and fermionic integrals over $[-\pi,\pi]$. The $\delta$-function is assumed to be periodic in $2\pi$ for $Q$, $Q'$ and in $4\pi$ for $K$. Only the physically interesting bosons $\hat R_1$, $\hat{\vec s}_2$ and $\hat\chi_1$ enter our formulation in this transcription. The other bosons are included as higher momentum modes of these three bosons.
\chapter{Renormalization group analysis}
The history of renormalization group approaches to the Hubbard model is a short one \cite{renhub}. Up to now all these investigations were performed in the framework of the purely fermionic model. In this section we will see how to apply the renormalization group formalism developed in chapter \ref{ch:renor} to our final partially bosonized form of the partition function \eqref{eq:partfunc}, where the action terms are given by \eqref{eq:action}. If convenient, we may rewrite the action terms by using \eqref{eq:action1}. The first task is then to write down a suitable truncation for the effective action. This truncation ansatz is then to be inserted in \eqref{eq:masterflow}, from which we can derive the flow equations for masses, couplings and wave function renormalization constants. Our renormalization group analysis will be focused on the properties of the bosonic effective potential, and the truncations proposed in the first section will be adjusted to this aim. 

\section{The truncation}

Since \eqref{eq:masterflow} cannot be solved exactly, we propose an ansatz for its solution, which is a truncated version of the effective average action. The guidelines for doing so are the following:
\begin{itemize}
\item Due to the property $\lim_{k\to\Lambda}\Gamma_k=S$ the effective average action will resemble the classical action as a function of expectation values of the fields at the beginning of the flow. The truncation will therefore include terms that look like the corresponding terms of the classical action and systematic generalizations of these.
\item The generalizations are limited by the fact that they should respect the symmetries of the theory.
\item We only keep terms that seem to be absolutely necessary to describe the behavior of quantities we want to calculate (in our case bosonic masses and quartic couplings). 
\end{itemize}
It is clear that by truncating the effective average action we introduce errors due to the approximation. The most difficult part in a successful renormalization group analysis is to decide which terms to include in the truncation. This requires physical intuition, some systematic expansion of the terms which possibly appear in the effective average action and a trial and error procedure --- at last, the only reliable way to estimate truncation errors is to include more terms in the systematic expansion and to see how they change the results. We want to stress that for theories with strong couplings (so that no perturbative expansion with respect to some small quantity is possible) there is no way to circumvent these approximation problems. One large advantage of the method of the effective average action we present here is that the property $\lim_{k\to\Lambda}\Gamma_k=S$ enhances our intuitive grip on possible truncation schemes.    

In general, the effective average action can be written in the form
\begin{equation}
\Gamma_k=\Gamma_k^B+\Gamma_k^F+\Gamma_k^{BF},
\end{equation}
where $\Gamma_k^B$ contains only bosonic fields, $\Gamma_k^F$ only fermionic fields and $\Gamma_k^{BF}$ coupling terms between bosonic and fermionic fields. Since we are mostly interested in the properties of the effective potential, which is part of $\Gamma_k^B$, we propose the following simple ansatz for $\Gamma_k^F$ and $\Gamma_k^{BF}$:
\begin{equation}
\label{eq:trunc:acf}
\Gamma_k^F=S_{kin}^F,\quad\Gamma_k^{BF}=S_Y.
\end{equation}
The reasoning behind this approximation is that we are mainly interested in the flow of the effective potential. Terms with more than two fermionic fields do not contribute to the flow of the effective potential. However, they do contribute to the flow of the Yukawa couplings --- that we keep constant in our truncation (see below). For constant couplings we can therefore ignore all terms with more than two fermionic fields. This leaves one fermionic propagator term which is part of $\Gamma_k^F$ and coupling terms of two fermionic fields to an arbitrary number of bosonic fields in $\Gamma_k^{BF}$. The propagator term can be written as the classical propagator term times a momentum dependent wave function renormalization constant. To arrive at \eqref{eq:trunc:acf}, we additionally make the following approximations:
\begin{itemize}
\item The fermionic wave function renormalization constant $Z_k^F$ is kept constant at its initial value $Z_k^F=1$.
\item All terms with two fermions and more than one boson are neglected. 
\item The Yukawa couplings are kept constant. This approximation would not be too good for most simpler theories, where the initial values of the couplings are known. It is even worse in our case, since the many different choices of the initial values of the couplings (although equivalent if everything is exact) lead to different results if approximations are made, as we already saw in the mean field case. However, to get a first impression of the properties of the effective potential, we will nevertheless take the Yukawa couplings as parameters and discuss the results as functions of these parameters. The inclusion of the flow of the couplings into the renormalization group analysis is subject to current work \cite{tobi}.
\end{itemize}
Again recall that with all these approximations the effective average action terms involving fermionic fields necessarily coincide with the corresponding terms of the classical action due to the property $\lim_{k\to\Lambda}\Gamma_k=S$. 

We now turn our attention to $\Gamma_k^B$. $\Gamma_k^B$ is a functional of the bosonic fields $b_i\in\{R_a,\vec s_a,\chi_a,\chi_a^*\}$ with $a\in\{1,2,3,4\}$. A systematic general expansion of $\Gamma_k^B$ in powers of the fields reads
\begin{align}
\Gamma_k^B[b_i]&=\sum_{K_1}\sum_{i_1}c_{i_1,k}(K_1)b_{i_1}(K_1)\nonumber\\
&\quad+\sum_{K_1K_2}\sum_{i_1i_2}c_{i_1i_2,k}(K_1,K_2)b_{i_1}(K_1)b_{i_2}(K_2)\nonumber\\
&\quad+\sum_{K_1K_2K_3}\sum_{i_1i_2i_3}c_{i_1i_2i_3,k}(K_1,K_2,K_3)b_{i_1}(K_1)b_{i_2}(K_2)b_{i_3}(K_3)\nonumber\\
&\quad+\cdots
\end{align}
with $k$-dependent coefficients ($k$ is the flow parameter from chapter \ref{ch:renor}). Each coefficient contains a momentum conserving $\delta$-function. From \eqref{eq:period:neven} we know that
\begin{equation}
b_i(K)=-b_i(k+4\pi\boldsymbol{e}_{1/2})
\end{equation}
which implies
\begin{equation}
c_{i_1\ldots i_n,k}(K_1,\ldots,K_j+4\pi\boldsymbol{e}_{1/2},\ldots,K_n)=(-1)^nc_{i_1\ldots i_n,k}(K_1,\ldots,K_n)
\end{equation}
to preserve the periodicity of the integrands. This tells us that we can expand each coefficient $c_{i_1\ldots i_n,k}(K_1,\ldots,K_n)$ with respect to the functions ($K_j=((\omega_m^B)_j,\boldsymbol{k}_j)$, $\boldsymbol{k}_j=((k_j)_1,(k_j)_2)$) 
\begin{equation}
\{1,\cos(m_j(k_j)_1/2)-1,\cos(m_j(k_j)_2/2)-1,\sin(m_j(k_j)_1/2),\sin(m_j(k_j)_2/2)\}
\end{equation}
 for $n$ even and 
\begin{equation}
\{\cos((2m_j+1)(k_j)_1/4),\cos((2m_j+1)(k_j)_2/4),\sin((2m_j+1)(k_j)_1/4),\sin((2m_j+1)(k_j)_2/4)\}
\end{equation}
 for $n$ odd and with $m\in\mathbbm{N}$ (we have chosen to expand with respect to $\cos(mk_{1/2}/2)-1$ instead of $\cos(mk_{1/2}/2)$ to achieve that all momentum dependent terms for even $n$ vanish for $\boldsymbol{k}=0$). The coefficients $\tilde c_{m,k}$ of this new expansion with respect to trigonometric functions only depend on the Matsubara frequency and the flow parameter $k$.

In our truncation of $\Gamma_k^B$ we include a momentum dependent propagator term (more on this later) and a second term containing all other bosonic terms
\begin{equation}
\Gamma_k^B=\Gamma_{kin,k}^B+U_k.
\end{equation}
We simplify the coefficients $\tilde c_{m,k}$ by neglecting their Matsubara mode dependence. Furthermore, for homogeneous fields we have $K_i=0\,\forall i$. Then \eqref{eq:Tsymm} tells us that all terms of the effective action with an {\em odd} number of fields have to vanish to preserve ``time reversal''-invariance\footnote{Note the tricky part of this argument: The symmetry transformation \eqref{eq:Tsymm} lived in the space of bosonic fields, which are all independent. If we switch to the effective action, the (linear) symmetry transformation carries over to the expectation values of the fields, {\em except for the charge density $R_1(0)\propto\rho$}. The charge density expectation value $\rho$ is regarded as a parameter controlled by the source $\mu$ that is no longer explicitly present in the effective action. The transformation $\rho\to-\rho$ does {\em not} follow from the transformation behavior of $\hat\rho$, since it is no longer a free field, but from $\mu\to-\mu$ in \eqref{eq:timereversal} and \eqref{eq:murho}.}. For the remaining terms with an even number of fields, only the terms with momentum independent coefficients survive, so that we truncate $U_k$ in the form
\begin{align}
\label{eq:genuk}
U_k[b_i]&=U_0{\cal V}+\sum_{K_1K_2}\sum_{i_1i_2}c_{i_1i_2,k}b_{i_1}(K_1)b_{i_2}(K_2)\delta(\ldots)\nonumber\\
&\quad+\sum_{K_1K_2K_3K_4}\sum_{i_1i_2i_3i_4}c_{i_1i_2i_3i_4,k}b_{i_1}(K_1)b_{i_2}(K_2)b_{i_3}(K_3)b_{i_4}(K_4)\delta(\ldots)\nonumber\\
&\quad+\cdots
\end{align}  
with homogeneous fields $b_i(K)=b_i\delta(K)$, coefficients that only depend on the flow parameter $k$ and appropriate momentum conserving $\delta$-functions $\delta(\ldots)$.

Due to $U(1)$- and $SU(2)$-symmetry the effective potential $U_k$ can only depend on the invariants
\begin{align}
\rho_{ab}(K_1,K_2)&=\frac{1}{2}R_a(K_1)R_b(K_2)\nonumber\\
\alpha_{ab}(K_1,K_2)&=\frac{1}{2}\vec s_a(K_1)\vec s_b(K_2)\nonumber\\
\Delta_{ai,bj}(K_1,K_2)&=\chi_{a_i}^*(K_1)\chi_{b_j}(K_2).
\end{align}
By use of \eqref{eq:action1} we can write the effective potential as a function of $\rho_{11}$, $\alpha_{22}$ and $\Delta_{11,11}$, $\Delta_{11,12}$, $\Delta_{12,11}$, $\Delta_{12,12}$ only. To simplify the notation, we write $\Delta_{ab}:=\Delta_{1a,1b}$. Recall that all other bosons are included as higher momentum modes of the bosons explicitly present in our truncation as discussed after \eqref{eq:action1}. This means that although we will only consider homogeneous modes of the invariants $\rho_{11}$, $\alpha_{22}$ and $\Delta_{11}$, $\Delta_{12}$, $\Delta_{21}$, $\Delta_{22}$ as external lines, the exchange of all virtual bosons is included as we also integrate over the higher virtual boson momentum modes representing the bosons  not explicitly present in the boson set we selected.  

The momentum dependent term in $\Gamma_{kin,k}^B$ is truncated to a propagator term only:
\begin{align}
\Gamma_{kin,k}^B &= \sum_{KK'}\Big[\frac{1}{2}\hat R_1(K)P_{11,k}^R(K,K')\hat R_1(K')+\frac{1}{2}\hat{\vec s}_2(K)P_{22,k}^{\vec s}(K,K')\hat{\vec s}_2(K')\nonumber\\
                 &  \hspace{2cm}+\hat\chi^\dagger_1(K)P_{11,k}^\chi(K,K')\hat\chi_1(K')\Big].
\end{align}
For the real bosons we set
\begin{align}
P_{11,k}^R(K,K')&=(2\pi)^2Z_k^R(P_1^R(K)-P_1^R(K=0)+(m\pi T)^2)\delta(K+K')\nonumber\\
P_{22,k}^{\vec s}(K,K')&=(2\pi)^2Z_k^{\vec s}(P_2^{\vec s}(K)-P_2^{\vec s}(K=0)+(m\pi T)^2)\delta(K+K').
\end{align}
This has to be compared to the classical action propagators
\begin{align}
P_{11}^R(K,K')&=(2\pi)^2P_1^R(K)\delta(K+K')\nonumber\\
P_{22}^{\vec s}(K,K')&=(2\pi)^2P_2^{\vec s}(K)\delta(K+K').
\end{align}
First note that the terms $P_1^R(K=0)$ and $P_2^{\vec s}(K=0)$ are momentum independent and therefore can be compensated by adjusting the coefficients in the effective potential $U_k$, so that we are free to add them. If we note that the propagators $P_1^R(K)$ and $P_2^{\vec s}(K)$ (for the Hubbard model values of the Yukawa couplings) become minimal in $K=0$, we see that (for $m=0$) the advantage of this addition is to make the propagator vanish at zero momentum and positive otherwise. This allows to define simple truncation schemes for the bosonic propagators. 

As we will see below, we use a temperature like cutoff in the fermionic sector. This means that during the beginning of the flow the system behaves as in the high temperature limit. The interesting physics emerges gradually as the temperature cutoff is lowered.

The last term we added involving the Matsubara frequency $m$ is needed to make loop Matsubara sums finite (since we will keep the Yukawa couplings constant, the complete $m$-dependence of loops will be provided by the propagators). By adding a term $\sim m^2$, we mimic the low temperature behavior of the one loop result as discussed in section \ref{sec:disoneloop}. We fit our truncation to the original action term by comparing the $m=0$-Matsubara mode (in the high temperature limit, this is the only one contributing in the one loop calculation). We therefore set $Z_k^R=1$ at the beginning of the flow, where $k$ is large. Note that we have truncated $Z_k^R$, $Z_k^{\vec s}$ to be momentum independent. 

In principle there is nothing that prevents us from replacing $(m\pi T)^2$ by $c_k(m\pi T)^2$, where $c_k$ is a $k$-dependent quantity. We have checked numerically that such a factor only has a small effect on our results, so that we set it equal to unity. 

For the complex bosons, there is no classical propagator that can be used as the main ingredient of the truncation, since for the Hubbard model values of the couplings the classical propagator of the complex bosons is momentum independent. We therefore propose a simple ansatz taking into account the first Fourier term in a general expansion of the effective action with respect to momentum
\begin{equation}
P_{11,k}^\chi(K,K')=(2\pi)^2Z_k^\chi\left(\frac{1}{2}\lambda_3(2-\cos(k_1/2)-\cos(k_2/2))+(m\pi T)^2\right)\delta(K-K').
\end{equation}
Since we know that the first momentum dependent contributions to the propagators of the complex bosons emerge at one loop level in ${\cal O}(1/T^4)$ (cf. \eqref{eq:hightemplim}), and our regularization scheme in the fermionic sector (as discussed below) will replace $m\pi T$ by $m\pi T_k=m\pi(T+k^2)$, the momentum dependence develops as ${\cal O}(1/k^8)$. We therefore approximately set $Z_{k=\Lambda}^\chi=1/\Lambda^8$ as the initial value of the wave function renormalization constant.

In conclusion, our truncation for the effective average action reads
\begin{align}
\label{eq:truncation}
\Gamma_k&=\Gamma_k^F+\Gamma_{kin,k}^B+U_k+\Gamma_k^{BF}\nonumber\\
\Gamma_k^F&=\sum_{QQ'}\psi^\dagger(Q)P^\psi(Q,Q')\psi(Q')\nonumber\\
\Gamma_{kin,k}^B &= \sum_{KK'}\Big[\frac{1}{2} R_1(K)P_{11,k}^R(K,K') R_1(K')+\frac{1}{2}{\vec s}_2(K)P_{22,k}^{\vec s}(K,K'){\vec s}_2(K')\nonumber\\
                 &  \hspace{2cm}+\chi^\dagger_1(K)P_{11,k}^\chi(K,K')\chi_1(K')\Big]\nonumber\\
U_k[b_i]&=U_k[\rho_{11},\alpha_{22},\Delta_{11},\Delta_{12},\Delta_{21},\Delta_{22}]\nonumber\\
\Gamma_k^{BF}&=-\sum_{KQQ'}\sum_{ab}\Big[\delta(K-Q+Q')\Big( R_1(K)\psi_a^\dagger(Q)V^R_{ab,1}(K)\psi_b(Q')\nonumber\\
          &                                           \hspace{5.5cm}+{\vec s}_2(K)\psi_a^\dagger(Q)V^{\vec s}_{ab,2}(K)\psi_b(Q')\Big)\nonumber\\
          &                             \hspace{2cm}+\delta(K-Q-Q')\Big(\chi^\dagger_1(K)\psi_a^T(Q)V^{\chi^*}_{ab,1}(Q,Q')\psi_b(Q')\nonumber\\
          &                                           \hspace{5.5cm}+\chi^T_1(K)\psi_a^\dagger(Q)V^\chi_{ab,1}(Q,Q')\psi^*_b(Q')\Big)\Big]
\end{align}
with
\begin{align}
P^\psi(Q,Q')&=(i\omega_n^F-2t(\cos(q_1/2)A_1+\cos(q_2/2)B_1))\delta(Q-Q')\nonumber\\
P_{11,k}^R(K,K')&=(2\pi)^2Z_k^R(P_1^R(K)-P_1^R(K=0)+(m\pi T)^2)\delta(K+K')\nonumber\\
P_{22,k}^{\vec s}(K,K')&=(2\pi)^2Z_k^{\vec s}(P_2^{\vec s}(K)-P_2^{\vec s}(K=0)+(m\pi T)^2)\delta(K+K')\nonumber\\
P_{11,k}^\chi(K,K')&=(2\pi)^2Z_k^\chi\left(\frac{1}{2}\lambda_3(2-\cos(k_1/2)-\cos(k_2/2))+(m\pi T)^2\right)\delta(K-K')
\end{align}
and the vertices given by \eqref{eq:vertexreal} and \eqref{eq:vertexcomplex}.

\section{The flow of the effective potential}

\subsection{The second derivative of $\Gamma_k$}
We will now start to derive the flow equations for the effective potential. In general, the flow equations for any interesting quantity can be derived from our master equation \eqref{eq:masterflow} for some particular truncation (in our case \eqref{eq:truncation}). As we see, we need the second derivative of $\Gamma_k$ with respect to the fields. Explicitly, we need
\begin{align}
\Gamma_k^{(2)}(K,K')&=
\begin{pmatrix}\frac{\delta^2\Gamma_k}{\delta u^\dagger(K)\delta u(K')}&\frac{\delta^2\Gamma_k}{\delta u^\dagger(K)\delta u^*(K')}&\frac{\delta^2\Gamma_k}{\delta u^\dagger(K)\delta w(K')}&-\frac{\delta^2\Gamma_k}{\delta u^\dagger(K)\delta\psi(K')}&\frac{\delta^2\Gamma_k}{\delta u^\dagger(K)\delta\psi^*(K')}\\
\frac{\delta^2\Gamma_k}{\delta u^T(K)\delta u(K')}&\frac{\delta^2\Gamma_k}{\delta u^T(K)\delta u^*(K')}&\frac{\delta^2\Gamma_k}{\delta u^T(K)\delta w(K')}&-\frac{\delta^2\Gamma_k}{\delta u^T(K)\delta\psi(K')}&\frac{\delta^2\Gamma_k}{\delta u^T(K)\delta\psi^*(K')}\\
\frac{\delta^2\Gamma_k}{\delta w^T(K)\delta u(K')}&\frac{\delta^2\Gamma_k}{\delta w^T(K)\delta u^*(K')}&\frac{\delta^2\Gamma_k}{\delta w^T(K)\delta w(K')}&-\frac{\delta^2\Gamma_k}{\delta w^T(K)\delta\psi(K')}&\frac{\delta^2\Gamma_k}{\delta w^T(K)\delta\psi^*(K')}\\
\frac{\delta^2\Gamma_k}{\delta \psi^\dagger(K)\delta u(K')}&\frac{\delta^2\Gamma_k}{\delta \psi^\dagger(K)\delta u^*(K')}&\frac{\delta^2\Gamma_k}{\delta \psi^\dagger(K)\delta w(K')}&-\frac{\delta^2\Gamma_k}{\delta \psi^\dagger(K)\delta\psi(K')}&\frac{\delta^2\Gamma_k}{\delta \psi^\dagger(K)\delta\psi^*(K')}\\
\frac{\delta^2\Gamma_k}{\delta \psi^T(K)\delta u(K')}&\frac{\delta^2\Gamma_k}{\delta \psi^T(K)\delta u^*(K')}&\frac{\delta^2\Gamma_k}{\delta \psi^T(K)\delta w(K')}&-\frac{\delta^2\Gamma_k}{\delta \psi^T(K)\delta\psi(K')}&\frac{\delta^2\Gamma_k}{\delta \psi^T(K)\delta\psi^*(K')}
\end{pmatrix},\nonumber\\
\end{align}
where $\Gamma_k^{(2)}(K,K')$ denotes the matrix of second derivatives occurring in \eqref{eq:masterflow} (we use this notation to avoid confusion between the complex bosonic fields $\chi$ we use here and the generalized fields $\chi$ in \eqref{eq:masterflow}). 

In our truncation, the kinetic terms yield
\begin{align}
(\Gamma_k^F)^{(2)}(K,K')&=\begin{pmatrix}0&0&0&0&0\\0&0&0&0&0\\0&0&0&0&0\\0&0&0&P^\psi&0\\0&0&0&0&P^\psi\end{pmatrix}(K,K')\nonumber\\
(\Gamma_{kin,k}^B)^{(2)}(K,K')&=\begin{pmatrix}P_{11,k}^\chi&0&0&0&0\\0&P_{11,k}^\chi&0&0&0\\0&0&\text{diag}(P_{11,k}^R,P_{22,k}^{\vec s})&0&0\\0&0&0&0&0\\0&0&0&0&0\end{pmatrix}(K,K'),
\end{align}
where we have used that the propagator matrices are symmetric. 

The contributions of the coupling terms are 
\begin{align}
\frac{\delta^2\Gamma_k^{BF}}{\delta\psi_a^\dagger(K)\delta\chi(K')}&=-\sum_Q\sum_b\delta(K-K'+Q)\tilde V_{ab,1}^\chi(K,Q)\psi^*_b(Q)\nonumber\\
\frac{\delta^2\Gamma_k^{BF}}{\delta\chi^T(K)\delta\psi_a^*(K')}&=-\sum_Q\sum_b\delta(K-K'-Q)\tilde V_{ab,1}^\chi(K',Q)\psi^\dagger_b(Q)\nonumber\\
\frac{\delta^2\Gamma_k^{BF}}{\delta\psi_a^T(K)\delta\chi^*(K')}&=-\sum_Q\sum_b\delta(K-K'+Q)\tilde V_{ab,1}^{\chi^*}(K,Q)\psi_b(Q)\nonumber\\
-\frac{\delta^2\Gamma_k^{BF}}{\delta\chi^\dagger(K)\delta\psi_a(K')}&=\sum_Q\sum_b\delta(K-K'-Q)\tilde V_{ab,1}^{\chi^*}(K',Q)\psi_b^T(Q)\nonumber\\
\frac{\delta^2\Gamma_k^{BF}}{\delta\psi_a^\dagger(K)\delta w_\gamma(K')}&=-\sum_Q\sum_b\delta(K-K'-Q)V_{ab}^{w_\gamma}(K')\psi_b(Q)\nonumber\\
\frac{\delta^2\Gamma_k^{BF}}{\delta w_\gamma^T(K)\delta\psi_a^*(K')}&=-\sum_Q\sum_b\delta(K-K'+Q)V_{ab}^{w_\gamma}(K)\psi_b^T(Q)\nonumber\\
\frac{\delta^2\Gamma_k^{BF}}{\delta\psi_a^T(K)\delta w_\gamma(K')}&=\sum_Q\sum_b\delta(K+K'-Q)\psi_b^*(Q)V_{ba}^{w_\gamma}(K')\nonumber\\
-\frac{\delta^2\Gamma_k^{BF}}{\delta w_\gamma^T(K)\delta\psi_a(K')}&=-\sum_Q\sum_b\delta(K+K'-Q)\psi_b^*(Q)V_{ba}^{w_\gamma}(K)\nonumber\\
-\frac{\delta^2\Gamma_k^{BF}}{\delta\psi_a^\dagger(K)\delta\psi_b(K')}&=-\sum_Q\sum_\gamma\delta(K-K'-Q)V_{ab}^{w_\gamma}(Q)w_\gamma(Q)\nonumber\\
-\frac{\delta^2\Gamma_k^{BF}}{\delta\psi_a^T(K)\delta\psi_b(K')}&=-\sum_Q\delta(K+K'-Q)\chi_1^\dagger(Q)\tilde V_{ab,1}^\chi(K,K')\nonumber\\
\frac{\delta^2\Gamma_k^{BF}}{\delta\psi_a^\dagger(K)\delta\psi_b^*(K')}&=\sum_Q\delta(K+K'-Q)\chi_1^T(Q)\tilde V_{ab,1}^\chi(K,K')\nonumber\\
\frac{\delta^2\Gamma_k^{BF}}{\delta\psi_a^T(K)\delta\psi_b^*(K')}&=-\sum_Q\sum_\gamma\delta(K-K'+Q)V_{ba}^{w_\gamma}(Q)w_\gamma(Q)
\end{align}
with
\begin{equation}
\tilde V(K,K')=V(K,K')-V^T(K',K)
\end{equation}
and $w_\gamma\in\{R_1,\vec s_2\}$.

The effective potential $U_k$ depends only on the invariants $\rho_{11}$, $\alpha_{22}$, $\Delta_{ij}$. It is therefore convenient to express the derivatives with respect to the fields by derivatives with respect to the invariants. Let
\begin{align}
D_{\rho}(K,K')&\equiv\frac{1}{2}\left(\frac{\delta}{\delta\rho_{11}(K,K')}+\frac{\delta}{\delta\rho_{11}(K',K)}\right)\nonumber\\
D_{\alpha}(K,K')&\equiv\frac{1}{2}\left(\frac{\delta}{\delta\alpha_{22}(K,K')}+\frac{\delta}{\delta\alpha_{22}(K',K)}\right)\nonumber\\
(D_{\Delta})_{ij}(K,K')&\equiv\frac{\delta}{\delta\Delta_{ij}(K,K')}.
\end{align}
Then we can write (generalized matrix notation!)
\begin{align}
U_k^{(2)}&=\text{diag}(D_\Delta,D_\Delta^T,D_\rho,D_\alpha,0,0)U_k\\
&+\begin{pmatrix}D_\Delta(\chi\chi^\dagger)D_\Delta&D_\Delta(\chi\chi^T)D_\Delta^T&D_\Delta(\chi R_1)D_\rho&D_\Delta(\chi \vec s_2^T)D_\alpha&0&0\\
D_\Delta^T(\chi^*\chi^\dagger)D_\Delta&(D_\Delta(\chi\chi^\dagger)D_\Delta)^T&D_\Delta^T(\chi^* R_1)D_\rho&D_\Delta(\chi^* \vec s_2^T)D_\alpha&0&0\\
D_\rho(R_1\chi^\dagger)D_\Delta&D_\rho(R_1\chi^T)D_\Delta^T&D_\rho(R_1R_1)D_\rho&D_\rho(R_1\vec s_2^T)D_\alpha&0&0\\
D_\alpha(\vec s_2\chi^\dagger)D_\Delta&D_\alpha(\vec s_2\chi^T)D_\Delta^T&D_\alpha(\vec s_2R_1)D_\rho&D_\alpha(\vec s_2\vec s_2^T)D_\alpha&0&0\\
0&0&0&0&0&0\\0&0&0&0&0&0
\end{pmatrix}U_k\nonumber.
\end{align}

\subsection{The flow of the effective potential}
In this section we will derive the flow of the effective potential from \eqref{eq:masterflow} or equivalently, the flow of the coefficients $c_{i_1\ldots i_n}$ in \eqref{eq:genuk}. In general, there is an infinite number of coefficients. To be able to calculate anything, there are two possible ways to render the number of flow equations finite:
\begin{itemize}
\item Truncate the effective potential at some finite power of the fields, e.g. at quartic order in the fields. This yields a finite set of flow equations for the remaining coefficients.
\item Instead of considering the flow of the coefficients, consider the flow of the full potential as a function of discretized homogeneous fields. For example, if $U[\phi]$ is an effective potential depending on the (continuous valued) field $\phi$, one could analyze the flow of the potential at given points $U[\phi=\phi_0]$, $U[\phi=\phi_1]$, $U[\phi=\phi_2]$ etc. A finite number of points yields a finite number of flow equations.
\end{itemize}
Both methods have their advantages and drawbacks. The first method is easier to apply, more stable and faster in the numerical treatment, but has the drawback that --- since it is nothing else than a polynomial series expansion around some given point to a given order --- the possible solutions for the effective potential are reliable only close to the point around which we expand. Usually, one expands around the minimum of the effective potential. Physical properties that can be inferred by looking at the flow of the potential close to this minimum can be well understood by this approach. However, phase transitions of first order typically can not be derived from properties of the flow of the potential near the minimum (cf. fig. \eqref{fig:mass}). 

The second method does not have this kind of problem and any possible shape of the potential can be described by it --- not only those corresponding to some finite power expansion in the fields. But on the other hand the numerical implementation of this second method is far from trivial. The reason for this is that we need the second derivatives of the effective potential on the right hand side of the flow equation. If we discretize the region the effective potential is defined on, these derivatives also have to be discretized. As is already known from more simple systems, it is a formidable task to achieve this discretization in a stable way. We will therefore use the first method, keeping in mind that we possibly face difficulties with first order phase transitions and postpone the more general treatment by the second method to future work. The implementation of the first method requires an explicit truncation of the effective potential. We will investigate different truncation schemes for the effective potential later on. In this section, we will derive the flow equations as far as possible without fixing some specific truncation.

The idea to extract the flow of the effective action from \eqref{eq:masterflow} is to note that
\begin{equation}
\lim_{\psi,\psi^*\to0}\lim_{b_i(Q)\to b_i\delta(Q)}\Gamma_k=U_k
\end{equation}
so that
\begin{equation}
\frac{d}{dk}U_k=\frac{1}{2}\lim_{\psi,\psi^*\to0}\lim_{b_i(Q)\to b_i\delta(Q)}\text{Tr}\,\left((\partial_kR_k)M\left(\Gamma_k^{(2)}+R_k\right)^{-1}\right).
\end{equation}
From the results of the last section, we have
\begin{equation}
\lim_{\psi,\psi^*\to0}\Gamma_k^{(2)}=\begin{pmatrix}B&0\\0&F\end{pmatrix}
\end{equation}
with
\begin{align}
\label{eq:matricesBF}
B&=\text{diag}(P_{11,k}^\chi+D_\Delta U_k,(P_{11,k}^\chi)^T+D_\Delta^TU_k,P_{11,k}^R+D_\rho U_k,P_{22,k}^{\vec s}+D_\alpha U_k)\nonumber\\
&\quad+\begin{pmatrix}
 \scriptstyle D_\Delta(\chi\chi^\dagger)D_\Delta U_k&\scriptstyle  D_\Delta(\chi\chi^T)D_\Delta^TU_k&\scriptstyle  D_\Delta(\chi R_1)D_\rho U_k&\scriptstyle  D_\Delta(\chi \vec s_2^T)D_\alpha U_k\\
\scriptstyle  D_\Delta^T(\chi^*\chi^\dagger)D_\Delta U_k&\scriptstyle  (D_\Delta(\chi\chi^\dagger)D_\Delta)^TU_k&\scriptstyle  D_\Delta^T(\chi^* R_1)D_\rho U_k&\scriptstyle  D_\Delta(\chi^* \vec s_2^T)D_\alpha U_k\\
\scriptstyle  D_\rho(R_1\chi^\dagger)D_\Delta U_k&\scriptstyle  D_\rho(R_1\chi^T)D_\Delta^T U_k&\scriptstyle  D_\rho(R_1R_1)D_\rho U_k&\scriptstyle  D_\rho(R_1\vec s_2^T)D_\alpha U_k\\
\scriptstyle  D_\alpha(\vec s_2\chi^\dagger)D_\Delta U_k&\scriptstyle  D_\alpha(\vec s_2\chi^T)D_\Delta^TU_k&\scriptstyle  D_\alpha(\vec s_2R_1)D_\rho U_k&\scriptstyle  D_\alpha(\vec s_2\vec s_2^T)D_\alpha U_k\\
\end{pmatrix}\nonumber\\
F(K,K')&=\text{diag}(P^\psi(K,K'),P^\psi(K,K'))\nonumber\\
&\quad+\begin{pmatrix}
\begin{matrix}-\sum_Q\delta(Q-K+K')\\\qquad\sum_\gamma V^{w_\gamma}(Q)w_\gamma(Q)\end{matrix}&\begin{matrix}\sum_Q\delta(Q-K-K')\\\qquad\chi^T\tilde V^\chi_{,1}(K,K')\end{matrix}\\\\
\begin{matrix}-\sum_Q\delta(Q-K-K')\\\qquad\chi^\dagger\tilde V^{\chi^*}_{,1}(K,K')\end{matrix}&\begin{matrix}-\sum_Q\delta(Q+K-K')\\\qquad\sum_\gamma V^{w_\gamma}(Q)^Tw_\gamma(Q)\end{matrix}
\end{pmatrix}.
\end{align}
We see that $\lim_{\psi,\psi^*\to0}\Gamma_k^{(2)}$ is block diagonal in the bosonic and fermionic sector. With
\begin{equation}
\tilde R_k^B=\text{diag}(R_k^\chi,(R_k^\chi)^T,R_k^{w_\gamma}),\quad \tilde R_k^F=\text{diag}(R_k^F,(R_k^F)^T)
\end{equation}
the flow equation now reads
\begin{align}
\frac{d}{dk}U_k&=\frac{1}{2}\lim_{b_i(Q)\to b_i\delta(Q)}\biggl[\text{Tr}\,\left((\partial_k\tilde R_k^B)\left(B+\tilde R_k^B\right)^{-1}\right)-\text{Tr}\,\left((\partial_k\tilde R_k^F)\left(F+\tilde R_k^F\right)^{-1}\right)\biggr].
\end{align}
We separate the flow equation by $U_k=U_k^B+U_k^F$, so that
\begin{align}
\frac{d}{dk}U_k^B&=\frac{1}{2}\lim_{b_i(Q)\to b_i\delta(Q)}\text{Tr}\,\left((\partial_k\tilde R_k^B)\left(B+\tilde R_k^B\right)^{-1}\right)\nonumber\\
\frac{d}{dk}U_k^F&=-\frac{1}{2}\lim_{b_i(Q)\to b_i\delta(Q)}\text{Tr}\,\left((\partial_k\tilde R_k^F)\left(F+\tilde R_k^F\right)^{-1}\right).
\end{align}
The right hand side of the flow equation for $U_k^F$ does not depend on $U_k$, so that we can further calculate it without specifying a truncation for $U_k$. 

It is very instructive to rewrite the flow equation for $U_k^F$ as
\begin{equation}
\frac{d}{dk}U_k^F=-\frac{1}{2}\lim_{b_i(Q)\to b_i\delta(Q)}\tilde\partial_k\ln\det(F+\tilde R_k^F),
\end{equation}
where
\begin{equation}
\tilde\partial_k=\frac{\partial\tilde R_k^F}{\partial k}\frac{\partial}{\partial\tilde R_k^F}.
\end{equation}
If the right hand side depended on $k$ only via $\tilde R_k^F$, we could replace $\tilde\partial_k$ by $\frac{d}{dk}$ and immediately integrate the equation to get $U_k$ as a function of $k$. For $k\to0$, we know that $\tilde R_k^F$ vanishes. In this case we have
\begin{equation}
U_{k=0}^F=U_{k\to\Lambda}^F-\frac{1}{2}\lim_{b_i(Q)\to b_i\delta(Q)}\ln\det F
\end{equation}
which is nothing else than the one loop corrected potential we calculated in our mean field analysis (cf. \eqref{eq:delUP}) --- besides the fact that we used a different set of bosons there. 

The calculation mainly goes through as in the mean field case. We set 
\begin{equation}
R_k^F(K)=i(2n+1)\pi k^2
\end{equation}
and finally arrive at
\begin{align}
\frac{d}{dk}U_k^F&=-2T{\cal V}\tilde\partial_k\sum_{\epsilon_1,\epsilon_2}\int_{-\pi}^\pi\frac{d^2q}{(2\pi)^2}\nonumber\\
&\quad\ln\cosh\left(\frac{1}{2T_k}\sqrt{\left(\sqrt{2h_\rho^4\rho_{11}}+\epsilon_2\sqrt{2h_a^4\alpha_{22}+4t^2(c_1+\epsilon_1c_2)^2}\right)^2+\bar\Delta(\epsilon_1)}\right)
\end{align}
with $c_i=\cos(q_i/2)$, $T_k=T+k^2$ and
\begin{equation}
\bar\Delta(\epsilon_1)=h_{e_y}^4\Delta_{11}\cos^2\frac{q_1}{2}+h_{e_x}^4\Delta_{22}\cos^2\frac{q_2}{2}+h_{e_y}^2h_{e_x}^2(\Delta_{12}+\Delta_{21})\epsilon_1\cos\frac{q_1}{2}\cos\frac{q_2}{2}.
\end{equation}
For $\Delta_{ij}=0$, this reproduces the mean field result \eqref{eq:effpot:U} with $\delta=0$, if we set $R_1=\rho/h_\rho$, $\vec s_2=\vec a/h_a$ and assume that the right hand side depends on $k$ only via $T_k$. This was to be expected from comparing the original coupling term in \eqref{eq:colpar:parmom} with \eqref{eq:couplhom} in the limit of homogeneous fields. The term involving the complex bosons looks somewhat different, which is due to our redefinition of the complex fields in \eqref{eq:exeydef}. The original field $e$ corresponds to $\chi_1+\chi_2$, whereas the $d$-wave corresponds to $\chi_1-\chi_2$. The corresponding invariants are
\begin{align}
e&:(\chi_1^*+\chi_2^*)(\chi_1+\chi_2)=\Delta_{11}+\Delta_{12}+\Delta_{21}+\Delta_{22}\nonumber\\
d&:(\chi_1^*-\chi_2^*)(\chi_1-\chi_2)=\Delta_{11}-\Delta_{12}-\Delta_{21}+\Delta_{22}.
\end{align}
If we assume that the original boson $\hat e$ always has a vanishing expectation value, we can set $\Delta_{12}+\Delta_{21}=-\Delta_{11}-\Delta_{22}$. If we further assume that rotational invariance is not broken in the sense that $\Delta_{11}\neq\Delta_{22}$, we can set
\begin{equation}
\Delta=2\Delta_{11}=2\Delta_{22}.
\end{equation}
Since we keep all couplings at their Hubbard values, we additionally have $h_{e_x}^2=h_{e_y}^2=2h_d^2$, so that
\begin{equation}
\bar\Delta(\epsilon_1)=2h_d^4\Delta(c_1-\epsilon_1 c_2).
\end{equation}
With these assumptions, we therefore are in complete agreement with the mean field result, if we additionally set $\delta=2h_d^2\Delta$.

\section{The flow of the bosonic wave function renormalization constant}
\label{flowwaveren}
In the following section we will also need the flow of the wave function renormalization constant $Z_k^{\vec s}$. We will derive the corresponding flow equation in this section and show that it can be analytically solved in our truncation. 

First note that $Z_k^{\vec s}$ can be extracted from our truncation of the effective action by
\begin{equation}
Z_k^{\vec s}=\frac{4}{\pi^4\lambda_3{\cal V}}\lim_{m\to0}\lim_{\boldsymbol{\epsilon}\to0}\frac{\partial}{\partial(\boldsymbol{\epsilon}^2)}\lim_{\boldsymbol{k}\to\boldsymbol{\epsilon}}\lim_{\vec s\to0}\frac{\delta^2}{\delta\vec s(K)\delta\vec s^T(K)}\lim_{\psi,\psi^*\to0}\Gamma_k\equiv O_Z^{\text{symm}}\Gamma_k
\end{equation}
in the symmetric phase and
\begin{equation}
Z_k^{\vec s}=\frac{4}{\pi^4\lambda_3{\cal V}}\lim_{m\to0}\lim_{\boldsymbol{\epsilon}\to0}\frac{\partial}{\partial(\boldsymbol{\epsilon}^2)}\lim_{\boldsymbol{k}\to\boldsymbol{\epsilon}}\lim_{\vec s\to\vec s_k^0}\frac{\delta^2}{\delta\vec s(K)\delta\vec s^T(K)}\lim_{\psi,\psi^*\to0}\Gamma_k\equiv O_Z^{\text{brok}}\Gamma_k
\end{equation}
in the broken phase, where $\vec s_k^0$ denotes the minimum of the effective potential, $K=(\omega_m^B,\boldsymbol{k})$ and by $\boldsymbol{\epsilon}$ we denote some small momentum $|\boldsymbol{\epsilon}|\ll1$. Note that the method of extracting $Z_k^{\vec s}$ is not uniquely determined by our truncation. Particularly, the limits $\vec s\to0$ and $\vec s\to\vec s_k^0$ are introduced only to yield more consistent equations when deriving the flow of $Z_k^{\vec s}$ from \eqref{eq:masterflow}. By substituting the momentum $\boldsymbol{k}$ by $\boldsymbol{\epsilon}$, expansions of the trigonometric functions with respect to $\boldsymbol{\epsilon}$ become possible. This is very convenient, since the quadratic expansion of $\lim_{\boldsymbol{k}\to\boldsymbol{\epsilon}}(c_1+c_2)$ is rotationally invariant and a function of $|\boldsymbol{\epsilon}|$ only. 

From \eqref{eq:masterflow} we immediately have 
\begin{equation}
\label{eq:flowZk1}
\partial_kZ_k^{\vec s}=O_Z^{\text{symm/brok}}\left(\frac{1}{2}\text{Tr}((\partial_kR_k)M(\Gamma_k^{(2)}+R_k)^{-1})\right).
\end{equation}

\subsection{The symmetric phase}
As we already discussed, the flow equation for the effective average action has the form of a one loop equation. The one loop $\vec s_2$-propagator term corrections have the diagrammatical form 
\[
\setlength{\unitlength}{1mm}
\begin{fmfgraph*}(80,50)
\fmfleft{i}
\fmfright{o}
\fmf{scalar,tension=3.0,label=$K$}{v1,i}
\fmf{scalar,tension=3.0,label=$K$}{o,v2}
\fmf{fermion,right,label=$Q$}{v1,v2}
\fmf{fermion,right,label=$K+Q$}{v2,v1}
\fmfv{label=$\vec s_2$, label.angle=180}{i}
\fmfv{label=$\vec s_2$, label.angle=0}{o}
\fmfdot{v1,v2}
\end{fmfgraph*}
\]
for fluctuations in the fermionic sector and
\[
\setlength{\unitlength}{1mm}
\begin{fmfgraph*}(80,50)
\fmfleft{i}
\fmfright{o}
\fmf{scalar,tension=3.0,label=$K$}{v1,i}
\fmf{scalar,tension=3.0,label=$K$}{o,v1}
\fmf{scalar,right}{v1,v1}
\fmfv{label=$\vec s_2$, label.angle=180}{i}
\fmfv{label=$\vec s_2$, label.angle=0}{o}
\fmfdot{v1}
\end{fmfgraph*}
\]
for fluctuations in the bosonic sector. Acting with $O_Z^{\text{symm}}$ on these expressions, the external legs are amputated. Since we have to differentiate with respect to $|\boldsymbol{\epsilon}|$ (which is nothing else than a small external momentum) and the loop momentum does not depend on $K$ in the second diagram, we conclude that in the symmetric phase we do not have any contribution to the flow of $Z_k^{\vec s}$ from the bosonic sector. The flow equation therefore reads
\begin{equation}
\label{eq:flowZksymm1}
\partial_kZ_k^{\vec s}=-O_Z^{\text{symm}}\left(\frac{1}{2}\text{Tr}((\partial_kR_k^F)(F+R_k^F)^{-1})\right),
\end{equation}
where we have already performed the limit $\psi,\psi^*\to0$. Inserting $F$ on the right hand side, performing the trace and all limits and derivatives, a lengthy but straightforward calculation yields
\begin{equation}
\label{eq:flowZksymm2}
\partial_kZ_k^{\vec s}=\frac{h_a^2}{\pi^4\lambda_3}\tilde\partial_k\int_{-\pi}^\pi\frac{d^2q}{(2\pi)^2}\sum_{\epsilon_1}\frac{T}{T_k^2}z_{\epsilon_1,k}^{\vec s}(\boldsymbol{q})
\end{equation}
with
\begin{align}
z_{\epsilon_1,k}^{\vec s}(\boldsymbol{q})&=\frac{h_a^2}{8\tilde t_{\epsilon_1}\cosh^2(\tilde t_{\epsilon_1}/2)}\left(\frac{1}{\tilde t_{\epsilon_1}^2}\left(\tilde t_{\epsilon_1}\left(1+\tilde t_{\epsilon_1}\tanh(\tilde t_{\epsilon_1}/2)\right)-\sinh\tilde t_{\epsilon_1}\right)\frac{s_1^2+s_2^2}{T_k^2}-\sinh\tilde t_{\epsilon_1}+\tilde t_{\epsilon_1}\right)\nonumber\\
&\quad-\pi^2\lambda_3\frac{\tanh\tilde t_{\epsilon_1}}{\tilde t_{\epsilon_1}},\nonumber\\
\tilde t_{\epsilon_1}&=\frac{2t(c_1+\epsilon_1c_2)}{T_k},\quad s_i=2t\sin(q_i/2).
\end{align}
Since in our truncation the right hand side of the flow equation depends on $k$ via $T_k$ only, we have $\tilde\partial_k=\partial_k$, so that we can directly integrate the flow equation to get 
\begin{equation}
\label{eq:Zksymm1}
Z_k^{\vec s}=1+\frac{h_a^2}{\pi^4\lambda_3}\int_{-\pi}^\pi\frac{d^2q}{(2\pi)^2}\sum_{\epsilon_1}\frac{T}{T_k^2}z_{\epsilon_1,k}^{\vec s}(\boldsymbol{q})
\end{equation}
From \eqref{eq:flowZksymm2} and \eqref{eq:Zksymm1}, we immediately find the anomalous dimension 
\begin{equation}
\eta_k=-k\partial_k\ln Z_k^{\vec s}=-\frac{k\partial_k Z_k^{\vec s}}{Z_k^{\vec s}}.
\end{equation}

\subsection{The broken phase}
In the broken phase, an additional diagram contributes to the one loop correction of the bosonic propagator term.
\[
\setlength{\unitlength}{1mm}
\begin{fmfgraph*}(80,50)
\fmfleft{i,i1}
\fmfright{o,o1}
\fmf{scalar,tension=3.0,label=$K$}{v1,i}
\fmf{scalar,tension=3.0,label=$K$}{o,v2}
\fmf{scalar,tension=3.0}{v1,i1}
\fmf{scalar,tension=3.0}{o1,v2}
\fmf{scalar,right,label=$Q$}{v2,v1}
\fmf{scalar,left,label=$K-Q$}{v2,v1}
\fmfv{label=$\vec s_2$, label.angle=180}{i}
\fmfv{label=$\vec s_2$, label.angle=0}{o}
\fmfv{label=$\vec s_2^0$, label.angle=180,decoration.shape=cross}{i1}
\fmfv{label=$\vec s_2^0$, label.angle=0,decoration.shape=cross}{o1}
\fmfdot{v1,v2}
\end{fmfgraph*}
\]
Here we have the possibility of a coupling to a non vanishing external field $\vec s_2^0$, so that we have a contribution from the bosonic sector in contrast to the symmetric case. The broken phase appears for small values of $k$. Since the flow in the fermionic sector goes as $k\partial_kZ_k^{\vec s}\propto k^2$, it is a good approximation to neglect the contribution from the fermionic sector in the broken phase. Furthermore, as we will see when discussing the flow equations for the effective potential below, only the Matsubara mode $m=0$ and momenta $\boldsymbol{q}$ close to zero contribute to the flow at small $k$. This means that --- with regard to the flow of $Z_k^{\vec s}$ --- we effectively face a two dimensional bosonic theory, which can be mapped to a simple $O(3)$-model in two dimensions. But for the $O(3)$-model the flow of the wave function renormalization constant has already been calculated \cite{wetton} to be
\begin{equation}
\eta_k=\frac{16v_2}{2}\alpha_{22,k}^0(\lambda_k^\alpha)^2m_{2,2}^2(2\lambda_k^\alpha\alpha_{22,k}^0,0)
\end{equation}
where $v_2=1/(8\pi)$ and $m_{2,2}^2(2\lambda_k^\alpha\alpha_{22,k}^0,0)$ is a so called threshold function whose exact form depends on the choice of the cutoff function $R_k^{\vec s}$ (again, for more details we refer to \cite{wetton}). If we choose $R_k^{\vec s}(\boldsymbol{q})=Z_k^{\vec s}(k^2-\boldsymbol{q}^2)\theta(k^2-\boldsymbol{q}^2)$, the threshold function simply reads \cite{nowak}
\begin{equation}
m_{2,2}^2(2\lambda_k^\alpha\alpha_{22,k}^0,0)=\frac{1}{(1+2\lambda_k^\alpha\alpha_{22,k}^0)^2}.
\end{equation}
With some minor adjustments reflecting different constant factors in our formulation as compared to the treatment of the $O(3)$-model in \cite{wetton}, the anomalous dimension finally reads
\begin{equation}
\label{eq:etak}
\eta_k=\frac{4T}{\pi^5\lambda_3}(\lambda_k^\alpha)^2\alpha_{22,k}^0\frac{1}{((2\pi)^2\pi^2+2\lambda_k^\alpha\alpha_{22,k}^0)^2}.
\end{equation}   
$Z_k^{\vec s}$ can then be derived from $\partial_kZ_k^{\vec s}=-Z_k^{\vec s}\eta_k$ by numerical integration.

\section{Results for different truncations of the effective potential}
A lot of interesting questions can be addressed by the formalism we developed so far. We will restrict our attention to two aspects:
\begin{itemize}
\item The Mermin-Wagner theorem \cite{mermin} tells us that in two dimensions at non vanishing temperature no long range order is present. However, in the mean field case we observed phase transitions for non vanishing temperature. Furthermore, in the effectively two dimensional cuprates antiferromagnetic and superconducting behavior is actually being observed. In a very simple setting (half filling, only antiferromagnetic degrees of freedom taken into account) we will see that this puzzle is solved by the renormalization group analysis. 
\item Mean field calculations tend to overemphasize symmetry breaking. It is interesting to study the influence of bosonic fluctuations on the phase diagram to better understand the basic mechanisms of the interplay of the different degrees of freedom.
\end{itemize}

\subsection{Antiferromagnetic behavior at half filling}
At half filling, we have $\rho=\rho_{11}=0$. Furthermore, we set $\Delta=0$, since we are only interested in the antiferromagnetic behavior.

\subsubsection{Fermionic contribution to the flow}
In this truncation, the flow equation for $U_k^F$ reads
\begin{align}
\label{eq:a22flowUkF}
\frac{d}{dk}U_k^F&=-2T{\cal V}\tilde\partial_k\sum_{\epsilon_1,\epsilon_2}\int_{-\pi}^\pi\frac{d^2q}{(2\pi)^2}\ln\cosh\left(\frac{1}{2T_k}\epsilon_2\sqrt{2{\hat h}_a^4\alpha_{22}+4t^2(c_1+\epsilon_1c_2)^2}\right)\nonumber\\
&=\frac{8kT{\cal V}}{T_k}\sum_{\epsilon_1}\int_{-\pi}^\pi\frac{d^2q}{(2\pi)^2}f_{\epsilon_1}(\alpha_{22})\tanh(f_{\epsilon_1}(\alpha_{22}))
\end{align}
with
\begin{equation}
f_{\epsilon_1}(\alpha_{22})=\frac{1}{2T_k}\sqrt{2{\hat h}_a^4\alpha_{22}+4t^2(c_1+\epsilon_1c_2)^2}.
\end{equation}
We have replaced $h_a$ by $\hat h_a$ to indicate that this coupling is not renormalized. 

\subsubsection{Truncation of the potential}
We work with a simple quartic truncation of the potential. We set 
\begin{align}
U_k&={\cal V}U_0+\sum_{K_1K_2}(\hat m_k^\alpha)^2\alpha_{22}(K_1,K_2)\delta(K_1+K_2)\nonumber\\
&\quad+\frac{1}{2}\sum_{K_1K_2K_3K_4}\hat\lambda_k^\alpha\alpha_{22}(K_1,K_2)\alpha_{22}(K_3,K_4)\delta(K_1+K_2+K_3+K_4)
\end{align}
in the symmetric phase and
\begin{align}
U_k&=\frac{1}{2}\sum_{K_1K_2K_3K_4}\hat\lambda_k^\alpha\delta(K_1+K_2+K_3+K_4)\nonumber\\
&\quad(\alpha_{22}(K_1,K_2)-\hat\alpha_{22,k}^0\delta(K_1)\delta(K_2))(\alpha_{22}(K_3,K_4)-\hat\alpha_{22,k}^0\delta(K_3)\delta(K_4))
\end{align}
in the broken phase. For homogeneous fields, these equations become
\begin{equation}
U_k={\cal V}\left(U_0+(\hat m_k^\alpha)^2\alpha_{22}+\frac{1}{2}\hat\lambda_k^\alpha\alpha_{22}^2\right)
\end{equation}
and 
\begin{equation}
U_k=\frac{1}{2}{\cal V}\hat\lambda_k^\alpha(\alpha_{22}-\hat\alpha_{22,k}^0)^2
\end{equation}
$\hat m_k^\alpha$ plays the role of a $k$-dependent mass, $\hat\alpha_{22,k}^0$ is the minimum of the potential if the symmetry is spontaneously broken and $\hat\lambda_k^\alpha$ is the quartic bosonic coupling.

\subsubsection{Bosonic contribution to the flow}
For our truncation, the matrix $B$ in \eqref{eq:matricesBF} becomes
\begin{equation}
B+\tilde R_k^B=\text{diag}(P_{11,k}^\chi+R_k^\chi,(P_{11,k}^\chi+R_k^\chi)^T,P_{11,k}^R+R_k^R,P_{22,k}^{\vec s}+R_k^{\vec s}+(D_\alpha+D_\alpha(\vec s_2\vec s_2^T)D_\alpha)U_k)
\end{equation}
If we neglect all terms in 
\begin{equation}
\frac{d}{dk}U_k^B=\frac{1}{2}\tilde\partial_k\lim_{b_i(Q)\to b_i\delta(Q)}\text{Tr}\ln\left(B+\tilde R_k^B\right)
\end{equation}
that do not depend on fields, we get
\begin{equation}
\frac{d}{dk}U_k^B=\frac{1}{2}\tilde\partial_k\lim_{b_i(Q)\to b_i\delta(Q)}\text{Tr}\ln\left(P_{22,k}^{\vec s}+R_k^{\vec s}+(D_\alpha+D_\alpha(\vec s_2\vec s_2^T)D_\alpha)U_k\right).
\end{equation}
If we now insert our truncation for $U_k$ and perform the trace, we arrive at
\begin{equation}
\label{eq:a22UkB1}
\frac{d}{dk}U_k^B=\frac{1}{2}{\cal V}\tilde\partial_k\sum_iT\sum_m\int_{-2\pi}^{2\pi}\frac{d^2q}{(2\pi)^2}\ln(P_{22,k}^{\vec s}(Q)+R_k^{\vec s}(Q)+c_k^i),
\end{equation}
where $Q=(\omega_m^B,\boldsymbol{q})$ and
\begin{equation}
\vec c_k=\left\{\begin{matrix}((\hat m_k^\alpha)^2+3\hat\lambda_k^\alpha\alpha_{22},(\hat m_k^\alpha)^2+\hat\lambda_k^\alpha\alpha_{22},(\hat m_k^\alpha)^2+\hat\lambda_k^\alpha\alpha_{22}),&\text{for the symmetric phase}\\\hat\lambda_k^\alpha(3\alpha_{22}-\hat\alpha_{22,k}^0,\alpha_{22}-\hat\alpha_{22,k}^0,\alpha_{22}-\hat\alpha_{22,k}^0),&\text{for the broken phase.}\end{matrix}\right.
\end{equation}

\subsubsection{Choice of the cutoff function}
We define the cutoff function $R_k^{\vec s}(Q)$ by
\begin{align}
R_k^{\vec s}(Q)&=(2\pi)^2\pi^2Z_k^{\vec s}\left(k^2-\frac{1}{2}\lambda_3(2-c_1-c_2)-m^2T^2\right)\nonumber\\
&\quad\theta\left(k^2-\frac{1}{2}\lambda_3(2-c_1-c_2)-m^2T^2\right)
\end{align}
where $\theta(x)$ is the usual step function\footnote{Explicitly, it is given by\begin{equation}\theta(x)=\left\{\begin{matrix}0&\quad\text{for $x<0$,}\\1&\quad\text{for $x\ge0$.}\end{matrix}\right.\end{equation}}. This cutoff function respects \eqref{eq:effavac:Rkconstr} and is therefore a viable choice. The step function cuts off large Matsubara frequencies and momenta. By lowering $k$, we average over larger and larger regions in position space. For this cutoff, we therefore can relate properties of the effective action at some given value of $k$ with long range order properties of the theory in position space at scales of order $1/k$. This will become important when interpreting the results of the renormalization group analysis with respect to the Mermin-Wagner theorem puzzle. Note however that from a purely numerical standpoint our choice of the cutoff function is not optimal, since the cutoff for the Matsubara sum is momentum dependent, so that the Matsubara sum cannot be carried out analytically. In the next section, where we consider a different truncation and are not dependent on the position space interpretation of the flow, we will use a different cutoff for which the Matsubara sum can be evaluated analytically. 

Inserting the cutoff function into \eqref{eq:a22UkB1}, we finally arrive at
\begin{align}
\label{eq:a22flowUkB}
\frac{d}{dk}U_k^B&=\frac{2T{\cal V}}{k}\sum_i\sum_{m=-M}^M\int^\theta\frac{d^2q}{(2\pi)^2}\nonumber\\
&\quad\frac{-\eta_k(k^2-\frac{1}{2}\lambda_3(2-\cos(q_1/2)-\cos(q_2/2))-(mT)^2)+2k^2}{k^2+\frac{c_k^i}{(2\pi)^2\pi^2Z_k^{\vec s}}},
\end{align}
where $M=\text{max}\{m\in\mathbbm{N}|m<k/T\}$ and the momentum integration $\int^\theta$ runs over $\{\{q_1,q_2\}|q_i\in[0,2\pi],k^2-(\lambda_3/2)(2-c_1-c_2)-m^2T^2>0\}$. $\eta_k=-k\partial_k\ln Z_k^{\vec s}$ is the anomalous dimension.

\subsubsection{Extraction of the coefficients}
The flow equations for the coefficients follow from the flow equations of the effective potential by
\begin{align}
\label{eq:excosymm}
\partial_k(\hat m_k^\alpha)^2&=\frac{1}{\cal V}\lim_{\alpha_{22}\to0}\frac{d}{d\alpha_{22}}\left(\frac{d}{dk}U_k\right)\nonumber\\
\partial_k\hat\lambda_k^\alpha&=\frac{1}{\cal V}\lim_{\alpha_{22}\to0}\frac{d^2}{d\alpha_{22}^2}\left(\frac{d}{dk}U_k\right)
\end{align}
for the symmetric phase and
\begin{align}
\label{eq:excobrok}
\partial_k\hat\alpha_{22,k}^0&=-\frac{1}{{\cal V}\hat\lambda_k^\alpha}\lim_{\alpha_{22}\to\hat\alpha_{22,k}^0}\frac{d}{d\alpha_{22}}\left(\frac{d}{dk}U_k\right)\nonumber\\
\partial_k\hat\lambda_k^\alpha&=\frac{1}{\cal V}\lim_{\alpha_{22}\to\hat\alpha_{22,k}^0}\frac{d^2}{d\alpha_{22}^2}\left(\frac{d}{dk}U_k\right)
\end{align}
for the broken phase. Note that if we set $\alpha_{22}=0$ in the symmetric phase expression of $\vec c_k$, we see that $\vec c_k\to((\hat m_k^\alpha)^2,(\hat m_k^\alpha)^2,(\hat m_k^\alpha)^2)$ describes three modes with equal mass. In contrast to this, if we set $\alpha_{22}=\hat\alpha_{22,k}^0$ in the broken phase expression of $\vec c_k$, we find $\vec c_k=(2\hat\lambda_k^\alpha\hat\alpha_{22,k}^0,0,0)$, that is, one massive mode and two massless modes. Of course, the massless modes are nothing else than the Goldstone modes which have to appear when breaking $SU(2)$ down to $U(1)$. 

\subsubsection{Introduction of rescaled, renormalized quantities}
It is very convenient to introduce rescaled, renormalized quantities defined by
\begin{equation}
\label{eq:scarenvar}
(m_k^\alpha)^2=\frac{(\hat m_k^\alpha)^2}{Z_k^{\vec s}k^2},\quad\alpha_{22,k}^0=Z_k^{\vec s}\hat\alpha_{22,k}^0,\quad\lambda_k^\alpha=\frac{\hat\lambda_k^\alpha}{(Z_k^{\vec s})^2k^2},\quad h_a^2=\frac{\hat h_a^2}{\sqrt{Z_k^{\vec s}}}.
\end{equation}
The advantage of rewriting the flow equations in terms of these variables is that they (at least for small $k$) no longer depend on $Z_k^{\vec s}$ and $k$ explicitly. 

\subsubsection{The flow equations}
Using \eqref{eq:excosymm} and \eqref{eq:excobrok}, we can derive the flow equations for the rescaled and renormalized quantities \eqref{eq:scarenvar} (except for the coupling $h_a$ that as we said will be kept constant) from \eqref{eq:a22flowUkF} and \eqref{eq:a22flowUkB}. They read
\begin{align}
\label{eq:flowa22symm}
\partial_t(m_k^\alpha)^2&=\frac{2Th_a^4}{T_k^3}\sum_{\epsilon_1}\int_{-\pi}^{\pi}\frac{d^2q}{(2\pi)^2}\left(\frac{\tanh f_{\epsilon_1}(0)}{f_{\epsilon_1}(0)}+\frac{1}{\cosh^2f_{\epsilon_1}(0)}\right)\nonumber\\
&\quad-\frac{10T\lambda_k^\alpha}{(2\pi)^2\pi^2}\sum_{m=-M}^M\int^\theta\frac{d^2q}{(2\pi)^2}\frac{1}{k^2}\frac{2-\eta_k\left(1-\frac{\lambda_3}{2k^2}(2-c_1-c_2)-(mT/k)^2\right)}{\left(1+\frac{(m_k^\alpha)^2}{(2\pi)^2\pi^2}\right)^2}\nonumber\\
&\quad-(m_k^\alpha)^2(2-\eta_k)\nonumber\\
\partial_t\lambda_k^\alpha&=\frac{Th_a^8}{2T_k^5}\sum_{\epsilon_1}\int_{-\pi}^\pi\frac{d^2q}{(2\pi)^2}\left(\frac{f_{\epsilon_1}(0)-\sinh f_{\epsilon_1}(0)\cosh f_{\epsilon_1}(0)}{f_{\epsilon_1}(0)^3\cosh^2f_{\epsilon_1}(0)}-\frac{2\tanh f_{\epsilon_1}(0)}{f_{\epsilon_1}(0)\cosh^2f_{\epsilon_1}(0)}\right)\nonumber\\
&\quad+\frac{44T(\lambda_k^\alpha)^2}{(2\pi)^4\pi^4}\sum_{m=-M}^M\int^\theta\frac{d^2q}{(2\pi)^2}\frac{1}{k^2}\frac{2-\eta_k\left(1-\frac{\lambda_3}{2k^2}(2-c_1-c_2)-(mT/k)^2\right)}{\left(1+\frac{(m_k^\alpha)^2}{(2\pi)^2\pi^2}\right)^3}\nonumber\\
&\quad-2\lambda_k^\alpha(1-\eta_k)
\end{align}
in the symmetric phase and
\begin{align}
\label{eq:flowa22brok}
\partial_t\alpha_{22,k}^0&=-\frac{2Th_a^4}{T_k^3\lambda_k^\alpha}\sum_{\epsilon_1}\int_{-\pi}^{\pi}\frac{d^2q}{(2\pi)^2}\left(\frac{\tanh f_{\epsilon_1}(\alpha_{22,k}^0)}{f_{\epsilon_1}(\alpha_{22,k}^0)}+\frac{1}{\cosh^2f_{\epsilon_1}(\alpha_{22,k}^0)}\right)\nonumber\\
&\quad+\frac{2T}{(2\pi)^2\pi^2}\sum_{m=-M}^M\int^\theta\frac{d^2q}{(2\pi)^2}\frac{1}{k^2}\left(2-\eta_k\left(1-\frac{\lambda_3}{2k^2}(2-c_1-c_2)-(mT/k)^2\right)\right)\nonumber\\
&\quad\qquad\left[2+\frac{3}{\left(1+\frac{2\lambda_k^\alpha\alpha_{22,k}^0}{(2\pi)^2\pi^2}\right)^2}\right]\nonumber\\
&\quad-\alpha_{22,k}^0\eta_k\nonumber\\
\partial_t\lambda_k^\alpha&=\frac{Th_a^8}{2T_k^5}\sum_{\epsilon_1}\int_{-\pi}^\pi\frac{d^2q}{(2\pi)^2}\left({\scriptstyle\frac{f_{\epsilon_1}(\alpha_{22,k}^0)-\sinh f_{\epsilon_1}(\alpha_{22,k}^0)\cosh f_{\epsilon_1}(\alpha_{22,k}^0)}{f_{\epsilon_1}(\alpha_{22,k}^0)^3\cosh^2f_{\epsilon_1}(\alpha_{22,k}^0)}-\frac{2\tanh f_{\epsilon_1}(\alpha_{22,k}^0)}{f_{\epsilon_1}(\alpha_{22,k}^0)\cosh^2f_{\epsilon_1}(\alpha_{22,k}^0)}}\right)\nonumber\\
&\quad+4T\left(\frac{\lambda_k^\alpha}{(2\pi)^2\pi^2}\right)^2\sum_{m=-M}^M\int^\theta\frac{d^2q}{(2\pi)^2}\frac{1}{k^2}\left(2-\eta_k\left(1-\frac{\lambda_3}{2k^2}(2-c_1-c_2)-(mT/k)^2\right)\right)\nonumber\\
&\quad\qquad\left[2+\frac{9}{\left(1+\frac{2\lambda_k^\alpha\alpha_{22,k}^0}{(2\pi)^2\pi^2}\right)^3}\right]\nonumber\\
&\quad-2\lambda_k^\alpha(1-\eta_k)
\end{align}
in the broken phase with $\partial_t=k\partial_k$ (that is, $t=\ln(k/\Lambda)$), $T_k=T+k^2$, $M=\text{max}\{m\in\mathbbm{N}|m<k/T\}$, $c_i=\cos(q_i/2)$, $\eta_k=-k\partial_k\ln Z_k^{\vec s}$,
\begin{equation}
f_{\epsilon_1}(\alpha_{22})=1/(2T_k)\sqrt{2h_a^4\alpha_{22}+4t^2(c_1+\epsilon_1c_2)^2},
\end{equation}
and the integration runs over 
\begin{equation}
\{\{q_1,q_2\}|q_i\in[0,2\pi],k^2-(\lambda_3/2)(2-c_1-c_2)-m^2T^2>0\}.
\end{equation}
The initial conditions are $(m_\Lambda^\alpha)^2=((2\pi)^2h_a^2)/\Lambda^2$, where $\Lambda$ is the initial value of $k$, $\lambda_\Lambda^\alpha=0$ and $Z_\Lambda^{\vec s}=1$. We did not write down a flow equation for $Z_k^{\vec s}$, since $Z_k^{\vec s}$ and the anomalous dimension $\eta_k$ can be calculated analytically in our truncation as we saw in sec \ref{flowwaveren}.

\subsubsection{Remarks}
For small $k$, we have $T_k\approx T$ and the integral in the bosonic sector
\begin{equation}
\int^\theta d^2q\,\frac{1}{k^2}\left(2-\eta_k\left(1-\frac{\lambda_3}{2k^2}(2-c_1-c_2)-(mT/k)^2\right)\right)
\end{equation}
can be evaluated analytically. First note that for small $k$ we have $M=0$, so that only the $m=0$-mode contributes to this integral. The integration region is then given by 
\begin{equation}
\{\{q_1,q_2\}|q_i\in[0,2\pi],k^2-(\lambda_3/2)(2-c_1-c_2)>0\}.
\end{equation}
Thus for small $k$, only momenta close to $\boldsymbol{q}=0$ contribute, and we can expand the trigonometric functions to quadratic order around $\boldsymbol{q}=0$. It is then simple to perform the integration and to see that all explicit $k$-dependences cancel. For small $k$, the right hand sides of the flow equations therefore indeed do not depend explicitly on $k$. This is a very nice property, since it allows to extend the numerical investigation to very small values of $k$ without the complications usually arising when dealing with very small numbers in numerical calculations. 

The possible zeroes in the denominators of the terms from the fermionic sector do not pose any problems, since they have the form
\begin{equation}
\lim_{x\to0}\frac{\tanh x}{x}=1,\quad\lim_{x\to0}\frac{x-\sinh x\cosh x}{x^3}=-\frac{2}{3}.
\end{equation}

Note that the emergence of the two different terms in the bosonic sector in the broken phase enclosed by the square brackets, one with denominator $1$ and one with denominator $(1+2\lambda_k^\alpha\alpha_{22,k}^0/((2\pi)^2\pi^2))^n$, is a consequence of the Goldstone bosons appearing in the broken phase. The first term stems from the two Goldstone modes and the second from the massive mode. 

It is illuminating to discuss qualitatively the flow in the broken phase. In the broken phase $k$ is very small. As can be numerically confirmed, the anomalous dimension $\eta_k$ is also small. In the flow equation for $\lambda_k^\alpha$ in the broken phase, we can therefore approximate $(1-\eta_k)\approx1$ and $2-\eta_k(\ldots)\approx2$. The quartic coupling becomes large, so that $\frac{9}{(1+\lambda_k^\alpha(\ldots))^3}\approx0$. If we keep $\hat h_a$ fixed, then (since $Z_k^{\vec s}$ becomes large) $h_a$ becomes very small, so that we can neglect the contributions from the fermionic sector. In this case, the flow equation for $\lambda_k^\alpha$ is 
\begin{equation}
\partial_t\lambda_k^\alpha=(\lambda_k^\alpha)^2(\ldots)-2\lambda_k^\alpha.
\end{equation} 
This means that for decreasing $k$ and initially increasing $\lambda_k^\alpha$, we eventually reach some value of $k$ where the right hand side vanishes, which means that we have reached a fixed point of $\lambda_k^\alpha$. Assuming that this fixed point has been reached, we can regard $\lambda_k^\alpha$ as constant with regard to the other flow equations. For small $k$ (so that the momentum integral can be performed), small $\eta_k$ and large $\lambda_k^\alpha\alpha_{22,k}^0$ ($\lambda_k^\alpha$ at its fixed point), the flow equation for the potential minimum reads
\begin{equation}
\partial_t\alpha_{22,k}^0=2\cdot\frac{T}{\lambda_3\pi^5}-\alpha_{22,k}^0\eta_k
\end{equation}
We emphasize the factor $2$ to indicate its origin from the presence of $2$ Goldstone modes. Inserting $\eta_k$ from \eqref{eq:etak}, we find
\begin{equation}
\partial_t\alpha_{22,k}^0=2\cdot\frac{T}{\lambda_3\pi^5}-\frac{T}{\lambda_3\pi^5}=\frac{T}{\lambda_3\pi^5}.
\end{equation}
The anomalous dimension correction exactly cancels the contribution of one of the two Goldstone modes! This is a well known feature which has already been discussed in the context of $O(N)$-models in \cite{wetton}. The right hand side of the flow equation in this simple approximation is positive. Thus the flow tends to lower $\alpha_{22,k}^0$ for decreasing $k$. If $\alpha_{22,k}^0$ becomes too small, our approximation $\lambda_k^\alpha\alpha_{22,k}^0\ll 1$ breaks down. However, if the minimum goes to zero, we can expect the system to return to the symmetric phase for small $k$. We will see that the numerical results confirm this expectation. 

The above discussion was based on fixing the unrenormalized coupling $\hat h_a$. Instead, we can fix the renormalized coupling $h_a$. In this case, the fermionic fluctuations will not be suppressed in the broken phase. As in the mean field case, the fermionic contributions will stabilize the symmetry breaking. We expect that in this case we will have true symmetry breaking and do not return to the symmetric phase even for very small $k$. Again, this expectation is confirmed by our numerical results.  

\subsubsection{Numerical results}
\begin{figure}
\centering
\psfrag{alpha0}{$\!\!\!\!\!\!\!\!\!\!\!\!\!\!10^{-3}(m_k^\alpha)^2$}
\psfrag{lambda}{$\!\!\!\!\!\!10^{-5}\lambda_k^\alpha$}
\psfrag{Zk}{$\!\!\!Z_k$}
\psfrag{eta}{$\!\eta_k$}
\psfrag{-t}{$-t$}
\includegraphics[scale=0.8]{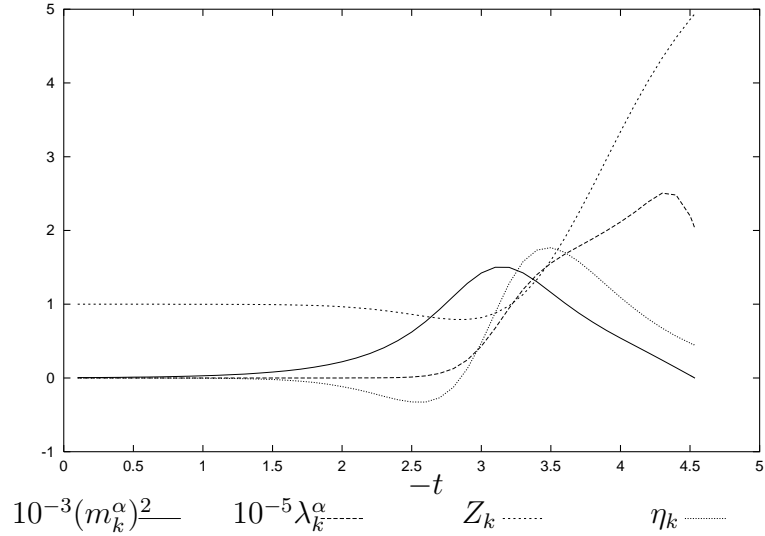}
\caption[The flow of the potential for $\vec s_2$ with fixed unrenormalized coupling in the symmetric regime]{The flow of the mass $(m_k^\alpha)^2$, the quartic coupling $\lambda_k^\alpha$, the wave function renormalization $Z_k^\alpha$ and the anomalous dimension $\eta_k$ in the symmetric regime for the antiferromagnetic boson $\vec s_2$ at half filling and temperature $T=0.15$. The Yukawa coupling is $h_{a}^2=10$. In this plot we keep the {\bf unrenormalized} Yukawa coupling fixed.}
\label{fig:alpha:symmhqfest}
\end{figure}

\begin{figure}
\centering
\psfrag{alpha0}{$\!\!\!\!\!\!\!\!\!\!\!\!\!\!10^{-3}(m_k^\alpha)^2$}
\psfrag{lambda}{$\!\!\!\!\!\!10^{-6}\lambda_k^\alpha$}
\psfrag{Zk}{$\!\!\!Z_k$}
\psfrag{eta}{$\!\eta_k$}
\psfrag{-t}{$-t$}
\includegraphics[scale=0.8]{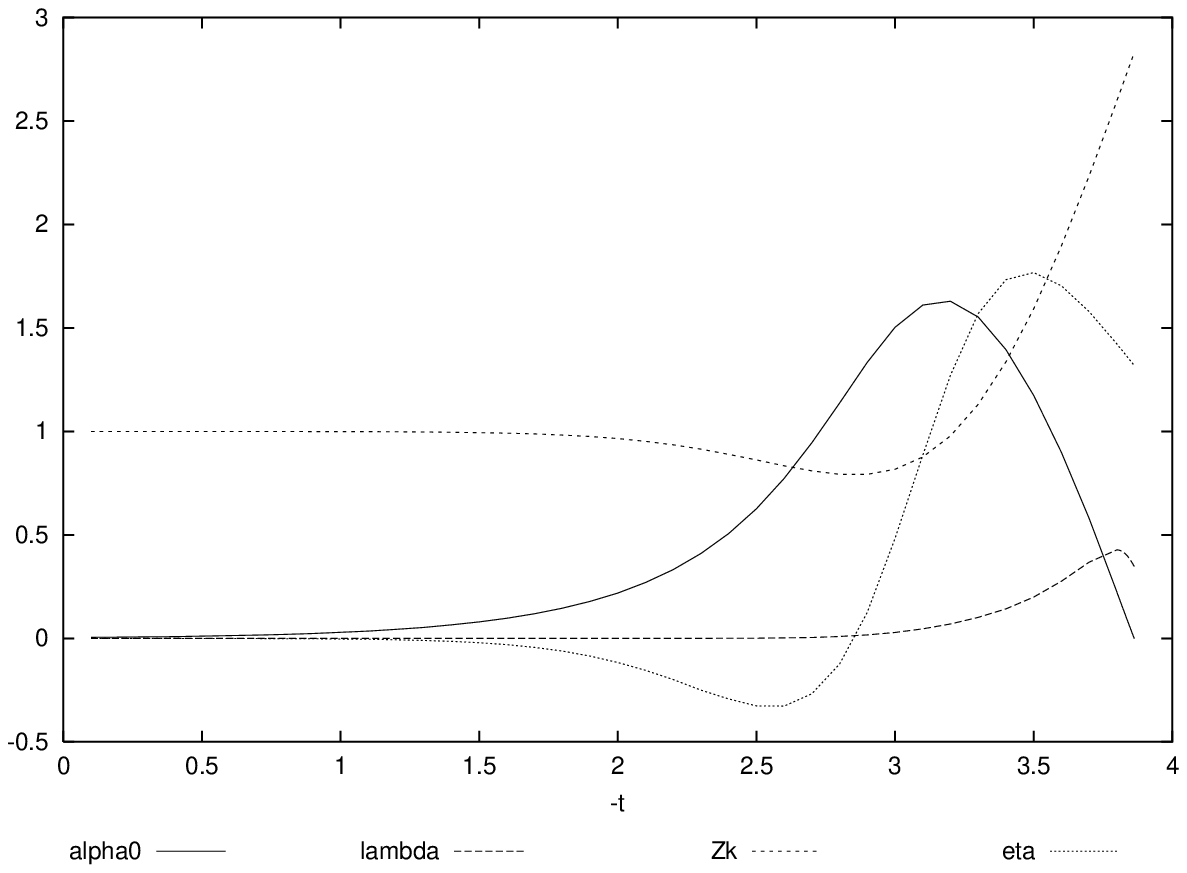}
\caption[The flow of the potential for $\vec s_2$ with fixed renormalized coupling in the symmetric regime]{The flow of the mass $(m_k^\alpha)^2$, the quartic coupling $\lambda_k^\alpha$, the wave function renormalization $Z_k^\alpha$ and the anomalous dimension $\eta_k$ in the symmetric regime for the antiferromagnetic boson $\vec s_2$ at half filling and temperature $T=0.15$. The Yukawa coupling is $h_{a}^2=10$. In this plot we keep the {\bf renormalized} Yukawa coupling fixed.}
\label{fig:alpha:symmhqrenfest}
\end{figure}

\begin{figure}
\centering
\psfrag{alpha0}{$\!\!\alpha_{0,k}$}
\psfrag{lambda}{$\!\!\!\!\!\!\!\!10^{-7}\lambda_k^\alpha$}
\psfrag{Zk}{$\!\!\!\!\!\!\!\!\!\!\!\!\!\!\!10^{-4}Z_k$}
\psfrag{-t}{$-t$}
\includegraphics[scale=0.8]{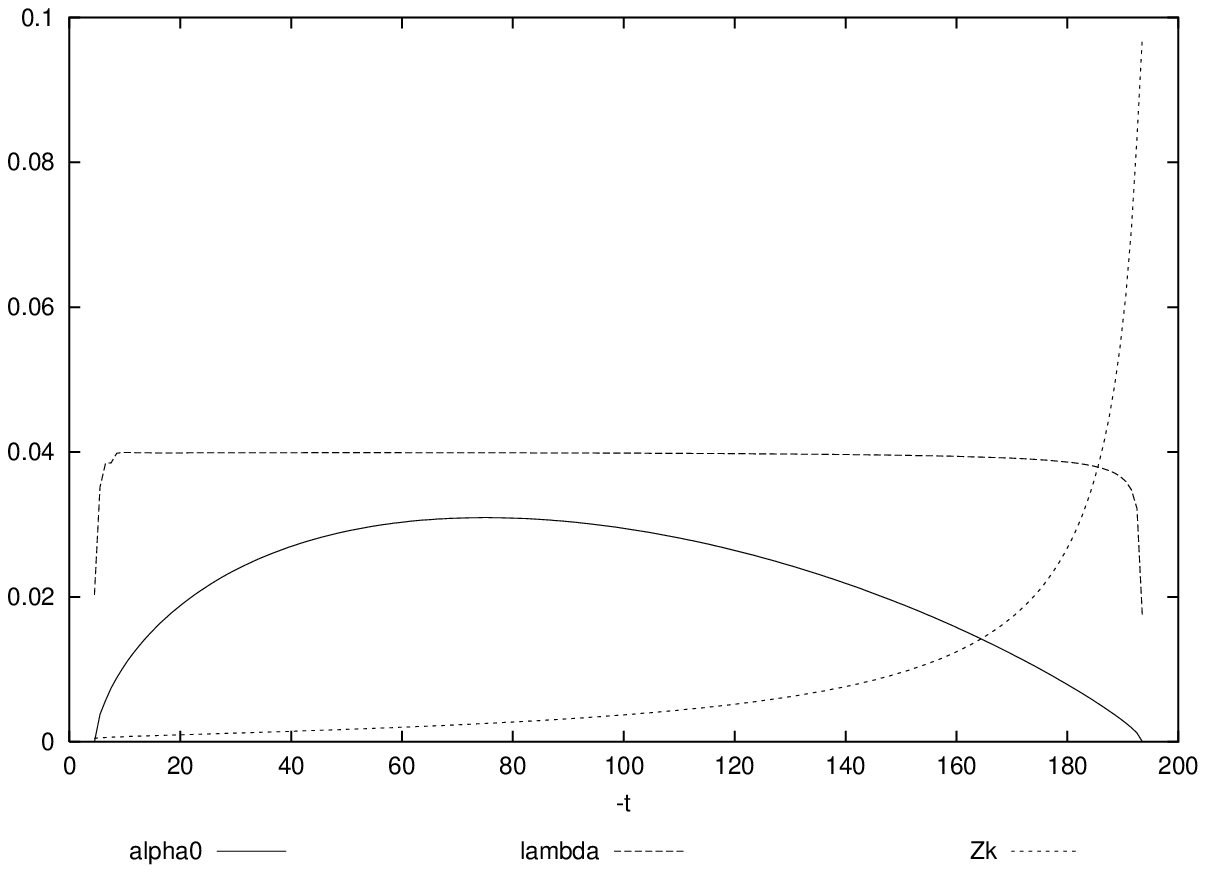}
\caption[The flow of the potential for $\vec s_2$ with fixed unrenormalized coupling in the regime of broken symmetry]{The flow of the minimum $\alpha_{0k}$, the quartic coupling $\lambda_k^\alpha$ and the wave function renormalization $Z_k^\alpha$ in the regime of broken symmetry for the antiferromagnetic boson $\vec s_2$ at half filling and temperature $T=0.15$. The Yukawa coupling is $h_{a}^2=10$. In this plot we keep the {\bf unrenormalized} Yukawa coupling fixed.}
\label{fig:alpha:brokenhqfest}
\end{figure}

\begin{figure}
\centering
\psfrag{alpha0}{$\!\!\!\!\alpha_{0,k}$}
\psfrag{lambda}{$\!\!\!\!\!\!\!\!10^{-4}\lambda_k^\alpha$}
\psfrag{Zk}{$\!\!\!\!\!\!\!\!\!\!\!\!\!\!\!10^{-6}Z_k$}
\psfrag{-t}{$-t$}
\includegraphics[scale=0.8]{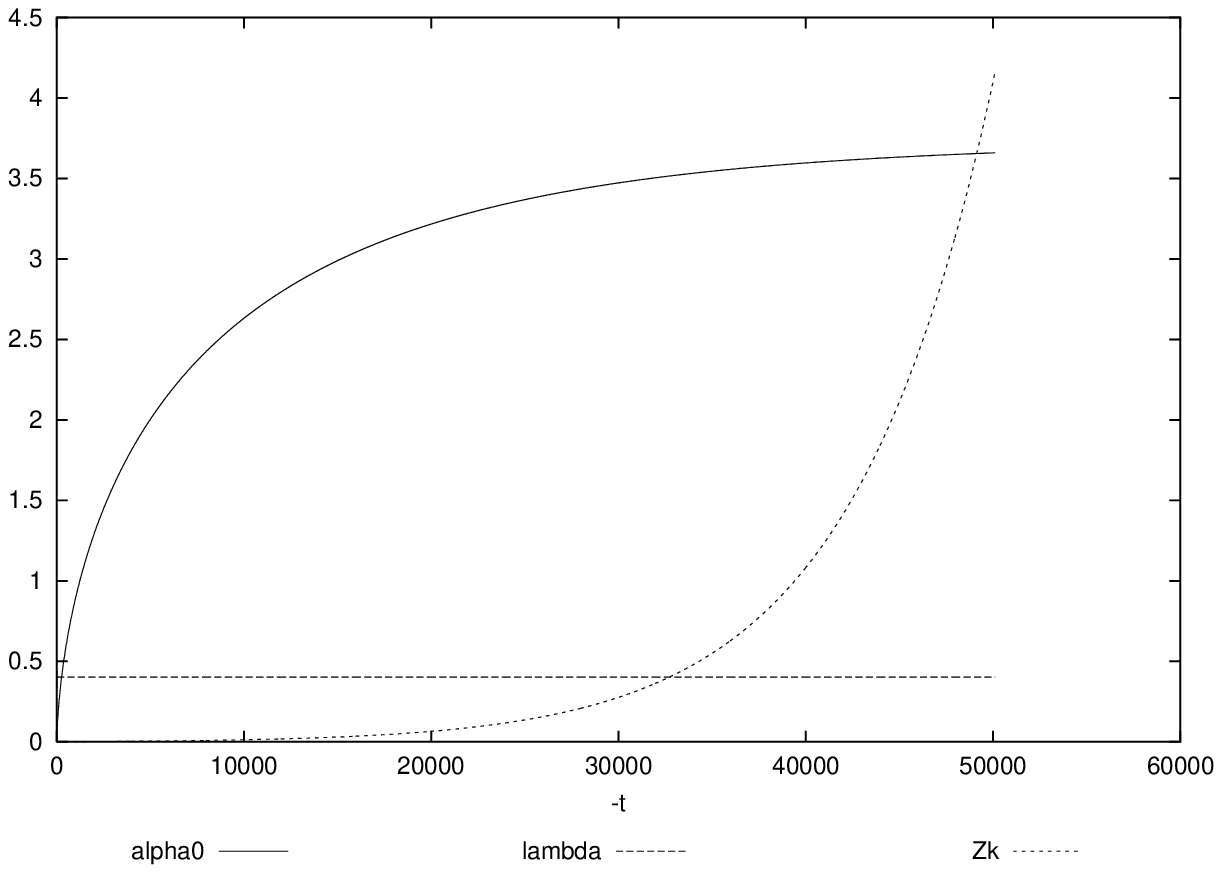}
\caption[The flow of the potential for $\vec s_2$ with fixed renormalized coupling in the regime of broken symmetry]{The flow of the mass $\alpha_{0k}$, the quartic coupling $\lambda_k^\alpha$ and the wave function renormalization $Z_k^\alpha$ in the regime of broken symmetry for the antiferromagnetic boson $\vec s_2$ at half filling and temperature $T=0.15$. The Yukawa coupling is $h_{a}^2=10$. In this plot we keep the {\bf renormalized} Yukawa coupling fixed.}
\label{fig:alpha:brokenhqrenfest}
\end{figure}

\begin{figure}
\centering
\psfrag{alpha0}{$\alpha_{0,k}$}
\psfrag{-t}{$-t$}
\psfrag{T003alpha0}{$\scriptstyle T=0.03$}
\psfrag{T007alpha0}{$\scriptstyle T=0.07$}
\psfrag{T011alpha0}{$\scriptstyle T=0.11$}
\psfrag{T015alpha0}{$\scriptstyle T=0.15$}
\includegraphics[scale=0.8]{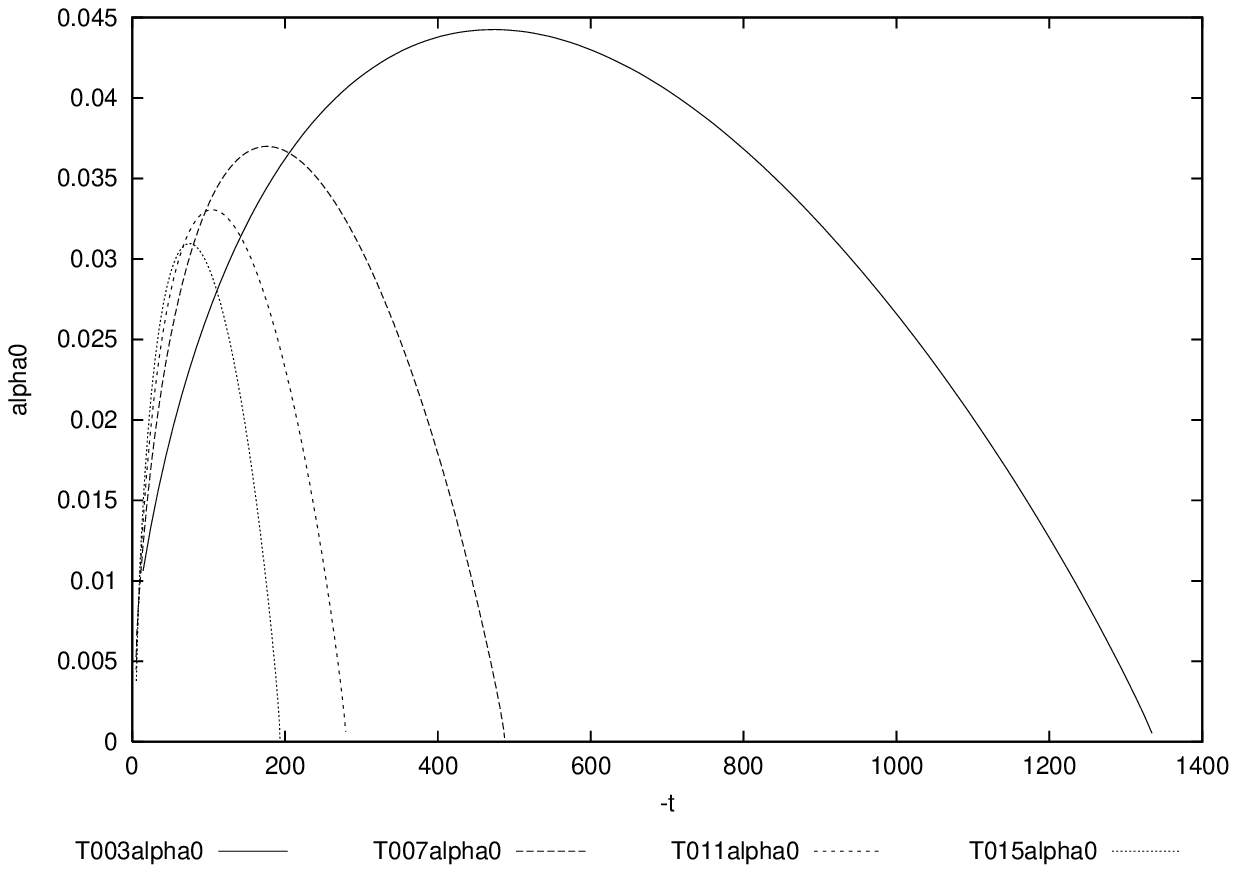}
\caption[The flow of the antiferromagnetic minimum with fixed unrenormalized coupling in the regime of broken symmetry for different temperatures.]{The flow of the minimum $\alpha_{0k}$ in the regime of broken symmetry for the antiferromagnetic boson $\vec s_2$ at half filling for different temperatures. The Yukawa coupling is $h_{a}^2=10$. In this plot we keep the {\bf unrenormalized} Yukawa coupling fixed.}
\label{fig:alpha:brokenhqfestdifftemp}
\end{figure}

\begin{figure}
\centering
\psfrag{alpha0}{$\alpha_{0,k}$}
\psfrag{-t}{$-t$}
\psfrag{T003alpha0}{$\scriptstyle T=0.03$}
\psfrag{T007alpha0}{$\scriptstyle T=0.07$}
\psfrag{T011alpha0}{$\scriptstyle T=0.11$}
\psfrag{T015alpha0}{$\scriptstyle T=0.15$}
\includegraphics[scale=0.8]{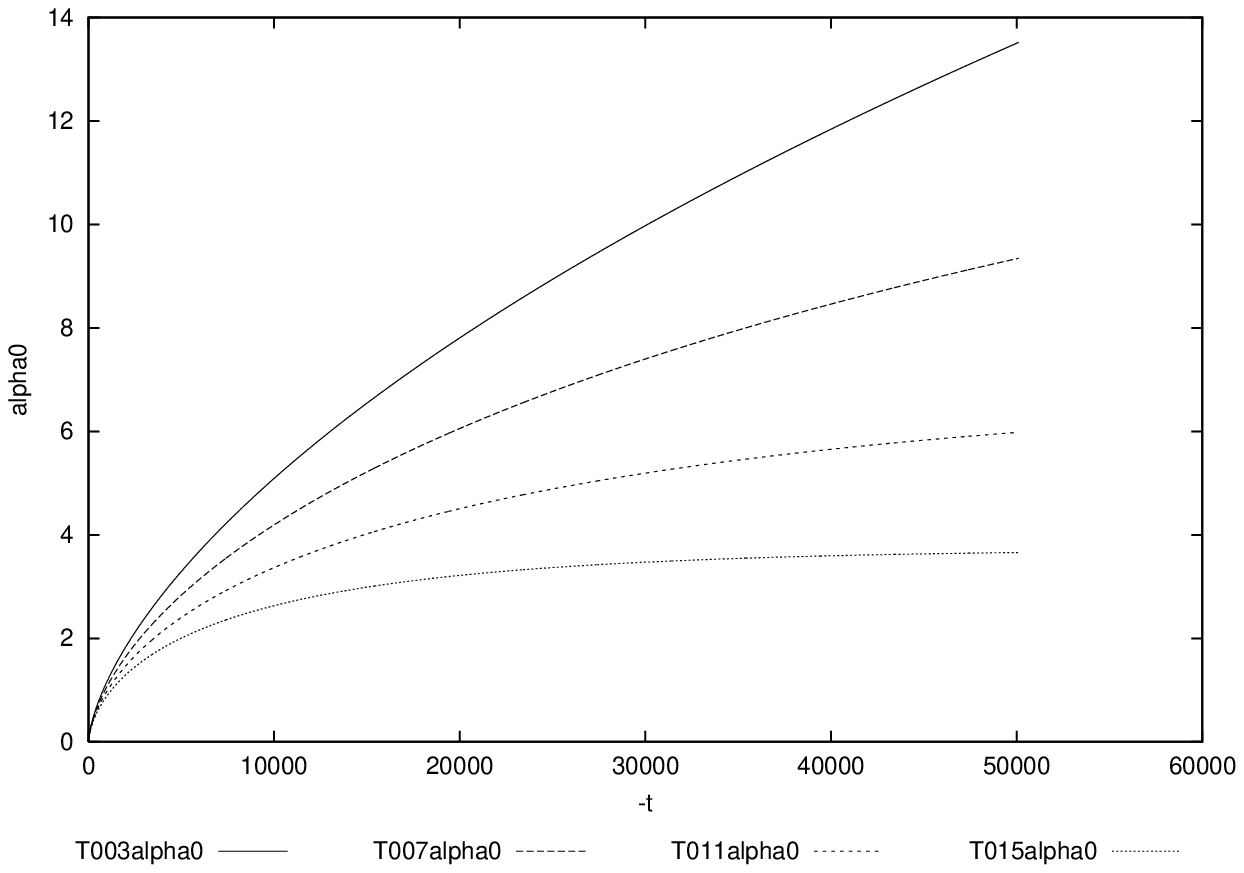}
\caption[The flow of the antiferromagnetic minimum with fixed renormalized coupling in the regime of broken symmetry for different temperatures]{The flow of the minimum $\alpha_{0k}$ in the regime of broken symmetry for the antiferromagnetic boson $\vec s_2$ at half filling for different temperatures. The Yukawa coupling is $h_{a}^2=10$. In this plot we keep the {\bf renormalized} Yukawa coupling fixed.}
\label{fig:alpha:brokenhqrenfestdifftemp}
\end{figure}

\begin{figure}
\centering
\psfrag{alpha0}{$\alpha_{0,k}$}
\psfrag{-t}{$-t$}
\psfrag{T003alpha0}{$\scriptstyle T=0.03$}
\psfrag{T007alpha0}{$\scriptstyle T=0.07$}
\psfrag{T011alpha0}{$\scriptstyle T=0.11$}
\psfrag{T015alpha0}{$\scriptstyle T=0.15$}
\includegraphics[scale=0.8]{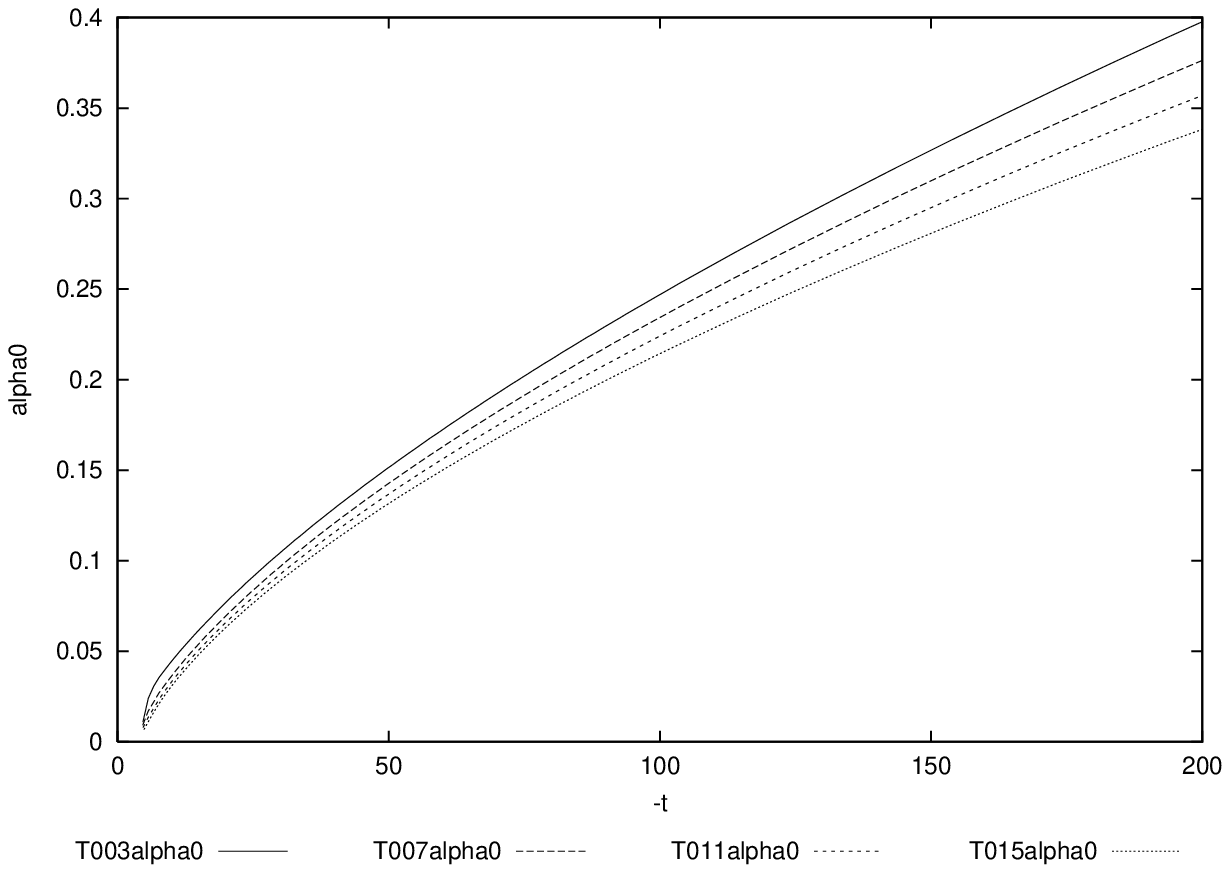}
\caption[The flow of the antiferromagnetic minimum with fixed renormalized coupling in the regime of broken symmetry for different temperatures at small $-t$]{The same plot as in fig. \ref{fig:alpha:brokenhqrenfestdifftemp}, but for smaller $-t$.}
\label{fig:alpha:brokenhqrenfestdifftemplowt}
\end{figure}

We solve the flow equations \eqref{eq:flowa22symm} and \eqref{eq:flowa22brok} numerically. We set $h_a^2=10$ and $\lambda_3=h_a^2/\pi^2$ at the beginning of the flow $k=\Lambda$. We choose $\Lambda$ large, so that the results no longer depend on the actual choice of $\Lambda$ in the limit $k\to0$. One finds that it actually suffices to set $\Lambda=10$. The Yukawa couplings are kept fixed. There are two ways to do so: Either we fix the unrenormalized coupling $\hat h_a$ or the renormalized coupling $h_a$. We give the results for both cases. We find that the critical temperature $T_c$ that describes the onset of symmetry breaking for $T<T_c$ is slightly decreased in comparison to the mean field case ($T_c\approx0.2$ here). In figs. \ref{fig:alpha:symmhqfest}--\ref{fig:alpha:brokenhqrenfest} we plot the results for the flowing variables at fixed temperature $T=0.15$. For this temperature, we find spontaneous symmetry breaking if $k$ is sufficiently small. In figs. \ref{fig:alpha:symmhqfest} and \ref{fig:alpha:symmhqrenfest} we plot the flow of the mass $(m_k^\alpha)^2$, the quartic coupling $\lambda_k^\alpha$, the wave function renormalization $Z_k^{\vec s}$ and the anomalous dimension $\eta_k$. We see that the mass becomes large if $k$ is lowered, reaching some maximum value and then drops to zero, indicating the phase transition. Note that we plotted $10^{-3}(m_k^\alpha)^2$, so that the initial value of the mass (which is $(m_\Lambda^\alpha)^2=(2\pi)^2h_a^2/\Lambda^2$) is not distinguishable from zero in the plot. The quartic coupling $\lambda_k^\alpha$ also reaches a maximum during the flow in the symmetric phase and begins to decay if $k$ is lowered towards the value where the phase transition occurs. Since the bosonic fluctuations (which enter the flow equation for the mass as a term $\propto \lambda_k^\alpha$) tend to prevent the phase transition, this behavior was to be expected. The wave function renormalization becomes large and remains large at the phase transition. This behavior will be important to explain the apparent contradiction to the Mermin-Wager theorem. In the symmetric phase the qualitative behavior of all quantities is independent of whether we fix the unrenormalized or the renormalized coupling. This is quite different in the broken phase. In  fig. \ref{fig:alpha:brokenhqfest} and fig. \ref{fig:alpha:brokenhqrenfest} we show the flow of the minimum $\alpha_{22,k}^0$, the quartic coupling $\lambda_k^\alpha$ and the wave function renormalization $Z_k^{\vec s}$ in the two cases. In both cases, the quartic coupling reaches some fixed point and $Z_k^{\vec s}$ diverges if $k$ is lowered. If we keep $\hat h_a$ fixed, the minimum reaches some maximum value and returns to zero. If we keep $h_a$ fixed, the maximum converges to some finite value. Both results are in agreement with our expectations based on analytical reasoning. 

In the case of fixing the unrenormalized coupling, no contradiction to the Mermin-Wagner theorem appears, since for $k\to0$ the symmetry becomes again unbroken. The symmetry breaking in some finite range of $k$ can then be interpreted as an antiferromagnetic order on large clusters that disappears if we average over even larger scales. Note the scale at which the broken symmetry becomes again unbroken. $t=-180$ corresponds to $k=\Lambda\exp(-180)$, which is extremely small, so that the symmetry becomes unbroken only when averaging over extremely large scales. For any probe of realistic size to be examined experimentally, we will find antiferromagnetic properties. In this interpretation, antiferromagnetic properties of superconductors do not contradict the Mermin-Wagner theorem, since they are finite size effects that would disappear if the probes were made large enough.

However, if we fix the unrenormalized coupling, the symmetry remains broken for $k\to0$. To see how this can be reconciled with the Mermin-Wagner theorem, we have to be careful in distinguishing the renormalized and unrenormalized minimum of the potential. The Mermin-Wagner theorem states that the {\em unrenormalized} minimum has to vanish for $k\to0$. But we have analyzed the flow of the renormalized minimum, and since the wave function renormalization diverges for $k\to0$, $\hat\alpha_{22,k}^0= \alpha_{22,k}^0/Z_k^{\vec s}$ actually vanishes in complete agreement with Mermin and Wagner. In fact, this is the mechanism how phase transitions can actually appear even in the case where they are forbidden by Mermin-Wagner (Kosterlitz-Thouless type phase transitions \cite{thouless}).

In figs. \ref{fig:alpha:brokenhqfestdifftemp}, \ref{fig:alpha:brokenhqrenfestdifftemp} we show how the flow of the minimum changes as a function of temperature. For fixed unrenormalized coupling, we see that the strength of the symmetry breaking and the scale at which the broken symmetry becomes unbroken increases if the temperature is lowered. This is what we intuitively expect, since the antiferromagnetic clusters should be larger for small temperature. For fixed renormalized coupling, the maximum value of the minimum also becomes larger for smaller temperature. Note that {\em all} the curves converge, but those for low $T$ on much larger scales than shown in the plot. Fig. \ref{fig:alpha:brokenhqrenfestdifftemplowt} enlarges the region of small $-t$ from \ref{fig:alpha:brokenhqrenfestdifftemp}, which is actually accessible by experiments. We see that the temperature dependence is weak in this region. 

In the framework of our truncation, we cannot decide which of the two possibilities of reconciling the occurrence of spontaneous symmetry breaking with the Mermin-Wagner theorem is realized. To do this, we have to include the flow of the Yukawa couplings.

\subsection{Charge density fluctuations and superconductivity}
In this section we will investigate the influence of charge density fluctuations on the superconducting properties of the theory. We set $\alpha_{22}=0$ and keep all wave function renormalizations constant.

\subsubsection{Fermionic contribution to the flow}
The flow equation for $U_k^F$ reads
\begin{align}
\frac{d}{dk}U_k^F&=-2T{\cal V}\tilde\partial_k\sum_{\epsilon_1,\epsilon_2}\int_{-\pi}^\pi\frac{d^2q}{(2\pi)^2}\nonumber\\
&\quad\ln\cosh\left(\frac{1}{2T_k}\sqrt{\left(\sqrt{2h_\rho^4\rho_{11}}+2t\epsilon_2(c_1+\epsilon_1c_2)\right)^2+2\hat h_d^4\Delta(c_1-\epsilon_1c_2)^2}\right)\nonumber\\
&=\frac{4kT{\cal V}}{T_k}\sum_{\epsilon_1\epsilon_2}\int_{-\pi}^\pi\frac{d^2q}{(2\pi)^2}f_{\epsilon_1,\epsilon_2}(\rho,\Delta)\tanh f_{\epsilon_1,\epsilon_2}(\rho,\Delta)
\end{align}
with
\begin{equation}
f_{\epsilon_1,\epsilon_2}(\rho,\Delta)=\frac{1}{2T_k}\sqrt{\left(\sqrt{2h_\rho^4\rho_{11}}+2t\epsilon_2(c_1+\epsilon_1c_2)\right)^2+2\hat h_d^4\Delta(c_1-\epsilon_1c_2)^2}.
\end{equation}
Although we keep all wave function renormalization constants fixed, we write $\hat h_d$ instead of $h_d$, since the initial value of $Z_k^\chi\neq1$.

\subsubsection{Truncation of the potential}
Our truncation for the potential reads
\begin{align}
U_k&={\cal V}U_0+\sum_{K_1K_2}({\hat m}_k^\Delta)^2\Delta(K_1,K_2)\delta(K_1-K_2)\nonumber\\
&\quad+\frac{1}{2}\sum_{K_1K_2K_3K_4}{\hat\lambda}_k^\Delta\Delta(K_1,K_2)\Delta(K_3,K_4)\delta(K_1-K_2+K_3-K_4)\nonumber\\
&\quad+\frac{1}{2}\sum_{K_1K_2K_3K_4}{\hat\lambda}_k^\rho\delta(K_1+K_2+K_3+K_4)\nonumber\\
&\quad\qquad(\rho_{11}(K_1,K_2)-\rho_{11}^0\delta(K_1)\delta(K_2))(\rho_{11}(K_3,K_4)-\rho_{11}^0\delta(K_3)\delta(K_4))\nonumber\\
&\quad+\frac{1}{2}\sum_{K_1K_2K_3K_4K_5K_6}{\hat\kappa}_k\delta(-K_1+K_2+K_3+K_4+K_5+K_6)\Delta(K_1,K_2)\nonumber\\
&\quad\qquad(\rho_{11}(K_3,K_4)-\rho_{11}^0\delta(K_3)\delta(K_4))(\rho_{11}(K_5,K_6)-\rho_{11}^0\delta(K_5)\delta(K_6))
\end{align}
in the symmetric phase. For homogeneous fields this becomes
\begin{equation}
U_k={\cal V}\left(({\hat m}_k^\delta)^2\Delta+\frac{1}{2}{\hat\lambda}_k^\Delta\Delta^2+\frac{1}{2}{\hat\lambda}_k^\rho(\rho_{11}-\rho_{11}^0)^2+\frac{1}{2}{\hat\kappa}_k\Delta(\rho_{11}-\rho_{11}^0)^2\right).
\end{equation}
We included terms up to quadratic order in $\Delta$ just as in the antiferromagnetic case. Additionally, we expanded $\rho_{11}$ around the minimum $\rho_{11}^0$. Recall that this minimum is not $k$-dependent, since the charge density is an external parameter controlled by the doping. The term $\propto{\hat\kappa}_k$ induces an interaction between the charge density and $\Delta$. 

We will restrict our attention to the flow in the symmetric regime, so that we do not write down a truncation for the potential in the broken phase. 

\subsubsection{Bosonic contribution to the flow}
In much the same way as in the antiferromagnetic case we find
\begin{align}
\label{eq:rhobosflow1}
\frac{d}{dk}U_k^B&=\frac{3}{2}{\cal V}\tilde\partial_kT\sum_m\int_{-2\pi}^{2\pi}\frac{d^2q}{(2\pi)^2}\ln\left(P_k^\chi(Q)+{\hat\lambda}_k^\Delta\Delta+\frac{1}{2}\kappa_k(\rho_{11}-\rho_{11}^0)^2\right)\nonumber\\
&\quad+\frac{1}{2}{\cal V}\tilde\partial_kT\sum_m\int_{-2\pi}^{2\pi}\frac{d^2q}{(2\pi)^2}\ln\nonumber\\
&\quad\biggl(\left(P_k^\chi(Q)+3{\hat\lambda}_k^\Delta\Delta+\frac{1}{2}{\hat\kappa}_k(\rho_{11}-\rho_{11}^0)^2\right)\left(\bar P_k^R(Q)+({\hat\lambda}_k^\rho+{\hat\kappa}_k\Delta)(3\rho_{11}-\rho_{11}^0)\right)\nonumber\\
&\quad\qquad-4{\hat\kappa}_k^2\Delta\rho_{11}(\rho_{11}-\rho_{11}^0)^2\biggr),
\end{align}
where 
\begin{equation}
P_k^\chi=P_{11,k}^\chi+R_k^\chi+({\hat m}_k^\Delta)^2,\quad \bar P_k^R=P_{11,k}^R+R_k^R.
\end{equation}

\subsubsection{Choice of the cutoff functions}
We define the cutoff functions to be
\begin{equation}
\frac{1}{Z_k^\chi}R_k^\chi(Q)=R_k^R(Q)=(2\pi)^2\pi^2(k^2-(mT)^2)\theta(k^2-(mT)^2).
\end{equation}
In contrast to the momentum dependent cutoff we chose in the last section, this cutoff allows to perform the Matsubara sum analytically, which speeds up the numerical calculation significantly. The drawback is that we can no longer interprete the flow as an averaging process over larger and larger clusters, but since in this section we will be mainly interested in whether a phase transition takes place for some temperature and $\rho_{11}^0$, we do not need this interpretation. 

Instead of inserting the cutoff function into \eqref{eq:rhobosflow1} immediately, it is more convenient to first extract the flow equations for the couplings and masses. 

\subsubsection{Extraction of the coefficients}
From the flow of the potential we can obtain the flow equations
\begin{align}
\partial_k(\hat m_k^\Delta)^2&=\frac{1}{\cal V}\lim_{\Delta\to0,\rho_{11}\to\rho_{11}^0}\frac{d}{d\Delta}\left(\frac{d}{dk}U_k\right)\nonumber\\
\partial_k\hat\lambda_k^\Delta&=\frac{1}{\cal V}\lim_{\Delta\to0,\rho_{11}\to\rho_{11}^0}\frac{d^2}{d\Delta^2}\left(\frac{d}{dk}U_k\right)\nonumber\\
\partial_k\hat\lambda_k^\rho&=\frac{1}{\cal V}\lim_{\Delta\to0,\rho_{11}\to\rho_{11}^0}\frac{d^2}{d\rho_{11}^2}\left(\frac{d}{dk}U_k\right)\nonumber\\
\partial_k\hat\kappa_k&=\frac{1}{\cal V}\lim_{\Delta\to0,\rho_{11}\to\rho_{11}^0}\frac{d^3}{d\rho_{11}^2d\Delta}\left(\frac{d}{dk}U_k\right).
\end{align}

\subsubsection{Introduction of rescaled quantities}
The rescaled quantities read
\begin{equation}
(m_k^\Delta)^2=\frac{(\hat m_k^\Delta)^2}{k^2},\quad\lambda_k^\Delta=\frac{\hat\lambda_k^\Delta}{k^2},\quad\lambda_k^\rho=\frac{\hat\lambda_k^\rho}{k^2},\quad\kappa_k=\frac{\hat\kappa_k}{k^2}.
\end{equation}

\subsubsection{The flow equations}
The flow equations are
\begin{align}
\label{eq:flowdeltarho}
\partial_t(m_k^\Delta)^2&=\frac{Th_d^4}{T_k^3}\sum_{\epsilon_1\epsilon_2}\int_{-\pi}^\pi\frac{d^2q}{(2\pi)^2}(c_1-\epsilon_1c_2)^2\left(\text{sech}^2f_{\epsilon_1\epsilon_2}+\frac{\tanh f_{\epsilon_1\epsilon_2}}{f_{\epsilon_1\epsilon_2}}\right)\nonumber\\
&\quad-(2\pi)^2T(2M+1)\int_0^{2\pi}\frac{d^2q}{k^2}\left(\frac{6Z_k^\chi\lambda_k^\Delta}{(P_k^\chi)^2}+\frac{2\kappa_k\rho_{11}^0}{(P_k^R)^2}\right)-2(m_k^\Delta)^2\nonumber\\
\partial_t\lambda_k^\Delta&=\frac{Th_d^8}{4T_k^5}\sum_{\epsilon_1\epsilon_2}\int_{-\pi}^\pi\frac{d^2q}{(2\pi)^2}(c_1-\epsilon_1c_2)^4\nonumber\\
&\quad\qquad\left(\frac{\text{sech}^2f_{\epsilon_1\epsilon_2}}{f_{\epsilon_1\epsilon_2}^2}-\frac{(1+2f_{\epsilon_1\epsilon_2}^2\text{sech}^2f_{\epsilon_1\epsilon_2})\tanh f_{\epsilon_1\epsilon_2}}{f_{\epsilon_1\epsilon_2}^3}\right)\nonumber\\
&\quad+(2\pi)^2T(2M+1)\int_0^{2\pi}\frac{d^2q}{k^2}\left(\frac{24Z_k^\chi(\lambda_k^\Delta)^2}{(P_k^\chi)^3}+\frac{8\kappa_k^2(\rho_{11}^0)^2}{(P_k^R)^3}\right)-2\lambda_k^\Delta\nonumber\\
\partial_t\lambda_k^\rho&=\frac{Th_\rho^8}{(h_\rho\rho)^2T_k^3}\sum_{\epsilon_1\epsilon_2}\int_{-\pi}^\pi\frac{d^2q}{(2\pi)^2}\left(\text{sech}^2f_{\epsilon_1\epsilon_2}-\frac{(1+2f_{\epsilon_1\epsilon_2}^2\text{sech}^2f_{\epsilon_1\epsilon_2})\tanh f_{\epsilon_1\epsilon_2}}{f_{\epsilon_1\epsilon_2}}\right)\nonumber\\
&\quad+(2\pi)^2T(2M+1)\int_0^{2\pi}\frac{d^2q}{k^2}\left(\frac{18(\lambda_k^\rho)^2}{(P_k^R)^3}-\frac{4Z_k^\chi\kappa_k}{(P_k^\chi)^2}\right)-2\lambda_k^\rho\nonumber\\
\partial_t\kappa_k&=\frac{Th_d^4h_\rho^8}{4(h_\rho\rho)^2T_k^5}\sum_{\epsilon_1\epsilon_2}\int_{-\pi}^\pi\frac{d^2q}{(2\pi)^2}(c_1-\epsilon_1c_2)^2\nonumber\\
&\quad\qquad\left(-\frac{3+2f_{\epsilon_1\epsilon_2}^2\text{sech}^2f_{\epsilon_1\epsilon_2}}{f_{\epsilon_1\epsilon_2}^2\cosh^2f_{\epsilon_1\epsilon_2}}+\frac{3\tanh f_{\epsilon_1\epsilon_2}}{f_{\epsilon_1\epsilon_2}^3}+4\text{sech}^2f_{\epsilon_1\epsilon_2}\tanh^2f_{\epsilon_1\epsilon_2}\right)\nonumber\\
&\quad+(2\pi)^2T(2M+1)\int_0^{2\pi}\frac{d^2q}{k^2}\nonumber\\
&\quad\qquad\left(\frac{12Z_k^\chi\lambda_k^\Delta\kappa_k}{(P_k^\chi)^3}+\frac{36\lambda_k^\rho\kappa_k}{(P_k^R)^3}-\frac{108(\lambda_k^\rho)^2\kappa_k\rho_{11}^0}{(P_k^R)^4}+\frac{8Z_k^\chi\kappa_k^2\rho_{11}^0}{(P_k^\chi)^2P_k^R}+\frac{8\kappa_k^2\rho_{11}^0}{P_k^\chi(P_k^R)^2}\right)\nonumber\\
&\quad-2\kappa_k
\end{align}
with $M=\text{max}\{m\in\mathbbm{N}|m<k/T\}$, 
\begin{align}
f_{\epsilon_1\epsilon_2}&=\frac{1}{2T_k}(h_\rho\rho+2t\epsilon_2(c_1+\epsilon_1c_2)),\nonumber\\
P_k^\chi&=Z_k^\chi\left(2\pi^4\lambda_3\frac{2-c_1-c_2}{k^2}+4\pi^4\right)+(m_k^\Delta)^2,\nonumber\\
P_k^R&=2\pi^4\lambda_3\frac{2-c_1-c_2}{k^2}+4\pi^4+2\lambda_k^\rho\rho_{11}^0,
\end{align}
where $\rho_{11}^0=\rho^2/(2h_\rho^2)$. The initial conditions are $(m_\Lambda^\Delta)^2=8\pi^2h_d^2/\Lambda^2$ and $\lambda_\Lambda^\Delta=\kappa_\Lambda=0$. $Z_k^\chi=1/\Lambda^8$ is kept fixed. The initial condition for $\lambda_k^\rho$ is set to $\lambda_\Lambda^\rho=1/\Lambda^8$, since the one loop correction to this coupling is of this order (we do not set $\lambda_\Lambda^\rho=0$ in order to have a potential minimum in $\rho_{11}^0$).

\subsubsection{Remarks}
As in the antiferromagnetic case, the zeroes of the denominators of the fermionic contributions to the flow do not cause any problems, since they can be canceled against numerator zeroes.

Note that the contributions in the bosonic sector have a simple diagrammatic interpretation. To see this, note that in our flow equations the derivatives $\tilde\partial_k$ have been carried out on the right hand side (which is of course necessary for the numerical investigation). For the diagrammatical interpretation, it is more useful to consider the expressions before taking the derivative. For example, the bosonic contribution to the flow of $(m_k^\Delta)^2$ is
\begin{equation}
\frac{1}{2}\tilde\partial_kT\sum_m\int_{-2\pi}^{2\pi}\frac{d^2q}{(2\pi)^2}\left(\frac{6\lambda_k^\Delta}{P_k^\chi}+\frac{2\kappa_k\rho_{11}^0}{P_k^R}\right).
\end{equation}  
and similarly for the other variables. The diagrammatic interpretation is then
\begin{gather}
m_k^\Delta:\parbox{20mm}{\setlength{\unitlength}{1mm}\begin{fmfgraph*}(20,15)
\fmfleft{i}
\fmfright{o}
\fmf{dashes}{v1,i}
\fmf{dashes}{o,v1}
\fmf{dashes}{v1,v1}
\fmfdot{v1}
\end{fmfgraph*}}
+\parbox{20mm}{\setlength{\unitlength}{1mm}\begin{fmfgraph*}(20,15)
\fmfleft{i}
\fmfright{o}
\fmftop{e}
\fmf{double,tension=0}{v1,e}
\fmf{dashes}{v1,i}
\fmf{dashes}{v1,o}
\fmf{plain}{v1,v1}
\fmfv{decoration.shape=cross}{e}
\fmfdot{v1}
\end{fmfgraph*}}\nonumber\\
\lambda_k^\Delta:\parbox{20mm}{\setlength{\unitlength}{1mm}\begin{fmfgraph*}(20,15)
\fmfleft{i1,i2}
\fmfright{o1,o2}
\fmf{dashes,tension=2}{i1,v1}\fmf{dashes,tension=2}{i2,v1}
\fmf{dashes,tension=2}{o1,v2}\fmf{dashes,tension=2}{o2,v2}
\fmf{dashes,left}{v1,v2}\fmf{dashes,right}{v1,v2}
\fmfdot{v1,v2}
\end{fmfgraph*}}
+\parbox{20mm}{\setlength{\unitlength}{1mm}\begin{fmfgraph*}(20,15)
\fmfleft{i1,i2,e1}
\fmfright{o1,o2,e2}
\fmf{dashes,tension=2}{i1,v1}\fmf{dashes,tension=2}{i2,v1}
\fmf{dashes,tension=2}{o1,v2}\fmf{dashes,tension=2}{o2,v2}
\fmf{plain,left}{v1,v2}\fmf{plain,right}{v1,v2}
\fmf{double,tension=2}{v1,e1}\fmf{double,tension=2}{v2,e2}
\fmfv{decoration.shape=cross}{e1,e2}
\fmfdot{v1,v2}
\end{fmfgraph*}},\qquad
\lambda_k^\rho:\parbox{20mm}{\setlength{\unitlength}{1mm}\begin{fmfgraph*}(20,15)
\fmfleft{i1,i2}
\fmfright{o1,o2}
\fmf{plain,tension=2}{i1,v1}\fmf{plain,tension=2}{i2,v1}
\fmf{plain,tension=2}{o1,v1}\fmf{plain,tension=2}{o2,v1}
\fmf{dashes}{v1,v1}
\fmfdot{v1}
\end{fmfgraph*}}
+\parbox{20mm}{\setlength{\unitlength}{1mm}\begin{fmfgraph*}(20,15)
\fmfleft{i1,i2}
\fmfright{o1,o2}
\fmf{plain,tension=2}{i1,v1}\fmf{plain,tension=2}{i2,v1}
\fmf{plain,tension=2}{o1,v2}\fmf{plain,tension=2}{o2,v2}
\fmf{plain,left}{v1,v2}\fmf{plain,right}{v1,v2}
\fmfdot{v1,v2}
\end{fmfgraph*}}\nonumber\\\nonumber\\
\kappa_k:\parbox{20mm}{\setlength{\unitlength}{1mm}\begin{fmfgraph*}(20,15)
\fmfleft{i1,i2,i3,i4}
\fmfright{o1,o2}
\fmf{plain,tension=2}{i1,v1}\fmf{plain,tension=2}{i2,v1}\fmf{plain,tension=2}{i3,v1}\fmf{plain,tension=2}{i4,v1}
\fmf{dashes,tension=2}{o1,v2}\fmf{dashes,tension=2}{o2,v2}
\fmf{dashes,left}{v1,v2}\fmf{dashes,right}{v1,v2}
\fmfdot{v1,v2}
\end{fmfgraph*}}
+\parbox{20mm}{\setlength{\unitlength}{1mm}\begin{fmfgraph*}(20,15)
\fmfleft{i1,i2,i3,i4}
\fmfright{o1,o2}
\fmf{plain,tension=2}{i1,v1}\fmf{plain,tension=2}{i2,v1}\fmf{dashes,tension=2}{i3,v1}\fmf{dashes,tension=2}{i4,v1}
\fmf{plain,tension=2}{o1,v2}\fmf{plain,tension=2}{o2,v2}
\fmf{plain,left}{v1,v2}\fmf{plain,right}{v1,v2}
\fmfdot{v1,v2}
\end{fmfgraph*}}
+\parbox{20mm}{\setlength{\unitlength}{1mm}\begin{fmfgraph*}(20,15)
\fmfsurround{i1,i2,i3,i4,e,i5,i6}
\fmf{plain,tension=2}{i1,v1}\fmf{plain,tension=2}{i2,v1}
\fmf{dashes,tension=2}{i3,v2}\fmf{dashes,tension=2}{i4,v2}
\fmf{plain,tension=2}{i5,v3}\fmf{plain,tension=2}{i6,v3}
\fmf{double,tension=2}{e,v2}
\fmfv{decoration.shape=cross}{e}
\fmf{plain}{v1,v2}\fmf{plain}{v2,v3}\fmf{plain}{v3,v1}
\fmfdot{v1,v2,v3}
\end{fmfgraph*}}
+\parbox{20mm}{\setlength{\unitlength}{1mm}\begin{fmfgraph*}(20,15)
\fmfleft{i1,i2,i3,i4}
\fmfright{o1,e,o2}
\fmf{plain,tension=2}{i1,v1}\fmf{plain,tension=2}{i2,v1}\fmf{plain,tension=2}{i3,v1}\fmf{dashes,tension=2}{i4,v1}
\fmf{plain,tension=2}{o1,v2}\fmf{dashes,tension=2}{o2,v2}
\fmf{double,tension=2}{e,v2}
\fmf{dashes,left}{v1,v2}\fmf{plain,right}{v1,v2}
\fmfv{decoration.shape=cross}{e}
\fmfdot{v1,v2}
\end{fmfgraph*}}
\end{gather}
where the dashed line represents the propagation of $\chi$, the solid line the propagation of $R_1$ and the double line the coupling to the external charge density $\rho_{11}^0$. 

\subsubsection{Numerical results}
\begin{figure}
\centering
\psfrag{T}{$T$}
\psfrag{hrho rho}{$\scriptstyle h_\rho\rho$}
\psfrag{hr1}{$\!\!\!\!\scriptstyle h_\rho^2=1$}
\psfrag{hr2}{$\!\!\!\!\!\!\!\!\scriptstyle h_\rho^2=20$}
\psfrag{hr4}{$\!\!\!\!\!\!\!\!\scriptstyle h_\rho^2=400$}
\includegraphics{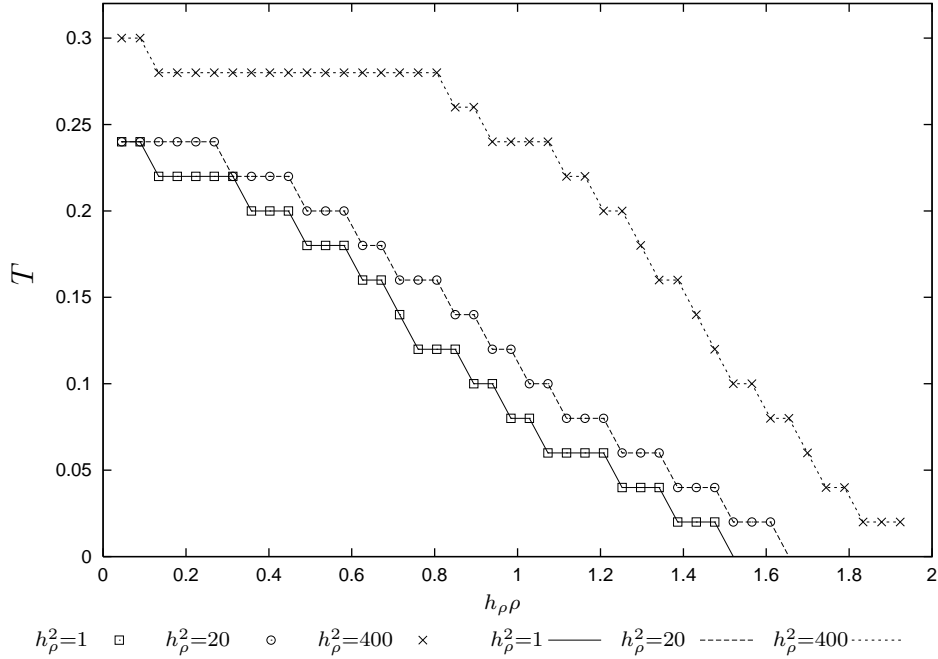}
\caption[Dependence of the superconducting region (including $\Delta$- and $\rho$- fluctuations) on the Yukawa coupling $h_\rho$]{Dependence of the superconducting region (including $\Delta$- and $\rho$- fluctuations) on the Yukawa coupling $h_\rho$. We take $h_d^2=20$.}
\label{fig:deltarho}
\end{figure}
We have solved the flow equations numerically for different temperatures, charge densities and Yukawa couplings $h_\rho$. We set $h_d^2=20$. We follow the flow of $(m_k^\Delta)^2$ until it either reaches zero or diverges. The first case indicates an instability in the $d$-wave superconducting channel (this should be taken with a grain of salt, since we know from our mean field discussion that phase transitions of first order render this criterion untrustworthy. However, in the superconducting regime we expect no phase transitions of first order to occur if no antiferromagnetic order is present). In fig. \ref{fig:deltarho} we show the borders of the regions where the mass happens to vanish during the flow. The comparison between different values of $h_\rho$ shows that a strong charge density coupling tends to enlarge the region of superconductivity. Note that we use the parameter $h_\rho\rho$ on the horizontal axis instead of $\rho$ in the mean field case. Recall that in the mean field case the results depended on $h_\rho$ only via $h_\rho\rho$. This means that if we ignored the bosonic fluctuations (which would reduce our flow equations to the mean field case), no difference between the phase borders for different $h_\rho$ would appear in our plot (a comparison of the mean field results and the results for the inclusion of various bosonic fluctuations can be found in the next section). The differences are consequences only of the inclusion of bosonic fluctuations.  The choice of the couplings in our plot is rather extreme; in the mean field case changing $h_a^2$ or $h_d^2$ only by a factor $2$ had a strong effect on the phase diagram. By changing $h_\rho^2$ by a factor of $2$ (not $20$ as in the plot), we see that the effect of charge density coupling is relatively small. We will see in the next section that --- in contrast to the charge density --- the phase diagram is very sensitive to the strength of antiferromagnetic fluctuations.    

\subsection{Antiferromagnetic fluctuations and superconductivity}
In this section, we will analyze the effect of antiferromagnetic fluctuations on the phase diagram. The derivation of the flow equations is very similar to the case of including charge density fluctuations in the flow. We set $\rho_{11}=\rho_{11}^0$ (that is, we do not consider charge density fluctuations in this section).

\subsubsection{Fermionic contribution to the flow}
The flow equation for $U_k^F$ reads
\begin{equation}
\frac{d}{dk}U_k^F=\frac{4kT{\cal V}}{T_k}\sum_{\epsilon_1\epsilon_2}\int_{-\pi}^\pi\frac{d^2q}{(2\pi)^2}f_{\epsilon_1,\epsilon_2}(\alpha_{22},\Delta)\tanh f_{\epsilon_1,\epsilon_2}(\alpha_{22},\Delta)
\end{equation}
with
\begin{equation}
f(\alpha_{22},\Delta)=\frac{1}{2T_k}\sqrt{\left(h_\rho\rho+\epsilon_2\sqrt{2h_a^4\alpha_{22}+4t^2(c_1+\epsilon_1c_2)^2}\right)^2+2h_d^4\Delta(c_1-\epsilon_1c_2)^2}.
\end{equation}

\subsubsection{Truncation of the potential}
We truncate the potential as 
\begin{align}
U_k&=U_0{\cal V}+\sum_{K_1K_2}(\hat m_k^\Delta)^2\Delta(K_1,K_2)\delta(K_1-K_2)\nonumber\\
&\quad+\frac{1}{2}\sum_{K_1K_2K_3K_4}\hat\lambda_k^\Delta\Delta(K_1,K_2)\Delta(K_3,K_4)\delta(K_1-K_2+K_3-K_4)\nonumber\\
&\quad+\sum_{K_1K_2}(\hat m_k^\alpha)^2\alpha_{22}(K_1,K_2)\delta(K_1+K_2)\nonumber\\
&\quad+\frac{1}{2}\sum_{K_1K_2K_3K_4}\hat\lambda_k^\alpha\alpha_{22}(K_1,K_2)\alpha_{22}(K_3,K_4)\delta(K_1+K_2+K_3+K_4)\nonumber\\
&\quad+\sum_{K_1K_2K_3K_4}\hat\gamma_k\alpha_{22}(K_1,K_2)\Delta(K_3,K_4)\delta(K_1+K_2-K_3+K_4).
\end{align}
This is an expansion up to quadratic order in $\alpha_{22}$ and $\Delta$. Again, we restrict our attention to the symmetric phase.

\subsubsection{Bosonic contribution to the flow}
The flow of the potential in the bosonic sector is given by
\begin{align}
\frac{d}{dk}U_k^B&=\frac{3}{2}{\cal V}\tilde\partial_kT\sum_m\int_{-2\pi}^{2\pi}\frac{d^2q}{(2\pi)^2}\ln(P_k^\chi(Q)+\lambda_k^\Delta\Delta+\gamma_k\alpha_{22})\nonumber\\
&\quad+{\cal V}\tilde\partial_kT\sum_m\int_{-2\pi}^{2\pi}\frac{d^2q}{(2\pi)^2}\ln(P_k^{\vec s}+\lambda_k^\alpha\alpha_{22}+\gamma_k\Delta)\nonumber\\
&\quad+\frac{1}{2}{\cal V}\tilde\partial_kT\sum_m\int_{-2\pi}^{2\pi}\frac{d^2q}{(2\pi)^2}\nonumber\\
&\quad\qquad\ln\left(\left((P_k^\chi+3\lambda_k^\Delta\Delta+\gamma_k\alpha_{22})(P_k^{\vec s}+3\lambda_k^\alpha\alpha_{22}+\gamma_k\Delta)\right)-4\gamma_k^2\Delta\alpha_{22}\right)
\end{align}
where 
\begin{equation}
P_k^\chi=P_{11,k}^\chi+R_k^\chi+(\hat m_k^\Delta)^2,\quad P_k^{\vec s}=P_{22,k}^{\vec s}+R_k^{\vec s}+(\hat m_k^\alpha)^2.
\end{equation}

\subsubsection{Choice of the cutoff function}
We choose the same cutoff function as in the last section
\begin{equation}
\frac{1}{Z_k^\chi}R_k^\chi(Q)=R_k^{\vec s}(Q)=(2\pi)^2\pi^2(k^2-(mT)^2)\theta(k^2-(mT)^2).
\end{equation}

\subsubsection{Extraction of the coefficients}
From the flow of the effective potential we obtain
\begin{align}
\label{eq:aldelex}
\partial_k(\hat m_k^\Delta)^2&=\frac{1}{\cal V}\lim_{\Delta,\alpha_{22}\to0}\frac{d}{d\Delta}\left(\frac{d}{dk}U_k\right)\nonumber\\
\partial_k\hat\lambda_k^\Delta&=\frac{1}{\cal V}\lim_{\Delta,\alpha_{22}\to0}\frac{d^2}{d\Delta^2}\left(\frac{d}{dk}U_k\right)\nonumber\\
\partial_k(\hat m_k^\alpha)^2&=\frac{1}{\cal V}\lim_{\Delta,\alpha_{22}\to0}\frac{d}{d\alpha_{22}}\left(\frac{d}{dk}U_k\right)\nonumber\\
\partial_k\hat\lambda_k^\alpha&=\frac{1}{\cal V}\lim_{\Delta,\alpha_{22}\to0}\frac{d^2}{d\alpha_{22}^2}\left(\frac{d}{dk}U_k\right)\nonumber\\
\partial_k\hat\gamma_k&=\frac{1}{\cal V}\lim_{\Delta,\alpha_{22}\to0}\frac{d^2}{d\alpha_{22}d\Delta}\left(\frac{d}{dk}U_k\right)\nonumber\\
\end{align}

\subsubsection{Introduction of rescaled quantities}
The rescaled quantities read
\begin{equation}
(m_k^\Delta)^2=\frac{(\hat m_k^\Delta)^2}{k^2},\quad\lambda_k^\Delta=\frac{\hat\lambda_k^\Delta}{k^2},\quad(m_k^\alpha)^2=\frac{(\hat m_k^\alpha)^2}{k^2},\quad\lambda_k^\alpha=\frac{\hat\lambda_k^\alpha}{k^2},\quad\gamma_k=\frac{\hat\gamma_k}{k^2}.
\end{equation}

\subsubsection{The flow equations}
We find the flow equations
\begin{align}
\label{eq:flowdeltaalpha}
\partial_t(m_k^\Delta)^2&=\frac{Th_d^4}{T_k^3}\sum_{\epsilon_1\epsilon_2}\int_{-\pi}^\pi\frac{d^2q}{(2\pi)^2}(c_1-\epsilon_1c_2)^2\left(\text{sech}^2f_{\epsilon_1\epsilon_2}+\frac{\tanh f_{\epsilon_1\epsilon_2}}{f_{\epsilon_1\epsilon_2}}\right)\nonumber\\
&\quad-(2\pi)^2T(2M+1)\int_0^{2\pi}\frac{d^2q}{k^2}\left(\frac{6Z_k^\chi\lambda_k^\Delta}{(P_k^\chi)^2}+\frac{3\gamma_k}{(P_k^{\vec s})^2}\right)-2(m_k^\Delta)^2\nonumber\\
\partial_t\lambda_k^\Delta&=\frac{Th_d^8}{4T_k^5}\sum_{\epsilon_1\epsilon_2}\int_{-\pi}^\pi\frac{d^2q}{(2\pi)^2}(c_1-\epsilon_1c_2)^4\nonumber\\
&\quad\qquad\left(\frac{\text{sech}^2f_{\epsilon_1\epsilon_2}}{f_{\epsilon_1\epsilon_2}^2}-\frac{(1+2f_{\epsilon_1\epsilon_2}^2\text{sech}^2f_{\epsilon_1\epsilon_2})\tanh f_{\epsilon_1\epsilon_2}}{f_{\epsilon_1\epsilon_2}^3}\right)\nonumber\\
&\quad+(2\pi)^2T(2M+1)\int_0^{2\pi}\frac{d^2q}{k^2}\left(\frac{24Z_k^\chi(\lambda_k^\Delta)^2}{(P_k^\chi)^3}+\frac{6\gamma_k^2}{(P_k^{\vec s})^3}\right)-2\lambda_k^\Delta\nonumber\\
\partial_t(m_k^\alpha)^2&=\frac{Th_a^4}{T_k^3}\sum_{\epsilon_1\epsilon_2}\int_{-\pi}^\pi\frac{d^2q}{(2\pi)^2}\left(1+\epsilon_2\frac{h_\rho\rho}{2t(c_1+\epsilon_1c_2)}\right)\left(\text{sech}^2f_{\epsilon_1\epsilon_2}+\frac{\tanh f_{\epsilon_1\epsilon_2}}{f_{\epsilon_1\epsilon_2}}\right)\nonumber\\
&\quad-(2\pi)^2T(2M+1)\int_0^{2\pi}\frac{d^2q}{k^2}\left(\frac{5\lambda_k^\alpha}{(P_k^{\vec s})^2}+\frac{4Z_k^\chi\gamma_k}{(P_k^\chi)^2}\right)-2(m_k^\alpha)^2\nonumber\\
\partial_t\lambda_k^\Delta&=\frac{Th_a^8}{4T_k^5}\sum_{\epsilon_1\epsilon_2}\int_{-\pi}^\pi\frac{d^2q}{(2\pi)^2}\nonumber\\
&\qquad\biggl[\left(1+\epsilon_2\frac{h_\rho\rho}{2t(c_1+\epsilon_1c_2)}\right)^2\left(\frac{\text{sech}^2f_{\epsilon_1\epsilon_2}}{f_{\epsilon_1\epsilon_2}^2}-\frac{(1+2f_{\epsilon_1\epsilon_2}^2\text{sech}^2f_{\epsilon_1\epsilon_2})\tanh f_{\epsilon_1\epsilon_2}}{f_{\epsilon_1\epsilon_2}^3}\right)\nonumber\\
&\qquad-\frac{4h_\rho\rho T_k^2\epsilon_2}{8t^3(c_1+\epsilon_1c_2)^3}\left(\text{sech}^2f_{\epsilon_1\epsilon_2}+\frac{\tanh f_{\epsilon_1\epsilon_2}}{f_{\epsilon_1\epsilon_2}}\right)\biggr]\nonumber\\
&\quad+(2\pi)^2T(2M+1)\int_0^{2\pi}\frac{d^2q}{k^2}\left(\frac{22(\lambda_k^\alpha)^2}{(P_k^{\vec s})^3}+\frac{8Z_k^\chi\gamma_k^2}{(P_k^\chi)^3}\right)-2\lambda_k^\alpha\nonumber\\
\partial_t\gamma_k&=\frac{Th_a^4h_d^4}{4T_k^5}\sum_{\epsilon_1\epsilon_2}\int_{-\pi}^\pi\frac{d^2q}{(2\pi)^2}(c_1-\epsilon_1c_2)^2\left(1+\epsilon_2\frac{h_\rho\rho}{2t(c_1+\epsilon_1c_2)}\right)\nonumber\\
&\quad\qquad\left(\frac{\text{sech}^2f_{\epsilon_1\epsilon_2}}{f_{\epsilon_1\epsilon_2}^2}-\frac{(1+2f_{\epsilon_1\epsilon_2}^2\text{sech}^2f_{\epsilon_1\epsilon_2})\tanh f_{\epsilon_1\epsilon_2}}{f_{\epsilon_1\epsilon_2}^3}\right)\nonumber\\
&\quad+(2\pi)^2T(2M+1)\int_0^{2\pi}\frac{d^2q}{k^2}\nonumber\\
&\quad\qquad\left(\frac{12Z_k^\chi\gamma_k\lambda_k^\Delta}{(P_k^\chi)^3}+\frac{10\gamma_k\lambda_k^\alpha}{(P_k^{\vec s})^3}+\frac{4Z_k^\chi\gamma_k^2}{(P_k^\chi)^2P_k^{\vec s}}+\frac{4\gamma_k^2}{P_k^\chi(P_k^{\vec s})^2}\right)-2\gamma_k
\end{align}
with $M=\text{max}\{m\in\mathbbm{N}|m<k/T\}$,
\begin{align}
f_{\epsilon_1\epsilon_2}&=\frac{1}{2T_k}(h_\rho\rho+2t\epsilon_2(c_1+\epsilon_1c_2)),\nonumber\\
P_k^\chi&=Z_k^\chi\left(2\pi^4\lambda_3\frac{2-c_1-c_2}{k^2}+4\pi^4\right)+(m_k^\Delta)^2,\nonumber\\
P_k^{\vec s}&=2\pi^4\lambda_3\frac{2-c_1-c_2}{k^2}+4\pi^4+(m_k^\alpha)^2.
\end{align}
The initial conditions are $(m_\Lambda^\Delta)^2=8\pi^2h_d^2/\Lambda^2$, $(m_\Lambda^\alpha)^2=4\pi^2h_a^2/\Lambda^2$, $\lambda_\Lambda^\Delta=\lambda_\Lambda^\alpha=\kappa_\Lambda=0$.

\subsubsection{Remarks}
Again the flow equations seem to be ill defined for $f_{\epsilon_1\epsilon_2}\to0$ and again all these denominator zeroes are canceled by corresponding numerator zeroes. However, we additionally face possible singularities in the flow equations which involve derivatives with respect to $\alpha_{22}$ in \eqref{eq:aldelex} for $c_1+\epsilon_1c_2\to0$. These singularities cancel if the sums over $\epsilon_j$ are performed. For example,
\begin{multline}
\sum_{\epsilon_2}\left(1+\epsilon_2\frac{h_\rho\rho}{2t(c_1+\epsilon_1c_2)}\right)\left(\text{sech}^2f_{\epsilon_1\epsilon_2}+\frac{\tanh f_{\epsilon_1\epsilon_2}}{f_{\epsilon_1\epsilon_2}}\right)\\
\to-4\,\text{sech}^2\frac{h_\rho\rho}{2T_k}\left(\frac{h_\rho\rho}{2T_k}\tanh\frac{h_\rho\rho}{2T_k}-1\right)
\end{multline}
for $c_1+\epsilon_1c_2\to0$, which is perfectly finite. In the same way, the sum over $\epsilon_2$ in the flow equations for $\lambda_k^\alpha$ and $\gamma_k$ can be carried out in the limit $c_1+\epsilon_1c_2\to0$.

Note that as in the last section, the contributions in the bosonic sector have simple diagrammatic interpretations:
\begin{gather}
m_k^\Delta:\parbox{20mm}{\setlength{\unitlength}{1mm}\begin{fmfgraph*}(20,15)
\fmfleft{i}
\fmfright{o}
\fmf{dashes}{v1,i}
\fmf{dashes}{o,v1}
\fmf{dashes}{v1,v1}
\fmfdot{v1}
\end{fmfgraph*}}
+\parbox{20mm}{\setlength{\unitlength}{1mm}\begin{fmfgraph*}(20,15)
\fmfleft{i}
\fmfright{o}
\fmf{dashes}{v1,i}
\fmf{dashes}{o,v1}
\fmf{plain}{v1,v1}
\fmfdot{v1}
\end{fmfgraph*}},\qquad
\lambda_k^\Delta:\parbox{20mm}{\setlength{\unitlength}{1mm}\begin{fmfgraph*}(20,15)
\fmfleft{i1,i2}
\fmfright{o1,o2}
\fmf{dashes,tension=2}{i1,v1}\fmf{dashes,tension=2}{i2,v1}
\fmf{dashes,tension=2}{o1,v2}\fmf{dashes,tension=2}{o2,v2}
\fmf{dashes,left}{v1,v2}\fmf{dashes,right}{v1,v2}
\fmfdot{v1,v2}
\end{fmfgraph*}}
+\parbox{20mm}{\setlength{\unitlength}{1mm}\begin{fmfgraph*}(20,15)
\fmfleft{i1,i2}
\fmfright{o1,o2}
\fmf{dashes,tension=2}{i1,v1}\fmf{dashes,tension=2}{i2,v1}
\fmf{dashes,tension=2}{o1,v2}\fmf{dashes,tension=2}{o2,v2}
\fmf{plain,left}{v1,v2}\fmf{plain,right}{v1,v2}
\fmfdot{v1,v2}
\end{fmfgraph*}},\nonumber\\
m_k^\alpha:\parbox{20mm}{\setlength{\unitlength}{1mm}\begin{fmfgraph*}(20,15)
\fmfleft{i}
\fmfright{o}
\fmf{plain}{v1,i}
\fmf{plain}{o,v1}
\fmf{plain}{v1,v1}
\fmfdot{v1}
\end{fmfgraph*}}
+\parbox{20mm}{\setlength{\unitlength}{1mm}\begin{fmfgraph*}(20,15)
\fmfleft{i}
\fmfright{o}
\fmf{plain}{v1,i}
\fmf{plain}{o,v1}
\fmf{dashes}{v1,v1}
\fmfdot{v1}
\end{fmfgraph*}},\qquad
\lambda_k^\alpha:\parbox{20mm}{\setlength{\unitlength}{1mm}\begin{fmfgraph*}(20,15)
\fmfleft{i1,i2}
\fmfright{o1,o2}
\fmf{plain,tension=2}{i1,v1}\fmf{plain,tension=2}{i2,v1}
\fmf{plain,tension=2}{o1,v2}\fmf{plain,tension=2}{o2,v2}
\fmf{plain,left}{v1,v2}\fmf{plain,right}{v1,v2}
\fmfdot{v1,v2}
\end{fmfgraph*}}
+\parbox{20mm}{\setlength{\unitlength}{1mm}\begin{fmfgraph*}(20,15)
\fmfleft{i1,i2}
\fmfright{o1,o2}
\fmf{plain,tension=2}{i1,v1}\fmf{plain,tension=2}{i2,v1}
\fmf{plain,tension=2}{o1,v2}\fmf{plain,tension=2}{o2,v2}
\fmf{dashes,left}{v1,v2}\fmf{dashes,right}{v1,v2}
\fmfdot{v1,v2}
\end{fmfgraph*}},\nonumber\\
\gamma_k^\alpha:\parbox{20mm}{\setlength{\unitlength}{1mm}\begin{fmfgraph*}(20,15)
\fmfleft{i1,i2}
\fmfright{o1,o2}
\fmf{dashes,tension=2}{i1,v1}\fmf{dashes,tension=2}{i2,v1}
\fmf{plain,tension=2}{o1,v2}\fmf{plain,tension=2}{o2,v2}
\fmf{dashes,left}{v1,v2}\fmf{dashes,right}{v1,v2}
\fmfdot{v1,v2}
\end{fmfgraph*}}
+\parbox{20mm}{\setlength{\unitlength}{1mm}\begin{fmfgraph*}(20,15)
\fmfleft{i1,i2}
\fmfright{o1,o2}
\fmf{dashes,tension=2}{i1,v1}\fmf{dashes,tension=2}{i2,v1}
\fmf{plain,tension=2}{o1,v2}\fmf{plain,tension=2}{o2,v2}
\fmf{plain,left}{v1,v2}\fmf{plain,right}{v1,v2}
\fmfdot{v1,v2}
\end{fmfgraph*}}
+\parbox{20mm}{\setlength{\unitlength}{1mm}\begin{fmfgraph*}(20,15)
\fmfleft{i1,i2}
\fmfright{o1,o2}
\fmf{dashes,tension=2}{i1,v1}\fmf{plain,tension=2}{i2,v1}
\fmf{dashes,tension=2}{o1,v2}\fmf{plain,tension=2}{o2,v2}
\fmf{plain,left}{v1,v2}\fmf{dashes,right}{v1,v2}
\fmfdot{v1,v2}
\end{fmfgraph*}}.
\end{gather}
The dashed line represents the propagation of $\chi$, whereas the solid line the propagation of $\vec s$. 

\subsubsection{Numerical results}
\begin{figure}
\centering
\psfrag{T}{$T$}
\psfrag{hrho rho}{$\scriptstyle h_\rho\rho$}
\psfrag{meanfield}{$\!\!\scriptstyle\text{meanfield}$}
\psfrag{delta}{$\scriptstyle\Delta$}
\psfrag{deltaalpha}{$\scriptstyle\Delta$, $\scriptstyle\alpha$}
\psfrag{deltarho}{$\scriptstyle \Delta$, $\scriptstyle\rho$}
\includegraphics[scale=0.9]{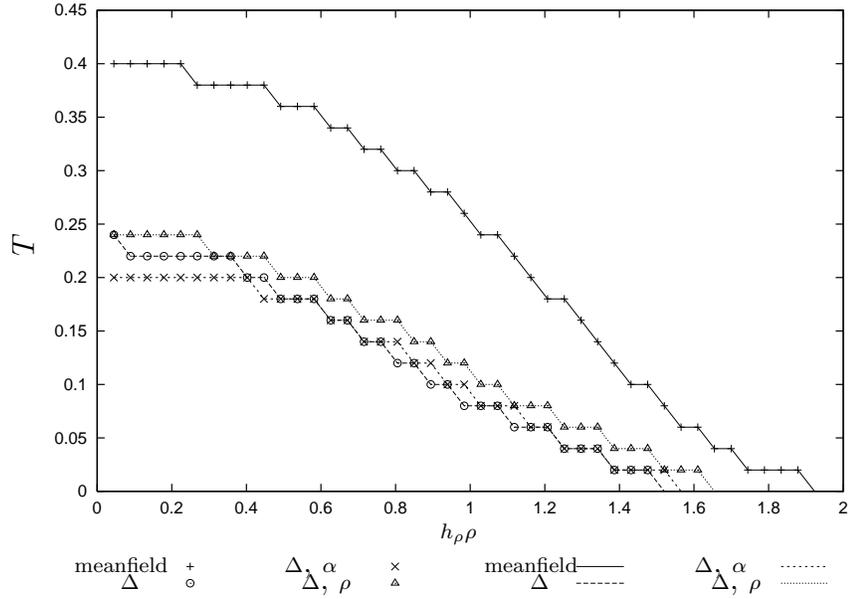}
\caption[Boundaries of the superconducting region for the mean field case, the inclusion of bosonic $\Delta$-fluctuations, $\Delta$- and $\alpha$-fluctuations as well as  $\Delta$- and $\rho$-fluctuations]{Boundaries of the superconducting region for the mean field case, the inclusion of bosonic $\Delta$-fluctuations, $\Delta$- and $\alpha$-fluctuations as well as  $\Delta$- and $\rho$-fluctuations. The Yukawa couplings are $h_\rho^2=20$, $h_a^2=10$ and $h_d^2=20$.}
\label{fig:delta}
\end{figure}
\begin{figure}
\centering
\psfrag{T}{$T$}
\psfrag{hrho rho}{$\scriptstyle h_\rho\rho$}
\psfrag{haq10}{$\!\!\scriptstyle h_a^2=10$}
\psfrag{haq15}{$\!\!\scriptstyle h_a^2=15$}
\psfrag{haq20}{$\!\!\scriptstyle h_a^2=20$}
\includegraphics[scale=0.9]{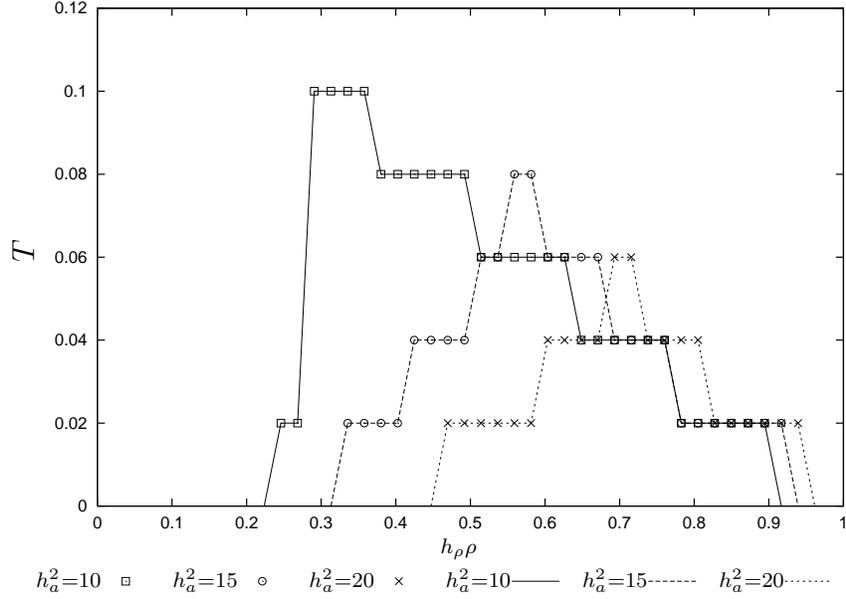}
\caption[Dependence of the superconducting region (including $\Delta$- and $\alpha$- fluctuations) on the Yukawa coupling $h_a$]{Dependence of the superconducting region (including $\Delta$- and $\alpha$- fluctuations) on the Yukawa coupling $h_a$. We take $h_d^2=15$.}
\label{fig:deltaalpha}
\end{figure}
In fig. \ref{fig:delta} we compare the effect of including different kinds of bosonic fluctuations into the flow with the mean field result. The plot for $\Delta$- and $\alpha$-fluctuations has been generated by using the flow equations of this section, the plot for $\Delta$- and $\rho$-fluctuations by using \eqref{eq:flowdeltarho}. By setting $(m_k^\alpha)^2=(m_\Lambda^\alpha)^2$, $\lambda_k^\alpha=0$ in \eqref{eq:flowdeltaalpha} we get the plot for the inclusion of $\Delta$-fluctuations only. We set $h_\rho^2=20$, $h_a^2=10$ and $h_d^2=20$. For this choice of the Yukawa couplings only the superconducting phase is present in the mean field result. The same is true if bosonic fluctuations are included, but compared to the mean field case, the superconducting region becomes smaller. This was to be expected, since the fermionic fluctuations favor the phase transition, whereas bosonic fluctuations have the opposite effect. In the mean field approximation we ignore the bosonic fluctuations, so that we overestimate the symmetry breaking behavior, which leads to a larger region of broken symmetry in the phase diagram. For this special choice of Yukawa couplings, the boundaries of the superconducting region approximately coincide. As we have seen in the last section, the phase boundary is not very sensitive to changes in $h_\rho$. However, fig. \ref{fig:deltaalpha} shows that we have a strong dependence on $h_a$ --- as we already had in the mean field case. In so far, the inclusion of bosonic couplings do not qualitatively alter the mean field results. The interesting feature of the plot is that the superconducting region is shifted to the right if $h_a$ is increased. Of course, this is a feature not present in the mean field approximation, since there $h_a$ entered only via the combination $h_a^2\alpha$, so that outside the antiferromagnetic phase (where $\alpha=0$) varying $h_a$ could not have any effect on the boundary of the superconducting region. This means that strong antiferromagnetic coupling tends to enlarge the superconducting region by means of antiferromagnetic fluctuations, even if there is no antiferromagnetic order present! In fact, it is suspected that antiferromagnetic fluctuations are crucial for the understanding why cuprates remain in the superconducting state even for large temperature.

Also note that --- as in the last section --- the phase boundaries have been inferred from an analysis of the masses only. Therefore our results have to be taken with some care --- possible phase transitions of first order will shift the phase boundaries. This is particularly true for the left boundaries of the superconducting regions in fig. \ref{fig:deltaalpha}, where the superconducting region is bordered by a region of antiferromagnetic behavior, since in our mean field calculation we found phase transitions of first order exactly at this boundary. However, at the right boundaries no phase transitions of first order appeared in the mean field approximation, so that hopefully our interpretation remains intact even for the more general case considered here.   

\subsection{Final remark}
Although many questions remain and a lot of work has to be invested to free ourselves from the limits of the problem of the coupling ambiguity and the possibility of first order phase transitions, the point we want to make is that we see that our formalism is in principle suitable for analyzing the properties of the model both in the symmetric and in the broken phase. Renormalization group approaches so far have been limited to the investigation of the flow in the symmetric phase and were not able to describe features of the flow beyond the point of spontaneous symmetry breaking. The problems we face in our analysis so far are not intrinsic to our formalism and can be cured by more refined truncation schemes (particularly including the flow of the couplings and some more general truncation for the effective potential). A systematical enhancement of the truncation scheme is straightforward and will be attacked in the future:
\begin{itemize}
\item In our approach so far, we approximated the effective potential by a polynomial in the fields, which raises the problem of dealing with phase transitions of first order. Calculations including the flow of the full potential are possible and have already been carried out successfully for a number of systems (cf. e. g. \cite{renormrev}). 
\item The phase diagrams we calculated in this chapter depend on the choice of the initial values of the Yukawa couplings that we kept constant. To get rid of this dependence, we have to include the flow of the Yukawa coupling into our sets of flow equations. This task is already worked on \cite{tobi}.
\item One qualitative feature of the phase diagram of high temperature superconductors fig. \ref{fig:cuprat} is the separation of the antiferromagnetic and superconducting region at low temperature and intermediate doping. In our calculations, the superconducting region is always bounded by the antiferromagnetic region towards small $\rho$. The reason for this shortcoming is our oversimplified homogeneous treatment of the charge density \cite{quantenphas}. More realistically, the charge density caused by doping is not homogeneously distributed, but forms stripe shaped regions of alternating high and low charge density. The width of the stripes depend on doping, and it turns out that for some intermediate doping the alternating charge density induces the formation of parallel spin ladders that decouple from each other. The effect is that long range order is lost, and antiferromagnetic order occurs only in the ladders. The material then becomes paramagnetically over large scales. This behavior should be reproducible with our formalism if we give up the assumption of homogeneous charge density and generalize it to stripe structures.
\item Although we introduced a large set of bosons in the formalism, we concentrated on only a small subset of them in our calculations. The choice of this subset was motivated by experiment: We know that high temperature superconductors exhibit antiferromagnetic and $d$-wave superconducting behavior and therefore chose exactly those bosons representing these properties. It is interesting to see whether a more unbiased choice, that is, taking into account more bosons representing e.g.  $s$-wave superconductivity or ferromagnetism, confirms that we have chosen those bosons actually leading to spontaneous symmetry breaking.        
\end{itemize}

Some words are in order comparing our formalism to the renormalization group approaches in \cite{renhub}. As already mentioned, all these calculations were performed in the purely fermionic theory. The authors basically investigate the flow of quartic fermionic couplings in different channels. The conventional way of these approaches was to introduce a regularization scheme that cut off momenta near the Fermi surface. In this scheme, following the flow to small $k$ corresponds to the inclusion of momentum modes increasingly close to the Fermi surface. More recently, temperature cutoffs resembling the regularization scheme used in our work have been applied (Honerkamp, Salmhofer and Rice 2002 in  \cite{renhub}). Independent of the regularization scheme, the flow of quartic fermionic couplings indicates instabilities in certain channels by the emergence of divergencies of the corresponding couplings. This means that in these approaches it is only possible to follow the flow until the point of symmetry breaking is reached, and no information about the behavior in the broken phase is available. By the bosonization and the investigation of the effective potential in this work, it becomes possible to analyze the flow in the broken phase. However, in contrast to \cite{renhub}, our formalism is plagued by the problem of the coupling ambiguity, and its success will be ultimately measured by our capability to overcome this complication. 

It should have become clear in the course of the explicit calculations presented over the last sections that our formalism provides an elegant and suitable starting point for the investigation of the Hubbard model and the properties of high temperature superconductors, and is also easily implemented for investigations of the behavior inside the broken phase. At this point, we have reproduced the gross qualitative features of the phase diagram of high temperature superconductors, with correctly placed regions of antiferromagnetic and superconducting behavior. We understand how the Mermin-Wagner theorem can be reconciled with the existence of antiferromagnetic long range order for non vanishing temperature, and we took a first glimpse on how antiferromagnetic fluctuations can enlarge regions of superconductivity. All these results let us suspect that we are on the right way and we hope to provide more insights into the nature of the Hubbard model and high temperature superconductors in the future.

\chapter*{Conclusion}
High temperature superconductors have a two dimensional layer structure with small interlayer coupling. They can be modeled by the two dimensional Hubbard model on a square lattice. This model describes electrons on a quadratic lattice that experience a local Coulomb interaction and are able to hop to adjacent lattice sites. The partition function for this model depends on the chemical potential (or equivalently the charge density), the temperature and the relative strength of Coulomb interaction and hopping amplitude. To compare the predictions of the model with experimental results for high temperature superconductors, one has to calculate the phase diagram of the model in the charge density-temperature plane. 

Our way to tackle this task is to identify the most prominent degrees of freedom of high temperature superconductors, which are antiferromagnetism and $d$-wave superconductivity, and to define a set of bosonic ``particles'', so that every particle corresponds to one degree of freedom of the model. We found an exact transcription of the partition function, which describes a Yukawa-like theory. The expectation values of the bosons in this rewritten theory indicate a possible long range order in the channel (antiferromagnetic, $d$-wave superconducting, etc.) to which the bosons correspond. 

A mean field calculation in this partially bosonized theory, neglecting all bosonic fluctuations and integrating out the fermions, already reveals the main features of the phase diagram of high temperature superconductors.

More refined calculations can be performed by means of exact renormalization group equations. We use the effective average action method. To simplify the definition of truncation schemes and to minimize the error induced by approximations, we rewrite the partially bosonized theory as a function of bosons which are eigenstates of translations on the lattice. Due to the lattice symmetries, these bosons are no longer mixed in the full effective action (e.g. the full propagator matrix becomes diagonal). 

We use this modified theory as a starting point for a renormalization group analysis. This analysis shows how the Mermin-Wagner theorem can be reconciled with the existence of antiferromagnetic long range order for non vanishing temperature and indicates that antiferromagnetic fluctuations tend to favor superconducting behavior in certain regions of the phase diagram.

The drawback of the present analysis is mainly the arbitrariness of the Yukawa couplings. Current work is dedicated to this problem. However, the cure is basically an improved truncation scheme for the effective action and does not intrinsically limit our formalism. Planned future work includes a more general treatment of the effective potential with special regard to first order phase transitions and the inclusion of more bosons in the truncation scheme, allowing to test whether e.g. superconducting channels other than that with spatial $d$-wave symmetry play a role.

In conclusion, we hope to have provided a formalism which is easy to implement for renormalization group studies, which introduces a convenient interpretation of non local fermionic order parameters as local expectation values of bosonic fields and that will continue to help investigating the rich and beautiful spectrum of properties of the Hubbard model and high temperature superconductors.
\begin{appendix}
\renewcommand{\chaptername}{Appendix}
\chapter{Conventions}

We use units for which $\hbar=c=k_B=1$. Bold symbols ($\boldsymbol{n}$, $\boldsymbol{x}$, $\boldsymbol{q}$, etc.) denote two dimensional vectors. Symbols with arrow ($\vec a$, $\vec m$, etc.) denote three dimensional vectors. Generalized momenta are called $Q$, $P$ and $K$, whereas $X$, $Y$ and $Z$ are generalized positions (see below for the definition of generalized quantities). By a $\,\hat{}\,$ we indicate fields. The same symbol without $\,\hat{}\,$ is the expectation value of the corresponding field. $\,\tilde{}\,$ is used to indicate fermion bilinears (to distinguish them from their bosonic counterparts).  

\section{Fourier transforms}
\label{sec:app:four}
We define
\begin{equation}
\begin{gathered}
Q=(\omega_n,\boldsymbol{q}),\quad X=(\tau,\boldsymbol{n}),\\
QX=\omega_n\tau+\boldsymbol{n}\boldsymbol{q},\\
\sum_X=\int_0^\beta d\tau\,\sum_{\boldsymbol{n}},\quad\sum_Q=T\sum_{n=-\infty}^\infty\int_{-\pi}^\pi\frac{d^2q}{(2\pi)^2},\\
\delta(Q-Q')=\beta\delta_{n,n'}(2\pi)^2\delta(\boldsymbol{q}-\boldsymbol{q'}),\\
\delta(X-X')=\delta(\tau-\tau')\delta_{\boldsymbol{n},\boldsymbol{n'}}.
\end{gathered}
\end{equation}
The Fourier transforms for both fermionic and bosonic fields are given by
\begin{equation}
\label{eq:app:fourtrans}
\begin{aligned}
{\hat\chi}_a(X)&=\sum_Q{\hat\chi}_a(Q)\exp(i(QX+\boldsymbol{z}_a\boldsymbol{q})),\\
{\hat\chi}_a^*(X)&=\sum_Q{\hat\chi}_a^*(Q)\exp(-i(QX+\boldsymbol{z}_a\boldsymbol{q})),
\end{aligned}
\end{equation}
where $\hat\chi$ stands for $\hat\psi$, $\hat w_\gamma$ or $\hat u_\beta$, whereas $\hat\chi^*$ stands for $\hat\psi^*$ or $\hat u_\beta^*$ and the $\boldsymbol{z}_a$ are given by
\begin{equation}
\begin{aligned}
\boldsymbol{z}_1&=(-1/4,1/4) & \boldsymbol{z}_2&=(1/4,1/4)\\
\boldsymbol{z}_4&=(-1/4,-1/4) & \boldsymbol{z}_3&=(1/4,-1/4).
\end{aligned}
\end{equation}

\section{Matrices}
\label{sec:matrices}
$\{\sigma_i\}$, $i\in\{1,2,3\}$ is the usual set of Pauli matrices. Additionally, we identify $\sigma_0$ with the unity matrix. We then define the matrices $\sigma_\mu\otimes\sigma_\nu$ by 
\begin{equation}
A_\mu=\begin{pmatrix}\sigma_\mu&0\\0&\sigma_\mu\end{pmatrix},\quad B_\mu=\begin{pmatrix}0&\sigma_\mu\\\sigma_\mu&0\end{pmatrix},\quad
C_\mu=\begin{pmatrix}0&-i\sigma_\mu\\i\sigma_\mu&0\end{pmatrix},\quad D_\mu=\begin{pmatrix}\sigma_\mu&0\\0&-\sigma_\mu\end{pmatrix}
\end{equation}
where $\mu\in\{0,1,2,3\}$. The matrices $A_\mu$ and $B_\mu$ have the properties
\begin{equation}
\begin{gathered}
\{A_i,B_j\}=2\delta_{ij}B_0,\quad\{A_i,A_j\}=\{B_i,B_j\}=2\delta_{ij},\\
[A_i,B_j]=2i\epsilon_{ijk}B_k,\quad[A_i,A_j]=[B_i,B_j]=2i\epsilon_{ijk}A_k,\\
B_0A_\mu=A_\mu B_0=B_\mu,\quad B_0B_\mu=B_\mu B_0=A_\mu
\end{gathered}
\end{equation}
with $i\in\{1,2,3\}$.

\section{Fermion bilinears}
\begin{equation}
\begin{aligned}
\tilde\sigma_{ab}(X)&=\hat\psi^\dagger_b(X)\hat\psi_a(X)\\
\vec{\tilde\varphi}_{ab}(X)&=\psi_b^\dagger(X)\vec\sigma\hat\psi_a(X)\\
\tilde\chi_{ab}(X)&=\hat\psi^T_b(X)i\sigma_2\hat\psi_a(X)\\
\tilde\chi_{ab}^*(X)&=-\hat\psi_b^\dagger(X)i\sigma_2\hat\psi_a^*(X).
\end{aligned}
\end{equation}

\begin{equation}
\begin{aligned}
\tilde\rho&=\tilde\sigma_{11}+\tilde\sigma_{22}+\tilde\sigma_{33}+\tilde\sigma_{44}
&\qquad\vec{\tilde m}&=\vec{\tilde\varphi}_{11}+\vec{\tilde\varphi}_{22}+\vec{\tilde\varphi}_{33}+\vec{\tilde\varphi}_{44}\\
\tilde p&=\tilde\sigma_{11}-\tilde\sigma_{22}+\tilde\sigma_{33}-\tilde\sigma_{44}
&\qquad\vec{\tilde a}&=\vec{\tilde\varphi}_{11}-\vec{\tilde\varphi}_{22}+\vec{\tilde\varphi}_{33}-\vec{\tilde\varphi}_{44}\\
\tilde{q}_y&=\tilde\sigma_{11}+\tilde\sigma_{22}-\tilde\sigma_{33}-\tilde\sigma_{44}
&\qquad\vec{\tilde g}_y&=\vec{\tilde\varphi}_{11}+\vec{\tilde\varphi}_{22}-\vec{\tilde\varphi}_{33}-\vec{\tilde\varphi}_{44}\\
\tilde{q}_x&=\tilde\sigma_{11}-\tilde\sigma_{22}-\tilde\sigma_{33}+\tilde\sigma_{44}
&\qquad\vec{\tilde g}_x&=\vec{\tilde\varphi}_{11}-\vec{\tilde\varphi}_{22}-\vec{\tilde\varphi}_{33}+\vec{\tilde\varphi}_{44}\\ \\
\tilde{s}&=\tilde\chi_{11}+\tilde\chi_{22}+\tilde\chi_{33}+\tilde\chi_{44}
&\qquad\tilde e&=\tilde\chi_{12}+\tilde\chi_{23}+\tilde\chi_{34}+\tilde\chi_{41}\\
\tilde{c}&=\tilde\chi_{11}-\tilde\chi_{22}+\tilde\chi_{33}-\tilde\chi_{44}
&\qquad\tilde d&=\tilde\chi_{12}-\tilde\chi_{23}+\tilde\chi_{34}-\tilde\chi_{41}\\
\tilde{t}_y&=\tilde\chi_{11}+\tilde\chi_{22}-\tilde\chi_{33}-\tilde\chi_{44}
&\qquad\tilde{v}_y&=\tilde\chi_{12}-\tilde\chi_{34}\\
\tilde{t}_x&=\tilde\chi_{11}-\tilde\chi_{22}-\tilde\chi_{33}+\tilde\chi_{44}
&\qquad\tilde{v}_x&=\tilde\chi_{23}-\tilde\chi_{41}\\
\tilde{s}^*&=\tilde\chi_{11}^*+\tilde\chi_{22}^*+\tilde\chi_{33}^*+\tilde\chi_{44}^*
&\qquad\tilde e^*&=\tilde\chi_{12}^*+\tilde\chi_{23}^*+\tilde\chi_{34}^*+\tilde\chi_{41}^*\\
\tilde{c}^*&=\tilde\chi_{11}^*-\tilde\chi_{22}^*+\tilde\chi_{33}^*-\tilde\chi_{44}^*
&\qquad\tilde d^*&=\tilde\chi_{12}^*-\tilde\chi_{23}^*+\tilde\chi_{34}^*-\tilde\chi_{41}^*\\
\tilde{t}_y^*&=\tilde\chi_{11}^*+\tilde\chi_{22}^*-\tilde\chi_{33}^*-\tilde\chi_{44}^*
&\qquad\tilde{v}_y^*&=\tilde\chi_{12}^*-\tilde\chi_{34}^*\\
\tilde{t}_x^*&=\tilde\chi_{11}^*-\tilde\chi_{22}^*-\tilde\chi_{33}^*+\tilde\chi_{44}^*
&\qquad\tilde{v}_x^*&=\tilde\chi_{23}^*-\tilde\chi_{41}^*\\ \\
\end{aligned}
\end{equation}

\chapter{Vertex factors for the Hubbard model}
\label{sec:vertexfac}
The vertices $V^w(Q',Q'')$ for the bosons $\hat\rho$, $\hat p$, $\hat q_{x,y}$ depend only on the momentum $Q=Q'-Q''$. With $\boldsymbol{e}_x=(1,0)$, $\boldsymbol{e}_y=(0,1)$ and $\boldsymbol{z}_a$, $a=1\ldots 4$, given in appendix \ref{sec:app:four}, they can be written in the form
\begin{gather}
  V^w_{ab,c}(Q',Q'') = V^w_{ab,c}(Q) = \frac{h_w}{4} e^{-i\boldsymbol{z}_a{\boldsymbol{q}}} e^{i\boldsymbol{z}_c{\boldsymbol{q}}} M^w_{ab,c}(Q)\otimes\mathbbm{1}_2^\mathrm{spin}, 
\end{gather}
The color matrices $M_c^w$ (with matrix elements $M_{ab,c}^w$) read
\begin{gather}
  \begin{aligned}
    M^\rho_1(Q) &= \text{diag}\{1,e^{i\boldsymbol{e}_x{\boldsymbol{q}}},e^{i(\boldsymbol{e}_x-\boldsymbol{e}_y){\boldsymbol{q}}},e^{-i\boldsymbol{e}_y{\boldsymbol{q}}}\},&\quad
    M^\rho_2(Q) &= \text{diag}\{1,1,e^{-i\boldsymbol{e}_y{\boldsymbol{q}}},e^{-i\boldsymbol{e}_y{\boldsymbol{q}}}\},\\
    M^\rho_3(Q) &= \text{diag}\{1,1,1,1\},&\quad
    M^\rho_4(Q) &= \text{diag}\{1,e^{i\boldsymbol{e}_x{\boldsymbol{q}}},e^{i\boldsymbol{e}_x{\boldsymbol{q}}},1\};
  \end{aligned}\nonumber\\
  M^p_c(Q) = (-1)^{c-1}\,\mathrm{diag}(1,-1,1,-1)\,M^\rho_c(Q); \nonumber\\
  \begin{aligned}
    M^{q_y}_1(Q) &= M^\rho_1(Q)\cdot\mathrm{diag}(-1,-1,1,1),&\quad
    M^{q_y}_2(Q) &= M^\rho_2(Q)\cdot\mathrm{diag}(-1,-1,1,1),\\
    M^{q_y}_3(Q) &= M^\rho_3(Q)\cdot\mathrm{diag}(1,1,-1,-1),&\quad
    M^{q_y}_4(Q) &= M^\rho_4(Q)\cdot\mathrm{diag}(1,1,-1,-1);
  \end{aligned}\nonumber\\
  \begin{aligned}
    M^{q_x}_1(Q) &= M^\rho_1(Q)\cdot\mathrm{diag}(-1,1,1,-1),&\quad
    M^{q_x}_2(Q) &= M^\rho_2(Q)\cdot\mathrm{diag}(1,-1,-1,1),\\
    M^{q_x}_3(Q) &= M^\rho_3(Q)\cdot\mathrm{diag}(1,-1,-1,1),&\quad
    M^{q_x}_4(Q) &= M^\rho_4(Q)\cdot\mathrm{diag}(-1,1,1,-1).
  \end{aligned}
\end{gather}
The same can be obtained for the bosons with spin index, $\vec{m}, \vec{a}, \vec{g}_{x,y}$, by substituting 
$\mathbbm{1}_2^\mathrm{spin} \to \vec\sigma^\mathrm{spin}$.

For the bosons $s,c,t_{x,y}$ one finds similarly: 
\begin{gather}
  V^{u^*}_{ab,c}(Q',Q'') = \frac{h_u}{4} 
  e^{i\boldsymbol{z}_a({\boldsymbol{q}}'+{\boldsymbol{q}}'')} e^{-i\boldsymbol{z}_c({\boldsymbol{q}}'+{\boldsymbol{q}}'')} M^{u^*}_{ab,c}(Q',Q'')\otimes i\sigma_2, \nonumber\\
  \begin{aligned}
    M^{s^*}_c(Q',Q'') &= M^\rho_c(-Q'-Q''),&\quad 
    M^{c^*}_c(Q',Q'') &= M^p_c(-Q'-Q''),\\
    M^{t_1^*}_c(Q',Q'') &= M^{q_1}_c(-Q'-Q''),&\quad 
    M^{t_2^*}_c(Q',Q'') &= M^{q_2}_c(-Q'-Q''),
  \end{aligned}  
\end{gather}
while $d,e,v_{x,y}$ are a bit more complicated. Let us define $e^{ij}=e^{i(\boldsymbol z_i\boldsymbol q'+\boldsymbol z_j\boldsymbol q'')}$ and a $*$-product $C=A*B$ by $C_{ij}:=A_{ij}B_{ij}$ (no sum over indices here!). One then obtains
\begin{gather}
  V^{e^*}_c(Q',Q'') = 
  \frac{h_e}{8} e^{-i\boldsymbol{z}_c({\boldsymbol{q}}'+{\boldsymbol{q}}'')} M^{e^*}_c(Q',Q'')\otimes i\sigma_2 \nonumber\\ 
  M^{e^*}_1(Q',Q'')=
  \left(\begin{array}{cccc}
      0&e^{12}e^{-i{\boldsymbol{q}}''\boldsymbol{e}_x}&0&e^{14}e^{i{\boldsymbol{q}}''\boldsymbol{e}_y}\\
      e^{21}e^{-i{\boldsymbol{q}}'\boldsymbol{e}_x}&0&e^{23}e^{i[{\boldsymbol{q}}''\boldsymbol{e}_y-({\boldsymbol{q}}'+{\boldsymbol{q}}'')\boldsymbol{e}_x]}&0\\
      0&e^{32}e^{i[{\boldsymbol{q}}'\boldsymbol{e}_y-({\boldsymbol{q}}'+{\boldsymbol{q}}'')\boldsymbol{e}_x]}&0&e^{34}e^{i[-\boldsymbol{e}_x{\boldsymbol{q}}'+\boldsymbol{e}_y({\boldsymbol{q}}'+{\boldsymbol{q}}'')]}\\
      e^{41}e^{i{\boldsymbol{q}}'\boldsymbol{e}_y}&0&e^{43}e^{i[-\boldsymbol{e}_x{\boldsymbol{q}}''+\boldsymbol{e}_y({\boldsymbol{q}}'+{\boldsymbol{q}}'')]}&0
    \end{array}\right), \nonumber\\
  M^{e^*}_2(Q',Q'')=
  \left(\begin{array}{cccc}
      0&e^{12}&0&e^{14}e^{i{\boldsymbol{q}}''\boldsymbol{e}_y}\\
      e^{21}&0&e^{23}e^{i{\boldsymbol{q}}''\boldsymbol{e}_y}&0\\
      0&e^{32}e^{i{\boldsymbol{q}}'\boldsymbol{e}_y}&0&e^{34}e^{i({\boldsymbol{q}}'+{\boldsymbol{q}}'')\boldsymbol{e}_y}\\
      e^{41}e^{i{\boldsymbol{q}}'\boldsymbol{e}_y}&0&e^{43}e^{i({\boldsymbol{q}}'+{\boldsymbol{q}}'')\boldsymbol{e}_y}&0
    \end{array}\right), \nonumber\\
  M^{e^*}_3(Q',Q'') = 
  \left(\begin{array}{cccc}
      0&e^{12}&0&e^{14} \\
      e^{21}&0&e^{23}&0 \\
      0&e^{32}&0&e^{34} \\
      e^{41}&0&e^{43}&0
    \end{array}\right), \nonumber\\
  M^{e^*}_4(Q',Q'')=
  \left(\begin{array}{cccc}
      0&e^{12}e^{-i{\boldsymbol{q}}''\boldsymbol{e}_x}&0&e^{14}\\
      e^{21}e^{-i{\boldsymbol{q}}'\boldsymbol{e}_x}&0&e^{23}e^{-i({\boldsymbol{q}}'+{\boldsymbol{q}}'')\boldsymbol{e}_x}&0\\
      0&e^{32}e^{-i({\boldsymbol{q}}'+{\boldsymbol{q}}'')\boldsymbol{e}_x}&0&e^{34}e^{-i{\boldsymbol{q}}'\boldsymbol{e}_x}\\
      e^{41}&0&e^{43}e^{-i{\boldsymbol{q}}''\boldsymbol{e}_x}&0
    \end{array}\right); 
\end{gather}
With the aid of the $*$-product the other vertices can now be obtained from these 
\begin{gather}
  M^{d^*}_c(Q',Q'')=
  \left(\begin{array}{cccc}
      0& 1& 0&-1\\
      1& 0&-1& 0\\
      0&-1& 0& 1\\
     -1& 0& 1& 0
    \end{array}\right)*M^{e^*}_c(Q',Q''); \nonumber\\
  M^{v_x^*}_c(Q',Q'')=
  \left(\begin{array}{cccc}
      0& 0& 0&-1\\
      0& 0& 1& 0\\
      0& 1& 0& 0\\
     -1& 0& 0& 0
    \end{array}\right)*M^{e^*}_c(Q',Q'')\cdot \lambda_c, \quad \lambda=(-1,1,1,-1); \nonumber\\
  M^{v_y^*}_c(Q',Q'')=
  \left(\begin{array}{cccc}
      0&1& 0& 0\\
      1&0& 0& 0\\
      0&0& 0&-1\\
      0&0&-1& 0
    \end{array}\right)*M^{e^*}_c(Q',Q'')\cdot \lambda_c, \quad \lambda=(-1,-1,1,1).
\end{gather}

The transition from the $\hat u^*\hat\psi\hat\psi$-vertices $V^{u^*}$ to the $\hat u\hat\psi^*\hat\psi^*$-vertices $V^u$ can be carried out by
\[V^{u^*}(Q',Q'')\to V^u(Q',Q'')=-(V^{u^*}(Q',Q''))^*=-V^{u^*}(-Q',-Q'').\]

\end{appendix}

\end{fmffile}




\end{document}